\documentclass[12pt,twoside,english]{extarticle}
\usepackage{lmodern}
\usepackage{helvet}
\usepackage[T1]{fontenc}
\usepackage[latin9]{inputenc}
\usepackage{geometry}
\usepackage{color}
\usepackage{babel}
\usepackage{float}
\usepackage{mathrsfs}
\usepackage{amsmath}
\usepackage{amsthm}
\usepackage{amssymb}
\usepackage{graphicx}
\usepackage{setspace}
\usepackage[unicode=true,pdfusetitle,
 bookmarks=true,bookmarksnumbered=false,bookmarksopen=false,
 breaklinks=true,pdfborder={0 0 0},pdfborderstyle={},backref=false,colorlinks=true]
 {hyperref}

\makeatletter

\providecommand{\tabularnewline}{\\}

\numberwithin{equation}{section}
\numberwithin{table}{section}
\numberwithin{figure}{section}
\usepackage[natbibapa]{apacite}
\theoremstyle{definition}
\newtheorem{defn}{\protect\definitionname}[section]
\theoremstyle{plain}
\newtheorem{assumption}{\protect\assumptionname}
\theoremstyle{plain}
\newtheorem{thm}{\protect\theoremname}[section]
\theoremstyle{definition}
\newtheorem{example}{\protect\examplename}[section]
\theoremstyle{plain}
\newtheorem{lem}{\protect\lemmaname}[section]

\usepackage{graphicx}
\usepackage{amsmath}
\usepackage{dcolumn}
\usepackage{listings}
\usepackage{color}
\usepackage{setspace}
\usepackage[normalem]{ulem}  
\definecolor{hellgelb}{rgb}{1,1,0.8}
\definecolor{colKeys}{rgb}{0,0,1}
\definecolor{colIdentifier}{rgb}{0,0,0}
\definecolor{colComments}{rgb}{1,0,0}
\definecolor{colString}{rgb}{0,0.5,0}

\usepackage{lmodern}
\usepackage[T1]{fontenc}
\usepackage{geometry}
\usepackage{fancyhdr}
\usepackage{amsthm}
\usepackage{thmtools}
\usepackage{amsmath}
\usepackage{amssymb}
\usepackage{esint}
\usepackage{multirow}
\usepackage{mathrsfs} 
\usepackage{textgreek}
\usepackage{amsfonts}
\usepackage{amssymb}
\usepackage{mathrsfs}

\usepackage[USenglish]{isodate}
\usepackage{chngcntr}

\counterwithout{figure}{section}
\counterwithout{table}{section}
\counterwithout{defn}{section}
\counterwithout{thm}{section}
\counterwithout{lem}{section}
\counterwithout{example}{section}
\counterwithout{equation}{section}

\makeatother

  \providecommand{\assumptionname}{Assumption}

  \providecommand{\definitionname}{Definition}
  \providecommand{\examplename}{Example}

  \providecommand{\lemmaname}{Lemma}

  \providecommand{\theoremname}{Theorem}
 
 \providecommand{\theoremname}{Theorem}
 
\usepackage{thmtools}

\newtheoremstyle{MyTheoremstyle}
  {\topsep} 
  {\topsep} 
  {} 
  {} 
  {\bfseries} 
  {.} 
  {.90em} 
  {} 
\theoremstyle{MyTheoremstyle} 
\theoremstyle{MyTheoremstyle} 
\theoremstyle{MyTheoremstyle} 
\theoremstyle{MyTheoremstyle} 
\theoremstyle{MyTheoremstyle}

\declaretheoremstyle[
    headfont=\bfseries,
    notefont=\normalfont,
    bodyfont=\itshape,
    headpunct=\newline,
    headformat={%
        \makebox{\NAME\ \NUMBER\ }{\NOTE}%
    },
]{theorem}

\newlength{\spacelength}
\declaretheoremstyle[
    headfont=\bfseries,
    notefont=\normalfont,
    bodyfont=\itshape,
    headpunct=\newline,
    headformat={%
        \makebox[0pt][l]{\NAME\ \NUMBER\ }\hskip-\spacelength{\NOTE}%
    },
]{theore}

 \usepackage{fancyhdr}
\usepackage{booktabs}
\usepackage{float}
\usepackage{graphicx}
\usepackage{epstopdf}
\usepackage{morefloats}
\usepackage[referable]{threeparttablex}
\usepackage{footnote}

\usepackage{setspace} 
\usepackage{graphicx,color}

\usepackage{listings}
\RequirePackage[T1]{fontenc}
\RequirePackage{ae,fancyvrb}
\DefineVerbatimEnvironment{Sinput}{Verbatim}{fontshape=sl}
\DefineVerbatimEnvironment{Soutput}{Verbatim}{}
\DefineVerbatimEnvironment{Scode}{Verbatim}{fontshape=sl}

\title{\bf Continuous Record Asymptotic Framework for Inference in Strucutral Change Models}
\author{
\textsc{\textcolor{MyBlue}{Alessandro Casini}}\thanks{Corresponding author at: Dep. of Economics and Finance, University of Rome Tor Vergata, Via Columbia 2, Rome, 00133, IT. 
Email: 
\texttt{\textcolor{MyBlue}{{alessandro.casini@uniroma2.it}}}.} 
\\
\small{\text{University of Rome Tor Vergata}}
\and
\textsc{\textcolor{MyBlue}{Taosong Deng}}\thanks{College of Finance and Statistics, Hunan University, 109 Shijiachong Road, Yuelu District, Changsha, Hunan 41006, China. 
Email: 
\texttt{\textcolor{MyBlue}{\mbox{tsdeng@hnu.edu.cn}}}.} 
\\
\small{\text{Hunan University}}
\and
\textsc{\textcolor{MyBlue}{Pierre Perron}}\thanks{Dep. of Economics, Boston University, 270 Bay State Road, Boston, MA 02215, US. 
Email: 
\texttt{\textcolor{MyBlue}{\mbox{perron@bu.edu}}}.} 
\\
\small{\text{Boston University}}
}

\date{\small{\today} \\} 



\usepackage{babel}
\addto\captionsenglish{%
}

\usepackage{color}
\usepackage{xcolor}

\definecolor{MyRed}{rgb}{0.8,0,0}
\definecolor{MyBlue}{rgb}{0,0,0.7}
\definecolor{Green}{rgb}{0,0.5,0}
\definecolor{hellgelb}{rgb}{1,1,0.8}
\definecolor{colKeys}{rgb}{0,0,1}
\definecolor{colIdentifier}{rgb}{0,0,0}
\definecolor{colComments}{rgb}{1,0,0}
\definecolor{colString}{rgb}{0,0.5,0}

\definecolor{MyLightRed}{rgb}{2.2,0.2,0.4} 
\definecolor{MyLightRed2}{rgb}{0.6,0.2,0.3} 
\definecolor{MyLightRed2temp}{rgb}{0.6,0.2,0.3}
\definecolor{MyLightRed3}{rgb}{0.8,0.1,0.1} 
\definecolor{MyRed}{rgb}{0.7,0.0,0}

\definecolor{MyLigthBlue13}{rgb}{0,0.2,0.7}
 \definecolor{MyLigthBlack}{rgb}{0.2,0.25,0.3} 

\hypersetup{%
bookmarks=true
  pdftitle = {Title Here},
  pdfsubject = {},
  pdfkeywords = {Keywords},
  pdfauthor = {Alessandro Casini},}
\hypersetup{ 
  colorlinks = {true}, 
  citecolor = {MyBlue},
  linkcolor = {MyRed},
  filecolor= {Green},	
  urlcolor = {MyRed},
  hyperindex = {true},
  linktocpage = {true},
  linkbordercolor ={1 0 0},
 citebordercolor ={0 1 1},
 urlbordercolor ={0 1 1}
}

\renewcommand*{\thesection}{\arabic{section}}

\usepackage{multibib}
\newcites{ReferencesSupp}{References}

\makeatother

\providecommand{\assumptionname}{Assumption}
\providecommand{\definitionname}{Definition}
\providecommand{\examplename}{Example}
\providecommand{\lemmaname}{Lemma}
\providecommand{\theoremname}{Theorem}

\begin{document}
\setcounter{page}{0}
\title{\textbf{\Large{}Theory of Low Frequency Contamination from Nonstationarity
and Misspecification: Consequences for HAR Inference}\textbf{}\thanks{We are grateful to Peter C.B. Phillips, Anna Mikusheva and the referees
for useful suggestions. We thank Whitney Newey and Tim Vogelsang for
discussions and Andrew Chesher, Adam McCloskey,  Zhongjun Qu and
Daniel Whilem for comments.}\textbf{ }}
\maketitle
\begin{abstract}
We establish theoretical results about the low frequency contamination
(i.e., long memory effects) induced by general nonstationarity for
estimates such as the sample autocovariance and the periodogram, and
deduce consequences for heteroskedasticity and autocorrelation robust
(HAR) inference. We present explicit expressions for the asymptotic
bias of these estimates. We show theoretically that nonparametric
smoothing over time is robust to low frequency contamination.  
Nonstationarity can have consequences for both the size and power
of HAR tests. Under the null hypothesis there are larger size distortions
than when data are stationary. Under the alternative hypothesis, existing
LRV estimators tend to be inflated and HAR tests can exhibit dramatic
power losses. Our theory indicates that long bandwidths or fixed-$b$
HAR tests suffer more from low frequency contamination relative to
HAR tests based on HAC estimators, whereas recently introduced double
kernel HAC estimators do not suffer from this problem. We present
second-order Edgeworth expansions under nonstationarity about the
distribution of HAC and DK-HAC estimators and about the corresponding
$t$-test in the regression model. The results show that the distortions
in the rejection rates can be induced by time variation in the second
moments even when there is no break in the mean. 
\end{abstract}

\onehalfspacing
\thispagestyle{empty}
\allowdisplaybreaks

\vfill{}
\pagebreak{}

\section{\label{Section Introduction Lap_BP}Introduction}

\begin{onehalfspace}
Many economic and financial time series have nonstationary characteristics
that need to be accounted for in inference {[}see, e.g., \citet{perron:89},
\citet{stock/watson:96}, \citet{ng/wright:13}, and \citet{giacomini/rossi:15}{]}.
We develop theoretical results about the behavior of the sample autocovariance
($\widehat{\Gamma}\left(k\right),\,k\in\mathbb{Z}$) and the periodogram
($I_{T}\left(\omega\right),\,\omega\in\left[-\pi,\,\pi\right]$) for
a short memory nonstationary process. This means processes that have
non-constant moments and whose sum of absolute autocovariances is
finite. The latter rules out processes with unbounded second moments
(e.g., unit root). We show that time-variation in the mean induces
low frequency contamination, meaning that the sample autocovariance
and the periodogram share features that are similar to those of a
long memory series. We present explicit expressions for the asymptotic
bias of these estimates, showing that it is always positive and increases
with the degree of heterogeneity in the data. 
\end{onehalfspace}

The low frequency contamination can be explained as follows. For a
short memory series, the autocorrelation function (ACF) displays exponential
decay and vanishes as the lag length $k\rightarrow\infty$, and the
periodogram is finite at the origin. Under general forms of nonstationarity
involving changes in the mean, we show theoretically that $\widehat{\Gamma}\left(k\right)=\lim_{T\rightarrow\infty}\Gamma_{T}\left(k\right)+d^{*},$
where $\Gamma_{T}\left(k\right)=T^{-1}\sum_{t=k+1}^{T}\mathbb{E}\left(V_{t}V_{t-k}\right)$,
$k\geq0$ and $d^{*}>0$ is independent of $k$. Assuming positive
dependence for simplicity (i.e., $\lim_{T\rightarrow\infty}\Gamma_{T}\left(k\right)>0$),
that means that each sample autocovariance overestimates the true
dependence in the data. The bias factor $d^{*}>0$ depends on the
type of nonstationarity and in general does not vanish as $T\rightarrow\infty$.
In addition, since short memory implies $\Gamma_{T}\left(k\right)\rightarrow0$
as $k\rightarrow\infty$, it follows that $d^{*}$ generates long
memory effects since $\widehat{\Gamma}\left(k\right)\thickapprox d^{*}>0$
as $k\rightarrow\infty$. As for the periodogram, $I_{T}\left(\omega\right)$,
we show that under nonstationarity $\mathbb{E}\left(I_{T}\left(\omega\right)\right)\rightarrow\infty$
as $\omega\rightarrow0$, a feature also shared by long memory processes. 

\begin{onehalfspace}
Several HAR inference problems in applied work (besides the $t$-
and $F$-test in regression models) are characterized by nonstationary
alternative hypotheses for which $d^{*}>0$ even asymptotically. This
class of  tests is very large. Tests for forecast evaluation {[}e.g.,
\citet{casini_CR_Test_Inst_Forecast}, \citet{diebold/mariano:95},
\citeauthor{giacomini/rossi:09} (\citeyear{giacomini/rossi:09},
\citeyear{giacomini/rossi:10}), \citet{giacomini/white:06}, \citet{perron/yamamoto:18}
and \citet{west:96}{]}, tests and inference for structural changes
{[}e.g., \citet{andrews:93}, \citet{bai/perron:98}, \citeauthor{casini/perron_CR_Single_Break}
(\citeyear{casini/perron_SC_BP_Lap}, \citeyear{casini/perron_Lap_CR_Single_Inf},
\citeyear{casini/perron_CR_Single_Break}), \citet{elliott/mueller:07},
and \citet{qu/perron:07}{]}, tests and inference in time-varying
parameters models {[}e.g., \citet{cai:07} and \citet{chen/hong:12}{]},
tests and inference for regime switching models {[}e.g., \citet{hamilton:89}
and \citet{qu/zhuo:2020}{]} and  others are part of this class.
 
\end{onehalfspace}

Recently, \citet{casini_hac} proposed a new HAC estimator that applies
nonparametric smoothing over time in order to account flexibly for
nonstationarity. We show theoretically that nonparametric smoothing
over time is robust to low frequency contamination and prove that
the resulting sample local autocovariance and the local periodogram
do not exhibit long memory features. Nonparametric smoothing avoids
mixing highly heterogeneous data coming from distinct nonstationary
regimes as opposed to what the sample autocovariance and the periodogram
do.  

Our work is different from the literature on spurious persistence
caused by the presence of level shifts or other deterministic trends.
\citet{perron:90} showed that the presence of breaks in mean often
induces spurious non-rejection of the unit root hypothesis, and that
the presence of a level shift asymptotically biases the estimate of
the AR coefficient towards one. \citet{bhattacharya/gupta/waymire:83}
demonstrated that certain deterministic trends can induce the spurious
presence of long memory. In other contexts, similar issues were discussed
by \citet{varneskov/christensen:17}, \citet{diebold/inoue:01}, \citet{demetrescu/salish:2020},
\citet{lamoureux/lastrapes:1990}, \citet{hillebrand:05}, \citet{granger/hyung:04},
\citet{mccloskey/hill:2017}, \citet{mikosh/starica:04}, \citet{muller/watson:2008}
and \citet{perron/qu:2010}. Our results are different from theirs
in that we consider a more general problem and we allow for more general
forms of nonstationarity using the segmented locally stationary framework
of \citet{casini_hac}. Importantly, we provide a general solution
to these problems and show theoretically its robustness to low frequency
contamination. Moreover, we discuss in detail the implications of
our theory for HAR inference.

HAR inference relies on estimation of the long-run variance (LRV).
The latter, from a time domain perspective, is equivalent to the sum
of all autocovariances while from a frequency domain perspective,
is equal to $2\pi$ times an integrated time-varying spectral density
at the zero frequency. From a time domain perspective, estimation
involves a weighted sum of the sample autocovariances, while from
a frequency domain perspective estimation is based on a weighted sum
of the periodogram ordinates near the zero frequency. Therefore, our
results on low frequency contamination for the sample autocovariances
and the periodogram can have important implications. 

There are two main approaches in HAR inference, one based on traditional
asymptotics and the other based on fixed-smoothing asymptotics. The
classical approach relies on an LRV estimator using a small bandwidth
{[}cf. the HAC estimators of \citeauthor{newey/west:87} (\citeyear{newey/west:87},
\citeyear{newey/west:94}) and \citet{andrews:91}{]}. Inference is
standard because HAR test statistics follow asymptotically standard
distributions. It was shown early that HAC standard errors can result
in oversized tests when there is substantial temporal dependence.
This stimulated a second approach  based on an LRV estimator that
keeps the bandwidth at a fixed fraction of the sample size and that
converges weakly to a random variable {[}cf. \citet{Kiefer/vogelsang/bunzel:00}{]}.
Inference is then based on a nonstandard reference distribution 
and it is shown that fixed-$b$ achieves high-order refinements {[}e.g.,
\citet{sun/phillips/jin:08}{]} and reduces the oversize problem of
HAR tests.\footnote{See \citet{dou:18}, \citet{hwang/sun:2017}, \citet{ibragimov/kattuman/skrobotov:2021},
\citet{ibragimov/muller:10}, \citet{jansson:04}, \citeauthor{Kiefer/vogelsang:02}
(\citeyear{Kiefer/vogelsang:02}, \citeyear{kiefer/vogelsang:05}),
\citet{lazarus/lewis/stock:17}, \textcolor{MyBlue}{Lazarus et al.}
\citeyearpar{lazarus/lewis/stock/watson:18} \citeauthor{muller:07}
(\citeyear{muller:07}, \citeyear{mueller:14}), \citet{phillips:05},
\citet{politis:11}, \citeauthor{potscher/preinerstorfer:18} (\citeyear{preinerstorfer/potscher:16},
\citeyear{potscher/preinerstorfer:18}, \citeyear{potscher/preinerstorfer:19}),
\citet{robinson:98}, \citeauthor{sun:14} (\citeyear{sun:13}, \citeyear{sun:14a},
\citeyear{sun:14}) and \citet{zhang/shao:13}.} However, unlike the classical approach, current fixed-$b$ HAR inference
is only valid under stationarity {[}cf. \citet{casini_fixed_b_erp}{]}
as the fixed-$b$ limiting distribution of the $t$/$F$ statistic
is non-pivotal under nonstationarity. More recently, a variant of
the fixed-$b$ approach {[}see, e.g., \citet{sun:14} and \citet{lazarus/lewis/stock/watson:18}{]}
considered the use of small-$b$ asymptotics in conjunction with
fixed-$b$ or $t/F$ critical values. These bandwidths are typically
larger than the MSE-optimal bandwidths used for the HAC estimators.

Recently, \citet{casini_hac} questioned the performance of HAR inference
under nonstationarity from a theoretical standpoint. Simulation evidence
of serious (e.g., non-monotonic) power or related issues in specific
HAR inference contexts were documented by \citet{altissimo/corradi:2003},
\citet{casini_CR_Test_Inst_Forecast}, \citeauthor{casini/perron_Lap_CR_Single_Inf}
(\citeyear{casini/perron:OUP-Breaks}, \citeyear{casini/perron_SC_BP_Lap},
\citeyear{casini/perron_Lap_CR_Single_Inf}), \citeauthor{chan:2020}
(\citeyear{chan:2022}, \citeyear{chan:2020}), \citet{crainiceanu/vogelsang:07},
\citet{deng/perron:06}, \citet{juhl/xiao:09}, \citet{kim/perron:09},
\citet{martins/perron:16}, \citet{otto/breitung:2021}, \citet{perron:1991},
\citet{perron/yamamoto:18}, \citet{shao/zhang:2010}, \citet{vogeslang:99}
and \citet{zhang/lavitas:2018} among others{]}. Our theoretical results
show that these issues occur because the unaccounted nonstationarity
alters the spectrum at low frequencies. Each sample autocovariance
is upward biased ($d^{*}>0$) and the resulting LRV estimators tend
to be inflated. When these estimators are used to normalize test statistics,
the latter lose power.  Interestingly,  $d^{*}$ is independent
of $k$ so that the more lags are included the more severe is the
problem. Further, by virtue of weak dependence, we have that $\Gamma_{T}\left(k\right)\rightarrow0$
as $k\rightarrow\infty$ but $d^{*}>0$ across $k$. We show formally
that long bandwidths/fixed-$b$ LRV estimators are expected to suffer
most from power losses  because they use many/all lagged autocovariances.

To precisely analyze the theoretical properties of the HAR tests under
the null hypothesis, we present second-order Edgeworth expansions
under nonstationarity for the distribution of the HAC and DK-HAC estimator
and for the distribution of the corresponding $t$-test in the linear
regression model. Under stationarity the results concerning the HAC
estimator were provided by \citet{velasco/robinson:01}. We show that
the order of the approximation error of the expansion is the same
as under stationarity from which it follows that the error in rejection
probability (ERP) is also the same. The ERP of the $t$-test based
on the DK-HAC estimator is slightly larger than that of the $t$-test
based on the HAC estimator due to the double smoothing. High-order
asymptotic expansions for spectral and other estimates were studied
by \citet{bhattacharya/ghosh:1978}, \citet{bentkus/rudzkis:1982},
\citet{janas:1994}, \citeauthor{phillips:1977} (\citeyear{phillips:1977},
\citeyear{phillips:1980}) and \citet{taniguchi/puri:1996}. The asymptotic
expansions of the fixed-$b$ HAR tests under stationarity were developed
by \citet{jansson:04} and \citet{sun/phillips/jin:08}. \citet{casini_fixed_b_erp}
showed that under nonstationarity the ERP of the fixed-$b$ HAR tests
can be larger than that of HAR tests based on HAC and DK-HAC estimators
thereby controverting the conclusion in the literature that the original
fixed-$b$ HAR tests have superior null rejection rates relative to
HAR tests based on traditional LRV estimators. \citet{casini_fixed_b_erp}
also developed fixed-$b$ methods that are valid under nonstationarity
and in fact provide better null rejection rates in finite-sample.

The Monte Carlo results suggest that under the null hypothesis nonstationarity
can generate larger size distortions than what one finds under stationarity.
In particular, fixed-smoothing methods can exhibit under-rejections
whereas HAC and DK-HAC methods can exhibit over-rejections when there
is strong persistence. For the latter problem, our second-order Edgeworth
expansions could be used to construct corrections to the standard
normal critical value. We relegate this opportunity to future research. 

  The paper is organized as follows. Section \ref{Section, Statistical Framework for Nonstationarity}
presents the statistical setting and Section \ref{Section Low Freq Cont - Theory}
establishes the theoretical results on low frequency contamination.
Section \ref{Section Edgeworth-Expansions-for} presents the Edgeworth
expansions of HAR tests based on the HAC and DK-HAC estimators. The
implications of our results for HAR inference are analyzed analytically
and computationally through simulations in Section \ref{Section Consequences for HAR}.
Section \ref{Section Conclusions} concludes. The supplemental materials
{[}cf. \citet{casini/perron_Low_Frequency_Contam_Nonstat:2020_supp}{]}
contain some additional examples and all mathematical proofs. 

\section{\label{Section, Statistical Framework for Nonstationarity}Statistical
Framework for Nonstationarity}

Suppose $\{V_{t,T}\}_{t=1}^{T}$ is defined on a probability space
$\left(\Omega,\,\mathscr{F},\,\mathbb{P}\right)$, where $\Omega$
is the sample space, $\mathscr{F}$ is the $\sigma$-algebra and $\mathbb{P}$
is a probability measure. In order to analyze time series models that
have a time-varying spectrum it is useful to introduce an infill asymptotic
setting whereby we rescale the original discrete time horizon $\left[1,\,T\right]$
by dividing each $t$ by $T.$ Letting $u=t/T$ we define a new time
scale $u\in\left[0,\,1\right]$ on which as $T\rightarrow\infty$
we observe more and more realizations of $V_{t,T}$ close to time
$t$. As a notion of nonstationarity, we use the concept of segmented
local stationarity (SLS)  introduced in \citet{casini_hac}.  This
extends the locally stationary processes {[}cf. \citet{dahlhaus:96}{]}
to allow for structural change and regime switching-type models. 
SLS processes allow for a finite number of discontinuities in the
spectrum over time. We collect the break dates in the set $\mathcal{T}\triangleq\left\{ T_{1}^{0},\,\ldots,\,T_{m}^{0}\right\} $.
 Let $i\triangleq\sqrt{-1}.$ A function $G\left(\cdot,\,\cdot\right):\,\left[0,\,1\right]\times\mathbb{R}\rightarrow\mathbb{C}$
is said to be left-differentiable at $u_{0}$ if $\partial G\left(u_{0},\omega\right)/\partial_{-}u\triangleq\lim_{u\rightarrow u_{0}^{-}}\left(G\left(u_{0},\,\omega\right)-G\left(u,\,\omega\right)\right)/\left(u_{0}-u\right)$
exists for any $\omega\in\mathbb{R}$.  Let $m_{0}\geq0$ be a finite
integer. 
\begin{defn}
\label{Definition Segmented-Locally-Stationary}A sequence of stochastic
processes $\{V_{t,T}\}_{t=1}^{T}$ is called segmented locally stationary
(SLS) with $m_{0}+1$ regimes, transfer function $A^{0}$  and trend
$\mu$ if there exists a representation 
\begin{align}
V_{t,T} & =\mu_{j}\left(t/T\right)+\int_{-\pi}^{\pi}\exp\left(i\omega t\right)A_{j,t,T}^{0}\left(\omega\right)d\xi\left(\omega\right),\qquad\qquad\left(t=T_{j-1}^{0}+1,\ldots,\,T_{j}^{0}\right),\label{Eq. Spectral Rep of SLS}
\end{align}
for $j=1,\ldots,\,m_{0}+1$, where by convention $T_{0}^{0}=0$ and
$T_{m_{0}+1}^{0}=T$. The following technical conditions are also
assumed to hold: (i) $\xi\left(\lambda\right)$ is a process on $\left[-\pi,\,\pi\right]$
with $\overline{\xi\left(\omega\right)}=\xi\left(-\omega\right)$
and 
\begin{align*}
\mathrm{cum}\left\{ d\xi\left(\omega_{1}\right),\ldots,\,d\xi\left(\omega_{r}\right)\right\}  & =\zeta\left(\sum_{j=1}^{r}\omega_{j}\right)g_{r}\left(\omega_{1},\ldots,\,\omega_{r-1}\right)d\omega_{1}\ldots d\omega_{r},
\end{align*}
 where $\mathrm{cum}\left\{ \cdots\right\} $ denotes the cumulant
spectra of $r$-th order, $g_{1}=0,\,g_{2}\left(\omega\right)=1$,
$\left|g_{r}\left(\omega_{1},\ldots,\,\omega_{r-1}\right)\right|\leq M_{r}$
for all $r$ with $M_{r}<\infty$ that may depend on $r$, and $\zeta\left(\omega\right)=\sum_{j=-\infty}^{\infty}\delta\left(\omega+2\pi j\right)$
is the period $2\pi$ extension of the Dirac delta function $\delta\left(\cdot\right)$;
 (ii) There exists a $C<\infty$  and a piecewise continuous function
$A:\,\left[0,\,1\right]\times\mathbb{R}\rightarrow\mathbb{C}$ such
that, for each $j=1,\ldots,\,m_{0}+1$, there exists a $2\pi$-periodic
function $A_{j}:\,(\lambda_{j-1}^{0},\,\lambda_{j}^{0}]\times\mathbb{R}\rightarrow\mathbb{C}$
with $A_{j}\left(u,\,-\omega\right)=\overline{A_{j}\left(u,\,\omega\right)}$,
$\lambda_{j}^{0}\triangleq T_{j}^{0}/T$ and for all $T,$
\begin{align}
A\left(u,\,\omega\right) & =A_{j}\left(u,\,\omega\right)\,\mathrm{\,for\,}\,\lambda_{j-1}^{0}<u\leq\lambda_{j}^{0},\label{Eq A(u) =00003D Ai}\\
\sup_{1\leq j\leq m_{0}+1} & \sup_{T_{j-1}^{0}<t\leq T_{j}^{0},\,\omega}\left|A_{j,t,T}^{0}\left(\omega\right)-A_{j}\left(t/T,\,\omega\right)\right|\leq CT^{-1};\label{Eq. 2.4 Smothenss Assumption on A}
\end{align}

(iii) $\mu_{\cdot}\left(\cdot\right)$ is piecewise Lipschitz continuous.
\end{defn}
Definition \ref{Definition Segmented-Locally-Stationary} states that
$V_{t,T}$ has a time-varying spectral representation where both the
mean $\mu_{\cdot}\left(\cdot\right)$ and transfer function $A_{\cdot,\cdot,T}^{0}\left(\omega\right)$
are piecewise continuous. Since the transfer function depends on the
parameters that enter the second moments of $V_{t,T}$, the smoothness
properties of $\mu_{\cdot}\left(\cdot\right)$ and $A$ guarantee
that $V_{t,T}$ has a piecewise locally stationary behavior. We require
additional smoothness properties for $A$ and an example is presented
at the end of this section.
\begin{assumption}
\label{Assumption Smothness of A (for HAC)}(i) $\left\{ V_{t,T}\right\} $
is an SLS process with $m_{0}+1$ regimes; (ii) $A\left(u,\,\omega\right)$
is \textcolor{red}{ }twice continuously differentiable in $u$ at
all $u\neq\lambda_{j}^{0}$, $j=1,\ldots,\,m_{0}+1,$ with bounded
derivatives $\left(\partial/\partial u\right)A\left(u,\,\cdot\right)$
and $\left(\partial^{2}/\partial u^{2}\right)A\left(u,\,\cdot\right)$;
(iii) $\left(\partial^{2}/\partial u^{2}\right)A\left(u,\,\cdot\right)$
is Lipschitz continuous at all $u\neq\lambda_{j}^{0}$ $(j=1,\ldots,\,m_{0}+1)$;
(iv) $A\left(u,\,\omega\right)$ is twice left-differentiable in $u$
at $u=\lambda_{j}^{0}$ $(j=1,\ldots,\,m_{0}+1)$ with  bounded derivatives
$\left(\partial/\partial_{-}u\right)A\left(u,\,\cdot\right)$ and
$\left(\partial^{2}/\partial_{-}u^{2}\right)A\left(u,\,\cdot\right)$
and has piecewise Lipschitz continuous derivative $\left(\partial^{2}/\partial_{-}u^{2}\right)A\left(u,\,\cdot\right)$;
(v) $A\left(u,\,\omega\right)$ is Lipschitz continuous in $\omega.$
\end{assumption}
 We define the time-varying spectral density as $f_{j}\left(u,\,\omega\right)\triangleq(2\pi)^{-1}|A_{j}\left(u,\,\omega\right)|^{2}$
for $T_{j-1}^{0}/T<u=t/T\leq T_{j}^{0}/T$.  Then we can define the
local covariance of $V_{t,T}$ at the rescaled time $u$ with $Tu\notin\mathcal{T}$
and lag $k\in\mathbb{Z}$ as $c\left(u,\,k\right)\triangleq\int_{-\pi}^{\pi}e^{i\omega k}f\left(u,\,\omega\right)d\omega$.
The same definition is also used when $Tu\in\mathcal{T}$ and $k\geq0$.
For $Tu\in\mathcal{T}$ and $k<0$ it is defined as $c\left(u,\,k\right)\triangleq\lim_{T\rightarrow\infty}\int_{-\pi}^{\pi}e^{i\omega k}A\left(u,\,\omega\right)A\left(u-k/T,\,-\omega\right)d\omega$.

Next, we impose conditions on the temporal dependence (we omit the
second subscript $T$ when it is clear from the context). Let 
\begin{align*}
\kappa_{V,t}^{\left(a_{1},a_{2},a_{3},a_{4}\right)} & \left(u,\,v,\,w\right)\\
 & \triangleq\kappa^{\left(a_{1},a_{2},a_{3},a_{4}\right)}\left(t,\,t+u,\,t+v,\,t+w\right)-\kappa_{\mathscr{N}}^{\left(a_{1},a_{2},a_{3},a_{4}\right)}\left(t,\,t+u,\,t+v,\,t+w\right)\\
 & \triangleq\mathbb{E}\left(V_{t}^{\left(a_{1}\right)}-\mathbb{E}V_{t}^{\left(a_{1}\right)}\right)\left(V_{t+u}^{\left(a_{2}\right)}-\mathbb{E}V_{t+u}^{\left(a_{2}\right)}\right)\left(V_{t+v}^{\left(a_{3}\right)}-\mathbb{E}V_{t+v}^{\left(a_{3}\right)}\right)\left(V_{t+w}^{\left(a_{4}\right)}-\mathbb{E}V_{t+w}^{\left(a_{4}\right)}\right)\\
 & \quad-\mathbb{E}\left(V_{\mathscr{N},t}^{\left(a_{1}\right)}-\mathbb{E}V_{\mathscr{N},t}^{\left(a_{1}\right)}\right)\left(V_{\mathscr{N},t+u}^{\left(a_{2}\right)}-\mathbb{E}V_{\mathscr{N},t+u}^{\left(a_{2}\right)}\right)\left(V_{\mathscr{N},t+v}^{\left(a_{3}\right)}-\mathbb{E}V_{\mathscr{N},t+v}^{\left(a_{3}\right)}\right)\left(V_{\mathscr{N},t+w}^{\left(a_{4}\right)}-\mathbb{E}V_{\mathscr{N},t+w}^{\left(a_{4}\right)}\right),
\end{align*}
where $\left\{ V_{\mathscr{N},t}\right\} $ is a Gaussian sequence
with the same mean and covariance structure as $\left\{ V_{t}\right\} $,
$\kappa_{V,t}^{\left(a_{1},a_{2},a_{3},a_{4}\right)}\left(u,\,v,\,w\right)$
is the time-$t$ fourth-order cumulant of $(V_{t}^{\left(a_{1}\right)},\,V_{t+u}^{\left(a_{2}\right)},\,V_{t+v}^{\left(a_{3}\right)},$
$\,V_{t+w}^{\left(a_{4}\right)})$ while $\kappa_{\mathscr{N}}^{\left(a_{1},a_{2},a_{3},a_{4}\right)}$
$(t,\,t+u,\,t+v,\,t+w)$ is the time-$t$ centered fourth moment of
$V_{t}$ if $V_{t}$ were Gaussian.
\begin{assumption}
\label{Assumption A - Dependence}(i) $\sum_{k=-\infty}^{\infty}\sup_{u\in\left[0,\,1\right]}$
$\left\Vert c\left(u,\,k\right)\right\Vert <\infty$ and $\sum_{k=-\infty}^{\infty}\sum_{j=-\infty}^{\infty}\sum_{l=-\infty}^{\infty}\sup_{u\in\left[0,\,1\right]}|\kappa_{V,\left\lfloor Tu\right\rfloor }^{\left(a_{1},a_{2},a_{3},a_{4}\right)}$
$\left(k,\,j,\,l\right)|<\infty$ for all $a_{1},a_{2},a_{3},a_{4}\leq p$.
(ii) For all $a_{1},a_{2},a_{3},a_{4}\leq p$ there exists a function
$\widetilde{\kappa}_{a_{1},a_{2},a_{3},a_{4}}:\,\left[0,\,1\right]\times\mathbb{Z}\times\mathbb{Z}\times\mathbb{Z}\rightarrow\mathbb{R}$
such that $\sup_{1\leq j\leq m_{0}+1}\sup_{\lambda_{j-1}^{0}<u\leq\lambda_{j}^{0}}|\kappa_{V,\left\lfloor Tu\right\rfloor }^{\left(a_{1},a_{2},a_{3},a_{4}\right)}\left(k,\,s,\,l\right)-\widetilde{\kappa}_{a_{1},a_{2},a_{3},a_{4}}$
$\left(u,\,k,\,s,\,l\right)|\leq LT^{-1}$ for some constant $L$;
the function $\widetilde{\kappa}_{a_{1},a_{2},a_{3},a_{4}}\left(u,\,k,\,s,\,l\right)$
is twice differentiable in $u$ at all $u\neq\lambda_{j}^{0}$ $(j=1,\ldots,\,m_{0}+1)$
with  bounded derivatives $\left(\partial/\partial u\right)\widetilde{\kappa}_{a_{1},a_{2},a_{3},a_{4}}$
$\left(u,\cdot,\cdot,\cdot\right)$ and $\left(\partial^{2}/\partial u^{2}\right)\widetilde{\kappa}_{a_{1},a_{2},a_{3},a_{4}}\left(u,\cdot,\cdot,\cdot\right)$,
and twice left-differentiable in $u$ with  bounded derivatives $\left(\partial/\partial_{-}u\right)\widetilde{\kappa}_{a_{1},a_{2},a_{3},a_{4}}\left(u,\cdot,\cdot,\cdot\right)$
and $\left(\partial^{2}/\partial_{-}u^{2}\right)\widetilde{\kappa}_{a_{1},a_{2},a_{3},a_{4}}\left(u,\cdot,\cdot,\cdot\right)$,
and piecewise Lipschitz continuous derivative $\left(\partial^{2}/\partial_{-}u^{2}\right)\widetilde{\kappa}_{a_{1},a_{2},a_{3},a_{4}}\left(u,\cdot,\cdot,\cdot\right)$.
\end{assumption}
If $\left\{ V_{t}\right\} $ is stationary then the cumulant condition
of Assumption \ref{Assumption A - Dependence}-(i) reduces to the
standard one used in the time series literature {[}see \citet{andrews:91}{]}.
Note that $\alpha$-mixing and some moment conditions imply that
the cumulant condition of Assumption \ref{Assumption A - Dependence}
holds.  Part (ii) extends the smoothness conditions on $A\left(u,\,\omega\right)$
in Assumption \ref{Assumption Smothness of A (for HAC)} to the fourth-order
cumulant. These smoothness conditions are not particularly restrictive. 

Consider the following time-varying AR(1) process with one break at
mid-sample $\lambda_{1}^{0}=0.5$, 
\begin{align}
V_{t,T} & =\rho\left(t/T\right)V_{t-1,T}+\sigma\left(t/T\right)u_{t},\label{Eq. Example TV AR(1)}\\
\rho\left(u\right) & =\begin{cases}
\rho_{1}\left(u\right), & u\leq0.5\\
\rho_{2}\left(u\right), & u>0.5
\end{cases},\nonumber 
\end{align}
 where $\rho_{1}\left(\cdot\right)$ and $\rho_{2}\left(\cdot\right)$
are Lipschitz continuous, $\sigma\left(\cdot\right)$ is piecewise
Lipschitz continuous and $\left\{ u_{t}\right\} $ are i.i.d. random
variables with mean zero and unit variance. Then, $V_{t,T}$ is an
SLS process with $A\left(u,\,\omega\right)=\sigma\left(u\right)\left(1+\rho\left(u\right)\exp\left(i\omega\right)\right)$.
If $\rho\left(u\right)$ and $\sigma\left(u\right)$ satisfy the same
smoothness conditions in $u$ required for $A\left(u,\,\omega\right)$
in Assumption \ref{Assumption Smothness of A (for HAC)}, $\sup_{u\in\left[0,\,1\right]}\left|\rho\left(u\right)\right|<1$
and $\sup_{u\in\left[0,\,1\right]}\sigma\left(u\right)<\infty$, then
$V_{t,T}$ fulfills Assumption \ref{Assumption Smothness of A (for HAC)}-\ref{Assumption A - Dependence}. 

\section{\label{Section Low Freq Cont - Theory}Theoretical Results on Low
Frequency Contamination}

In this section we establish theoretical results about the low frequency
contamination induced by nonstationarity, misspecification and outliers.
We first consider the asymptotic proprieties of two key  quantities
for inference in time series contexts, i.e., the sample autocovariance
 and the periodogram. These are defined, respectively, by 
\begin{align}
\widehat{\Gamma}\left(k\right) & =T^{-1}\sum_{t=|k|+1}^{T}\left(V_{t}-\overline{V}\right)\left(V_{t-|k|}-\overline{V}\right),\label{Eq. Definition of Gamma(k)}
\end{align}
where $\overline{V}$ is the sample mean and 
\begin{align*}
I_{T}\left(\omega\right) & =\left|\frac{1}{\sqrt{T}}\sum_{t=1}^{T}\exp\left(-i\omega t\right)V_{t}\right|^{2},\qquad\qquad\omega\in\left[0,\,\pi\right],
\end{align*}
 which is evaluated at the Fourier frequencies $\omega_{j}=\left(2\pi j\right)/T\in[0,\,\pi]$.
In the context of autocorrelated data, hypotheses testing and construction
of confidence intervals require estimation of the so-called long-run
variance. Traditional HAC estimators are weighted sums of sample autocovariances
while frequency domain estimators are weighted sums of the periodograms.
\citet{casini_hac} considered an alternative estimate for the sample
autocovariance to be used in the DK-HAC estimators, defined in Section
\ref{Subsection HAR inference methods}, namely,
\begin{align*}
\widehat{\Gamma}_{\mathrm{DK}}\left(k\right) & \triangleq\frac{n_{T}}{T}\sum_{r=1}^{\left\lfloor T/n_{T}\right\rfloor }\widehat{c}_{T}\left(rn_{T}/T,\,k\right),
\end{align*}
 where $k\in\mathbb{Z},$ $n_{T}\rightarrow\infty$ satisfying the
conditions given below, and 
\begin{align}
\widehat{c}_{T}\left(rn_{T}/T,\,k\right) & =n_{2,T}^{-1}\sum_{s=0}^{n_{2,T}-1}\left(V_{rn_{T}+\left\lfloor |k/2|\right\rfloor -n_{2,T}/2+s+1}-\overline{V}{}_{rn_{T},T}\right)\left(V_{rn_{T}-\left\lfloor |k/2|\right\rfloor -n_{2,T}/2+s+1}-\overline{V}{}_{rn_{T},T}\right),\label{Eq. chat}
\end{align}
with $\overline{V}{}_{rn_{T},T}=n_{2,T}^{-1}\sum_{s=0}^{n_{2,T}-1}V_{rn_{T}-n_{2,T}/2+s+1}$
and $n_{2,T}\rightarrow\infty$ such that $n_{2,T}/T\rightarrow0$.
For notational simplicity we assume that $n_{T}$ and $n_{2,T}$ are
even. $\widehat{c}_{T}\left(rn_{T}/T,\,k\right)$ is an estimate of
the autocovariance at time $rn_{T}$ and lag $k$, i.e., $\mathrm{cov}(V_{rn_{T}},\,V_{rn_{T}-k})$.
One could use a smoothed or tapered version; the estimate $\widehat{\Gamma}_{\mathrm{DK}}\left(k\right)$
is an integrated local sample autocovariance. It extends $\widehat{\Gamma}\left(k\right)$
to better account for nonstationarity. Similarly, the DK-HAC estimator
does not relate to the periodogram but to the local periodogram defined
by
\begin{align*}
I_{\mathrm{L},T}\left(u,\,\omega\right) & \triangleq\left|\frac{1}{\sqrt{n_{T}}}\sum_{s=0}^{n_{T}-1}V_{\left\lfloor Tu\right\rfloor -n_{T}/2+s+1,T}\exp\left(-i\omega s\right)\right|^{2},
\end{align*}
where $I_{\mathrm{L},T}\left(u,\,\omega\right)$ is the (untapered)
periodogram over a segment of length $n_{T}$ with midpoint $\left\lfloor Tu\right\rfloor $.
 We also consider the statistical properties of both $\widehat{\Gamma}_{\mathrm{DK}}\left(k\right)$
and $I_{\mathrm{L},T}\left(u,\,\omega\right)$ under nonstationarity.
Define $r_{j}=(\lambda_{j}^{0}-\lambda_{j-1}^{0})$ for $j=1,\ldots,\,m_{0}+1$
with $\lambda_{0}^{0}=0$ and $\lambda_{m_{0}+1}^{0}=1$.  Note that
$\lambda_{j}^{0}=\sum_{s=0}^{j}r_{s}.$ 

The low frequency bias is generated by breaks in the mean function.
For the sample autocovariance, the bias factor is given by $d^{*}=2^{-1}\sum_{j_{1}\neq j_{2}}r_{j_{1}}r_{j_{2}}(\overline{\mu}_{j_{2}}-\overline{\mu}_{j_{1}})^{2}$
where 
\begin{align*}
\overline{\mu}_{j} & =r_{j}^{-1}\int_{\lambda_{j-1}^{0}}^{\lambda_{j}^{0}}\mu_{j}\left(u\right)du,\qquad\mathrm{for}\,j=1,\ldots,\,m_{0}+1,
\end{align*}
with $\mu_{j}\left(\cdot\right)$ defined in \eqref{Eq. Spectral Rep of SLS}
and we use $\sum_{j_{1}\neq j_{2}}$ as a shorthand for  $\sum_{\left\{ j_{1},\,j_{2}=1,\ldots,\,m_{0}+1,\,j_{1}\neq j_{2}\right\} }.$
When the mean is constant in each regime $\mu_{j}\left(t/T\right)=\mu_{j}$.
Then, $\overline{\mu}_{j}=\mu_{j}$ and $d^{*}=2^{-1}\sum_{j_{1}\neq j_{2}}r_{j_{1}}r_{j_{2}}(\mu_{j_{2}}-\mu_{j_{1}})^{2}.$
If the mean is constant across regimes, then there is no low frequency
bias and $d^{*}=0.$ 

In Section \ref{Subsection Extension Sample Autocovariance and Periodogram}
we generalize the results in the literature on low frequency contamination
for the sample autocovariance and the periodogram. In Section \ref{Subsection Local Autocovariance and Local Periodogram}
we show that the local sample autocovariance and the local periodogram
are in general robust to low frequency contamination. 

\subsection{\label{Subsection Extension Sample Autocovariance and Periodogram}The
Sample Autocovariance and the Periodogram Under Nonstationarity }

\citet*{mikosh/starica:04} established some results on the low frequency
bias for the sample autocovariance and  periodogram under the assumption
that $V_{t}$ is stationary in each regime and that the regimes are
independent. In Section \ref{Section Results on Low Frequency Contamination}
in the supplement we extend these results by allowing time-varying
mean and autocovariace function in each regime and weak dependence
across regimes. Here we present a brief summary of these results.
Theorem \ref{Theorem ACF Nonstat} shows that for $\left\{ V_{t,T}\right\} $
that satisfies Definition \ref{Definition Segmented-Locally-Stationary}
and Assumption \ref{Assumption Smothness of A (for HAC)}-\ref{Assumption A - Dependence},
 we have
\begin{align}
\widehat{\Gamma}\left(k\right)\geq & \int_{0}^{1}c\left(u,\,k\right)du+d^{*}+o_{\mathrm{a.s}.}\left(1\right),\label{Eq. Gamma_hat(k) Ineq Theorem-1}
\end{align}
and as $k\rightarrow\infty,$ $\widehat{\Gamma}\left(k\right)\geq d^{*}$
$\mathbb{P}$-a.s. This suggests that $\widehat{\Gamma}\left(k\right)$
is asymptotically the sum of two terms. The first is the  autocovariance
of $\left\{ V_{t}\right\} $ at lag $k$.  The second, $d_{\mathrm{}}^{*}$,
is  always positive and increases  with the difference in the mean
across regimes. Thus, the time-varying mean induces a positive bias.
The result that $\widehat{\Gamma}\left(k\right)\geq d^{*}$ $\mathbb{P}$-a.s.
as $k\rightarrow\infty$ implies that unaccounted nonstationarity
generates long memory effects. The intuition is straightforward. A
long memory SLS process satisfies $\sum_{k=-\infty}^{\infty}|\Gamma\left(u,\,k\right)|\rightarrow\infty$
for some $u\in\left(0,\,1\right)$, similar to a stationary long memory
process.\footnote{In Section \ref{Subsection Long-Memory-Segmented-Locally} in the
supplement we define long memory SLS processes that are characterized
by the property $\sum_{k=-\infty}^{\infty}\left|\rho_{V}\left(u,\,k\right)\right|=\infty$
for some $u\in\left[0,\,1\right]$ where $\rho_{V}\left(u,\,k\right)\triangleq\mathrm{Corr}(V_{\left\lfloor Tu\right\rfloor },\,V_{\left\lfloor Tu\right\rfloor +k})$
and $\vartheta\left(u\right)\in\left(0,\,1/2\right)$ is the long
memory parameter at time $u$. } The theorem shows that $\widehat{\Gamma}\left(k\right)$ exhibits
a similar property and $\widehat{\Gamma}\left(k\right)$ decays more
slowly than for a short memory stationary process for small lags and
approaches a constant $d^{*}>0$ for large lags. 

Theorem \ref{Theorem Periodogram Long Memory Effects} in the supplement
analyzes the properties of the periodogram $I_{T}\left(\omega_{l}\right)$
as $\omega\rightarrow0$ when the mean is time-varying. The result
states that as $\omega\rightarrow0$ $\mathbb{E}\left(I_{T}\left(\omega\right)\right)$
generally takes  unbounded values except for some $\omega$ for which
$\mathbb{E}\left(I_{T}\left(\omega\right)\right)$ is bounded below
by $2\pi\int_{0}^{1}f\left(u,\,\omega\right)du>0.$  An SLS process
with long memory has an unbounded local spectral density $f\left(u,\,\omega\right)$
as $\omega\rightarrow0$ for some $u\in\left[0,\,1\right]$. Since
$f\left(\cdot,\,\cdot\right)$ cannot be negative, it follows that
$\int_{0}^{1}f\left(u,\,\omega\right)du$ is also unbounded as $\omega\rightarrow0$.
Theorem \ref{Theorem Periodogram Long Memory Effects} suggests that
nonstationarity consisting of time-varying first moment results in
a periodogram sharing features of a long memory series. 

This discussion suggests that certain deviations from stationarity
can generate a long memory component that leads to overestimation
of the true autocovariance. It follows that the LRV is also overestimated.
Since the LRV is used to normalize test statistics, this has important
consequences for many HAR inference tests  characterized by deviations
from stationarity under the alternative hypothesis. These include
tests for forecast evaluation, tests and inference for structural
change models, time-varying parameters models and regime-switching
models. In the linear regression model, $V_{t}$ corresponds to the
regressors multiplied by the fitted residuals. Unaccounted nonlinearities
and outliers can contaminate the mean of $V_{t}$ and therefore contribute
to $d^{*}$.

\subsection{\label{Subsection Local Autocovariance and Local Periodogram}The
Sample Local Autocovariance and Local Periodogram Under Nonstationarity }

We now consider the behavior of $\widehat{c}_{T}\left(rn_{T}/T,\,k\right)$
 defined in \eqref{Eq. chat} for fixed $k$ as well as for $k\rightarrow\infty$.
For notational simplicity we assume that $k$ is even. For $u\in\left(0,\,1\right)$
define $\mathbf{S}\left(u,\,k,\,n_{2,T}\right)=\{\left\lfloor Tu\right\rfloor +k/2-n_{2,T}/2+1,\ldots,\,\left\lfloor Tu\right\rfloor +k/2+n_{2,T}/2\}$,
$n_{j,L}\left(u,\,k,\,n_{2,T}\right)=(T_{j}^{0}-(\left\lfloor Tu\right\rfloor +k/2-n_{2,T}/2+1)),$
and $n_{j,R}\left(u,\,k,\,n_{2,T}\right)=((\left\lfloor Tu\right\rfloor +k/2+n_{2,T}/2+1)-T_{j}^{0})$.
$\mathbf{S}\left(u,\,k,\,n_{2,T}\right)$ denotes a window of length
$n_{2,T}$ around $\left\lfloor Tu\right\rfloor $, $n_{j,L}\left(u,\,k,\,n_{2,T}\right)$
(resp. $n_{j,R}\left(u,\,k,\,n_{2,T}\right)$) denotes the distance
between the left (resp. right) end point of $\mathbf{S}\left(u,\,k,\,n_{2,T}\right)$
and $T_{j}^{0}$.
\begin{thm}
\label{Theorem Local ACF Nonstat}Assume that $\left\{ V_{t,T}\right\} $
satisfies Definition \ref{Definition Segmented-Locally-Stationary},
$n_{T},\,n_{2,T}\rightarrow\infty$ with $n_{T}/T\rightarrow0$, $n_{2,T}/T\rightarrow0$
and $n_{T}/n_{2,T}\rightarrow0.$ Under Assumption \ref{Assumption Smothness of A (for HAC)}-\ref{Assumption A - Dependence},

(i) for $u\in\left(0,\,1\right)$ such that $T_{j}^{0}\notin\mathbf{S}\left(u,\,k,\,n_{2,T}\right)$
for all $j=1,\ldots,\,m_{0}$, $\widehat{c}_{T}\left(u,\,k\right)=c\left(u,\,k\right)+o_{\mathbb{P}}\left(1\right)$; 

(ii) for $u\in\left(0,\,1\right)$ such that $T_{j}^{0}\in\mathbf{S}\left(u,\,k,\,n_{2,T}\right)$
for some $j=1,\ldots,\,m_{0}$, we have two sub-cases: (a) if $n_{j,L}\left(u,\,k,\,n_{2,T}\right)/n_{2,T}\rightarrow\gamma$
or $n_{j,R}\left(u,\,k,\,n_{2,T}\right)/n_{2,T}\rightarrow\gamma$
with $\gamma\in\left(0,\,1\right)$, then 
\begin{align*}
\widehat{c}_{T}\left(u,\,k\right) & \geq\gamma c\left(\lambda_{j}^{0},\,k\right)+\left(1-\gamma\right)c\left(u,\,k\right)+\gamma\left(1-\gamma\right)\left(\mu_{j}\left(\lambda_{j}^{0}\right)-\mu_{j+1}\left(u\right)\right)^{2}+o_{\mathbb{P}}\left(1\right).
\end{align*}
 (b) if $n_{j,L}\left(u,\,k,\,n_{2,T}\right)/n_{2,T}\rightarrow0$
or $n_{j,R}\left(u,\,k,\,n_{2,T}\right)/n_{2,T}\rightarrow0$, then
$\widehat{c}_{T}\left(u,\,k\right)=c\left(u,\,k\right)+o_{\mathbb{P}}\left(1\right)$.

Further, if there exists an $r=1,\ldots,\,\left\lfloor T/n_{T}\right\rfloor $
such that there exists a $j=1,\ldots,\,m_{0}$ with $T_{j}^{0}\in\mathbf{S}\left(rn_{T},\,k,\,n_{2,T}\right)$
satisfying (ii-a), then, as $k\rightarrow\infty$, $\widehat{\Gamma}_{\mathrm{DK}}\left(k\right)\geq d_{T}^{*}$
$\mathbb{P}$-a.s., where $d_{T}^{*}=\left(n_{2,T}/T\right)\gamma\left(1-\gamma\right)$
$(\mu_{j}(\lambda_{j}^{0})-\mu_{j+1}\left(u\right))^{2}>0$ and $d_{T}^{*}\rightarrow0$
as $T\rightarrow\infty$.
\end{thm}
The theorem shows that the behavior of  $\widehat{c}_{T}\left(u,\,k\right)$
depends on whether a change in mean is present, and if so whether
it is close enough to $\left\lfloor Tu\right\rfloor $. For a given
$u\in\left(0,\,1\right)$ and $k\in\mathbb{Z}$, if the condition
of part (i) of the theorem holds, then $\widehat{c}_{T}\left(u,\,k\right)$
is consistent for $\mathrm{cov}(V_{\left\lfloor Tu\right\rfloor }V_{\left\lfloor Tu\right\rfloor -k})=c\left(u,\,k\right)+O\left(T^{-1}\right)$
{[}see \citet{casini_hac}{]}. If a change-point falls close to either
boundary of the window $\mathbf{S}\left(u,\,k,\,n_{2,T}\right)$,
as specified in case (ii-b), then $\widehat{c}_{T}\left(u,\,k\right)$
remains consistent. The only case in which a non-negligible bias
arises is when the change-point falls in a neighborhood around $\left\lfloor Tu\right\rfloor $
sufficiently far from either boundary. This represents case (ii-a),
for which a biased estimate results. However, the bias vanishes asymptotically.
Since $\widehat{\Gamma}_{\mathrm{DK}}\left(k\right)$ is an average
of $\widehat{c}_{T}\left(rn_{T},\,k\right)$ over blocks $r=1,\ldots,\,\left\lfloor T/n_{T}\right\rfloor $,
if case (ii-a) holds then $\widehat{\Gamma}_{\mathrm{DK}}\left(k\right)\geq d_{T}^{*}$
as $k\rightarrow\infty$ but $d_{T}^{*}\rightarrow0$ as $T\rightarrow\infty$.
Thus, comparing this result with the discussion above on $\widehat{\Gamma}\left(k\right)$
(see also Theorem \ref{Theorem ACF Nonstat}), in practice the long
memory effects are unlikely to occur when using $\widehat{\Gamma}_{\mathrm{DK}}\left(k\right)$.
Furthermore, one can reduce this problem by appropriately choosing
the blocks $r=1,\ldots,\,\left\lfloor T/n_{T}\right\rfloor $. A 
procedure was proposed  in \citet{casini_hac} using the methods
developed in \citet{casini/perron:change-point-spectra}. 

We now study the asymptotic properties of $I_{\mathrm{L},T}\left(u,\,\omega\right)$
as $\omega\rightarrow0$ for $u\in\left[0,\,1\right]$.  We consider
the Fourier frequencies $\omega_{l}=2\pi l/n_{T}\in(-\pi,\,\pi)$
for an integer $l\neq0$ (mod $n_{T}$). We need the following high-level
conditions. Part (i) corresponds to Assumption \ref{Assumption Means for Periodogram},
part (ii) is satisfied if $\left\{ V_{t}\right\} $ is strong mixing
with mixing parameters of size $-2\nu/\left(\nu-1/2\right)$ for
some $\nu>1$ such that $\sup_{t\geq1}\mathbb{E}\left|V_{t}\right|^{4\nu}<\infty,$
while part (iii) requires additional smoothness. 
\begin{assumption}
\label{Assumption Means for Local Periodogram}(i) For each $\omega_{l}$
and $u\in\left[0,\,1\right]$ with $T_{j}^{0}\in\mathbf{S}\left(u,\,0,\,n_{T}\right)$
there exist $B_{j}\in\mathbb{R}$, $j=1,\ldots,\,m_{0}$ with $B_{j_{1}}\neq B_{j_{2}}$
for $j_{1}\neq j_{2}$ such that
\begin{align*}
\left|\sum_{s=0}^{n_{T}-1}\mu\left(\left(\left\lfloor Tu\right\rfloor -n_{T}/2+s+1\right)/T\right)\exp\left(-i\omega_{l}s\right)\right|^{2}\geq\\
\quad\left|B_{j}\sum_{s=0}^{T_{j}^{0}-\left(\left\lfloor Tu\right\rfloor -n_{T}/2+1\right)}\exp\left(-i\omega_{l}s\right)+B_{j+1}\sum_{s=T_{j}^{0}-\left(\left\lfloor Tu\right\rfloor -n_{T}/2\right)}^{n_{T}-1}\exp\left(-i\omega_{l}s\right)\right|^{2}.
\end{align*}
(ii) $\left|\Gamma\left(u,\,k\right)\right|=C_{u,k}k^{-m}$ for all
$u\in\left[0,\,1\right]$ and all $k\geq C_{3}T^{\kappa}$ for some
$C_{3}<\infty$ , $C_{u,k}<\infty$ (which depends on $u$ and $k$),
$0<\kappa<1/2$, and $m>2$. (iii) $\sup_{u\in\left[0,\,1\right],\,u\neq\lambda_{0}^{j},\,j=1,\ldots,\,m_{0}}\left(\partial^{2}/\partial u^{2}\right)f\left(u,\,\omega\right)$
is continuous in $\omega.$  
\end{assumption}
\begin{thm}
\label{Theorem Local Periodogram Long Memory Effects}Assume that
$\left\{ V_{t,T}\right\} $ satisfies Definition \ref{Definition Segmented-Locally-Stationary}
and that $n_{T}\rightarrow\infty$ with $n_{T}/T\rightarrow0$. Under
Assumption \ref{Assumption Smothness of A (for HAC)}-\ref{Assumption A - Dependence},
and \ref{Assumption Means for Local Periodogram}, 

(i) for any $u\in\left(0,\,1\right)$ such that $T_{j}^{0}\notin\mathbf{S}\left(u,\,0,\,n_{T}\right)$
for all $j=1,\ldots,\,m_{0}$, $\mathbb{E}\left(I_{\mathrm{L},T}\left(u,\,\omega_{l}\right)\right)\geq f\left(u,\,\omega_{l}\right)$
as $\omega_{l}\rightarrow0$;

(ii) for any $u\in\left(0,\,1\right)$ such that $T_{j}^{0}\in\mathbf{S}\left(u,\,0,\,n_{T}\right)$
for some $j=1,\ldots,\,m_{0}$ we have two sub-cases: (a) if $n_{j,L}\left(u,\,0,\,n_{T}\right)/n_{T}\rightarrow\gamma$
or $n_{j,R}\left(u,\,0,\,n_{T}\right)/n_{T}\rightarrow\gamma$ with
$\gamma\in\left(0,\,1\right),$ and $n_{T}\omega_{l}^{2}\rightarrow0$
as $T\rightarrow\infty$, then $\mathbb{E}\left(I_{\mathrm{L},T}\left(u,\,\omega\right)\right)\rightarrow\infty$
for many values in the sequence $\left\{ \omega_{l}\right\} $ as
$\omega_{l}\rightarrow0$; (b) if $n_{j,L}\left(u,\,0,\,n_{T}\right)/n_{T}\rightarrow0$
or $n_{j,R}\left(u,\,0,\,n_{T}\right)/n_{T}\rightarrow0$, then $\mathbb{E}\left(I_{\mathrm{L},T}\left(u,\,\omega_{l}\right)\right)\geq f\left(u,\,\omega_{l}\right)$
as $\omega_{l}\rightarrow0$.
\end{thm}
It is useful to compare Theorem \ref{Theorem Local Periodogram Long Memory Effects}
with the discussion above about the periodogram (see also Theorem
\ref{Theorem Periodogram Long Memory Effects}). Unlike the periodogram,
the asymptotic behavior of the local periodogram as $\omega_{l}\rightarrow0$
depends on the vicinity of $u$ to  $\lambda_{j}^{0}$ $\left(j=1,\ldots,\,m_{0}\right)$.
Since $I_{\mathrm{L},T}\left(u,\,\omega_{l}\right)$ uses observations
in the window $\mathbf{S}\left(u,\,0,\,n_{T}\right)$, if no discontinuity
in the mean occurs in this window then $I_{\mathrm{L},T}\left(u,\,\omega_{l}\right)$
is asymptotically unbiased for the spectral density $f\left(u,\,\omega_{l}\right)$.
More complex is its behavior if some $T_{j}^{0}$ falls in $\mathbf{S}\left(u,\,0,\,n_{T}\right)$.
The theorem shows that if $T_{j}^{0}$ is close to the boundary, as
indicated in case (ii-b), then $I_{\mathrm{L},T}\left(u,\,\omega_{l}\right)$
is bounded below by $f\left(u,\,\omega_{l}\right)$, similarly to
case (i). If instead $T_{j}^{0}$ falls sufficiently close to the
mid-point $\left\lfloor Tu\right\rfloor ,$ as indicated in case (ii-a),
then $\mathbb{E}\left(I_{\mathrm{L},T}\left(u,\,\omega\right)\right)\rightarrow\infty$
for many values in the sequence $\left\{ \omega_{l}\right\} $ as
$\omega_{l}\rightarrow0$ provided it satisfies $n_{T}\omega_{l}^{2}\rightarrow0$
as $T\rightarrow\infty$. Hence, unless $T\lambda_{j}^{0}$ is close
to $\left\lfloor Tu\right\rfloor ,$ the local periodogram $I_{\mathrm{L},T}\left(u,\,\omega_{l}\right)$
behaves very differently from the periodogram $I_{T}\left(\omega_{l}\right)$.
Accordingly, nonstationarity is unlikely to generate long memory effects
if one uses the local periodogram. As for $\widehat{c}_{T}\left(u,\,k\right)$,
if one uses preliminary inference procedures {[}cf. \citet{casini:change-point-spectra}{]}
for the detection and estimation of the discontinuities in the spectrum
and for the estimation of their locations, then one can construct
the window efficiently and avoid $T_{j}^{0}$ being too close to $\left\lfloor Tu\right\rfloor .$ 

\section{\label{Section Edgeworth-Expansions-for}Edgeworth Expansions for
 HAR Tests Under Nonstationarity}

We now consider Edgeworth expansions for the distribution of the
$t$-statistic in the location model based on the HAC and DK-HAC estimator
where $\left\{ V_{t}\right\} $ is assumed to have zero-mean and time-varying
second moments. This is useful for analyzing the theoretical properties
of the null rejection probabilities of the HAR tests under nonstationarity.
As in the literature, we make use of the Gaussianity assumption for
mathematical convenience.\footnote{This can be relaxed by considering distributions with Gram-Charlier
representations at the expense of more complex derivations. } We relax the stationarity assumption used in the literature {[}cf.
\citet{jansson:04}, \citet{sun/phillips/jin:08} and \citet{velasco/robinson:01}{]}
which has important consequences for the nature of the results. The
results concerning the $t$-test based on the HAC estimator are presented
in Section \ref{Subsection: HAC-based-HAR-Tests} while those based
on the DK-HAC estimator are presented in Section \ref{Subsection: DK-HAC-based-HAR-Tests}. 

Let $\left\{ V_{t}\right\} $ be a zero-mean Gaussian SLS process
satisfying Assumption \ref{Assumption Smothness of A (for HAC)}-(i-iv).
Let
\begin{align}
h_{1} & \triangleq\frac{\sqrt{T}\,\overline{V}}{\sqrt{J_{T}}}\sim\mathscr{N}\left(0,\,1\right),\label{Eq. (h1)}
\end{align}
which is valid for all $T$ such that $J_{T}>0$ where $J_{T}=T^{-1}\sum_{s=1}^{T}\sum_{t=1}^{T}\mathbb{E}(V_{s}V_{t})$.

\subsection{\label{Subsection: HAC-based-HAR-Tests}HAC-based HAR Tests}

The classical HAC estimator is defined as
\begin{align*}
\widehat{J}_{\mathrm{HAC,}T}\triangleq\sum_{k=-T+1}^{T-1}K_{1}\left(b_{1,T}k\right)\widehat{\Gamma}\left(k\right), & \qquad\widehat{\Gamma}\left(k\right)=T^{-1}\sum_{t=|k|+1}^{T}V_{t}V_{t-|k|},
\end{align*}
 where $K_{1}\left(\cdot\right)$ is a kernel and $b_{1,T}$ a bandwidth
parameter. Under appropriate conditions on $b_{1,T},$ we have $\widehat{J}_{\mathrm{\mathrm{HAC},}T}-J_{T}\overset{\mathbb{P}}{\rightarrow}0$
from which it follows that
\begin{align*}
Z_{T} & \triangleq\frac{\sqrt{T}\,\overline{V}}{\sqrt{\widehat{J}_{\mathrm{HAC,}T}}}\overset{d}{\rightarrow}\mathscr{N}\left(0,\,1\right).
\end{align*}
Let $\mathbf{V}=(V_{1},\ldots,\,V_{T})'$. Note that $\widehat{J}_{\mathrm{HAC,}T}=\mathbf{V}'W_{b_{1}}\mathbf{V}/T$
where $W_{b_{1}}$  has $\left(r,\,s\right)$th element 
\begin{align}
W_{b_{1}}^{(r,s)} & =w\left(b_{1,T}(r-s)\right)=\int_{\Pi}\widetilde{K}_{b_{1}}\left(\omega\right)e^{i\left(r-s\right)\omega}d\omega,\label{Eq. Definition W_b1}
\end{align}
such that $\widetilde{K}_{b_{1}}\left(\omega\right)$ is a kernel
with smoothing number $b_{1,T}^{-1}$ and $\Pi=(-\pi,\,\pi]$. For
an even function $K$ that integrates to one, we define 
\begin{align*}
\widetilde{K}_{b_{1}}\left(\omega\right) & =b_{1,T}^{-1}\sum_{j=-\infty}^{\infty}K\left(b_{1,T}^{-1}(\omega+2\pi j)\right).
\end{align*}
Note that $\widetilde{K}_{b_{1}}\left(\omega\right)$ is periodic
of period $2\pi$, even and satisfies $\smallint_{-\pi}^{\pi}\widetilde{K}_{b_{1}}\left(\omega\right)d\omega=1$.
It follows that $w\left(r\right)=\int_{-\infty}^{\infty}e^{irx}K\left(x\right)dx$
and $\widehat{J}_{\mathrm{HAC,}T}=2\pi\int_{\Pi}\widetilde{K}_{b_{1}}\left(\omega\right)I_{T}\left(\omega\right)d\omega$.
$\widetilde{K}_{b_{1}}\left(\omega\right)$ is the so-called spectral
window generator. We refer to \citet{brillinger:75} for a review
of these introductory concepts. 

We now analyze the joint distribution of $\overline{V}$ and $\widehat{J}_{\mathrm{HAC,}T}$.
Let $\mathsf{B}_{T}=\mathbb{E}(\widehat{J}_{\mathrm{HAC,}T})/J_{T}-1$
and $\mathsf{V}_{T}^{2}=\mathrm{Var}(\sqrt{Tb_{1,T}}\widehat{J}_{\mathrm{HAC,}T}/J_{T})$
denote the relative bias and variance, respectively, of $\widehat{J}_{\mathrm{HAC,}T}$.
It is convenient to work with standardized statistics with zero mean
and unit variance. Write 
\begin{align*}
Z_{T} & =Z_{T}\left(\mathbf{h}\right)=h_{1}\left(1+\mathrm{\mathsf{B}}_{T}+\mathsf{V}_{T}h_{2}\left(Tb_{1,T}\right)^{-1/2}\right)^{-1/2},\qquad h_{2}=\sqrt{Tb_{1,T}}\left(\frac{\widehat{J}_{\mathrm{HAC,}T}-\mathbb{E}\left(\widehat{J}_{\mathrm{HAC,}T}\right)}{J_{T}\mathsf{V}_{T}}\right),
\end{align*}
where $\mathbf{h}=\left(h_{1},\,h_{2}\right)'$. Note that $h_{2}=\mathbf{V}'Q_{T}\mathbf{V}-$$\mathbb{E}\left(\mathbf{V}'Q_{T}\mathbf{V}\right)$
is a centered quadratic form in a Gaussian vector where $Q_{T}=W_{b_{1}}(\sqrt{T/b_{1,T}}\mathsf{V}_{T}J_{T})^{-1}$.
The joint characteristic function of $\mathbf{h}$ is
\begin{align*}
\psi_{T}\left(\mathbf{t}\right)=\psi_{T}\left(t_{1},\,t_{2}\right) & =\left|I-2it_{2}\Sigma_{V}Q_{T}\right|^{-1/2}\exp\left(-2^{-1}t_{1}^{2}\xi'_{T}\left(I-2it_{2}\Sigma_{V}Q_{T}\right)^{-1}\Sigma_{V}\xi_{T}-it_{2}\Upsilon_{T}\right),
\end{align*}
where $\Upsilon_{T}=\mathbb{E}\left(\mathbf{V}'Q_{T}\mathbf{V}\right)=\mathrm{Tr}\left(\Sigma_{V}Q_{T}\right),$
$\Sigma_{V}=\mathbb{E}\left(\mathbf{VV}'\right)$, and $\xi_{T}=\mathbf{1}/\sqrt{TJ_{T}}$
with $\mathbf{1}$ being the $T\times1$ vector $\left(1,1,\ldots,\,1\right)'$.
 The cumulant generating function of $\mathbf{h}$ is
\begin{align*}
\mathrm{K}_{T}\left(t_{1},\,t_{2}\right) & =\log\psi_{T}\left(t_{1},\,t_{2}\right)=\sum_{r=0}^{\infty}\sum_{s=0}^{\infty}\kappa_{T}\left(r,\,s\right)\frac{\left(it_{1}\right)^{r}}{r!}\frac{\left(it_{2}\right)^{r}}{s!},
\end{align*}
where $\kappa_{T}\left(r,\,s\right)$ is the cumulant of $\mathbf{h}.$
\citet{phillips:1980} considered the distribution of linear and quadratic
forms under Gaussianity. From his derivations, the  nonzero bivariate
cumulants are 
\begin{align*}
\kappa_{T}\left(0,\,s\right) & =2^{s-1}\left(s-1\right)!\mathrm{Tr}\left(\left(\Sigma_{V}Q_{T}\right)^{s}\right),\qquad s>1,\\
\kappa_{T}\left(2,\,s\right) & =2^{s}s!\xi_{T}'\left(\Sigma_{V}Q_{T}\right)^{s}\Sigma_{V}\xi_{T},\qquad\qquad s>0.
\end{align*}
We introduce the following assumptions about $\left\{ V_{t}\right\} $
and $f\left(u,\,0\right)$.
\begin{assumption}
\label{Assumption 1 in VR}For all $u\in\left[0,\,1\right]$, $0<f\left(u,\,0\right)<\infty$
and $f\left(u,\,\omega\right)$ has $d_{f}$ continuous derivatives
$\left(d_{f}\geq2\right)$ $f^{\left(d_{f}\right)}\left(u,\,\omega\right)$
in a neighborhood of $\omega=0$ and the $d_{f}$th derivative satisfies
a Lipschitz condition of order $\varrho$ with $\varrho\in(0,\,1]$. 
\end{assumption}
\begin{assumption}
\label{Assumption: Assumption 2 in VR }For all $u$, $f\left(u,\,\omega\right)\in L_{p}$
for some $p>1$, i.e., $\left\Vert f\left(u,\cdot\right)\right\Vert _{p}^{p}=\int_{\Pi}f^{p}\left(u,\,\omega\right)d\omega<\infty.$ 
\end{assumption}
\begin{assumption}
\label{Assumption 3 VR}$|K\left(x\right)|<\infty$, $K\left(x\right)=K\left(-x\right)$,
$K\left(x\right)=0$ for $x\notin\Pi$ and $\int_{\Pi}K\left(x\right)dx=1$. 
\end{assumption}
\begin{assumption}
\label{Assumption 4 in VR}$K\left(x\right)$ satisfies a uniform
Lipschitz condition of order 1 in $\left[-\pi,\,\pi\right]$. 
\end{assumption}
\begin{assumption}
\label{Assumption 5 VR}For $j=0,\,1,\ldots,\,d_{f}$, $d_{f}\geq2$
and $r=1,\,2,\ldots$ 
\begin{align*}
\mu_{j}\left(K^{r}\right) & \triangleq\int_{\Pi}x^{j}\left(K\left(x\right)\right)^{r}dx=\begin{cases}
=0, & j<d_{f},\,r=1;\\
\neq0, & j=d_{f},\,r=1.
\end{cases}
\end{align*}
\end{assumption}
\begin{assumption}
\label{Assumption 6 VR} $b_{1,T}+(Tb_{1,T})^{-1}\rightarrow0$ as
$T\rightarrow\infty$.
\end{assumption}
\begin{assumption}
\label{Assumption 7 VR}$b_{1,T}=CT^{-q}$ where $0<q<1$ and $0<C<\infty$. 
\end{assumption}
Assumptions \ref{Assumption 3 VR}-\ref{Assumption 7 VR} about the
kernel and bandwidth are the same as in \citet{velasco/robinson:01}
in which a discussion can be found. They are satisfied by most kernels
used in practice. The bandwidth condition in Assumption \ref{Assumption 6 VR}
is sufficient for the consistency of $\widehat{J}_{\mathrm{HAC,}T}$
and is strengthened in Assumption \ref{Assumption 7 VR}, for some
parts of the proofs, which is satisfied by popular MSE-optimal bandwidths
{[}cf. \citet{andrews:91}, \citet{casini_comment_andrews91}, \textcolor{MyBlue}{Belotti et al.}
\citeyearpar{belotti/casini/catania/grassi/perron_HAC_Sim_Bandws}
and \citet{whilelm:2015}{]}. 

Assumptions \ref{Assumption 1 in VR}-\ref{Assumption: Assumption 2 in VR }
impose conditions on the smoothness and boundedness of the spectral
density.  Assumption \ref{Assumption 1 in VR} is implied by $\sum_{k=-\infty}^{\infty}\left|k\right|^{d_{f}+\varrho}\sup_{t}|\mathbb{E}V_{t}V_{t-k}|<\infty$
but it is stronger than necessary because it extends the smoothness
restriction to all frequencies. Assumption \ref{Assumption: Assumption 2 in VR }
does impose some restrictions on $f\left(u,\,\cdot\right)$ beyond
the origin, though it is not particularly restrictive since any $p>1$
arbitrarily close to 1 will suffice.

We now analyze the asymptotic distribution of $\widehat{J}_{\mathrm{HAC,}T}$.
Under stationarity this was discussed by \citet{bentkus/rudzkis:1982}
and \citet{velasco/robinson:01}. From Lemmas \ref{Lemma 4.1 in VR}-\ref{Lemma 2 in VR }
in the supplement we obtain
\begin{align}
\mathsf{B}_{T}=\overline{c}_{1}b_{1,T}^{d_{f}}+O\left(b_{1,T}^{d_{f}+\varrho}+T^{-1}\log T\right), & \qquad\mathrm{where}\qquad\overline{c}_{1}=\frac{\mu_{d_{f}}\left(K\right)\int_{0}^{1}f^{\left(d_{f}\right)}\left(u,\,0\right)du}{d_{f}!\int_{0}^{1}f\left(u,\,0\right)du}.\label{Eq. (c1)}
\end{align}
The order of the asymptotic bias $b_{1,T}^{d_{f}}$ depends on the
smoothness of the spectral density at $\omega=0$ {[}cf. Assumption
\ref{Assumption 1 in VR}{]}. The constant $\overline{c}_{1}$ depends
on the moment of order $d_{f}$ of the kernel $K$ and on the smoothness
of $f\left(u,\,\omega\right)$ at $\omega=0$. For example, for the
time-varying AR(1) in \eqref{Eq. Example TV AR(1)},
\begin{align}
f^{\left(2\right)}\left(u,\,0\right) & =-\frac{\sigma^{2}\left(u\right)\rho\left(u\right)}{\pi\left(1+\rho\left(u\right)^{2}-2\rho\left(u\right)\right)^{2}}.\label{Eq.(f(2) (u,0) TV AR(1)}
\end{align}
If there is positive dependence at time $u$, then $\rho\left(u\right)>0$
and $f^{\left(2\right)}\left(u,\,0\right)<0$. Suppose $K\left(x\right)\geq0$
for all $x$ so that $\mu_{2}\left(K\right)>0$. Then the sign of
the bias is determined by the sign of $\int_{0}^{1}f^{\left(2\right)}\left(u,\,0\right)du$.
A positive local AR(1) coefficient contributes negative bias which
corresponds to the well-known downward bias of the LRV estimator when
there is positive dependence. Conversely, with anti-persistence $\rho\left(u\right)<0$
and $f^{\left(2\right)}\left(u,\,0\right)>0$. Since $\rho\left(\cdot\right)$
is time-varying, whether the bias is positive or negative depends
on the path of $\rho\left(\cdot\right)$. The smoother the spectral
density is at frequency zero, the smoother the kernel and the slower
$b_{1,T}$ can be. The factor $\int_{0}^{1}f\left(u,\,0\right)du$
in the denominator follows by definition because $\mathsf{B}_{T}$
is the relative bias. 

We present a second-order Edgeworth expansion to approximate the
distribution of $\mathbf{h}$, with error $o((Tb_{1,T})^{-1/2})$
and including terms up to order $(Tb_{1,T})^{-1/2}$ to correct the
asymptotic normal distribution. This will imply the validity of that
expansion for the distribution of $\widehat{J}_{\mathrm{HAC},T}$.
For $\mathbf{B}\in\mathscr{B}^{2}$, where $\mathscr{B}^{2}$ is any
class of Borel sets in $\mathbb{R}^{2}$, let $\mathbb{Q}_{T}^{\left(2\right)}\left(\mathbf{B}\right)=\int_{\mathbf{B}}\varphi_{2}\left(\mathbf{h}\right)q_{T}^{\left(2\right)}\left(\mathbf{h}\right)d\mathbf{h},$
where $\varphi_{2}\left(\mathbf{h}\right)=\left(2\pi\right)^{-1}\exp\{-\left(1/2\right)\left\Vert \mathbf{h}\right\Vert ^{2}\}$
is the density of the bivariate standard normal distribution, 
\begin{align*}
q_{T}^{\left(2\right)}\left(\mathbf{h}\right) & =1+(1/3!)\left(Tb_{1,T}\right)^{-1/2}\left(\Xi_{0}(0,\,3)\mathcal{H}_{3}\left(h_{2}\right)+\Xi_{0}(2,\,1)\mathcal{H}_{2}\left(h_{1}\right)\mathcal{H}_{1}\left(h_{2}\right)\right),
\end{align*}
where $\mathcal{H}_{j}\left(\cdot\right)$ are the univariate Hermite
polynomials of order $j$, and $\Xi_{0}\left(0,\,3\right)=\left(4\pi\right)^{1/2}2!\int_{\Pi}K^{3}\left(\omega\right)$
$d\omega\left\Vert K\right\Vert _{2}^{-3}$ and $\Xi_{0}(2,\,1)=\left(4\pi\right)^{1/2}K\left(0\right)\left\Vert K\right\Vert _{2}^{-1}$
(see Lemmas \ref{Lemma 3 in VR}-\ref{Lemma: Lemma 4 in VR}). Let
$\left(\partial\mathbf{B}\right)^{\phi}$denote a neighborhood of
radius $\phi$ of the boundary of a set $\mathrm{\mathbf{B}}.$ Let
$\mathbb{P}_{T}$ denote the probability measure of $\mathbf{h}.$
\begin{thm}
\label{Theorem: Theorem 1 in VR} Let Assumptions \ref{Assumption 1 in VR},
\ref{Assumption: Assumption 2 in VR } $\left(p>1\right)$, \ref{Assumption 3 VR}-\ref{Assumption 4 in VR}
and \ref{Assumption 7 VR} $\left(0<q<1\right)$ hold. For $\phi_{T}=(Tb_{1,T})^{-\varpi}$
with $1/2<\varpi<1$, we have
\begin{align}
\sup_{\mathbf{B}\in\mathscr{B}^{2}}\left|\mathbb{P}_{T}\left(\mathbf{B}\right)-\mathbb{Q}_{T}^{\left(2\right)}\left(\mathbf{B}\right)\right| & =o\left(\left(Tb_{1,T}\right)^{-1/2}\right)+(4/3)\sup_{\mathbf{B}\in\mathscr{B}^{2}}\mathbb{Q}_{T}^{\left(2\right)}\left(\left(\partial\mathbf{B}\right)^{2\phi_{T}}\right).\label{Eq. in Th. 1 in VR}
\end{align}
\end{thm}
Theorem \ref{Theorem: Theorem 1 in VR} shows that $\mathbb{Q}_{T}^{\left(2\right)}$
is a valid second-order Edgeworth expansion for the measure $\mathbb{P}_{T}$.
 The method of proof is the same as in \citet{velasco/robinson:01}.
We first approximate the true characteristic function and then apply
a smoothing lemma {[}cf. Lemma \ref{Lemma Bhattacharya and Rao 1975}
in the supplement which is from \citet{bhattacharya/rao:1975}{]}.
The leading term of the approximation error is of order $o((Tb_{1,T})^{-1/2})$
as the second term on the right hand side of \eqref{Eq. in Th. 1 in VR}
is negligible if $\mathbf{B}$ is convex because $\phi_{T}$ decreases
as a power of $T$. This is the same order obtained for the corresponding
leading term under stationarity. Since the higher-order correction
terms in $q_{T}^{\left(2\right)}$ depend only on $K\left(\cdot\right)$
but not on $f\left(\cdot,\,\cdot\right)$, they are equal to the one
obtained under stationarity. 

Next, we focus on  $Z_{T}$, i.e., a $t$-statistic for the mean.
 Proceeding as in \citet{velasco/robinson:01}, we first derive a
linear stochastic approximation to $Z_{T}\left(\mathbf{h}\right)$
and show that its distribution is the same as that of $Z_{T}$ up
to order $o((Tb_{1,T}){}^{-1/2})$. Then, we show that the asymptotic
approximation for the distribution of the linear stochastic approximation
is valid also for $Z_{T}$ with the same error $o((Tb_{1,T}){}^{-1/2})$.
Using Lemmas \ref{Lemma 3 in VR}-\ref{Lemma: Lemma 4 in VR} in the
supplement we can substitute out $\mathsf{B}_{T}$ and $\mathsf{V}_{T}$
in  $Z_{T}$ and, by only focusing on the leading terms, we define
the following linear stochastic approximation,
\begin{align*}
\widetilde{Z}_{T} & \triangleq h_{1}\left(1-2^{-1}\overline{c}_{1}b_{1,T}^{d_{f}}-2^{-1}\sqrt{4\pi}\left\Vert K_{2}\right\Vert h_{2}\left(Tb_{1,T}\right)^{-1/2}\right).
\end{align*}
The next theorem presents a valid Edgeworth expansion for the distribution
of $\widetilde{Z}_{T}$ from that of $\mathbf{h}.$ 
\begin{thm}
\label{Theorem: Theorem 2 in VR}Let Assumptions \ref{Assumption 1 in VR},
\ref{Assumption: Assumption 2 in VR } $\left(p>1\right)$, \ref{Assumption 3 VR}-\ref{Assumption 5 VR}
and \ref{Assumption 7 VR} $(q=1/\left(1+2d_{f}\right))$ hold. For
a convex Borel set $\mathbf{C}$, we have, for $r_{2}\left(x\right)=-\overline{c}_{1}\left(x^{2}-1\right)/2$,
\begin{align}
\sup_{\mathbf{C}}\left|\mathbb{P}\left(Z_{T}\in\mathbf{C}\right)-\int_{\mathbf{C}}\varphi\left(x\right)\left(1+r_{2}\left(x\right)b_{1,T}^{d_{f}}\right)dx\right| & =o\left(\left(Tb_{1,T}\right)^{-1/2}\right).\label{Eq. (3) in VR}
\end{align}
\end{thm}
Theorem \ref{Theorem: Theorem 2 in VR} shows the form of the correction
term to the standard normal distribution, i.e., $b_{1,T}^{d_{f}}\int_{\mathbf{C}}\varphi\left(x\right)r_{2}\left(x\right)dx$.
The error of the approximation is of order $o((Tb_{1,T})^{-1/2})$
which is the same as the one obtained under stationarity by \citet{velasco/robinson:01}.

Let $\Phi\left(\cdot\right)$ denote the distribution function of
the standard normal. Setting $\mathbf{C}=(-\infty,\,z]$, integrating
and Taylor expanding $\Phi\left(\cdot\right)$, we obtain, uniformly
in $z$,
\begin{align}
\mathbb{P}\left(Z_{T}\leq z\right) & =\Phi\left(z\right)+\frac{1}{2}\overline{c}_{1}z\varphi\left(z\right)b_{1,T}^{d_{f}}+o\left(\left(Tb_{1,T}\right)^{-1/2}\right)\label{Eq. (4) in VR}\\
 & =\Phi\left(z\left(1+\frac{1}{2}\overline{c}_{1}b_{1,T}^{d_{f}}\right)\right)+o\left(\left(Tb_{1,T}\right)^{-1/2}\right)=\Phi\left(z\right)+O\left(\left(Tb_{1,T}\right)^{-1/2}\right).\nonumber 
\end{align}
This shows that under the conditions of Theorem \ref{Theorem: Theorem 2 in VR},
 the standard normal approximation is correct up to order $O((Tb_{1,T})^{-1/2})$.
Eq. \eqref{Eq. (4) in VR} has an immediate interpretation. Consider
the time-varying AR(1) example in \eqref{Eq. Example TV AR(1)} and
suppose $K\left(x\right)\geq0$ for all $x$ so that $\mu_{2}\left(K\right)\geq0$.
Given \eqref{Eq.(f(2) (u,0) TV AR(1)} we know that with local positive
persistence (i.e., $\rho\left(u\right)>0$) $f\left(u,\,\omega\right)$
has a peak at $\omega=0$. If the pattern of $\rho\left(u\right)$
is such that $\int_{0}^{1}f^{\left(2\right)}\left(u,\,0\right)du<0$
so that the positive persistence dominates, then $\overline{c}_{1}<0$
and as is well-known the HAC estimator underestimates the true LRV
and the corresponding HAC-based test over-rejects. The approximation
in \eqref{Eq. (4) in VR} tends to correct this problem as it follows
that one uses $\Phi\left(z\left(1+\gamma_{T}\right)\right)$ where
$\gamma_{T}\leq0$, so for a given significance level the critical
value $z$ is larger in absolute value than the corresponding standard
normal critical value. Conversely, if there is anti-persistence, then
$\overline{c}_{1}>0$ and the implied critical value is smaller than
the corresponding standard normal critical value. For $d_{f}>2$ the
reasoning is the same but one has to take into account the sign of
$\mu_{d_{f}}\left(K\right)$.

Consider the location model $y_{t}=\beta+V_{t}$ $\left(t=1,\ldots,\,T\right).$
For the null hypothesis $\mathbb{H}_{0}:\,\beta=\beta_{0}$, consider
the following $t$-test, 
\begin{align*}
t_{\mathrm{HAC}} & =\frac{\sqrt{T}\left(\widehat{\beta}-\beta_{0}\right)}{\sqrt{\widehat{J}_{\mathrm{HAC},T}}},
\end{align*}
where $\widehat{\beta}$ is the least-squares estimator of $\beta$.
Theorem \ref{Theorem: Theorem 2 in VR} and \eqref{Eq. (4) in VR}
imply that 
\begin{align}
\mathbb{P}\left(t_{\mathrm{HAC}}\leq z\right) & =\Phi\left(z\right)+p\left(z\right)\left(Tb_{1,T}\right)^{-1/2}+o\left(\left(Tb_{1,T}\right)^{-1/2}\right),\label{Eq. ERP t-hac}
\end{align}
for any $z\in\mathbb{R},$ where $p\left(z\right)$ is an odd function.
When $q=1/\left(1+2d_{f}\right)$ we have $p\left(z\right)=2^{-1}\overline{c}_{1}z\varphi\left(z\right)C^{d_{f}+1/2}$
where $C$ is defined in Assumption \ref{Assumption 7 VR}. Thus,
the error in rejection probability (ERP) of $t_{\mathrm{HAC}}$ is
of order $O((Tb_{1,T})^{-1/2})$. If $\left\{ V_{t}\right\} $ is
second-order stationary, the results in \citet{velasco/robinson:01}
imply that the ERP of $t_{\mathrm{HAC}}$ is also of order $O((Tb_{1,T})^{-1/2})$.
Below we establish the corresponding ERP when the $t$-statistic is
instead normalized by $\widehat{J}_{\mathrm{DK},T}$ and also discuss
the ERP of the $t$-test under fixed-$b$ asymptotics. 

\subsection{\label{Subsection: DK-HAC-based-HAR-Tests}DK-HAC-based HAR Tests }

We now consider the Edgeworth expansion for tests based on the DK-HAC
estimator. In order to simplify some parts of the proof here we
consider an asymptotically equivalent version of the DK-HAC estimator
discussed in Section \ref{Section Consequences for HAR}. Let 
\begin{align*}
\widehat{J}_{\mathrm{DK},T}^{*}=\sum_{k=-T+1}^{T-1}K_{1}\left(b_{1,T}k\right)\widehat{\Gamma}_{\mathrm{DK}}^{*}\left(k\right), & \qquad\widehat{\Gamma}_{\mathrm{DK}}^{*}\left(k\right)\triangleq\int_{0}^{1}\widehat{c}_{\mathrm{DK,}T}\left(r,\,k\right)dr,
\end{align*}
 where $b_{1,T}$ is a bandwidth sequence and  
\begin{align*}
\widehat{c}_{\mathrm{DK,}T}\left(r,\,k\right) & =\left(Tb_{2,T}\right)^{-1}\sum_{s=|k|+1}^{T}K_{2}\left(\frac{\left(Tr-\left(s-|k|/2\right)\right)/T}{b_{2,T}}\right)V_{s}V{}_{s-|k|},
\end{align*}
with $K_{2}$ a kernel and $b_{2,T}$ a bandwidth. Note that $\widehat{\Gamma}_{\mathrm{DK}}\left(k\right)$
and $\widehat{\Gamma}_{\mathrm{DK}}^{*}\left(k\right)$ are asymptotically
equivalent and $\widehat{c}_{T}$ is a special case of $\widehat{c}_{\mathrm{DK,}T}$
with $K_{2}$ being a rectangular kernel and $n_{2,T}=Tb_{2,T}$. 
\begin{assumption}
\label{Assumption K2 and b2}$K_{2}\left(\cdot\right):\,\mathbb{R}\rightarrow\left[0,\,\infty\right]$,
$K_{2}\left(x\right)=K_{2}\left(1-x\right)$, ${\textstyle \int_{0}^{1}}K_{2}\left(x\right)dx=1$,
$K_{2}\left(x\right)=0$ for $x\notin\left[0,\,1\right]$ and $K_{2}\left(\cdot\right)$
is continuous. The bandwidth sequence $\{b_{2,T}\}$ satisfies $b_{2,T}\rightarrow0$,
$b_{2,T}^{2}/b_{1,T}^{q_{2}}\rightarrow\overline{b}\in[0,\,\infty)$
and $1/Tb_{1,T}b_{2,T}\rightarrow0$ where $q_{2}$ is the index of
smoothness of $K_{1}\left(\cdot\right)$ at 0. 
\end{assumption}
Under Assumptions \ref{Assumption 3 VR}-\ref{Assumption 4 in VR},
\ref{Assumption 6 VR} and \ref{Assumption K2 and b2} it holds that
$\widehat{J}_{\mathrm{DK},T}^{*}-J_{T}\overset{\mathbb{P}}{\rightarrow}0$
{[}cf. \citet{casini_hac}{]} and 
\begin{align}
U_{T} & \triangleq\frac{\sqrt{T}\,\overline{V}}{\sqrt{\widehat{J}_{\mathrm{DK},T}^{*}}}\overset{d}{\rightarrow}\mathscr{N}\left(0,\,1\right).\label{Eq. U_T}
\end{align}
Note that $\widehat{J}_{\mathrm{DK},T}^{*}=\int_{0}^{1}\mathbf{\widetilde{V}}\left(r\right)'W_{b_{1}}\mathbf{\widetilde{V}}\left(r\right)dr/(Tb_{2,T})$
where $\mathbf{\widetilde{V}}\left(r\right)=(\widetilde{V}_{1}\left(r\right),\,\widetilde{V}_{2}\left(r\right),\ldots,\,\widetilde{V}_{T}\left(r\right))'$
with $\widetilde{V}_{j}\left(r\right)=\sqrt{K_{2}\left(\left(r-j\right)/Tb_{2,T}\right)}V_{j}$
and $W_{b_{1}}$ defined in \eqref{Eq. Definition W_b1}. Let
\begin{align*}
\widetilde{I}_{T}\left(r,\,\omega\right) & =\frac{1}{2\pi Tb_{2,T}}\left|\sum_{t=1}^{T}\exp\left(-i\omega t\right)\widetilde{V}_{t}\left(r\right)\right|^{2}.
\end{align*}
$\widetilde{I}_{T}\left(r,\,\omega\right)$ is the local periodogram
of $\{\mathbf{\widetilde{V}}\left(r\right)\}$. Then, $\widehat{J}_{\mathrm{DK},T}^{*}=2\pi\int_{0}^{1}\int_{\Pi}\widetilde{K}_{b_{1}}\left(\omega\right)\widetilde{I}_{T}\left(r,\,\omega\right)d\omega dr$. 

We begin by analyzing the joint distribution of $\overline{V}$ and
$\widehat{J}_{\mathrm{DK},T}^{*}$. Let $\mathsf{B}_{\mathrm{2},T}=\mathbb{E}(\widehat{J}_{\mathrm{DK},T}^{*})/J_{T}-1$
and $\mathsf{V}_{2,T}^{2}=\mathrm{Var}(\sqrt{Tb_{1,T}b_{2,T}}\widehat{J}_{\mathrm{DK},T}^{*}/J_{T})$
denote the relative bias and variance of $\widehat{J}_{\mathrm{DK},T}^{*}$,
respectively. It is convenient to work with standardized statistics
with zero mean and unit variance. Write 
\begin{align*}
U_{T} & =U_{T}\left(\mathbf{v}\right)=v_{1}\left(1+\mathrm{\mathsf{B}}_{2,T}+\mathsf{V}_{2,T}v_{2}\left(Tb_{1,T}b_{2,T}\right)^{-1/2}\right)^{-1/2},\quad v_{2}=\sqrt{Tb_{1,T}b_{2,T}}\left(\frac{\widehat{J}_{\mathrm{DK},T}^{*}-\mathbb{E}\left(\widehat{J}_{\mathrm{DK},T}^{*}\right)}{J_{T}\mathsf{V}_{2,T}}\right),
\end{align*}
where $\mathbf{v}=\left(v_{1},\,v_{2}\right)'$ with $v_{1}=h_{1}.$
Note that $v_{2}=\int_{0}^{1}(\mathbf{\widetilde{V}}\left(r\right)'Q_{2,T}\mathbf{\widetilde{V}}\left(r\right)-$$\mathbb{E}(\mathbf{\widetilde{V}}\left(r\right)'Q_{2,T}\mathbf{\widetilde{V}}\left(r\right)))dr$
is a centered quadratic form in a Gaussian vector where $Q_{2,T}=W_{b_{1}}(\sqrt{Tb_{2,T}/b_{1,T}}\mathsf{V}_{2,T}J_{T})^{-1}$.
The joint characteristic function of $\mathbf{v}$ is
\begin{align*}
\psi_{2,T}\left(t_{1},\,t_{2}\right) & =\left|I-2it_{2}\Sigma_{\widetilde{V}}Q_{2,T}\right|^{-1/2}\exp\left\{ -2^{-1}t_{1}^{2}\xi'_{2,T}\left(I-2it_{2}\Sigma_{\widetilde{V}}Q_{2,T}\right)^{-1}\Sigma_{\widetilde{V}}\xi_{2,T}-it_{2}\Upsilon_{2,T}\right\} ,
\end{align*}
where $\Upsilon_{2,T}=\mathbb{E}(\int_{0}^{1}(\mathbf{\widetilde{V}}\left(r\right)'Q_{2,T}\mathbf{\widetilde{V}}\left(r\right))dr)=\mathrm{Tr}(\Sigma_{\widetilde{V}}Q_{2,T}),$
$\Sigma_{\widetilde{V}}=\mathbb{E}(\int_{0}^{1}(\mathbf{\widetilde{V}}\left(r\right)\mathbf{\widetilde{V}}\left(r\right)')dr)$
and $\xi_{2,T}=\mathbf{1}/\sqrt{Tb_{2,T}J_{T}}$.  The cumulant generating
function of $\mathbf{v}$ is
\begin{align*}
\mathrm{K}_{2,T}\left(t_{1},\,t_{2}\right) & =\log\psi_{2,T}\left(t_{1},\,t_{2}\right)=\sum_{r=0}^{\infty}\sum_{s=0}^{\infty}\kappa_{2,T}\left(r,\,s\right)\frac{\left(it_{1}\right)^{r}}{r!}\frac{\left(it_{2}\right)^{r}}{s!},
\end{align*}
 where $\kappa_{2,T}\left(r,\,s\right)$ is the cumulant of $\mathbf{v}.$
To obtain more precise bounds in some parts of the proofs we use
the following assumption on the cross-partial derivatives of $f\left(u,\,\omega\right)$.
Let $\widetilde{\mathbf{C}}$ denote the set of continuity points
of $f\left(u,\,\omega\right)$ in $u$, i.e., $\widetilde{\mathbf{C}}=\{\left[0,\,1\right]/\{\lambda_{j}^{0},\,j=1,\ldots,\,m_{0}\}\}$.
Define 
\begin{align*}
\Delta_{f}\left(\omega\right) & =\sum_{j=1}^{m_{0}}\int_{0}^{1}\left(\frac{\partial}{\partial u_{-}}f\left(\lambda_{j}^{0},\,\omega\right)\int_{0}^{1-s}xK_{2}\left(x\right)dx+\frac{\partial}{\partial u_{+}}f\left(\lambda_{j}^{0},\,\omega\right)\int_{1-s}^{1}xK_{2}\left(x\right)dx\right)ds,
\end{align*}
where 
\begin{align*}
\frac{\partial}{\partial u_{-}}f\left(\lambda_{j}^{0},\,\omega\right)=\underset{h\uparrow0}{\lim}\frac{f\left(\lambda_{j}^{0}+h,\,\omega\right)-f\left(\lambda_{j}^{0},\,\omega\right)}{h}, & \qquad\frac{\partial}{\partial u_{+}}f\left(\lambda_{j}^{0},\,\omega\right)=\underset{h\downarrow0}{\lim}\frac{f\left(\lambda_{j}^{0}+h,\,\omega\right)-f\left(\lambda_{j}^{0},\,\omega\right)}{h}.
\end{align*}

\begin{assumption}
\label{Assumption Lip of d2 f(u,w)}For $u\in\widetilde{\mathbf{C}},$
$\left(\partial^{2}/\partial u^{2}\right)f\left(u,\,\omega\right)$
has $d_{f}$ continuous derivatives in $\omega$ in a neighborhood
of $\omega=0,$ the $d_{f}$ derivative satisfying a Lipschitz condition
of order $\varrho_{2}\in(0,\,1]$. \\
For $u\notin\widetilde{\mathbf{C}},$ $\left(\partial/\partial u_{-}\right)f\left(u,\,\omega\right)$
and $\left(\partial/\partial u_{+}\right)f\left(u,\,\omega\right)$
have $d_{f}$ continuous derivatives in $\omega$ in a neighborhood
of $\omega=0,$ the $d_{f}$ derivative satisfying a Lipschitz condition
of order $\varrho_{2}\in(0,\,1]$.
\end{assumption}
From Lemmas \ref{Lemma 4.1 in VR} and \ref{Lemma 2 in VR DK-HAC},
the relative bias of $\widehat{J}_{\mathrm{DK},T}^{*}$ is 
\begin{align*}
\mathsf{B}_{2,T}=\overline{c}_{1}b_{1,T}^{d_{f}}+\overline{c}_{2}b_{2,T}^{2}+O\left(b_{1,T}^{d_{f}+\varrho}+T^{-1}\log T+\left(Tb_{2,T}\right)^{-1}\right)+o\left(b_{2,T}^{2}\right) & ,
\end{align*}
 where 
\begin{align*}
\overline{c}_{1}=\frac{\mu_{d_{f}}\left(K\right)\int_{0}^{1}f^{\left(d_{f}\right)}\left(u,\,0\right)du}{d_{f}!\int_{0}^{1}f\left(u,\,0\right)du}, & \qquad\overline{c}_{2}=\frac{2^{-1}\int_{0}^{1}x^{2}K_{2}\left(x\right)dx\int_{\widetilde{\mathbf{C}}}\frac{\partial^{2}}{\partial u^{2}}f\left(u,\,0\right)du+\Delta_{f}\left(0\right)}{\int_{0}^{1}f\left(u,\,0\right)du}.
\end{align*}

The factor $\overline{c}_{1}$ in the relative bias $\mathsf{B}_{2,T}$
also enters $\mathsf{B}_{T}$ and we already discussed it. The second
factor, $\overline{c}_{2}$, includes two elements. The first depends
on the second moment of the kernel $K_{2}$ and on the smoothness
over time of the spectral density $f\left(u,\,0\right)$. The second
element in $\overline{c}_{2}$ is $\Delta_{f}\left(0\right)$ which
depends on the right and left first partial derivatives of $f\left(u,\,0\right)$
with respect to $u$ at the discontinuity points. The more nonstationary
is the data the more complex is $\overline{c}_{2}$, and in fact the
larger in magnitude are $\partial^{2}f\left(u,\,0\right)/\partial u^{2}$
and $\Delta_{f}\left(0\right)$. For the special case of stationary
data, $\overline{c}_{2}=0$.  The more nonstationary is the data,
the smaller $b_{2,T}$ should be chosen so as to weight more the data
locally. The smoothing over sample autocovariances is needed to achieve
consistency while the time-smoothing is introduced to more flexibly
account for the time-varying properties of the data. The disadvantage
of the time-smoothing is that it reduces the effective sample size
thereby making accounting for strong dependence more difficult. 

We now present a second-order Edgeworth expansion to approximate
the distribution of $\mathbf{v}$ with error $o((Tb_{1,T}b_{2,T})^{-1/2})$.
 The expansion includes terms up to order $(Tb_{1,T}b_{2,T})^{-1/2}$
to correct the asymptotic normal distribution. This implies the validity
of that expansion for the distribution of $\widehat{J}_{\mathrm{DK},T}^{*}$.
For $\mathbf{B}\in\mathscr{B}^{2}$,  let $\mathbb{Q}_{2,T}^{\left(2\right)}(\mathbf{B})=\int_{\mathbf{B}}\varphi_{2}\left(\mathbf{v}\right)q_{2,T}^{\left(2\right)}\left(\mathbf{v}\right)d\mathbf{v},$
where  
\begin{align*}
q_{2,T}^{\left(2\right)}\left(\mathbf{v}\right) & =1+(1/3!)\left(Tb_{1,T}b_{2,T}\right)^{-1/2}\left\{ \Xi_{2,0}(0,\,3)\mathcal{H}_{2,3}\left(v_{2}\right)+\Xi_{2,0}(2,\,1)\mathcal{H}_{2,2}\left(v_{1}\right)\mathcal{H}_{2,1}\left(v_{1}\right)\right\} ,
\end{align*}
 $\mathcal{H}_{2,j}\left(\cdot\right)$ are the univariate Hermite
polynomials of order $j$ and $\Xi_{2,0}(0,\,3)$ and $\Xi_{2,0}(2,\,1)$
are bounded and depend on $K,\,K_{2}$ and on $f\left(u,\,0\right)$
(see Lemmas \ref{Lemma: Proposition 1 in VR DK-HAC}-\ref{Lemma: Proposition 2 in VR DK-HAC}).

\begin{thm}
\label{Theorem: Theorem 1 in VR DK-HAC}Let Assumptions \ref{Assumption 1 in VR},
\ref{Assumption: Assumption 2 in VR } $\left(p>1\right)$, \ref{Assumption 3 VR}-\ref{Assumption 4 in VR},
\ref{Assumption 7 VR} $\left(0<q<1\right)$, \ref{Assumption K2 and b2}-\ref{Assumption Lip of d2 f(u,w)}
hold. For $\phi_{T}=(Tb_{1,T}b_{2,T})^{-\varpi}$ with $1/2<\varpi<1$,
and every class $\mathscr{B}^{2}$ of Borel sets in $\mathbb{R}^{2}$,
we have
\begin{align}
\sup_{\mathbf{B}\in\mathscr{B}^{2}}\left|\mathbb{P}_{T}\left(\mathbf{B}\right)-\mathbb{Q}_{2,T}^{\left(2\right)}\left(\mathbf{B}\right)\right| & =o\left(\left(Tb_{1,T}b_{2,T}\right)^{-1/2}\right)+(4/3)\sup_{\mathbf{B}\in\mathscr{B}^{2}}\mathbb{Q}_{2,T}^{\left(2\right)}\left(\left(\partial\mathbf{B}\right)^{2\phi_{T}}\right).\label{Eq. in Th. 1 in VR DK-HAC}
\end{align}
\end{thm}
Theorem \ref{Theorem: Theorem 1 in VR DK-HAC} shows that $\mathbb{Q}_{2,T}^{\left(2\right)}$
is a valid second-order Edgeworth expansion for the probability measure
$\mathbb{P}_{T}$ of $\mathbf{v}.$  The correction $q_{2,T}^{\left(2\right)}\left(\mathbf{v}\right)$
differs from $q_{T}^{\left(2\right)}\left(\mathbf{h}\right)$  in
Theorem \ref{Theorem: Theorem 1 in VR}. This difference depends on
the smoothing over time, i.e., on $b_{2,T}$ and $K_{2}\left(\cdot\right)$.
The theorem also suggests that the leading term of the error of the
approximation is of order $o((Tb_{1,T}b_{2,T})^{-1/2})$. 

Next, we focus on  $U_{T}$ defined in \eqref{Eq. U_T}, i.e., a
$t$-statistic based on $\widehat{J}_{\mathrm{DK},T}^{*}$,  and
present the Edgeworth expansion. We need the following assumption,
replacing Assumptions \ref{Assumption 6 VR}-\ref{Assumption 7 VR},
that controls the rate of smoothing over lagged autocovariances and
time implied by the bandwidths $b_{1,T}$ and $b_{2,T}$, respectively.
It requires that the bias due to smoothing over frequency and over
time is of the same order as the correction term obtained in $\mathbb{Q}_{2,T}^{\left(2\right)}\left(\mathbf{B}\right)$
or as the standard deviation of $\widehat{J}_{\mathrm{DK},T}^{*}$.
The assumption is satisfied by, for example, the MSE-optimal DK-HAC
estimators proposed by \textcolor{MyBlue}{Belotti et al.} \citeyearpar{belotti/casini/catania/grassi/perron_HAC_Sim_Bandws}
and \citet{casini_hac}. 
\begin{assumption}
\label{Assumption condition Repalce Assumption 7}The bandwidths $b_{1,T}\rightarrow0$
and $b_{2,T}\rightarrow0$ satisfy $0<b_{1,T}^{d_{f}}\left(Tb_{1,T}b_{2,T}\right)^{-1/2}<\infty$
and $0<b_{2,T}^{2}\left(Tb_{1,T}b_{2,T}\right)^{-1/2}<\infty$.
\end{assumption}
\begin{thm}
\label{Theorem: Theorem 2 in VR DK-HAC}Let Assumptions \ref{Assumption 1 in VR},
\ref{Assumption: Assumption 2 in VR } $\left(p>1\right)$, \ref{Assumption 3 VR}-\ref{Assumption 5 VR},
and \ref{Assumption K2 and b2}-\ref{Assumption condition Repalce Assumption 7}
hold. For convex Borel sets $\mathbf{C}$, we have, for $r_{2}\left(x\right)=-\overline{c}_{1}\left(x^{2}-1\right)/2$
and $r_{3}\left(x\right)=-\overline{c}_{2}\left(x^{2}-1\right)/2$,
\begin{align}
\sup_{\mathbf{C}}\left|\mathbb{P}\left(U_{T}\in\mathbf{C}\right)-\int_{\mathbf{C}}\varphi\left(x\right)\left(1+r_{2}\left(x\right)b_{1,T}^{d_{f}}+r_{3}\left(x\right)b_{2,T}^{2}\right)dx\right| & =o\left(\left(Tb_{1,T}b_{2,T}\right)^{-1/2}\right).\label{Eq. (3) in VR-1}
\end{align}
\end{thm}
Theorem \ref{Theorem: Theorem 2 in VR DK-HAC} shows that the correction
term to the standard normal distribution, i.e., $\int_{\mathbf{C}}\varphi\left(x\right)$
$(r_{2}\left(x\right)b_{1,T}^{d_{f}}+r_{3}\left(x\right)b_{2,T}^{2})dx$,
depends on both smoothing directions. The error of the approximation
is of order $o((Tb_{1,T}$ $b_{2,T})^{-1/2})$ which can be larger
than that obtained in Theorem \ref{Theorem: Theorem 2 in VR} for
the HAC estimators. Similar to \eqref{Eq. (4) in VR}, we obtain uniformly
in $z$, 
\begin{align}
\mathbb{P}\left(U_{T}\leq z\right) & =\Phi\left(z\left(1+\frac{1}{2}\overline{c}_{1}b_{1,T}^{d_{f}}+\frac{1}{2}\overline{c}_{2}b_{2,T}^{2}\right)\right)+O\left(\left(Tb_{1,T}b_{2,T}\right)^{-1/2}\right),\label{Eq. (4) in VR DK-HAC}
\end{align}
where $\mathbf{C}=(-\infty,\,z]$, which suggests that the standard
normal approximation is correct up to order $O((Tb_{1,T}b_{2,T})^{-1/2})$.
Eq. \eqref{Eq. (4) in VR DK-HAC} has a similar interpretation to
\eqref{Eq. (4) in VR}. Consider the time-varying AR(1) example in
\eqref{Eq. Example TV AR(1)} and suppose $\rho\left(u\right)>0$
for all $u.$ Then, $\overline{c}_{1}<0$. However, the sign of $\overline{c}_{2}$
is not easily determined even for this simple model. For the special
case $\rho\left(u\right)=\sin(u\pi/10)$, no break and $\sigma^{2}\left(u\right)=\sigma^{2}$
we have $\overline{c}_{2}<0$. Then, the implied critical value from
the approximation is larger than the standard normal critical value.
In general, however, the correction to strong persistence might be
either attenuated or strengthened by the correction to nonstationarity
depending on the true data-generating process. 

Returning to the location model, consider the $t$-statistic based
on $\widehat{J}_{\mathrm{DK},T}^{*}$,
\begin{align*}
t_{\mathrm{DK}} & =\frac{\sqrt{T}\left(\widehat{\beta}-\beta_{0}\right)}{\sqrt{\widehat{J}_{\mathrm{DK},T}^{*}}}.
\end{align*}
Theorem \ref{Theorem: Theorem 2 in VR DK-HAC} and \eqref{Eq. (4) in VR DK-HAC}
imply that 
\begin{align}
\mathbb{P}\left(t_{\mathrm{DK}}\leq z\right) & =\Phi\left(z\right)+p_{2}\left(z\right)\left(Tb_{1,T}b_{2,T}\right)^{-1/2}+o\left(\left(Tb_{1,T}b_{2,T}\right)^{-1/2}\right),\label{Eq. ERP t-DK}
\end{align}
for any $z\in\mathbb{R},$ where $p_{2}\left(z\right)$ is an odd
function. Under the conditions of Theorem \ref{Theorem: Theorem 2 in VR DK-HAC}
$p_{2}\left(z\right)=2^{-1}((C^{d_{f}+1/2}\overline{c}_{1}+C_{2}\overline{c}_{2})z\varphi\left(z\right))$
where $C$ is defined in Assumption \ref{Assumption 7 VR}, $C_{2}=(\overline{b}C^{d_{f}+1/2})^{1/2}$
and $\overline{b}$ is defined in Assumption \ref{Assumption K2 and b2}.
 Thus, the ERP of $t_{\mathrm{DK}}$ can be larger than that of $t_{\mathrm{HAC}}$,
though the margin is small. This follows from the fact that $\widehat{J}_{\mathrm{DK},T}^{*}$
applies smoothing over two directions. The smoothing over time is
useful to flexibly account for nonstationarity. Its benefits  appear
explicitly under the alternative hypothesis as we show in Section
\ref{Section Consequences for HAR} whereas the ERP refers to the
null hypothesis. One can show that the ERP of $t_{\mathrm{DK}}$ and
$t_{\mathrm{HAC}}$ remain unchanged if prewhitening is applied, though
the proofs are omitted since they are similar. 

We can further compare the ERP of $t_{\mathrm{HAC}}$ and $t_{\mathrm{DK}}$
to that of the corresponding $t$-test under the fixed-$b$ asymptotics.
\citet{casini_fixed_b_erp} showed that the limiting distribution
of the original fixed-$b$ HAR test statistics under nonstationarity
is not pivotal as it depends on the true data-generating process of
the errors and regressors. This contrasts to the stationarity case
for which the fixed-$b$ limiting distribution is pivotal and the
ERP is of order $O(T^{-1})$ {[}see \citet{jansson:04} and \citet{sun/phillips/jin:08}{]}.
Based on an ERP of smaller magnitude relative to that of HAR tests
based on HAC estimators {[}cf. $O(T^{-1})<O((Tb_{1,T})^{-1/2})${]},
the literature has long suggested that the original fixed-$b$ HAR
tests are superior to HAR tests based on HAC estimators. However,
this breaks down under nonstationarity as shown by \citet{casini_fixed_b_erp}
who established that (i) the ERP of the original fixed-$b$ HAR tests
does not converge to zero because under nonstationarity the fixed-$b$
limiting distribution is different; (ii) for fixed-$b$ HAR tests
that use the critical values from the non-pivotal fixed-$b$ limiting
distribution the ERP increases by an order of magnitude relative to
the stationary case {[}i.e., from $O(T^{-1})$ to $O(T^{-\eta})$
with $\eta\in(0,\,1/2)${]}. Therefore, fixed-$b$ HAR tests can have
an ERP larger than that of $t_{\mathrm{HAC}}$ and $t_{\mathrm{DK}}$.
Overall, the results based on Edgeworth expansions show that the distortions
on the null rejection rates of the HAR tests can arise from time variation
in the second moments even when the mean is constant. Thus, these
results complement the asymptotic bias results induced by breaks in
the mean function.

\section{\label{Section Consequences for HAR}Consequences for HAR Inference}

In this section, we discuss the implications of the theoretical results
from Section \ref{Section Low Freq Cont - Theory}-\ref{Section Edgeworth-Expansions-for}.
In Section \ref{Subsection HAR inference methods}, we first present
a review of HAR inference methods and their connection to the estimates
considered in Section \ref{Section Low Freq Cont - Theory}. In Section
\ref{Subsec Finite-Sample-Low-Frequency} we present evidence that
the HAR inference tests can suffer from larger size distortions under
nonstationarity than under stationarity. In Section \ref{Subsec General-Low-Frequency}
we show the consequences of low frequency contamination for the power
of the HAR tests and we provide the corresponding theoretical results
in Section \ref{Subsec: Theoretical-Results-on Power}.

\subsection{\label{Subsection HAR inference methods}HAR Inference Methods}

There are two main approaches for HAR inference. Classical HAC standard
errors {[}cf. \citeauthor{newey/west:87} (\citeyear{newey/west:87},
\citeyear{newey/west:94}) and \citet{andrews:91}{]} require estimation
of the LRV defined as $J\triangleq\mathrm{lim}_{T\rightarrow\infty}J_{T}$
where $J_{T}$ is defined after \eqref{Eq. (h1)}. The form of $\left\{ V_{t}\right\} $
depends on the specific problem under study. For example, for a $t$-test
on a regression coefficient in the linear model $y_{t}=x{}_{t}\beta_{0}+e_{t}$
$\left(t=1,\ldots,\,T\right)$ we have $V_{t}=x_{t}e_{t}$. Classical
HAC estimators take the following form, 
\begin{align*}
\widehat{J}_{\mathrm{HAC,}T}=\sum_{k=-T+1}^{T-1}K_{1}\left(b_{1,T}k\right)\widehat{\Gamma}\left(k\right) & ,
\end{align*}
where $\widehat{\Gamma}\left(k\right)$ is given in \eqref{Eq. Definition of Gamma(k)}
with $\widehat{V}_{t}=x_{t}\widehat{e}_{t}$ where $\left\{ \widehat{e}_{t}\right\} $
are the least-squares residuals, $K_{1}\left(\cdot\right)$ is a kernel
and $b_{1,T}$ is bandwidth. One can use the the Bartlett kernel,
advocated by \citet{newey/west:87}, the quadratic spectral kernel
as suggested by \citet{andrews:91}, or any other kernel suggested
in the literature, see e.g. \citet{dejong/davidson:00} and \citet{ng/perron:1996}.
Under $b_{1,T}\rightarrow0$ at an appropriate rate, we have $\widehat{J}_{\mathrm{HAC,}T}\overset{\mathbb{P}}{\rightarrow}J.$
Hence, equipped with $\widehat{J}_{\mathrm{HAC,}T}$, HAR inference
is standard and simple because HAR test statistics follow asymptotically
standard distributions. 

 HAC standard errors can result in oversized tests when there is
substantial temporal dependence {[}e.g., \citet{andrews:91}{]}. This
stimulated a second approach based on LRV estimators that keeps the
bandwidth at some fixed fraction of $T$ {[}cf. \citet{Kiefer/vogelsang/bunzel:00}{]},
e.g., using all autocovariances, so that $\widehat{J}_{\mathrm{\mathrm{KVB},}T}\triangleq T^{-1}\sum_{t=1}^{T}\sum_{s=1}^{T}\left(1-\left|t-s\right|/T\right)$
$\widehat{V}_{t}\widehat{V}_{s}$ which is equivalent to the Newey-West
estimator with $b_{1,T}=T^{-1}$.  Under fixed-$b$ asymptotics
the reference distribution of HAR test statistics is nonstandard.
The validity of fixed-$b$  inference rests on stationarity {[}cf.
\citet{casini_fixed_b_erp}{]}.  Many authors have considered various
versions of $\widehat{J}_{\mathrm{\mathrm{KVB},}T}$. However, the
one that leads to HAR inference tests that are least oversized is
the original $\widehat{J}_{\mathrm{\mathrm{KVB},}T}$ {[}see \citet{casini/perron_PrewhitedHAC}
for simulation results{]}. For comparison we also report the equally-weighted
cosine (EWC) estimator of \citet{lazarus/lewis/stock:17}. It is an
orthogonal series estimators that use long bandwidths,
\begin{align*}
\widehat{J}_{\mathrm{\mathrm{EWC},}T} & \triangleq B^{-1}\sum_{j=1}^{B}\Lambda_{j}^{2},\qquad\mathrm{where}\quad\Lambda_{j}=\sqrt{\frac{2}{T}}\sum_{t=1}^{T}\widehat{V}_{t}\cos\left(\pi j\left(\frac{t-1/2}{T}\right)\right)
\end{align*}
with $B$ some fixed integer. Assuming $B$ satisfies some conditions,
under fixed-$b$ asymptotics a $t$-statistic normalized by $\widehat{J}_{\mathrm{\mathrm{EWC},}T}$
follows a $t_{B}$ distribution where $B$ is the degree of freedom. 

Recently, a new HAC estimator was proposed in \citet{casini_hac}.
Motivated by the power impact of low frequency contamination of existing
LRV estimators, he proposed a double kernel HAC (DK-HAC) estimator,
defined by 
\begin{align*}
\widehat{J}_{\mathrm{DK},T}\triangleq\sum_{k=-T+1}^{T-1}K_{1}\left(b_{1,T}k\right)\widehat{\Gamma}_{\mathrm{DK}}\left(k\right) & ,
\end{align*}
 where $b_{1,T}$ is a bandwidth sequence and $\widehat{\Gamma}_{\mathrm{DK}}\left(k\right)$
defined in Section \ref{Section Low Freq Cont - Theory} with $\widehat{c}_{T}\left(\cdot,\,k\right)$
replaced by 
\begin{align*}
\widehat{c}_{\mathrm{DK,}T}\left(rn_{T}/T,\,k\right) & =\left(Tb_{2,T}\right)^{-1}\sum_{s=|k|+1}^{T}K_{2}\left(\frac{\left(rn_{T}-\left(s-|k|/2\right)\right)/T}{b_{2,T}}\right)\widehat{V}_{s}\widehat{V}{}_{s-|k|},
\end{align*}
with $K_{2}$ a kernel and $b_{2,T}$ a bandwidth. Note that $\widehat{c}_{\mathrm{DK,}T}$
and $\widehat{c}_{T}$ are asymptotically equivalent and the results
of Section \ref{Section Low Freq Cont - Theory} continue to hold
for $\widehat{c}_{\mathrm{DK,}T}$. More precisely, $\widehat{c}_{T}$
is a special case of $\widehat{c}_{\mathrm{DK,}T}$ with $K_{2}$
being a rectangular kernel and $n_{2,T}=Tb_{2,T}$. This approach
falls in the first category of standard inference  $\widehat{J}_{\mathrm{DK},T}\overset{\mathbb{P}}{\rightarrow}J$
and HAR test statistics normalized by $\widehat{J}_{\mathrm{DK},T}$
follows standard distribution asymptotically. The DK-HAC estimator
involves two kernels: $K_{1}$ smooths the lagged sample autocovariances,
akin to the classical HAC estimators, while $K_{2}$ applies smoothing
over time. The latter feature is useful to avoid the low frequency
contamination. Additionally, \citet{casini/perron_PrewhitedHAC} proposed
prewhitened DK-HAC $(\widehat{J}_{\mathrm{\mathrm{pw},DK},T})$ estimator
that improves the size control of HAR tests and enjoys the same
asymptotic properties of $\widehat{J}_{\mathrm{DK},T}$. \citet{casini_hac}
and \citet{casini/perron_PrewhitedHAC} demonstrated via simulations
that tests based on $\widehat{J}_{\mathrm{DK},T}$ and $\widehat{J}_{\mathrm{pw,DK},T}$
have superior power properties relative to tests based on the other
estimators. In terms of size, the simulation results showed that
tests based on $\widehat{J}_{\mathrm{pw,DK},T}$ perform better than
those based on $\widehat{J}_{\mathrm{HAC,}T}$ and $\widehat{J}_{\mathrm{DK},T}$,
and is competitive with $\widehat{J}_{\mathrm{\mathrm{KVB},}T}$ when
the latter works well.  We include $\widehat{J}_{\mathrm{DK},T}$
and $\widehat{J}_{\mathrm{pw,DK},T}$ in our simulations below. We
report the results only for the DK-HAC estimators that do not use
the pre-test for discontinuities in the spectrum {[}cf. \citet{casini/perron:change-point-spectra}{]}
because we do not want the results to be affected by such pre-test.

\subsection{\label{Subsec Finite-Sample-Low-Frequency}Null Rejection Rates and
Power in Finite-Sample}

In order to better understand the effect of nonstationarity on the
null rejection rates of HAR tests we first conduct a Monte Carlo analysis
where we compare a nonstationary model with a stationary one that
has either the same spectral density at frequency zero or the same
average dependence. Consider the following four AR(1) data-generating
processes (DGPs). DGP 1 is given by 
\begin{align*}
V_{t} & =0.26V_{t-1}+e_{t},\qquad t=1,\ldots,\,T,
\end{align*}
 where $e_{t}\sim\mathscr{N}\left(0,\,1\right)$ for all $t$. The
LRV of DGP 1 is $J=1.826$. DGP 2 is 
\begin{align*}
V_{t} & =0.7817V_{t-1}+e_{t},\qquad t=1,\ldots,\,T,
\end{align*}
 where $e_{t}\sim\mathscr{N}\left(0,\,1\right)$ for all $t$. Its
LRV is $J=20.988$. We now introduce two nonstationary DGPs. DGP 3
takes the following form 
\begin{align*}
V_{t} & =\begin{cases}
0.9V_{t-1}+e_{t}, & 1\leq t\leq0.2T\\
0.1V_{t-1}+e_{t}, & 0.2T<t\leq T,
\end{cases}
\end{align*}
where $e_{t}\sim\mathscr{N}\left(0,\,1\right)$. Note that the spectral
density at frequency zero of $V_{t}$ is given by the weighted average
of the spectral densities of $V_{t}$ in the two regimes: 
\begin{align*}
f\left(0\right)=\int_{0}^{1}f\left(u,\,0\right)du & =0.2\frac{1}{2\pi\left(1-2\cdot0.9+0.9^{2}\right)}+0.8\frac{1}{2\pi\left(1-2\cdot0.1+0.1^{2}\right)}=3.342.
\end{align*}
Thus, the LRV of $V_{t}$ is $J=2\pi\int_{0}^{1}f\left(u,\,0\right)du=20.988$
which takes the same value as the LRV of DGP 2. Further, DGP 3 has
the same average dependence as DGP 1, meaning that the AR(1) coefficient
in DGP 1 is equal to the weighted average of the AR(1) coefficients
of DGP 3 in the two regimes, i.e., $\overline{\rho}=0.2\cdot0.9+0.8\cdot0.1=0.26$.
We also want to verify whether the location of the break in persistence
in DGP 3 is important for the bias. Thus, we consider DGP 4: 
\begin{align*}
V_{t} & =\begin{cases}
0.1V_{t-1}+e_{t}, & 1\leq t\leq0.5T\\
0.9V_{t-1}+e_{t}, & 0.5T<t\leq0.5T+0.2T\\
0.1V_{t-1}+e_{t}, & 0.5T+0.2T<t\leq T,
\end{cases}
\end{align*}
 where $e_{t}\sim\mathscr{N}\left(0,\,1\right)$ for all $t$. While
in DGP 3 the regime with strong persistence occurs in the first 20\%
of the sample, in DGP 4 it occurs between the 50\% and 70\% of the
sample. The LRV of DGP 4 is the same as that of DGP 3. 

For each DGP we consider three different initial conditions: (a) $V_{0}=0$;
(b) $V_{0}\sim\mathscr{N}\left(0,\,1\right)$; (c) $V_{0}\sim\mathscr{N}\left(0,\,4\right)$.
This is useful in order to verify whether the initial condition has
any effect on the bias generated by changes in the second-order properties.
DGP 3(a) should exhibit a smaller bias due to nonstationarity than
DGP 3(b,c) and 4. To see this, note that in DGP 3(a) the initial condition
is $V_{0}=0.$ Thus, the process starts from zero. Since there is
strong persistence in the first 20\% of the sample, the process is
more likely to stay close to zero in the first regime than when the
initial condition is $V_{0}\sim\mathscr{N}\left(0,\,1\right)$ or
$V_{0}\sim\mathscr{N}\left(0,\,4\right)$. In DGP 4 the different
specifications of the initial condition should not lead to any differences
in the bias due to nonstationarity because the regime with strong
dependence occurs about mid-sample.

To summarize, we have four DGPs. DGP 1 and 2 are stationary while
DGP 3 and 4 are nonstationary. Since DGP 2 has a LRV that takes the
same value as that of DGP 3 and 4, this allows us to better separate
the effect of persistence from that of nonstationarity in the second
moments on the following quantities: $\widehat{J}_{\mathrm{HAC}}$,
$-\widehat{c}_{1}b_{1,T}$ and $\widehat{\Gamma}\left(k\right)$ for
$k=0,1,\,5,\,10.$ In the simulations below $\widehat{J}_{\mathrm{HAC}}$
is the Newey-West estimator based on a predetermined number of lagged
sample autocovariances following the rule $4\left(T/100\right)^{2/9}$
{[}cf. \citet{lazarus/lewis/stock/watson:18}{]}. We compare $\widehat{\Gamma}\left(k\right)$
to the theoretical value $\Gamma_{T}\left(k\right)$ corresponding
to each DGP which can be computed by hand given the simple form of
the DGPs. In fact, for the nonstationary DGPs, $\Gamma_{T}\left(k\right)$
is a weighed average of the  theoretical autocovariances corresponding
to each regime. Here, $\widehat{c}_{1}$ is an estimate of $\overline{c}_{1}$
in \eqref{Eq. (c1)} that enters the asymptotic bias of $\widehat{J}_{\mathrm{HAC}}$.
In order to compute $\widehat{c}_{1}$ we recall that the asymptotic
bias of the LRV estimator based on the Bartlett kernel is given by
\begin{align*}
\lim_{T\rightarrow\infty}b_{1,T}^{-1}\mathbb{E}\left(\widehat{J}_{\mathrm{HAC}}-J_{T}\right) & =-2\pi K_{\mathrm{BT},1}\int_{0}^{1}f^{\left(1\right)}\left(u,\,0\right)du,
\end{align*}
 where 
\begin{align*}
K_{\mathrm{BT},q} & =\lim_{x\rightarrow0}\frac{1-K_{\mathrm{BT}}\left(x\right)}{\left|x\right|^{q}}
\end{align*}
 denotes the index of smoothness of the kernel at zero and $f^{\left(1\right)}\left(u,\,0\right)$
is the index of smoothness of the local spectral density at time $u$
and frequency zero. For the Bartlett kernel $K_{\mathrm{BT},q}=0$
if $q<1$, $K_{\mathrm{BT},q}=1$ if $q=1$ and $K_{\mathrm{BT},q}=\infty$
if $q>1.$ The Parzen characteristic exponent is the largest $q$
such that $K_{\mathrm{BT},q}$ is finite. Thus, the relative bias
is
\begin{align*}
\lim_{T\rightarrow\infty}b_{1,T}^{-1}\mathbb{E}\left(\widehat{J}_{\mathrm{HAC}}/J_{T}-1\right) & =-K_{\mathrm{BT},1}\frac{\int_{0}^{1}f^{\left(1\right)}\left(u,\,0\right)du}{\int_{0}^{1}f\left(u,\,0\right)du}=-\overline{c}_{1},
\end{align*}
 using $K_{\mathrm{BT},1}=1.$ The index of smoothness of $f\left(u,\,\omega\right)$
at $\omega=0$ is defined as 
\begin{align*}
f^{\left(1\right)}\left(u,\,0\right) & =\frac{1}{2\pi}\sum_{k=-\infty}^{\infty}|k|\Gamma\left(u,\,k\right).
\end{align*}
 For an AR(1) process with parameters $\rho\left(u\right)$ and $\sigma_{e}^{2}\left(u\right)$,
we have $\Gamma\left(u,\,k\right)=\sigma_{e}^{2}\left(u\right)\rho\left(u\right)^{|k|}/(1-\rho\left(u\right)^{2}).$
It follows that 
\begin{align*}
f^{\left(1\right)}\left(u,\,0\right) & =-\frac{1}{2\pi}\frac{2\rho\left(u\right)\sigma_{e}^{2}\left(u\right)}{\left(\rho\left(u\right)-1\right)^{3}\left(1+\rho\left(u\right)\right)}.
\end{align*}
 Based on this result we can obtain $\overline{c}_{1}$ for each model.
In particular, for model DGP 1, 2, 3 and 4 we have $\overline{c}_{1}=0.55$,
3.92, 9.04 and 9.05, respectively. 

We estimate $\overline{c}_{1}$ as follows. For DGP 1, we obtain the
OLS residuals $\widehat{V}_{t}$ and estimate $\rho$ and $\sigma_{e}^{2}$
from the autoregression
\begin{align*}
\widehat{V}_{t} & =\rho\widehat{V}_{t-1}+e_{t},\qquad\qquad t=1,\ldots,\,T,
\end{align*}
where $\sigma_{e}^{2}$ is the variance of $e_{t}$. Let these estimates
be denoted by $\widehat{\rho}$ and $\widehat{\sigma}_{e}^{2}$, respectively.
Then, the estimate of $\overline{c}_{1}$ is defined as 
\begin{align*}
\widehat{c}_{1} & =-\frac{2\widehat{\rho}\widehat{\sigma}_{e}^{2}}{\widehat{J}_{\mathrm{HAC}}\left(\widehat{\rho}-1\right)^{3}\left(1+\widehat{\rho}\right)}.
\end{align*}
The same applies to DGP 2. For DGP 3, we obtain the estimate of the
autoregressive coefficient of $V_{t}$ and of the variance of the
innovations by estimating the autoregression in the two regimes separately.
That is, we obtain
\begin{align*}
\widehat{V}_{t} & =\begin{cases}
\widehat{\rho}_{1}\widehat{V}_{t-1}+\widehat{e}_{t}, & 1\leq t\leq0.2T\\
\widehat{\rho}_{2}\widehat{V}_{t-1}+\widehat{e}_{t}, & 0.2T<t\leq T,
\end{cases}
\end{align*}
where we also compute $\widehat{\sigma}_{1,e}^{2}$ and $\widehat{\sigma}_{2,e}^{2}$
which are the sample variances of the residuals $\widehat{e}_{t}$
in the two regimes, respectively. Then, the estimate of $\overline{c}_{1}$
is defined as 
\begin{align*}
\widehat{c}_{1} & =-0.2\frac{2\widehat{\rho}_{1}\widehat{\sigma}_{1,e}^{2}}{\widehat{J}_{\mathrm{HAC}}\left(\widehat{\rho}_{1}-1\right)^{3}\left(1+\widehat{\rho}_{1}\right)}-0.8\frac{2\widehat{\rho}_{2}\widehat{\sigma}_{2,e}^{2}}{\widehat{J}_{\mathrm{HAC}}\left(\widehat{\rho}_{2}-1\right)^{3}\left(1+\widehat{\rho}_{2}\right)}.
\end{align*}
 The same applies to DGP 4 with the difference that the autoregressive
coefficient and the variance of the innovations are estimated separately
in each of the three distinct regimes. 

We consider the sample size $T=100,\,200$ and 1000, and 50,000 repetitions
were used for each DGP. The results are reported in Table \ref{Table T=00003D100, 200, 1000}.
Let us first discuss the finite-sample properties of $\widehat{J}_{\mathrm{HAC}}$.
The results clearly suggest that $\widehat{J}_{\mathrm{HAC}}$ deviates
substantially from $J$ when the data are nonstationary. $\widehat{J}_{\mathrm{HAC}}$
underestimates $J$ for all DGPs but it does so much more when the
DGP is nonstationary. The difference between the values of $\widehat{J}_{\mathrm{HAC}}$
in DGP 2 and those in DGP 3-4 is about one half, e.g., $\widehat{J}_{\mathrm{HAC}}=6.775$
in DGP 2(a) and $\widehat{J}_{\mathrm{HAC}}=3.142$ in DGP 3(a). As
the sample size increases the downward bias becomes smaller, though
$\widehat{J}_{\mathrm{HAC}}$ still underestimates $J$ for $T=1000$.
The downward bias continues to remain larger in DGP 3-4 than in DGP
2 even when $T=1000$. Thus, this evidence based on $\widehat{J}_{\mathrm{HAC}}$
already points out that basic forms of nonstationarity generate bias
in the LRV estimator. This bias adds to the well-known bias generated
by strong persistence in stationary data documented in the literature. 

Let us discuss the relative bias $-\overline{c}_{1}b_{1,T}$ and its
estimate $-\widehat{c}_{1}b_{1,T}$. First note that $-\overline{c}_{1}b_{1,T}<0$
and $-\widehat{c}_{1}b_{1,T}<0$ for all DGPs and sample sizes considered.
This confirms the downward bias of $\widehat{J}_{\mathrm{HAC}}$ observed
above. For a given model, the asymptotic relative bias $-\overline{c}_{1}b_{1,T}$
and its estimate increase with the sample size. The downward bias
is much larger for the nonstationary DGP 3-4 than for the stationary
DGP 1-2. The estimates $-\widehat{c}_{1}b_{1,T}$ of the relative
bias $-\overline{c}_{1}b_{1,T}$ significantly underestimate $-\overline{c}_{1}b_{1,T}$
in DGP 3-4 while in DGP 1-2 the deviations are much smaller. The large
deviations of $-\widehat{c}_{1}b_{1,T}$ from $-\overline{c}_{1}b_{1,T}$
continue to hold even for $T=1000.$ 

We now move to discuss the finite-sample properties of $\widehat{\Gamma}\left(k\right)$.
When the data are stationary, $\widehat{\Gamma}\left(k\right)$ is
close to $\Gamma_{T}\left(k\right)$ even when $T=100$ and it approaches
$\Gamma_{T}\left(k\right)$ when $T=1000$. For nonstationary data,
$\widehat{\Gamma}\left(k\right)$ is much farther from $\Gamma_{T}\left(k\right)$.
For example, in DGP 2(a) $\widehat{\Gamma}\left(0\right)=2.507$ and
$\Gamma_{T}\left(0\right)=2.571$ whereas in DGP 3(a) $\widehat{\Gamma}\left(0\right)=1.589$
and $\Gamma_{T}\left(0\right)=1.861$. Thus, $\widehat{\Gamma}\left(k\right)$
has larger bias (in general downward) when the data are nonstationary.
This result is present even when $T=200$.  As $T$ increases, $\widehat{\Gamma}\left(k\right)$
approaches $\Gamma_{T}\left(k\right)$ for all DGPs, though the downward
bias remains larger in DGP 3-4 than in DGP 1-2.

We repeated this exercise for other DGPs and the conclusions were
the same. The results suggest that under nonstationarity the bias
in the LRV estimator is affected by multiple factors. In addition
to the downward bias arising from strong persistence which is also
present under stationarity there is bias generated by the time-varying
properties of the process. Under the null hypothesis this time variation
occurs in the autocovariance structure of the process. For example,
in DGP 3 one has $0.2T$ observations to estimate $2\pi\int_{0}^{0.2}f\left(u,\,0\right)du=0.4\pi f\left(0\right)$
where $f\left(0\right)=1/(2\pi\left(1-2\rho+\rho^{2}\right))$ with
$\rho=0.9$, and $0.8T$ observations to estimate $2\pi\int_{0.2}^{1}f\left(u,\,0\right)du=1.6\pi f\left(0\right)$
where $f\left(0\right)=1/(2\pi\left(1-2\rho+\rho^{2}\right))$ with
$\rho=0.1$. This is more difficult than estimating $2\pi f\left(0\right)=1/(2\pi\left(1-2\rho+\rho^{2}\right))$
with $\rho=0.7817$ using $T$ observations, which applies to DGP
2. Even if the total sample size is $T$ in both DGP 2 and 3, nonstationarity
reduces the effective sample size making the estimation of the LRV
in DGP 3 effectively based on a smaller number of observations. For
example, $\widehat{\Gamma}\left(k\right)$ involves an average on
$\{\widehat{V}_{t}\widehat{V}_{t-k}\}$ for $t=k+1,\ldots,\,T$. Some
of these pairs $\{\widehat{V}_{t}\widehat{V}_{t-k}\}$ are such that
$\widehat{V}_{t}$ and $\widehat{V}_{t-k}$ belong to two different
regimes, and so contribute bias to the estimation of $\Gamma_{T}\left(k\right)$.
Under stationarity all the pairs $\{\widehat{V}_{t}\widehat{V}_{t-k}\}$
are such that $\widehat{V}_{t}$ and $\widehat{V}_{t-k}$ belong to
the same regime leading to more precise estimates of $\widehat{\Gamma}\left(k\right)$
and LRV. In addition, changes in persistence over short regimes share
features similar to shifts in the mean, at least graphically. While
the former is consistent with the null hypothesis, the latter is not.
This is likely to generate some bias where changes in persistence
are confounded with shifts in the mean even when the unconditional
mean of the series has not changed. The downward bias due to strong
persistence and the bias due to time-varying second-order properties
are likely to influence each other making the estimation problem even
harder. 

We now investigate the consequence of nonstationarity for HAR inference.
We obtain the empirical size and power for a two-tailed $t$-test
on the intercept normalized by several LRV estimators for the model
$y_{t}=\delta+V_{t}$ with $\delta=0$ under the null  and $\delta>0$
under the alternative hypothesis. Model M1 involves an SLS process:
$V_{t}=0.9V_{t-1}+u_{t}$, $V_{0}\sim\mathscr{N}\left(0,\,1\right)$,
$u_{t}\sim\mathrm{i.i.d.}\,\mathscr{N}\left(0,\,1\right)$ for $t=1,\ldots,\,T_{1}^{0}$
with $T_{1}^{0}=T\lambda_{1}^{0}$, and $V_{t}=\rho\left(t/T\right)V_{t-1}+u_{t}$,
$\rho\left(t/T\right)=0.3\left(\cos\left(1.5-\cos\left(t/T\right)\right)\right)$,
$u_{t}\sim\mathscr{\mathrm{i.i.d.}\,\mathscr{N}}\left(0,\,0.5\right)$
for $t=T_{1}^{0}+1,\ldots,\,T$. Note that $\rho\left(\cdot\right)$
varies between 0.172 and 0.263. We set $\lambda_{1}^{0}=0.1$. In
addition to M1, we consider other models: M2 involves a time-varying
AR(1) with a break in volatility $V_{t}=\rho\left(t/T\right)V_{t-1}+u_{t}$,
$\rho\left(t/T\right)=0.7(\cos\left(1.5t/T\right))$, $u_{t}\sim\mathscr{N}\left(0,\,\sigma_{t}^{2}\right)$,
$\sigma_{t}^{2}=5$ for $t\leq4$ and $\sigma_{t}^{2}=0.25$ for $t>4$,
$V_{0}\sim\mathscr{N}\left(0,\,5\right)$; M3 involves $V_{t}=\rho\left(t/T\right)V_{t-1}+u_{t}$,
$\rho\left(t/T\right)=0.8(\cos\left(1.5t/T\right))$, $u_{t}\sim\mathscr{N}\left(0,\,0.25\right)$,
$V_{0}=0$ with outliers $V_{t}\sim\mathrm{Uniform}\left(\underline{c},\,5\underline{c}\right)$
for $t=T/2,\,3T/4$ where $\underline{c}=-1/(\sqrt{2}\mathrm{erfc^{-1}\left(3/2\right))}\mathrm{med}\left(\left|V-\mathrm{med}\left(V\right)\right|\right)$
with $\mathrm{erfc}^{-1}$ the inverse complementary error function,
$\mathrm{med}\left(\cdot\right)$ is the median and $V=\left(V_{t}\right)_{t=1}^{T}$;\footnote{In this literature, values smaller than $\underline{c}$ are not classified
as outliers. } M4 involves a time varying AR(1) with periods of strong persistence
where $V_{t}=\rho\left(t/T\right)V_{t-1}+u_{t}$, $\rho\left(t/T\right)=0.95(\cos\left(1.5t/T\right))$,
$u_{t}\sim\mathscr{\mathrm{i.i.d.}\,\mathscr{N}}\left(0,\,0.4\right)$
and $V_{0}\sim\mathscr{N}\left(0,\,4\right)$. $\rho\left(\cdot\right)$
varies between 0.7 and 0.05 in M2, between 0.05 and 0.8 in M3 and
between 0.95 and 0.07 in M4.

We consider the DK-HAC estimators with and without prewhitening ($\widehat{J}_{\mathrm{DK},T}$,
$\widehat{J}_{\mathrm{DK,pw},\mathrm{SLS},T}$, $\widehat{J}_{\mathrm{DK,pw},\mathrm{SLS},\mu,T}$)
of \citet{casini_hac} and \citet{casini/perron_PrewhitedHAC}, respectively;
\citeauthor{andrews:91}' \citeyearpar{andrews:91} HAC estimator
with and without the prewhitening procedure of \citet{andrews/monahan:92};
\citeauthor{newey/west:87}'s \citeyearpar{newey/west:87} HAC estimator
with the popular rule to select the number of lags (i.e., $b_{1,T}=(4(T/100)^{2/9})^{-1}$;
Newey-West with the fixed-$b$ method of \citet{Kiefer/vogelsang/bunzel:00}
with $b=1$ (labeled KVB); and the Equally-Weighted Cosine (EWC) of
\citet{lazarus/lewis/stock/watson:18} with the bandwidth choice recommended
by the authors. For the DK-HAC estimators we use the data-dependent
methods for the bandwidths, kernels and choice of $n_{T}$ as proposed
in \citet{casini_hac} and \citet{casini/perron_PrewhitedHAC}, which
are optimal under mean-squared error (MSE). Let $\widehat{V}_{t}$
denote the least-squares residual based on $\widehat{\delta}$ where
the latter is the least-squares estimate of $\delta$. We set $\widehat{b}_{1,T}=0.6828(\widehat{\phi}\left(2\right)T\widehat{\overline{b}}_{2,T})^{-1/5}$
where 
\begin{align*}
\widehat{\phi}\left(2\right) & =\left(18\left(\frac{n_{T}}{T}\sum_{j=0}^{\left\lfloor T/n_{3,T}\right\rfloor -1}\frac{\left(\widehat{\sigma}\left(\left(jn_{T}+1\right)/T\right)\widehat{a}_{1}\left(\left(jn_{T}+1\right)/T\right)\right)^{2}}{\left(1-\widehat{a}_{1}\left(\left(jn_{T}+1\right)/T\right)\right)^{4}}\right)^{2}\right)/\\
 & \quad\left(\frac{n_{T}}{T}\sum_{j=0}^{\left\lfloor T/n_{3,T}\right\rfloor -1}\frac{\left(\widehat{\sigma}\left(\left(jn_{T}+1\right)/T\right)\right)^{2}}{\left(1-\widehat{a}_{1}\left(\left(jn_{T}+1\right)/T\right)\right)^{2}}\right)^{2},
\end{align*}
 with 
\begin{align*}
\widehat{a}_{1}\left(u\right)=\frac{\sum_{j=t-n_{T}+1}^{t}\widehat{V}_{j}\widehat{V}_{j-1}}{\sum_{j=t-n_{T}+1}^{t}(\widehat{V}_{j-1})^{2}}, & \qquad\mathrm{and}\qquad\widehat{\sigma}\left(u\right)=(\sum_{j=t-n_{T}+1}^{t}(\widehat{V}_{j}-\widehat{a}_{1}\left(u\right)\widehat{V}_{j-1})^{2})^{1/2},
\end{align*}
and $\widehat{\overline{b}}_{2,T}=\left(n_{T}/T\right)\sum_{r=1}^{\left\lfloor T/n_{T}\right\rfloor -1}$
$\widehat{b}_{2,T}\left(rn_{T}/T\right)$, $\widehat{b}_{2,T}\left(u\right)=1.6786(\widehat{D}_{1}\left(u\right)){}^{-1/5}(\widehat{D}_{2}\left(u\right))^{1/5}T^{-1/5}$
where $\widehat{D}_{2}\left(u\right)\triangleq2\sum_{l=-\left\lfloor T^{4/25}\right\rfloor }^{\left\lfloor T^{4/25}\right\rfloor }\widehat{c}_{\mathrm{DK,}T}\left(u,\,l\right)^{2}$
and
\begin{align*}
\widehat{D}_{1}\left(u\right) & \triangleq(\left[S_{\omega}\right]^{-1}\sum_{s\in S_{\omega}}[3\pi^{-1}(1+0.8(\cos1.5+\cos4\pi u)\exp(-i\omega_{s}))^{-4}(0.8(-4\pi\sin(4\pi u)))\exp(-i\omega_{s})\\
 & \quad-\pi^{-1}\left|1+0.8(\cos1.5+\cos4\pi u)\exp(-i\omega_{s})\right|^{-3}(0.8(-16\pi^{2}\cos(4\pi u)))\exp(-i\omega_{s})])^{2},
\end{align*}
 with $\left[S_{\omega}\right]$ being the cardinality of $S_{\omega}$
and $\omega_{s+1}>\omega_{s}$, $\omega_{1}=-\pi,\,\omega_{\left[S_{\omega}\right]}=\pi.$
We set $n_{T}=T^{0.6}$, $S_{\omega}=\{-\pi,\,-3,\,-2,\,-1,\,0,\,1,\,2,\,3,\,\pi\}.$
$K_{1}\left(\cdot\right)$ is the QS kernel and $K_{2}\left(x\right)=6x\left(1-x\right)$
for $x\in\left[0,\,1\right].$ 

Table \ref{Table S1 - Figure 1} reports the results using 5,000 replications.
The $t$-test based on \citeauthor{newey/west:87}'s \citeyearpar{newey/west:87}
and \citeauthor{andrews:91}' \citeyearpar{andrews:91} prewhitened
HAC estimators are excessively oversized. \citeauthor{andrews:91}'
\citeyearpar{andrews:91} HAC-based test is slightly undersized while
the KVB's fixed-$b$ and EWC-based tests are severely undersized.
The fact that the KVB's fixed-$b$ and EWC-based tests have larger
size distortions than other tests is consistent with the results in
Section \ref{Section Edgeworth-Expansions-for} which suggest that
they have a larger ERP. For the $t$-test on the intercept, $\widehat{J}_{\mathrm{DK},T}$
can lead to tests that are oversized when there is strong dependence.
However, the prewhitened DK-HAC estimators $\widehat{J}_{\mathrm{DK,pw},\mathrm{SLS},T}$
and $\widehat{J}_{\mathrm{DK,pw},\mathrm{SLS},\mu,T}$ lead to tests
having more accurate rejection rates. Nonstationarity affects the
power of the tests based on LRV estimators that rely on  $\widehat{\Gamma}\left(k\right)$
or equivalently on $I_{T}\left(\omega\right)$ (e.g., the EWC). The
KVB's fixed-$b$ and EWC-based tests suffer from relatively large
power losses. The power of tests normalized by \citeauthor{newey/west:87}'s
(1987) and \citeauthor{andrews:91}' \citeyearpar{andrews:91} prewhitened
HAC are not comparable because they are significantly oversized. The
DK-HAC-based tests have the best power, the second best being \citeauthor{andrews:91}'
\citeyearpar{andrews:91} HAC-based test. 

Turning to M2, Table \ref{Table S1 - Figure 1} shows some size distortions
and power losses for KVB's fixed-$b$ and EWC-based tests. The prewhitened
DK-HAC-based tests display accurate size control and good power. \citeauthor{newey/west:87}'s
(1987) and \citeauthor{andrews:91}' \citeyearpar{andrews:91} prewhitened
HAC-based tests are again excessively oversized. \citeauthor{andrews:91}'
\citeyearpar{andrews:91} HAC-based test and the DK-HAC-based test
show a similar performance. For model M3-M4, Table \ref{Table S1 - Figure 1}
shows that all methods lead to oversized tests except prewhitened
DK-HAC and KVB's fixed-$b$. However, the KVB's fixed-$b$-based tests
show substantial unde-rejection that has consequences for power whereas
the prewhitened DK-HAC-based-tests show accurate null rejection rates
and good power. Finally, the simulations show that the null rejection
rates of HAC- and DK-HAC-based tests are not very far from each other,
thereby confirming that their respective ERP are close as shown in
Section \ref{Section Edgeworth-Expansions-for}.

\subsection{\label{Subsec General-Low-Frequency}General Low Frequency Contamination}

We now discuss HAR inference tests for which the low frequency contamination
results of Section \ref{Section Low Freq Cont - Theory} hold asymptotically.
This means that $d^{*}>0$ for all $T$ and as $T\rightarrow\infty$.
This comprises the class of HAR tests that admit a nonstationary alternative
hypothesis. This class is very large and includes most HAR tests 
as discussed in the Introduction. Here we consider the Diebold-Mariano
test for the sake of illustration and remark that similar issues apply
to other HAR tests.  

The Diebold-Mariano test statistic is defined as $t_{\mathrm{DM}}\triangleq T_{n}^{1/2}\overline{d}_{L}/\sqrt{\widehat{J}_{d_{L},T}}$,
where $\overline{d}_{L}$ is the average of the loss differentials
between two competing forecast models, $\widehat{J}_{d_{L},T}$ is
an estimate of the LRV of the loss differential series and $T_{n}$
is the number of observations in the out-of-sample. We use the quadratic
loss. We consider an out-of-sample forecasting exercise with a fixed
forecasting scheme where, given a sample of $T$ observations, $0.5T$
observations are used for the in-sample and the remaining half is
used for prediction {[}see \citet{perron/yamamoto:18} for recommendations
on using a fixed scheme in the presence of breaks{]}. The DGP under
the null hypothesis is given by $y_{t}=1+\beta_{0}x_{t-1}^{(0)}+e_{t}$
where $x_{t-1}^{(0)}\sim\mathrm{i.i.d.}\,\mathscr{N}\left(1,\,1\right)$,
$e_{t}=0.3e_{t-1}+u_{t}$ with $u_{t}\sim\mathrm{i.i.d.\,}\mathscr{N}\left(0,\,1\right)$,
and we set $\beta_{0}=1$ and $T=400.$ The two competing models both
involve an intercept but differ with respect to the predictor used
in place of $x_{t}^{(0)}$. The first forecast model uses $x_{t}^{(1)}$
while the second uses $x_{t}^{(2)}$ where $x_{t}^{(1)}$ and $x_{t}^{(2)}$
are independent $\mathrm{i.i.d.}\,\mathscr{N}\left(1,\,1\right)$
sequences, both independent from $x_{t}^{(0)}$. Each forecast model
generates a sequence of $\tau\left(=1\right)$-step ahead out-of-sample
losses $L_{t}^{(j)}$ $\left(j=1,\,2\right)$ for $t=T/2+1,\ldots,\,T-\tau.$
Then $d_{t}\triangleq L_{t}^{(2)}-L_{t}^{(1)}$ denotes the loss differential
at time $t$. The Diebold-Mariano test rejects the null hypothesis
of equal predictive ability when $\overline{d}_{L}$ is sufficiently
far from zero. Under the alternative hypothesis, the two competing
forecast models are as follows: the first uses $x_{t}^{(1)}=x_{t}^{(0)}+u_{X_{1},t}$
where $u_{X_{1},t}\sim\mathrm{i.i.d.\,}\mathscr{N}\left(0,\,1\right)$
while the second uses $x_{t}^{(2)}=x_{t}^{(0)}+0.2z_{t}+2u_{X_{2},t}$
for $t\in\left[1,\ldots,\,3T/4-1,\,3T/4+21,\ldots T\right]$ and $x_{t}^{(2)}=\delta\left(t/T\right)+0.2z_{t}+2u_{X_{2},t}$
for $t=3T/4,\ldots,\,3T/4+20$ with $u_{X_{2},t}\sim\mathrm{i.i.d.\,}\mathscr{N}\left(0,\,1\right)$,
where $z_{t}$ has the same distribution as $x_{t}^{(0)}.$ 

We consider four specifications for $\delta\left(\cdot\right).$ In
the first $x_{t}^{(2)}$ is subject to an abrupt break in the mean
$\delta\left(t/T\right)=\delta>0$; in the second $x_{t}^{(2)}$ is
locally stationary with time-varying mean $\delta\left(t/T\right)=\delta\left(\sin\left(t/T-3/4\right)\right)$;
in the third specification $x_{t}^{(2)}=x_{t}^{(0)}+0.2z_{t}+2u_{X_{2},t}$
for $t\in[1,\ldots,\,T/2-30,\,T/2$ $+21,\ldots T]$ and $x_{t}^{(2)}=\delta\left(t/T\right)+0.2z_{t}+2u_{X_{2},t}$
for $t=T/2-30,\ldots,\,T/2+20$ with $\delta\left(t/T\right)=\delta(\sin(t/T-1/2$
$-30/T))$; in the fourth $x_{t}^{(2)}$ is the same as in the second
with in addition two outliers $x_{t}^{(2)}\sim\mathrm{Uniform}\left(\left|\underline{c}\right|,\,5\left|\underline{c}\right|\right)$
for $t=6T/10,\,8T/10$ where $\underline{c}=-1/(\sqrt{2}\mathrm{erfc^{-1}\left(3/2\right))}\mathrm{med}(|x^{(2)}-\mathrm{med}$
$(x^{(2)})|)$ where $x^{(2)}=(x_{t}^{(2)})_{t=1}^{T}$. That is,
in the second model $x_{t}^{(2)}$ is locally stationary only in the
out-of-sample, in the third it is locally stationary in both the in-sample
and out-of sample and in the fourth model $x_{t}^{(2)}$ has two outliers
in the out-of-sample. The location of the outliers is irrelevant for
the results; they can also occur in the in-sample.

Table \ref{Table Power DM Test} reports the null rejection rate and
the power of the various tests for all models. We begin with the case
$\delta\left(t/T\right)=\delta>0$ (top panel). The null rejection
rate of the test using the DK-HAC estimators is accurate while the
tests using other LRV estimators are oversized with the exception
of the KVB's fixed-$b$ method for which the rejection rate is equal
to zero. The HAR tests using existing LRV estimators have lower power
relative to that obtained with the DK-HAC estimators for small values
of $\delta$. When $\delta$ increases the tests standardized by the
HAC estimators of \citet{andrews:91} and \citet{newey/west:87},
and by the KVB's fixed-$b$ and EWC LRV estimators display non-monotonic
power gradually converging to zero as the alternative gets further
away from the null value. In contrast, when using the DK-HAC estimators
the test has monotonic power that reaches and maintains unit power.
The results for the other models are even stronger. In general, except
when using the DK-HAC estimators, all tests display serious power
problems. Thus, either form of nonstationarity or outliers leads to
similar implications, consistent with our theoretical results. 

\begin{center}
\begin{figure}[H]
\includegraphics[width=18cm,height=17cm]{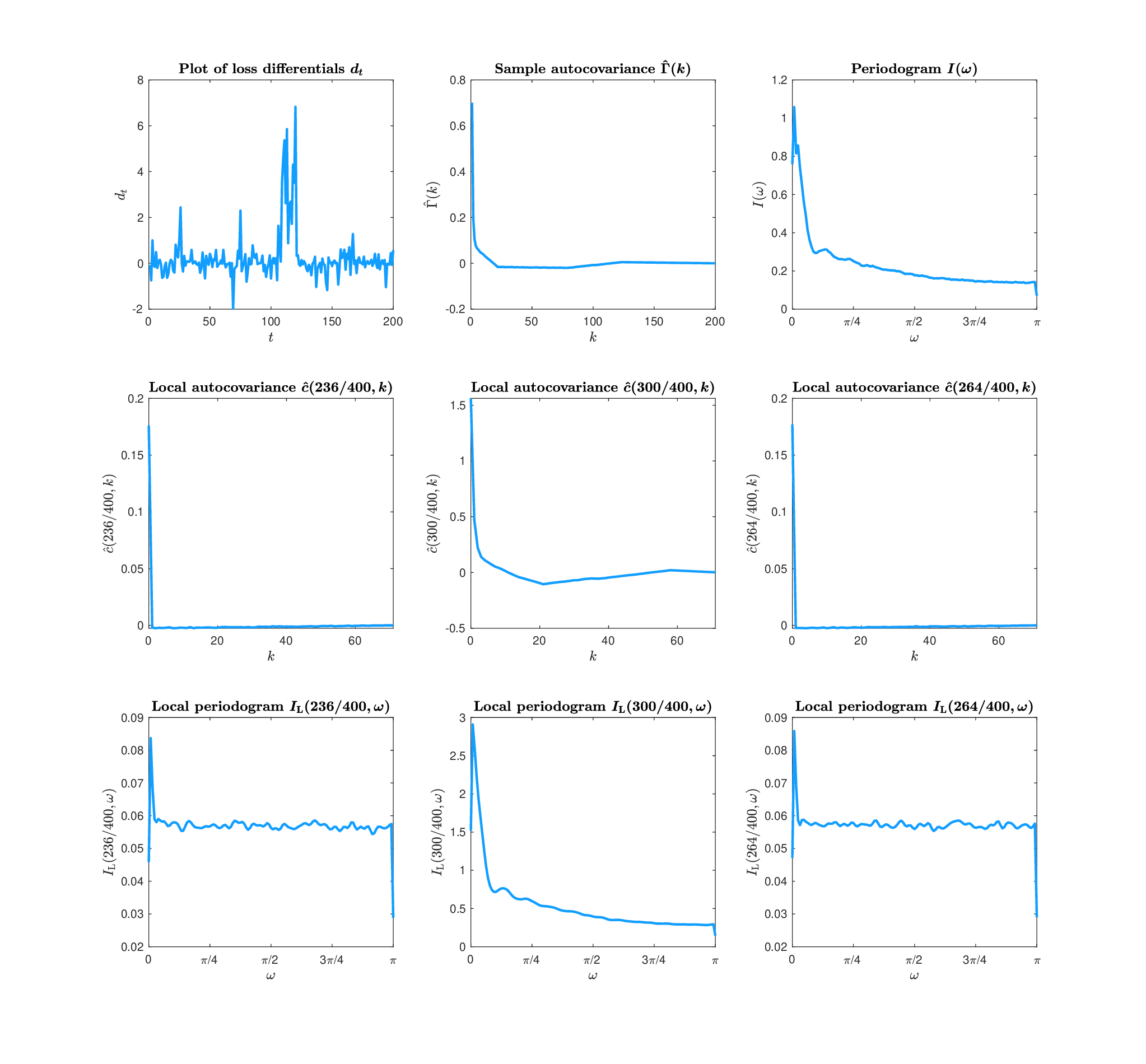}

{\footnotesize{}\caption{{\scriptsize{}\label{Fig_1_ET}Plots of loss differentials $d_{t}$,
sample autocovariance $\widehat{\Gamma}\left(k\right)$, periodogram
$I\left(\omega\right)$, sample local autocovariance $\widehat{c}(u,\,k)$
and local periodogram $I_{\mathrm{L}}(u,\,\omega)$.  In all panels
$\delta=1.$}}
}{\footnotesize\par}
\end{figure}
\end{center}

In order to further assess the theoretical results from Section \ref{Section Low Freq Cont - Theory},
Figure \ref{Fig_1_ET} (top panel) reports the plots of $d_{t}$,
its sample autocovariances and its periodogram, for $\delta=1$.
Figures \ref{Fig_2_ET}-\ref{Fig_3_ET} (top panels) in the supplement
report the corresponding plots for $\delta=2,\,5$, respectively.
We only consider the case $\delta_{t}=\delta>0$. The other cases
lead to the same conclusions. For $\delta=1$, Figure \ref{Fig_1_ET}
(top panel) shows that $\widehat{\Gamma}\left(k\right)$ decays slowly.
As $\delta$ increases, from Figures \ref{Fig_2_ET} and \ref{Fig_3_ET}
(top panels), $\widehat{\Gamma}\left(k\right)$ decays even more slowly
at a rate far from the typical exponential decay of short memory processes.
 This suggests evidence of long memory. However, the data are short
memory with small temporal dependence. What is generating the spurious
long memory effect is the nonstationarity present under the alternative
hypothesis. This is visible in the top panels which present plots
of $d_{t}$ for the first specification. The shift in the mean of
$d_{t}$ for $t=3T/4,\ldots,\,3T/4+20$ is responsible for the long
memory effect. This corresponds to the second term of \eqref{Eq. Gamma_hat(k) Ineq Theorem}
in Theorem \ref{Theorem ACF Nonstat}. The overall behavior of the
sample autocovariance is as predicted by Theorem \ref{Theorem ACF Nonstat}.
For small lags, $\widehat{\Gamma}\left(k\right)$ shows a power-like
decay and it is positive. As $k$ increases to medium lags, the autocovariances
turn negative because the sum of all sample autocovariances has to
be equal to zero {[}cf. \citet{percival:1992}{]}.    Next, we
move to the bottom panels which plot the periodogram of $\{d_{t}\}$.
It is unbounded at frequencies close to $\omega=0$ as predicted by
Theorem \ref{Theorem Periodogram Long Memory Effects} and as would
occur if long memory was present. It also explains why the Diebold-Mariano
test normalized by Newey-West's, Andrews', KVB's fixed-$b$ and EWC's
LRV estimators have serious power problems. These LRV estimators are
inflated and consequently the  tests lose power. The figures show
that as we raise $\delta$ the more severe these issues and the 
power losses so that the power eventually reaches zero. This is consistent
with our theory since $d^{*}$ is increasing in $\delta$ (cf. $d^{*}\thickapprox0.1\cdot0.9\delta^{2}$).

We now verify the results about the local sample autocovariance $\widehat{c}_{T}\left(u,\,k\right)$
and the local periodogram from Theorems \ref{Theorem Local ACF Nonstat}-\ref{Theorem Local Periodogram Long Memory Effects}.
We set $n_{2,T}=T^{0.6}=36$ following the MSE criterion of \citet{casini_hac}.
We consider (i) $u=236/T$, (ii-a) $u=T_{1}^{0}/T=3/4$ and (ii-b)
$u=264/T$. Note that cases (i)-(ii-b) correspond to parts (i)-(ii-b)
in Theorems \ref{Theorem Local ACF Nonstat}-\ref{Theorem Local Periodogram Long Memory Effects}.
 We consider $\delta=1,\,2$ and $5$. According to Theorems \ref{Theorem Local ACF Nonstat}-\ref{Theorem Local Periodogram Long Memory Effects},
we should expect long memory features only for case (ii-a). Figures
\ref{Fig_1_ET} and \ref{Fig_2_ET}-\ref{Fig_3_ET} in the supplement
confirm this. The results pertaining to case (ii-a) are plotted in
the middle panels. They show that the local autocovariance displays
slow decay similar to the pattern discussed above for $\widehat{\Gamma}\left(k\right)$
and that this problem becomes more severe as $\delta$ increases.
Such long memory features also appear for $I_{\mathrm{L}}\left(3/4,\,\omega\right)$.
The bottom panels in Figures \ref{Fig_1_ET} and \ref{Fig_2_ET}-\ref{Fig_3_ET}
show that the local periodogram at $u=3/4$ and at a frequency close
to $\omega=0$ are extremely large. The latter result is consistent
with Theorem \ref{Theorem Local Periodogram Long Memory Effects}-(ii-a)
which suggests that $I_{\mathrm{L,}T}\left(3/4,\,\omega\right)\rightarrow\infty$
as $\omega\rightarrow0$. For case (i) and (ii-b) both figures show
that the local autocovariance and the local periodogram do not display
long memory features. Indeed, they have forms similar to those of
a short memory process, a result consistent with Theorems \ref{Theorem Local ACF Nonstat}-\ref{Theorem Local Periodogram Long Memory Effects}
also for cases (i) and (ii-b). 

It is noteworthy to explain why HAR inference based on the DK-HAC
estimators does not suffer from the low frequency contamination even
for case (ii-a). The DK-HAC estimator computes an average of the local
spectral density over time blocks. If one of these blocks contains
a discontinuity in the spectrum, then as in case (ii-a) some bias
would arise for the local spectral density estimate corresponding
to that block. However, by virtue of the time-averaging over blocks
that bias becomes negligible. Hence, nonparametric smoothing over
time asymptotically cancels the bias, so that inference based on the
DK-HAC estimators is robust to nonstationarity. 

\subsection{\label{Subsec: Theoretical-Results-on Power}Theoretical Results
about the Power}

We present theoretical results about the power of $t_{\mathrm{DM}}$
for the case of general low frequency contamination discussed in Section
\ref{Subsec General-Low-Frequency}. In particular, we focus on specification
(1) (i.e., $\delta>0$). The same intuition and qualitative theoretical
results apply to the other specifications of $\delta\left(\cdot\right)$.

Let $t_{\mathrm{DM},i}=T_{n}^{1/2}\overline{d}_{L}/\sqrt{\widehat{J}_{d_{L},i,T}}$
denote the DM test statistic where $i=\mathrm{DK},\,\mathrm{pwDK},\,\mathrm{KVB},\,\mathrm{EWC},$
$\mathrm{A91},\,\mathrm{pwA}91$, $\mathrm{NW87}$ and $\mathrm{pwNW87}$
with $\widehat{J}_{\mathrm{A91},T}$ and $\widehat{J}_{\mathrm{NW87},T}$
being $\widehat{J}_{\mathrm{HAC},T}$ using the quadratic spectral
and Bartlett kernel, respectively. Define the power of $t_{\mathrm{DM},i}$
as $\mathbb{P}_{\delta}(|t_{\mathrm{DM},i}|>z_{1-\alpha/2})$ where
$z_{1-\alpha/2}$ is the $1-\alpha/2$ quantile of the standard normal
for a two-sided test with significance level $\alpha\in\left(0,\,1\right)$.
To avoid repetitions we present the results only for $i=\mathrm{DK},\,\mathrm{KVB}$
and $\mathrm{NW87}$. The results concerning the prewhitening DK-HAC
estimator are the same as those corresponding to the DK-HAC estimator
while the results concerning the EWC estimator are similar to those
corresponding to the KVB's fixed-$b$ estimator, though for the latter
the non-monotonic power is more pronounced. The results pertaining
to \citeauthor{andrews:91}' \citeyearpar{andrews:91} HAC estimator
(with and without prewhitening) are the same as those corresponding
to \citeauthor{newey/west:87}'s \citeyearpar{newey/west:87} estimator.
Let $n_{\delta}=T-T_{b}-2$ denote the length of the regime in which
$x_{t}^{(2)}$ exhibits a shift $\delta$ in the mean. The deviation
from the null hypothesis depends on the shift magnitude $\delta$
and on $n_{\delta}$.
\begin{thm}
\label{Theorem Power DM HAR Tests}Let $\left\{ d_{t}-\mathbb{E}(d_{t})\right\} _{t=1}^{T_{n}}$
be an SLS process satisfying Assumption \ref{Assumption Smothness of A (for HAC)}-(i-iv)
and \ref{Assumption A - Dependence}. Let Assumptions \ref{Assumption 3 VR}-\ref{Assumption 4 in VR}
hold and $n_{\delta}=O(T_{n}^{1/2+\zeta})$ where $\zeta\in\left(0,\,1/2\right)$
such that $T_{n}^{\zeta}b_{1,T}^{1/2}\rightarrow0$ and $T_{n}^{\zeta}(\widehat{b}_{1,T})^{1/2}\rightarrow0$.
Then, we have:

(i) Under Assumption \ref{Assumption 6 VR}, $\mathbb{P}_{\delta}(|t_{\mathrm{DM},\mathrm{NW87}}|>z_{\alpha})\rightarrow0$$.$
If Assumption \ref{Assumption 6 VR} is replaced by Assumption \ref{Assumption 7 VR}
with $q=1/3$, then $|t_{\mathrm{DM},\mathrm{NW87}}|=O_{\mathbb{P}}(T_{n}^{\zeta-1/6})$
and $\mathbb{P}_{\delta}(|t_{\mathrm{DM},\mathrm{NW87}}|>z_{\alpha})$$\rightarrow0.$ 

(ii) If $b_{1,T}=T^{-1}$, then $|t_{\mathrm{DM},\mathrm{KVB}}|=O_{\mathbb{P}}(T_{n}^{\zeta-1/2})$
and $\mathbb{P}_{\delta}(|t_{\mathrm{DM},\mathrm{KVB}}|>z_{\alpha})$$\rightarrow0.$

(iii) Under Assumption \ref{Assumption K2 and b2}, $|t_{\mathrm{DM},\mathrm{DK}}|=\delta^{2}O_{\mathbb{P}}(T_{n}^{\zeta})$
and $\mathbb{P}_{\delta}(|t_{\mathrm{DM},\mathrm{DK}}|>z_{\alpha})$$\rightarrow1$.
\end{thm}
 Note that Assumption \ref{Assumption 7 VR} with $q=1/3$ refers
to the MSE-optimal bandwidth for the \citeauthor{newey/west:87}'s
\citeyearpar{newey/west:87} estimator. The conditions $T_{n}^{\zeta}b_{1,T}^{1/2}\rightarrow0$
and $T_{n}^{\zeta}(\widehat{b}_{1,T})^{1/2}\rightarrow0$ mean that
the length of the regime in which $x_{t}^{(2)}$ exhibits a shift
$\delta$ in the mean increases to infinity at a slower rate than
$T$. Theorem \ref{Theorem Power DM HAR Tests} shows that when the
HAC estimators or the fixed-$b$ LRV estimators are used, the DM test
is not consistent and its power approaches zero. The theorem also
implies that the power functions corresponding to tests based on HAC
estimators lie above the power functions corresponding to those based
on fixed-$b$/EWC LRV estimators. This follows from $|t_{\mathrm{DM},\mathrm{KVB}}|\ll|t_{\mathrm{DM},\mathrm{NW87}}|.$
Another interesting feature is that $|t_{\mathrm{DM},\mathrm{NW87}}|$
and $|t_{\mathrm{DM},\mathrm{KVB}}|$ do not increase in magnitude
with $\delta$ because $\delta$ appears in both the numerator and
denominator ($\delta$ enters the denominator through the low frequency
contamination term $d^{*}$ that accounts for the bias in the HAC
and fixed-$b$ estimators (cf. Theorem \ref{Theorem ACF Nonstat})).
Part (iii) of the theorem suggests that these issues do not occur
when the DK-HAC estimator is used since the test is consistent and
its power increases with $\delta$ and with the sample size as it
should be. These results match the empirical results in Table \ref{Table Power DM Test}
discussed above, thereby confirming the relevance of Theorem \ref{Theorem Power DM HAR Tests}.

\section{\label{Section Conclusions}Conclusions}

Economic time series often display nonstationary features that are
usefully addressed in testing by allowing for some misspecification
in standard model formulations. If nonstationarity is not accounted
for properly, parameter estimates and, in particular, asymptotic LRV
estimates can be largely biased. We establish results on the low
frequency contamination induced by nonstationarity and misspecification
for the sample autocovariance and the periodogram under general conditions.
These estimates can exhibit features akin to long memory when the
data are nonstationary short memory. We show, using theoretical arguments,
that nonparametric smoothing is robust. Since the autocovariances
and the periodogram are basic elements for HAR inference, our results
allow a  better understanding of  LRV estimation. Under the
null hypothesis there are larger size distortions than when the data
are stationary. Under the alternative hypothesis, existing LRV estimators
tend to be inflated and HAR tests can exhibit dramatic power losses.
Long bandwidths/fixed-$b$ HAR tests suffer more from low frequency
contamination relative to HAR tests based on HAC estimators, whereas
the DK-HAC estimators do not suffer from this problem.

\section*{Supplemental Materials}

Casini, A., T. Deng and P. Perron (2024): Supplement to ``Theory
of low frequency contamination from nonstationarity and misspecification:
consequences for HAR inference\textquotedbl , Econometric Theory
Supplementary Material.  

\newpage{}

\bibliographystyle{elsarticle-harv}
\bibliography{References_JoE}

\addcontentsline{toc}{section}{References}

\newpage{}

\clearpage 
\appendix


\setlength{\belowcaptionskip}{-0pt}

\section{Appendix}

\begin{singlespace}

\end{singlespace}

\pagebreak{}

\newpage{}

\thispagestyle{empty}

\begin{table}[H]
\caption{{\small{}\label{Table T=00003D100, 200, 1000}Average estimates of
$\widehat{J}_{\mathrm{HAC}},$ $\widehat{c}_{1}$ and $\widehat{\Gamma}\left(k\right)$,
$k=0,\,1,\,5,\,10$}}

\centering{}{\footnotesize{}}%
\begin{tabular}{ccccccccccccc}
\hline 
\multicolumn{13}{c}{{\footnotesize{}$T=100$}}\tabularnewline
\hline 
{\footnotesize{}DGP} & {\footnotesize{}$J$} & {\footnotesize{}$\widehat{J}_{\mathrm{HAC}}$} & {\footnotesize{}$-\overline{c}_{1}b_{1,T}$} & {\footnotesize{}$-\widehat{c}_{1}b_{1,T}$} & {\footnotesize{}$\Gamma_{T}\left(0\right)$} & {\footnotesize{}$\widehat{\Gamma}\left(0\right)$} & {\footnotesize{}$\Gamma_{T}\left(1\right)$} & {\footnotesize{}$\widehat{\Gamma}\left(1\right)$} & {\footnotesize{}$\Gamma_{T}\left(5\right)$} & {\footnotesize{}$\widehat{\Gamma}\left(5\right)$} & {\footnotesize{}$\Gamma_{T}\left(10\right)$} & {\footnotesize{}$\widehat{\Gamma}\left(10\right)$}\tabularnewline
\hline 
\hline 
{\footnotesize{}1(a)} & {\footnotesize{}1.826} & {\footnotesize{}1.483} & {\footnotesize{}-0.138} & {\footnotesize{}-0.169} & {\footnotesize{}1.072} & {\footnotesize{}1.062} & {\footnotesize{}0.279} & {\footnotesize{}0.273} & {\footnotesize{}0.001} & {\footnotesize{}0.002} & {\footnotesize{}0.000} & {\footnotesize{}0.000}\tabularnewline
{\footnotesize{}1(b)} & {\footnotesize{}1.826} & {\footnotesize{}1.499} & {\footnotesize{}-0.138} & {\footnotesize{}-0.165} & {\footnotesize{}1.072} & {\footnotesize{}1.072} & {\footnotesize{}0.279} & {\footnotesize{}0.276} & {\footnotesize{}0.001} & {\footnotesize{}0.001} & {\footnotesize{}0.000} & {\footnotesize{}0.000}\tabularnewline
{\footnotesize{}1(c)} & {\footnotesize{}1.826} & {\footnotesize{}1.549} & {\footnotesize{}-0.138} & {\footnotesize{}-0.160} & {\footnotesize{}1.072} & {\footnotesize{}1.105} & {\footnotesize{}0.279} & {\footnotesize{}0.285} & {\footnotesize{}0.001} & {\footnotesize{}0.001} & {\footnotesize{}0.000} & {\footnotesize{}0.000}\tabularnewline
{\footnotesize{}2(a)} & {\footnotesize{}20.988} & {\footnotesize{}6.755} & {\footnotesize{}-0.980} & {\footnotesize{}-2.685} & {\footnotesize{}2.571} & {\footnotesize{}2.507} & {\footnotesize{}2.009} & {\footnotesize{}1.940} & {\footnotesize{}0.751} & {\footnotesize{}0.696} & {\footnotesize{}0.219} & {\footnotesize{}0.195}\tabularnewline
{\footnotesize{}2(b)} & {\footnotesize{}20.988} & {\footnotesize{}6.830} & {\footnotesize{}-0.980} & {\footnotesize{}-2.617} & {\footnotesize{}2.571} & {\footnotesize{}2.533} & {\footnotesize{}2.009} & {\footnotesize{}1.961} & {\footnotesize{}0.751} & {\footnotesize{}0.702} & {\footnotesize{}0.219} & {\footnotesize{}0.195}\tabularnewline
{\footnotesize{}2(c)} & {\footnotesize{}20.988} & {\footnotesize{}7.038} & {\footnotesize{}-0.980} & {\footnotesize{}-2.622} & {\footnotesize{}2.571} & {\footnotesize{}2.609} & {\footnotesize{}2.009} & {\footnotesize{}2.019} & {\footnotesize{}0.751} & {\footnotesize{}0.725} & {\footnotesize{}0.219} & {\footnotesize{}0.206}\tabularnewline
{\footnotesize{}3(a)} & {\footnotesize{}20.988} & {\footnotesize{}3.142} & {\footnotesize{}-2.260} & {\footnotesize{}-40.480} & {\footnotesize{}1.861} & {\footnotesize{}1.589} & {\footnotesize{}1.028} & {\footnotesize{}0.736} & {\footnotesize{}0.622} & {\footnotesize{}0.312} & {\footnotesize{}0.367} & {\footnotesize{}0.100}\tabularnewline
{\footnotesize{}3(b)} & {\footnotesize{}20.988} & {\footnotesize{}3.301} & {\footnotesize{}-2.260} & {\footnotesize{}-38.312} & {\footnotesize{}1.861} & {\footnotesize{}1.635} & {\footnotesize{}1.028} & {\footnotesize{}0.781} & {\footnotesize{}0.622} & {\footnotesize{}0.338} & {\footnotesize{}0.367} & {\footnotesize{}0.113}\tabularnewline
{\footnotesize{}3(c)} & {\footnotesize{}20.988} & {\footnotesize{}3.761} & {\footnotesize{}-2.260} & {\footnotesize{}-35.695} & {\footnotesize{}1.861} & {\footnotesize{}1.790} & {\footnotesize{}1.028} & {\footnotesize{}0.920} & {\footnotesize{}0.622} & {\footnotesize{}0.427} & {\footnotesize{}0.367} & {\footnotesize{}0.161}\tabularnewline
{\footnotesize{}4(a)} & {\footnotesize{}20.988} & {\footnotesize{}3.437} & {\footnotesize{}-2.260} & {\footnotesize{}-37.756} & {\footnotesize{}1.861} & {\footnotesize{}1.670} & {\footnotesize{}1.028} & {\footnotesize{}0.829} & {\footnotesize{}0.622} & {\footnotesize{}0.373} & {\footnotesize{}0.367} & {\footnotesize{}0.133}\tabularnewline
{\footnotesize{}4(b)} & {\footnotesize{}20.988} & {\footnotesize{}3.448} & {\footnotesize{}-2.260} & {\footnotesize{}-37.145} & {\footnotesize{}1.861} & {\footnotesize{}1.680} & {\footnotesize{}1.028} & {\footnotesize{}0.830} & {\footnotesize{}0.622} & {\footnotesize{}0.373} & {\footnotesize{}0.367} & {\footnotesize{}0.134}\tabularnewline
{\footnotesize{}4(c)} & {\footnotesize{}20.988} & {\footnotesize{}3.472} & {\footnotesize{}-2.260} & {\footnotesize{}-35.472} & {\footnotesize{}1.861} & {\footnotesize{}1.711} & {\footnotesize{}1.028} & {\footnotesize{}0.834} & {\footnotesize{}0.622} & {\footnotesize{}0.373} & {\footnotesize{}0.367} & {\footnotesize{}0.134}\tabularnewline
\hline 
\multicolumn{13}{c}{{\footnotesize{}$T=200$}}\tabularnewline
\hline 
{\footnotesize{}DGP} & {\footnotesize{}$J$} & {\footnotesize{}$\widehat{J}_{\mathrm{HAC}}$} & {\footnotesize{}$-\overline{c}_{1}b_{1,T}$} & {\footnotesize{}$-\widehat{c}_{1}b_{1,T}$} & {\footnotesize{}$\Gamma_{T}\left(0\right)$} & {\footnotesize{}$\widehat{\Gamma}\left(0\right)$} & {\footnotesize{}$\Gamma_{T}\left(1\right)$} & {\footnotesize{}$\widehat{\Gamma}\left(1\right)$} & {\footnotesize{}$\Gamma_{T}\left(5\right)$} & {\footnotesize{}$\widehat{\Gamma}\left(5\right)$} & {\footnotesize{}$\Gamma_{T}\left(10\right)$} & {\footnotesize{}$\widehat{\Gamma}\left(10\right)$}\tabularnewline
\hline 
\hline 
{\footnotesize{}1(a)} & {\footnotesize{}1.826} & {\footnotesize{}1.569} & {\footnotesize{}-0.110} & {\footnotesize{}-0.127} & {\footnotesize{}1.072} & {\footnotesize{}1.067} & {\footnotesize{}0.279} & {\footnotesize{}0.276} & {\footnotesize{}0.001} & {\footnotesize{}0.001} & {\footnotesize{}0.000} & {\footnotesize{}0.000}\tabularnewline
{\footnotesize{}1(b)} & {\footnotesize{}1.826} & {\footnotesize{}1.577} & {\footnotesize{}-0.110} & {\footnotesize{}-0.128} & {\footnotesize{}1.072} & {\footnotesize{}1.071} & {\footnotesize{}0.279} & {\footnotesize{}0.277} & {\footnotesize{}0.001} & {\footnotesize{}0.001} & {\footnotesize{}0.000} & {\footnotesize{}0.000}\tabularnewline
{\footnotesize{}1(c)} & {\footnotesize{}1.826} & {\footnotesize{}1.602} & {\footnotesize{}-0.110} & {\footnotesize{}-0.124} & {\footnotesize{}1.072} & {\footnotesize{}1.089} & {\footnotesize{}0.279} & {\footnotesize{}0.281} & {\footnotesize{}0.001} & {\footnotesize{}0.001} & {\footnotesize{}0.000} & {\footnotesize{}0.000}\tabularnewline
{\footnotesize{}2(a)} & {\footnotesize{}20.988} & {\footnotesize{}8.388} & {\footnotesize{}-0.784} & {\footnotesize{}-1.862} & {\footnotesize{}2.571} & {\footnotesize{}2.539} & {\footnotesize{}2.009} & {\footnotesize{}1.975} & {\footnotesize{}0.751} & {\footnotesize{}0.722} & {\footnotesize{}0.219} & {\footnotesize{}0.207}\tabularnewline
{\footnotesize{}2(b)} & {\footnotesize{}20.988} & {\footnotesize{}8.449} & {\footnotesize{}-0.784} & {\footnotesize{}-1.839} & {\footnotesize{}2.571} & {\footnotesize{}2.553} & {\footnotesize{}2.009} & {\footnotesize{}1.988} & {\footnotesize{}0.751} & {\footnotesize{}0.728} & {\footnotesize{}0.219} & {\footnotesize{}0.207}\tabularnewline
{\footnotesize{}2(c)} & {\footnotesize{}20.988} & {\footnotesize{}8.555} & {\footnotesize{}-0.784} & {\footnotesize{}-1.821} & {\footnotesize{}2.571} & {\footnotesize{}2.588} & {\footnotesize{}2.009} & {\footnotesize{}2.013} & {\footnotesize{}0.751} & {\footnotesize{}0.737} & {\footnotesize{}0.219} & {\footnotesize{}0.211}\tabularnewline
{\footnotesize{}3(a)} & {\footnotesize{}20.988} & {\footnotesize{}4.354} & {\footnotesize{}-1.808} & {\footnotesize{}-30.914} & {\footnotesize{}1.861} & {\footnotesize{}1.723} & {\footnotesize{}1.028} & {\footnotesize{}0.883} & {\footnotesize{}0.622} & {\footnotesize{}0.465} & {\footnotesize{}0.367} & {\footnotesize{}0.229}\tabularnewline
{\footnotesize{}3(b)} & {\footnotesize{}20.988} & {\footnotesize{}4.459} & {\footnotesize{}-1.808} & {\footnotesize{}-30.284} & {\footnotesize{}1.861} & {\footnotesize{}1.749} & {\footnotesize{}1.028} & {\footnotesize{}0.903} & {\footnotesize{}0.622} & {\footnotesize{}0.479} & {\footnotesize{}0.367} & {\footnotesize{}0.237}\tabularnewline
{\footnotesize{}3(c)} & {\footnotesize{}20.988} & {\footnotesize{}4.771} & {\footnotesize{}-1.808} & {\footnotesize{}-30.321} & {\footnotesize{}1.861} & {\footnotesize{}1.823} & {\footnotesize{}1.028} & {\footnotesize{}0.978} & {\footnotesize{}0.622} & {\footnotesize{}0.526} & {\footnotesize{}0.367} & {\footnotesize{}0.265}\tabularnewline
{\footnotesize{}4(a)} & {\footnotesize{}20.988} & {\footnotesize{}4.548} & {\footnotesize{}-1.808} & {\footnotesize{}-28.901} & {\footnotesize{}1.861} & {\footnotesize{}1.766} & {\footnotesize{}1.028} & {\footnotesize{}0.929} & {\footnotesize{}0.622} & {\footnotesize{}0.496} & {\footnotesize{}0.367} & {\footnotesize{}0.247}\tabularnewline
{\footnotesize{}4(b)} & {\footnotesize{}20.988} & {\footnotesize{}4.552} & {\footnotesize{}-1.808} & {\footnotesize{}-29.944} & {\footnotesize{}1.861} & {\footnotesize{}1.770} & {\footnotesize{}1.028} & {\footnotesize{}0.931} & {\footnotesize{}0.622} & {\footnotesize{}0.496} & {\footnotesize{}0.367} & {\footnotesize{}0.248}\tabularnewline
{\footnotesize{}4(c)} & {\footnotesize{}20.988} & {\footnotesize{}4.569} & {\footnotesize{}-1.808} & {\footnotesize{}-29.132} & {\footnotesize{}1.861} & {\footnotesize{}1.786} & {\footnotesize{}1.028} & {\footnotesize{}0.932} & {\footnotesize{}0.622} & {\footnotesize{}0.499} & {\footnotesize{}0.367} & {\footnotesize{}0.248}\tabularnewline
\hline 
\multicolumn{13}{c}{{\footnotesize{}$T=1000$}}\tabularnewline
\hline 
{\footnotesize{}DGP} & {\footnotesize{}$J$} & {\footnotesize{}$\widehat{J}_{\mathrm{HAC}}$} & $-\overline{c}_{1}b_{1,T}$ & $-\widehat{c}_{1}b_{1,T}$ & {\footnotesize{}$\Gamma_{T}\left(0\right)$} & {\footnotesize{}$\widehat{\Gamma}\left(0\right)$} & {\footnotesize{}$\Gamma_{T}\left(1\right)$} & {\footnotesize{}$\widehat{\Gamma}\left(1\right)$} & {\footnotesize{}$\Gamma_{T}\left(5\right)$} & {\footnotesize{}$\widehat{\Gamma}\left(5\right)$} & {\footnotesize{}$\Gamma_{T}\left(10\right)$} & {\footnotesize{}$\widehat{\Gamma}\left(10\right)$}\tabularnewline
\hline 
\hline 
{\footnotesize{}1(a)} & {\footnotesize{}1.826} & {\footnotesize{}1.667} & {\footnotesize{}-0.079} & {\footnotesize{}-0.088} & {\footnotesize{}1.072} & {\footnotesize{}1.071} & {\footnotesize{}0.279} & {\footnotesize{}0.278} & {\footnotesize{}0.001} & {\footnotesize{}0.001} & {\footnotesize{}0.000} & {\footnotesize{}0.000}\tabularnewline
{\footnotesize{}1(b)} & {\footnotesize{}1.826} & {\footnotesize{}1.669} & {\footnotesize{}-0.079} & {\footnotesize{}-0.087} & {\footnotesize{}1.072} & {\footnotesize{}1.073} & {\footnotesize{}0.279} & {\footnotesize{}0.279} & {\footnotesize{}0.001} & {\footnotesize{}0.000} & {\footnotesize{}0.000} & {\footnotesize{}0.000}\tabularnewline
{\footnotesize{}1(c)} & {\footnotesize{}1.826} & {\footnotesize{}1.673} & {\footnotesize{}-0.079} & {\footnotesize{}-0.087} & {\footnotesize{}1.072} & {\footnotesize{}1.076} & {\footnotesize{}0.279} & {\footnotesize{}0.279} & {\footnotesize{}0.001} & {\footnotesize{}0.002} & {\footnotesize{}0.000} & {\footnotesize{}0.000}\tabularnewline
{\footnotesize{}2(a)} & {\footnotesize{}20.988} & {\footnotesize{}10.904} & {\footnotesize{}-0.560} & {\footnotesize{}-1.097} & {\footnotesize{}2.571} & {\footnotesize{}2.565} & {\footnotesize{}2.009} & {\footnotesize{}2.003} & {\footnotesize{}0.751} & {\footnotesize{}0.743} & {\footnotesize{}0.219} & {\footnotesize{}0.216}\tabularnewline
{\footnotesize{}2(b)} & {\footnotesize{}20.988} & {\footnotesize{}10.934} & {\footnotesize{}-0.560} & {\footnotesize{}-1.084} & {\footnotesize{}2.571} & {\footnotesize{}2.571} & {\footnotesize{}2.009} & {\footnotesize{}2.008} & {\footnotesize{}0.751} & {\footnotesize{}0.749} & {\footnotesize{}0.219} & {\footnotesize{}0.219}\tabularnewline
{\footnotesize{}2(c)} & {\footnotesize{}20.988} & {\footnotesize{}10.935} & {\footnotesize{}-0.560} & {\footnotesize{}-1.084} & {\footnotesize{}2.571} & {\footnotesize{}2.574} & {\footnotesize{}2.009} & {\footnotesize{}2.009} & {\footnotesize{}0.751} & {\footnotesize{}0.746} & {\footnotesize{}0.219} & {\footnotesize{}0.217}\tabularnewline
{\footnotesize{}3(a)} & {\footnotesize{}20.988} & {\footnotesize{}6.510} & {\footnotesize{}-1.291} & {\footnotesize{}-20.845} & {\footnotesize{}1.861} & {\footnotesize{}1.834} & {\footnotesize{}1.028} & {\footnotesize{}1.001} & {\footnotesize{}0.622} & {\footnotesize{}0.592} & {\footnotesize{}0.367} & {\footnotesize{}0.339}\tabularnewline
{\footnotesize{}3(b)} & {\footnotesize{}20.988} & {\footnotesize{}6.541} & {\footnotesize{}-1.291} & {\footnotesize{}-20.449} & {\footnotesize{}1.861} & {\footnotesize{}1.841} & {\footnotesize{}1.028} & {\footnotesize{}1.001} & {\footnotesize{}0.622} & {\footnotesize{}0.595} & {\footnotesize{}0.367} & {\footnotesize{}0.343}\tabularnewline
{\footnotesize{}3(c)} & {\footnotesize{}20.988} & {\footnotesize{}6.629} & {\footnotesize{}-1.291} & {\footnotesize{}-20.475} & {\footnotesize{}1.861} & {\footnotesize{}1.857} & {\footnotesize{}1.028} & {\footnotesize{}1.021} & {\footnotesize{}0.622} & {\footnotesize{}0.605} & {\footnotesize{}0.367} & {\footnotesize{}0.349}\tabularnewline
{\footnotesize{}4(a)} & {\footnotesize{}20.988} & {\footnotesize{}6.543} & {\footnotesize{}-1.291} & {\footnotesize{}-20.854} & {\footnotesize{}1.861} & {\footnotesize{}1.840} & {\footnotesize{}1.028} & {\footnotesize{}0.838} & {\footnotesize{}0.622} & {\footnotesize{}0.595} & {\footnotesize{}0.367} & {\footnotesize{}0.344}\tabularnewline
{\footnotesize{}4(b)} & {\footnotesize{}20.988} & {\footnotesize{}6.555} & {\footnotesize{}-1.291} & {\footnotesize{}-20.361} & {\footnotesize{}1.861} & {\footnotesize{}1.843} & {\footnotesize{}1.028} & {\footnotesize{}1.009} & {\footnotesize{}0.622} & {\footnotesize{}0.598} & {\footnotesize{}0.367} & {\footnotesize{}0.347}\tabularnewline
{\footnotesize{}4(c)} & {\footnotesize{}20.988} & {\footnotesize{}6.559} & {\footnotesize{}-1.291} & {\footnotesize{}-20.551} & {\footnotesize{}1.861} & {\footnotesize{}1.846} & {\footnotesize{}1.028} & {\footnotesize{}1.011} & {\footnotesize{}0.622} & {\footnotesize{}0.598} & {\footnotesize{}0.367} & {\footnotesize{}0.347}\tabularnewline
\hline 
\end{tabular}{\footnotesize\par}
\end{table}

\begin{table}[H]
\caption{\label{Table S1 - Figure 1}Empirical small-sample null rejection
rates and power of $t$-test for model M1-M4}

\begin{centering}
{\footnotesize{}}%
\begin{tabular}{lccccc}
\multicolumn{6}{c}{{\footnotesize{}M1}}\tabularnewline
{\footnotesize{}$\alpha=0.05,\,T=200$} & {\footnotesize{}$\delta=0$ (null rejection)} & {\footnotesize{}$\delta=0.05$} & {\footnotesize{}$\delta=0.1$} & {\footnotesize{}$\delta=0.25$} & {\footnotesize{}$\delta=1.5$}\tabularnewline
\hline 
\hline 
{\footnotesize{}$\widehat{J}_{\mathrm{DK,}T}$} & {\footnotesize{}0.068} & {\footnotesize{}0.189} & {\footnotesize{}0.286} & {\footnotesize{}0.661} & {\footnotesize{}1.000}\tabularnewline
{\footnotesize{}$\widehat{J}_{\mathrm{DK,pw},\mathrm{SLS},T}$ } & {\footnotesize{}0.045} & {\footnotesize{}0.085} & {\footnotesize{}0.199} & {\footnotesize{}0.612} & {\footnotesize{}1.000}\tabularnewline
{\footnotesize{}$\widehat{J}_{\mathrm{DK,pw},\mathrm{SLS},\mu,T}$} & {\footnotesize{}0.046} & {\footnotesize{}0.090} & {\footnotesize{}0.202} & {\footnotesize{}0.613} & {\footnotesize{}1.000}\tabularnewline
{\footnotesize{}Andrews (1991)} & {\footnotesize{}0.039} & {\footnotesize{}0.095} & {\footnotesize{}0.185} & {\footnotesize{}0.623} & {\footnotesize{}0.999}\tabularnewline
{\footnotesize{}Andrews (1991), prewhite} & {\footnotesize{}0.115} & {\footnotesize{}0.168} & {\footnotesize{}0.304} & {\footnotesize{}0.650} & {\footnotesize{}0.999}\tabularnewline
{\footnotesize{}Newey-West (1987)} & {\footnotesize{}0.209} & {\footnotesize{}0.272} & {\footnotesize{}0.398} & {\footnotesize{}0.689} & {\footnotesize{}1.000}\tabularnewline
{\footnotesize{}KVB fixed-$b$} & {\footnotesize{}0.004} & {\footnotesize{}0.018} & {\footnotesize{}0.063} & {\footnotesize{}0.301} & {\footnotesize{}0.969}\tabularnewline
{\footnotesize{}EWC} & {\footnotesize{}0.011} & {\footnotesize{}0.038} & {\footnotesize{}0.137} & {\footnotesize{}0.539} & {\footnotesize{}0.999}\tabularnewline
\hline 
\multicolumn{6}{c}{{\footnotesize{}M2}}\tabularnewline
{\footnotesize{}$\alpha=0.05,\,T=200$} & {\footnotesize{}$\delta=0$ (null rejection)} & {\footnotesize{}$\delta=0.05$} & {\footnotesize{}$\delta=0.1$} & {\footnotesize{}$\delta=0.3$} & {\footnotesize{}$\delta=1$}\tabularnewline
\hline 
\hline 
{\footnotesize{}$\widehat{J}_{\mathrm{DK,}T}$} & {\footnotesize{}0.080} & {\footnotesize{}0.132} & {\footnotesize{}0.257} & {\footnotesize{}0.842} & {\footnotesize{}1.000}\tabularnewline
{\footnotesize{}$\widehat{J}_{\mathrm{DK,pw},\mathrm{SLS},T}$} & {\footnotesize{}0.059} & {\footnotesize{}0.098} & {\footnotesize{}0.190} & {\footnotesize{}0.736} & {\footnotesize{}1.000}\tabularnewline
{\footnotesize{}$\widehat{J}_{\mathrm{DK,pw},\mathrm{SLS},\mu,T}$} & {\footnotesize{}0.055} & {\footnotesize{}0.088} & {\footnotesize{}0.187} & {\footnotesize{}0.735} & {\footnotesize{}1.000}\tabularnewline
{\footnotesize{}Andrews (1991)} & {\footnotesize{}0.081} & {\footnotesize{}0.133} & {\footnotesize{}0.266} & {\footnotesize{}0.838} & {\footnotesize{}1.000}\tabularnewline
{\footnotesize{}Andrews (1991), prewhite} & {\footnotesize{}0.094} & {\footnotesize{}0.141} & {\footnotesize{}0.268} & {\footnotesize{}0.842} & {\footnotesize{}1.000}\tabularnewline
{\footnotesize{}Newey-West (1987)} & {\footnotesize{}0.137} & {\footnotesize{}0.190} & {\footnotesize{}0.336} & {\footnotesize{}0.881} & {\footnotesize{}1.000}\tabularnewline
{\footnotesize{}KVB fixed-$b$} & {\footnotesize{}0.014} & {\footnotesize{}0.036} & {\footnotesize{}0.078} & {\footnotesize{}0.561} & {\footnotesize{}0.990}\tabularnewline
{\footnotesize{}EWC} & {\footnotesize{}0.032} & {\footnotesize{}0.064} & {\footnotesize{}0.157} & {\footnotesize{}0.712} & {\footnotesize{}1.000}\tabularnewline
\hline 
\multicolumn{6}{c}{{\footnotesize{}M3}}\tabularnewline
{\footnotesize{}$\alpha=0.05,\,T=200$} & {\footnotesize{}$\delta=0$ (null rejection)} & {\footnotesize{}$\delta=0.1$} & {\footnotesize{}$\delta=0.15$} & {\footnotesize{}$\delta=0.3$} & {\footnotesize{}$\delta=1$}\tabularnewline
\hline 
\hline 
{\footnotesize{}$\widehat{J}_{\mathrm{DK,}T}$} & {\footnotesize{}0.117} & {\footnotesize{}0.363} & {\footnotesize{}0.537} & {\footnotesize{}0.928} & {\footnotesize{}1.000}\tabularnewline
{\footnotesize{}$\widehat{J}_{\mathrm{DK,pw},\mathrm{SLS},T}$} & {\footnotesize{}0.049} & {\footnotesize{}0.227} & {\footnotesize{}0.384} & {\footnotesize{}0.865} & {\footnotesize{}1.000}\tabularnewline
{\footnotesize{}$\widehat{J}_{\mathrm{DK,pw},\mathrm{SLS},\mu,T}$} & {\footnotesize{}0.052} & {\footnotesize{}0.223} & {\footnotesize{}0.374} & {\footnotesize{}0.855} & {\footnotesize{}1.000}\tabularnewline
{\footnotesize{}Andrews (1991)} & {\footnotesize{}0.106} & {\footnotesize{}0.334} & {\footnotesize{}0.515} & {\footnotesize{}0.917} & {\footnotesize{}1.000}\tabularnewline
{\footnotesize{}Andrews (1991), prewhite} & {\footnotesize{}0.122} & {\footnotesize{}0.351} & {\footnotesize{}0.524} & {\footnotesize{}0.928} & {\footnotesize{}1.000}\tabularnewline
{\footnotesize{}Newey-West (1987)} & {\footnotesize{}0.169} & {\footnotesize{}0.412} & {\footnotesize{}0.596} & {\footnotesize{}0.948} & {\footnotesize{}1.000}\tabularnewline
{\footnotesize{}KVB fixed-$b$} & {\footnotesize{}0.024} & {\footnotesize{}0.165} & {\footnotesize{}0.309} & {\footnotesize{}0.712} & {\footnotesize{}0.999}\tabularnewline
{\footnotesize{}EWC} & {\footnotesize{}0.058} & {\footnotesize{}0.245} & {\footnotesize{}0.400} & {\footnotesize{}0.858} & {\footnotesize{}1.000}\tabularnewline
\hline 
\multicolumn{6}{c}{{\footnotesize{}M4}}\tabularnewline
{\footnotesize{}$\alpha=0.05,\,T=200$} & {\footnotesize{}$\delta=0$ (null rejection)} & {\footnotesize{}$\delta=0.1$} & {\footnotesize{}$\delta=0.3$} & {\footnotesize{}$\delta=0.5$} & {\footnotesize{}$\delta=3$}\tabularnewline
\hline 
\hline 
{\footnotesize{}$\widehat{J}_{\mathrm{DK,}T}$} & {\footnotesize{}0.154} & {\footnotesize{}0.146} & {\footnotesize{}0.496} & {\footnotesize{}0.706} & {\footnotesize{}1.000}\tabularnewline
{\footnotesize{}$\widehat{J}_{\mathrm{DK,pw},\mathrm{SLS},T}$} & {\footnotesize{}0.037} & {\footnotesize{}0.050} & {\footnotesize{}0.168} & {\footnotesize{}0.459} & {\footnotesize{}1.000}\tabularnewline
{\footnotesize{}$\widehat{J}_{\mathrm{DK,pw},\mathrm{SLS},\mu,T}$} & {\footnotesize{}0.041} & {\footnotesize{}0.079} & {\footnotesize{}0.198} & {\footnotesize{}0.477} & {\footnotesize{}1.000}\tabularnewline
{\footnotesize{}Andrews (1991)} & {\footnotesize{}0.127} & {\footnotesize{}0.162} & {\footnotesize{}0.398} & {\footnotesize{}0.623} & {\footnotesize{}0.999}\tabularnewline
{\footnotesize{}Andrews (1991), prewhite} & {\footnotesize{}0.197} & {\footnotesize{}0.226} & {\footnotesize{}0.439} & {\footnotesize{}0.653} & {\footnotesize{}1.000}\tabularnewline
{\footnotesize{}Newey-West (1987)} & {\footnotesize{}0.397} & {\footnotesize{}0.423} & {\footnotesize{}0.584} & {\footnotesize{}0.758} & {\footnotesize{}1.000}\tabularnewline
{\footnotesize{}KVB fixed-$b$} & {\footnotesize{}0.005} & {\footnotesize{}0.012} & {\footnotesize{}0.135} & {\footnotesize{}0.339} & {\footnotesize{}0.964}\tabularnewline
{\footnotesize{}EWC} & {\footnotesize{}0.115} & {\footnotesize{}0.147} & {\footnotesize{}0.367} & {\footnotesize{}0.681} & {\footnotesize{}0.999}\tabularnewline
\hline 
\end{tabular}{\footnotesize\par}
\par\end{centering}
\end{table}

\begin{table}
\caption{\label{Table Power DM Test}Empirical small-sample null rejection
rates and power of the DM (1995) test}

\begin{centering}
{\footnotesize{}}%
\begin{tabular}{lcccccc}
 &  & \multicolumn{5}{c}{{\footnotesize{}(1) $\delta>0$}}\tabularnewline
{\footnotesize{}$\alpha=0.05,\,T=200$} & {\footnotesize{}(null rejection)} & {\footnotesize{}$\delta=0.2$} & {\footnotesize{}$\delta=0.5$} & {\footnotesize{}$\delta=2$} & {\footnotesize{}$\delta=5$} & {\footnotesize{}$\delta=10$}\tabularnewline
\hline 
\hline 
{\footnotesize{}$\widehat{J}_{\mathrm{DK,}T}$} & {\footnotesize{}0.033} & {\footnotesize{}0.312} & {\footnotesize{}0.551} & {\footnotesize{}0.997} & {\footnotesize{}1.000} & {\footnotesize{}1.000}\tabularnewline
{\footnotesize{}$\widehat{J}_{\mathrm{DK,pw},\mathrm{SLS},T}$} & {\footnotesize{}0.042} & {\footnotesize{}0.322} & {\footnotesize{}0.563} & {\footnotesize{}0.999} & {\footnotesize{}1.000} & {\footnotesize{}1.000}\tabularnewline
{\footnotesize{}$\widehat{J}_{\mathrm{DK,pw},\mathrm{SLS},\mu,T}$} & {\footnotesize{}0.046} & {\footnotesize{}0.348} & {\footnotesize{}0.573} & {\footnotesize{}0.998} & {\footnotesize{}1.000} & {\footnotesize{}1.000}\tabularnewline
{\footnotesize{}Andrews (1991)} & {\footnotesize{}0.085} & {\footnotesize{}0.254} & {\footnotesize{}0.305} & {\footnotesize{}0.114} & {\footnotesize{}0.000} & {\footnotesize{}0.000}\tabularnewline
{\footnotesize{}Andrews (1991), prewhite} & {\footnotesize{}0.085} & {\footnotesize{}0.246} & {\footnotesize{}0.293} & {\footnotesize{}0.401} & {\footnotesize{}0.045} & {\footnotesize{}0.000}\tabularnewline
{\footnotesize{}Newey-West (1987)} & {\footnotesize{}0.083} & {\footnotesize{}0.246} & {\footnotesize{}0.299} & {\footnotesize{}0.612} & {\footnotesize{}0.817} & {\footnotesize{}0.782}\tabularnewline
{\footnotesize{}KVB fixed-$b$} & {\footnotesize{}0.002} & {\footnotesize{}0.212} & {\footnotesize{}0.185} & {\footnotesize{}0.000} & {\footnotesize{}0.000} & {\footnotesize{}0.000}\tabularnewline
{\footnotesize{}EWC} & {\footnotesize{}0.083} & {\footnotesize{}0.252} & {\footnotesize{}0.268} & {\footnotesize{}0.045} & {\footnotesize{}0.000} & {\footnotesize{}0.000}\tabularnewline
\hline 
 &  & \multicolumn{5}{c}{{\footnotesize{}(2) $\delta\left(t/T\right)$ locally stationary}}\tabularnewline
{\footnotesize{}$\alpha=0.05,\,T=200$} &  & {\footnotesize{}$\delta=0.2$} & {\footnotesize{}$\delta=0.5$} & {\footnotesize{}$\delta=2$} & {\footnotesize{}$\delta=5$} & {\footnotesize{}$\delta=10$}\tabularnewline
\hline 
\hline 
{\footnotesize{}$\widehat{J}_{\mathrm{DK,}T}$} &  & {\footnotesize{}0.278} & {\footnotesize{}0.297} & {\footnotesize{}0.592} & {\footnotesize{}0.889} & {\footnotesize{}1.000}\tabularnewline
{\footnotesize{}$\widehat{J}_{\mathrm{DK,pw},\mathrm{SLS},T}$} &  & {\footnotesize{}0.301} & {\footnotesize{}0.363} & {\footnotesize{}0.634} & {\footnotesize{}0969} & {\footnotesize{}1.000}\tabularnewline
{\footnotesize{}$\widehat{J}_{\mathrm{DK,pw},\mathrm{SLS},\mu,T}$} &  & {\footnotesize{}0.327} & {\footnotesize{}0.368} & {\footnotesize{}0.642} & {\footnotesize{}0.969} & {\footnotesize{}1.000}\tabularnewline
{\footnotesize{}Andrews (1991)} &  & {\footnotesize{}0.255} & {\footnotesize{}0.259} & {\footnotesize{}0.255} & {\footnotesize{}0.110} & {\footnotesize{}0.005}\tabularnewline
{\footnotesize{}Andrews (1991), prewhite} &  & {\footnotesize{}0.249} & {\footnotesize{}0.243} & {\footnotesize{}0.268} & {\footnotesize{}0.188} & {\footnotesize{}0.031}\tabularnewline
{\footnotesize{}Newey-West (1987)} &  & {\footnotesize{}0.281} & {\footnotesize{}0.282} & {\footnotesize{}0.313} & {\footnotesize{}0.268} & {\footnotesize{}0.078}\tabularnewline
{\footnotesize{}KVB fixed-$b$} &  & {\footnotesize{}0.203} & {\footnotesize{}0.202} & {\footnotesize{}0.178} & {\footnotesize{}0.025} & {\footnotesize{}0.000}\tabularnewline
{\footnotesize{}EWC} &  & {\footnotesize{}0.244} & {\footnotesize{}0.252} & {\footnotesize{}0.219} & {\footnotesize{}0.045} & {\footnotesize{}0.000}\tabularnewline
\hline 
\multicolumn{2}{l}{} & \multicolumn{5}{c}{{\footnotesize{}(3) $\delta\left(t/T\right)$ segmented locally stationary}}\tabularnewline
{\footnotesize{}$\alpha=0.05,\,T=200$} &  & {\footnotesize{}$\delta=0.2$} & {\footnotesize{}$\delta=1$} & {\footnotesize{}$\delta=2$} & {\footnotesize{}$\delta=5$} & {\footnotesize{}$\delta=10$}\tabularnewline
\hline 
\hline 
{\footnotesize{}$\widehat{J}_{\mathrm{DK,}T}$} &  & {\footnotesize{}0.540} & {\footnotesize{}0.862} & {\footnotesize{}0.992} & {\footnotesize{}1.000} & {\footnotesize{}1.000}\tabularnewline
{\footnotesize{}$\widehat{J}_{\mathrm{DK,pw},\mathrm{SLS},T}$} &  & {\footnotesize{}0.396} & {\footnotesize{}0.664} & {\footnotesize{}0.988} & {\footnotesize{}1.000} & {\footnotesize{}1.000}\tabularnewline
{\footnotesize{}$\widehat{J}_{\mathrm{DK,pw},\mathrm{SLS},\mu,T}$} &  & {\footnotesize{}0.412} & {\footnotesize{}0.724} & {\footnotesize{}0.987} & {\footnotesize{}1.000} & {\footnotesize{}1.000}\tabularnewline
{\footnotesize{}Andrews (1991)} &  & {\footnotesize{}0.328} & {\footnotesize{}0.234} & {\footnotesize{}0.235} & {\footnotesize{}0.241} & {\footnotesize{}0.777}\tabularnewline
{\footnotesize{}Andrews (1991), prewhite} &  & {\footnotesize{}0.342} & {\footnotesize{}0.315} & {\footnotesize{}0.512} & {\footnotesize{}0.296} & {\footnotesize{}0.882}\tabularnewline
{\footnotesize{}Newey-West (1987)} &  & {\footnotesize{}0.381} & {\footnotesize{}0.384} & {\footnotesize{}0.720} & {\footnotesize{}0.972} & {\footnotesize{}0.999}\tabularnewline
{\footnotesize{}KVB fixed-$b$} &  & {\footnotesize{}0.100} & {\footnotesize{}0.032} & {\footnotesize{}0.000} & {\footnotesize{}0.002} & {\footnotesize{}0.040}\tabularnewline
{\footnotesize{}EWC} &  & {\footnotesize{}0.312} & {\footnotesize{}0.152} & {\footnotesize{}0.142} & {\footnotesize{}0.296} & {\footnotesize{}0.852}\tabularnewline
\hline 
 &  & \multicolumn{5}{c}{{\footnotesize{}(4) case (2) with outliers}}\tabularnewline
{\footnotesize{}$\alpha=0.05$, $T=400$} &  & {\footnotesize{}$\delta=0.5$} & {\footnotesize{}$\delta=1$} & {\footnotesize{}$\delta=2$} & {\footnotesize{}$\delta=5$} & {\footnotesize{}$\delta=10$}\tabularnewline
\hline 
\hline 
{\footnotesize{}$\widehat{J}_{\mathrm{DK,}T}$} &  & {\footnotesize{}0.694} & {\footnotesize{}0.733} & {\footnotesize{}0.822} & {\footnotesize{}0.981} & {\footnotesize{}1.000}\tabularnewline
{\footnotesize{}$\widehat{J}_{\mathrm{DK,pw},\mathrm{SLS},T}$} &  & {\footnotesize{}0.724} & {\footnotesize{}0.777} & {\footnotesize{}0.846} & {\footnotesize{}0.982} & {\footnotesize{}1.000}\tabularnewline
{\footnotesize{}$\widehat{J}_{\mathrm{DK,pw},\mathrm{SLS},\mu,T}$} &  & {\footnotesize{}0.727} & {\footnotesize{}0.771} & {\footnotesize{}0.847} & {\footnotesize{}0.981} & {\footnotesize{}1.000}\tabularnewline
{\footnotesize{}Andrews (1991)} &  & {\footnotesize{}0.192} & {\footnotesize{}0.242} & {\footnotesize{}0.245} & {\footnotesize{}0.203} & {\footnotesize{}0.022}\tabularnewline
{\footnotesize{}Andrews (1991), prewhite} &  & {\footnotesize{}0.182} & {\footnotesize{}0.233} & {\footnotesize{}0.243} & {\footnotesize{}0.288} & {\footnotesize{}0.114}\tabularnewline
{\footnotesize{}Newey-West (1987)} &  & {\footnotesize{}0.222} & {\footnotesize{}0.271} & {\footnotesize{}0.245} & {\footnotesize{}0.345} & {\footnotesize{}0.225}\tabularnewline
{\footnotesize{}KVB fixed-$b$} &  & {\footnotesize{}0.203} & {\footnotesize{}0.222} & {\footnotesize{}0.212} & {\footnotesize{}0.075} & {\footnotesize{}0.000}\tabularnewline
{\footnotesize{}EWC} &  & {\footnotesize{}0.186} & {\footnotesize{}0.221} & {\footnotesize{}0.174} & {\footnotesize{}0.062} & {\footnotesize{}0.000}\tabularnewline
\hline 
\end{tabular}{\footnotesize\par}
\par\end{centering}
\end{table}

\clearpage{}

\pagebreak{}

\newpage{}

\pagebreak{}

\section*{}
\addcontentsline{toc}{part}{Supplemental Material}

\begin{center}
\title{\textbf{\Large{Supplement to ``Theory of Low Frequency Contamination from Nonstationarity and Misspecification: Consequences for HAR Inference"}}} 
\maketitle
\end{center}
\medskip{} 
\medskip{} 
\medskip{} 
\thispagestyle{empty}

\begin{center}
\quad  \quad \author{\textsc{\textcolor{MyBlue}{Alessandro Casini}}
\quad \quad \quad  \,
\textsc{\textcolor{MyBlue}{Taosong Deng}}
\quad \quad \quad \quad 
\textsc{\textcolor{MyBlue}{Pierre Perron}}}
\\
\small{University of Rome Tor Vergata} 
\quad \quad \, 
\small{Hunan University}
\quad \quad \quad \quad \, 
\small{Boston University}
\\
\medskip{}
\medskip{} 
\medskip{} 
\medskip{} 
\date{\small{\today}} 
\medskip{} 
\medskip{} 
\medskip{}
\end{center}
\begin{abstract}
{\footnotesize{}This supplemental material is for online publication
only. Section \ref{Section Results on Low Frequency Contamination}
introduces the notion of long memory segmented locally stationary
processes and presents the theoretical results referenced in Section
\ref{Section Low Freq Cont - Theory}. Section \ref{Supp Math App}
contains the proofs of the results in the paper and Section \ref{Section Figures Supp}
contains additional figures. }{\footnotesize\par}
\end{abstract}
\setcounter{page}{0}
\setcounter{section}{0}
\setcounter{equation}{0}
\setcounter{thm}{0}
\setcounter{lem}{0}
\setcounter{defn}{0}
\setcounter{assumption}{0}
\renewcommand*{\theHsection}{\the\value{section}}

\newpage{}

\begin{singlespace} 
\noindent 
\small

\allowdisplaybreaks


\renewcommand{\thepage}{S-\arabic{page}}   
\renewcommand{\thesection}{S.\Alph{section}}   
\renewcommand{\theequation}{S.\arabic{equation}}
\renewcommand{\theassumption}{S.\arabic{assumption}}
\renewcommand{\thethm}{S.\arabic{thm}}
\renewcommand{\theexample}{S.\arabic{example}}
\renewcommand{\thedefn}{S.\arabic{defn}}
\renewcommand{\thelem}{S.\arabic{lem}}




\section{\label{Section Results on Low Frequency Contamination}Results on
Low Frequency Bias for the Sample Autocovariance and the Periodogram}

In Section \ref{Subsection Long-Memory-Segmented-Locally} we define
the long memory SLS processes. In Section \ref{Subsection:  The Sample Autocovariance under Nonstationarity}
and \ref{Subsec The-Periodogram under Nonstationary} we present results
on the low frequency bias for the sample autocovariance and the periodogram,
respectively. 

\subsection{\label{Subsection Long-Memory-Segmented-Locally}Long Memory Segmented
Locally Stationary Processes}

 Define the backward difference operator $\Delta V_{t}=\Delta^{1}V_{t}=V_{t}-V_{t-1}$
and $\Delta^{l}V_{t}$ recursively. Long memory features can be expressed
as a ``pole'' in the spectral density at frequency zero. That is,
for a stationary process, long memory implies that $f\left(\omega\right)\thicksim\omega^{-2\vartheta}$
as $\omega\rightarrow0$ where $\vartheta\in\left(0,\,1/2\right)$
is the  long memory parameter. In what follows, $l$ is some non-negative
integer. 
\begin{defn}
\label{Definition Long Memory Segmented-Locally-Stationary}A sequence
of stochastic processes $\left\{ V_{t,T}\right\} $ is called long
memory segmented locally stationary with $m_{0}+1$ regimes, transfer
function $A^{0}$  and trend $\mu_{\cdot}$ if there exists a representation
\begin{align}
\Delta^{l}V_{t} & =\mu_{j}\left(t/T\right)+\int_{-\pi}^{\pi}\exp\left(i\omega t\right)A_{j,t,T}^{0}\left(\omega\right)d\xi\left(\omega\right),\qquad\qquad\left(t=T_{j-1}^{0}+1,\ldots,\,T_{j}^{0}\right),\label{Eq. Spectral Rep of Long Memory SLS}
\end{align}
for $j=1,\ldots,\,m_{0}+1$, where by convention $T_{0}^{0}=0$ and
$T_{m_{0}+1}^{0}=T$, (i) and (iii) of Definition \ref{Definition Segmented-Locally-Stationary}
hold, and (ii) of Definition \ref{Definition Segmented-Locally-Stationary}
is replaced by

(ii) There exist two constants $L_{2}>0$ and $D<1/2$ (which depend
on $j$) and a piecewise continuous function $A:\,\left[0,\,1\right]\times\mathbb{R}\rightarrow\mathbb{C}$
such that, for each $j=1,\ldots,\,m_{0}+1$, there exists a $2\pi$-periodic
function $A_{j}:\,(\lambda_{j-1}^{0},\,\lambda_{j}^{0}]\times\mathbb{R}\rightarrow\mathbb{C}$
with $A_{j}\left(u,\,-\omega\right)=\overline{A_{j}\left(u,\,\omega\right)}$,
 
\begin{align}
A\left(u,\,\omega\right) & =A_{j}\left(u,\,\omega\right)\,\mathrm{\,for\,}\,\lambda_{j-1}^{0}<u\leq\lambda_{j}^{0},\label{Eq A(u) =00003D Ai Long Memory}\\
\sup_{1\leq j\leq m_{0}+1} & \sup_{T_{j-1}^{0}<t\leq T_{j}^{0},\,\omega}\left|A_{j,t,T}^{0}\left(\omega\right)-A_{j}\left(t/T,\,\omega\right)\right|\leq L_{2}T^{-1}\left|\omega\right|^{-D},\label{Eq. (3) Reouf and von Sasch}
\end{align}
and 
\begin{align}
\sup_{0\leq v\leq u\leq1,\,u\neq\lambda_{j}^{0}\,\left(j=1,\ldots,\,m_{0}+1,\right),\,\omega}\left|A\left(u,\,\omega\right)-A\left(v,\,\omega\right)\right| & \leq L_{2}\left|u-v\right|\left|\omega\right|^{-D}.\label{Eq. (4) Reouf and von Sasch}
\end{align}
 The spectral density of $\{V_{t,T}\}$ is given by $f_{j}\left(u,\,\omega\right)=|1-\exp\left(-i\omega\right)|^{-2l}|A_{j}\left(u,\,\omega\right)|^{-2}$
for $j=1,\ldots,\,m_{0}+1$. We say that the process $\left\{ V_{t,T}\right\} $
has local memory parameter $\vartheta\left(u\right)\in\left(-\infty,\,l+1/2\right)$
at time $u\in\left[0,\,1\right]$ if it satisfies \eqref{Eq. Spectral Rep of Long Memory SLS}-\eqref{Eq. (4) Reouf and von Sasch},
and its generalized spectral density $f_{j}\left(u,\,\omega\right)$
$(j=1,\ldots,$ $m_{0}+1)$ satisfies the following condition, 
\begin{align}
f_{j}\left(u,\,\omega\right) & =\left|1-e^{-i\omega}\right|^{-2\vartheta_{j}\left(u\right)}f_{j}^{*}\left(u,\,\omega\right),\label{Eq. (6) Reuf and von Sachs}
\end{align}
 with $f_{j}^{*}\left(u,\,\omega\right)>0$ and
\begin{align}
\left|f_{j}^{*}\left(u,\,\omega\right)-f_{j}^{*}\left(u,\,0\right)\right| & \leq L_{4}f_{j}^{*}\left(u,\,\omega\right)\left|\omega\right|^{\nu},\qquad\omega\in\left[-\pi,\,\pi\right],\label{Eq. (7) Reuf and von Sachs}
\end{align}
 where $L_{4}>0$ and $\nu\in(0,\,2]$. 
\end{defn}
Definition \ref{Definition Long Memory Segmented-Locally-Stationary}
extends Definition \ref{Definition Segmented-Locally-Stationary}
and Assumption \ref{Assumption Smothness of A (for HAC)} by requiring
the bound on the smoothness of $A\left(\cdot,\,\omega\right)$ to
depend also on $\left|\omega\right|^{-D}$ thereby allowing a singularity
at $\omega=0.$ \citet{casini_hac} showed that $f_{j}\left(u,\,\omega\right)=\left|A_{j}\left(u,\,\omega\right)\right|^{2}$
for $j=1,\ldots,\,m_{0}+1$. Using similar arguments, we obtain the
form $f_{j}\left(u,\,\omega\right)$ given in \eqref{Eq. (6) Reuf and von Sachs}.
See \citeReferencesSupp{roueff/vonsachs:2011} for a definition of
 long memory local stationarity. Definition \ref{Definition Long Memory Segmented-Locally-Stationary}
extends their definition to allow for $m_{0}$ discontinuities. We
have assumed that breaks in the long memory parameter occur at the
same locations as the breaks in the spectrum. This can be relaxed
but would provide no added value in this paper.
\begin{example}
A time-varying AR fractionally integrated moving average $(p,\,\vartheta,\,q)$
process with $m_{0}$ structural breaks satisfies Definition \ref{Definition Long Memory Segmented-Locally-Stationary}
with $\vartheta_{j}:\,\left[0,\,1\right]\rightarrow\left(-\infty,\,l+1/2\right)$,
$\sigma_{j}:\,\left[0,\,1\right]\rightarrow\mathbb{R}_{+}$, $\phi_{j}=\left[\phi_{1},\ldots,\,\phi_{p}\right]':\,\left[0,\,1\right]\rightarrow\mathbb{R}^{q}$
and $\theta_{j}=\left[\theta_{1},\ldots,\,\theta_{q}\right]'\,:\,\left[0,\,1\right]\rightarrow\mathbb{R}^{p}$
are left-Lipschitz functions for each $j=1,\ldots,\,m_{0}+1$ such
that $1-\sum_{k=1}^{p}\phi_{j,k}\left(u\right)z^{k}$ does not vanish
for all $u\in\left[0,\,1\right]$ and $z\in\mathbb{C}$ such that
$\left|z\right|\leq1$. Using the latter condition, the local transfer
function $A_{j}\left(u;\,\cdot\right)$ defines for each $j$ a causal
autoregressive fractionally integrated moving average (ARFIMA$\left(p,\,\vartheta\left(u\right)-l,\,q\right)$
process whose spectral density satisfies the conditions \eqref{Eq. (6) Reuf and von Sachs}
and \eqref{Eq. (7) Reuf and von Sachs} with $\nu=2$. Using Lemma
3 in \citeReferencesSupp{roueff/vonsachs:2011}, condition \eqref{Eq. (4) Reouf and von Sasch}
holds with $D>\sup_{1\leq j\leq m_{0}+1}\sup_{\lambda_{j-1}^{0}<u\leq\lambda_{j}^{0},\,\omega}\vartheta_{j}\left(u\right)-l$. 
\end{example}
Definition \ref{Definition Long Memory Segmented-Locally-Stationary}
implies that $\rho_{V}\left(u,\,k\right)\triangleq\mathrm{Corr}(V_{\left\lfloor Tu\right\rfloor },\,V_{\left\lfloor Tu\right\rfloor +k})\sim Ck^{2\vartheta_{j}\left(u\right)-1}$
for $\lambda_{j-1}^{0}<u<\lambda_{j}^{0}$ and large $k$ where $C>0.$
This means that the rescaled time-$u$ autocorrelation function (ACF$\left(u\right)$)
has a power law decay which implies $\sum_{k=-\infty}^{\infty}\left|\rho_{V}\left(u,\,k\right)\right|=\infty$
if $\vartheta_{j}\left(u\right)\in\left(0,\,1/2\right)$. 

\subsection{\label{Subsection:  The Sample Autocovariance under Nonstationarity}The
Sample Autocovariance Under Nonstationarity }

We now establish some asymptotic properties of the sample autocovariance
under nonstationarity. We consider the case $k\geq0$ only; the case
$k<0$ is similar. 
\begin{thm}
\label{Theorem ACF Nonstat}Assume that $\left\{ V_{t,T}\right\} $
satisfies Definition \ref{Definition Segmented-Locally-Stationary}.
Under Assumptions \ref{Assumption Smothness of A (for HAC)}-\ref{Assumption A - Dependence},
\begin{align}
\widehat{\Gamma}\left(k\right)\geq & \int_{0}^{1}c\left(u,\,k\right)du+d^{*}+o_{\mathrm{a.s}.}\left(1\right),\label{Eq. Gamma_hat(k) Ineq Theorem}
\end{align}
where $d^{*}=2^{-1}\sum_{j_{1}\neq j_{2}}r_{j_{1}}r_{j_{2}}(\overline{\mu}_{j_{2}}-\overline{\mu}_{j_{1}})^{2}$.
Further, as $k\rightarrow\infty,$ $\widehat{\Gamma}\left(k\right)\geq d^{*}$
$\mathbb{P}$-a.s. If in addition it holds that $\mu_{j}\left(t/T\right)=\mu_{j}$
for $j=1,\ldots,\,m_{0}+1$, then
\begin{align*}
\widehat{\Gamma}\left(k\right)= & \int_{0}^{1}c\left(u,\,k\right)du+d_{\mathrm{Sta}}^{*}+o_{\mathrm{a.s}.}\left(1\right),
\end{align*}
where $d_{\mathrm{Sta}}^{*}=2^{-1}\sum_{j_{1}\neq j_{2}}r_{j_{1}}r_{j_{2}}\left(\mu_{j_{2}}-\mu_{j_{1}}\right)^{2}$
and, as $k\rightarrow\infty$, $\widehat{\Gamma}\left(k\right)=d_{\mathrm{Sta}}^{*}+o_{\mathrm{a.s.}}\left(1\right)$. 
\end{thm}

\subsection{\label{Subsec The-Periodogram under Nonstationary}The Periodogram
Under Nonstationarity }

Classical LRV estimators are weighted averages of periodogram ordinates
around the zero frequency. Thus, it is useful to study the behavior
of the periodogram as the frequency $\omega$ approaches zero. We
now establish some properties of the asymptotic bias of the periodogram
under nonstationarity. We consider the Fourier frequencies $\omega_{l}=2\pi l/T\in(-\pi,\,\pi)$
for an integer $l\neq0$ (mod $T$) and exclude $\omega_{l}=0$ for
mathematical convenience.   
\begin{assumption}
\label{Assumption Means for Periodogram}(i) For each $j=1,\ldots,\,m_{0}+1$
there exists a $B_{j}\in\mathbb{R}$ such that
\begin{align*}
\left|\sum_{j=1}^{m_{0}+1}\sum_{t=\left\lfloor T\lambda_{j-1}^{0}\right\rfloor +1}^{\left\lfloor T\lambda_{j}^{0}\right\rfloor }\mu_{j}\left(t/T\right)\exp\left(-i\omega_{l}t\right)\right|^{2} & \geq\left|\sum_{j=1}^{m_{0}+1}B_{j}\sum_{t=\left\lfloor T\lambda_{j-1}^{0}\right\rfloor +1}^{\left\lfloor T\lambda_{j}^{0}\right\rfloor }\exp\left(-i\omega_{l}t\right)\right|^{2},\qquad\omega_{l}\in\left(-\pi,\,\pi\right),
\end{align*}
where $B_{j_{1}}\neq B_{j_{2}}$ for $j_{1}\neq j_{2}$; (ii) $\left|\Gamma\left(u,\,k\right)\right|=C_{u,k}k^{-m}$
for all $u\in\left[0,\,1\right]$ and all $k\geq C_{3}T^{\kappa}$
for some $C_{3}<\infty$ , $C_{u,k}<\infty$ (which depends on $u$
and $k$), $0<\kappa<1/2$, and $m>2$.
\end{assumption}
Part (i) is easily satisfied (e.g., the special case with $\mu_{j}\left(t/T\right)=\mu_{j}$).
Part (ii) is satisfied if $\left\{ V_{t}\right\} $ is strong mixing
with mixing parameters of size $-2\nu/\left(\nu-1/2\right)$ for
some $\nu>1$ such that $\sup_{t\geq1}\mathbb{E}\left|V_{t}\right|^{4\nu}<\infty.$
This is less stringent than the size condition $-3\nu/\left(\nu-1\right)$
for some $\nu>1$ sufficient for Assumption \ref{Assumption A - Dependence}-(i).
\begin{thm}
\label{Theorem Periodogram Long Memory Effects}Assume that $\left\{ V_{t,T}\right\} $
satisfies Definition \ref{Definition Segmented-Locally-Stationary}.
Under Assumptions \ref{Assumption Smothness of A (for HAC)}-\ref{Assumption A - Dependence}
and \ref{Assumption Means for Periodogram}, 
\begin{align}
\mathbb{E}\left(I_{T}\left(\omega_{l}\right)\right) & =2\pi\int_{0}^{1}f\left(u,\,\omega_{l}\right)du\label{Eq. (E(I_T(wl))) Theorem B in Mikosh and Starica}\\
 & \quad+\frac{1}{T\omega_{l}^{2}}\left|\left[B_{1}-B_{m_{0}+1}-\sum_{j=1}^{m_{0}}\left(B_{j}-B_{j+1}\right)\exp\left(-2\pi il\lambda_{l}^{0}\right)\right]\right|^{2}+o\left(1\right).\nonumber 
\end{align}
 Under Assumptions \ref{Assumption Smothness of A (for HAC)}-\ref{Assumption A - Dependence}
and \ref{Assumption Means for Periodogram}-(ii), if $\mu_{j}\left(t/T\right)=\mu_{j}$
for each $j=1,\ldots,\,m_{0}+1$, then
\begin{align*}
\mathbb{E}\left(I_{T}\left(\omega_{l}\right)\right) & =2\pi\int_{0}^{1}f\left(u,\,\omega_{l}\right)du\\
 & \quad+\frac{1}{T\omega_{l}^{2}}\left|\left[\mu_{j}-\mu_{m_{0}+1}-\sum_{j=1}^{m_{0}}\left(\mu_{j}-\mu_{j+1}\right)\exp\left(-2\pi il\lambda_{j}^{0}\right)\right]\right|^{2}+o\left(1\right).
\end{align*}
 In either case, if $T\omega_{l}^{2}\rightarrow0$ as $T\rightarrow\infty$
then $\mathbb{E}\left(I_{T}\left(\omega_{l}\right)\right)\rightarrow\infty$
for many values in $\left\{ \omega_{l}\right\} $ as $\omega_{l}\rightarrow0.$ 
\end{thm}
The theorem suggests that for small frequencies $\omega_{l}$ close
to $0,$ the periodogram attains very large values. This follows because
the first term of \eqref{Eq. (E(I_T(wl))) Theorem B in Mikosh and Starica}
is bounded for all $\omega_{j}$.  Since $B_{1},\ldots,\,B_{m_{0}+1}$
are fixed, the order of the second term of \eqref{Eq. (E(I_T(wl))) Theorem B in Mikosh and Starica}
is $O((T\omega_{j}^{2})^{-1})$. Note that as $\omega_{l}\rightarrow0$
there are some values $l$ for which the corresponding term involving
$\left|\cdot\right|^{2}$ on the right-hand side of \eqref{Eq. (E(I_T(wl))) Theorem B in Mikosh and Starica}
is equal to zero. In such cases, $\mathbb{E}\left(I_{T}\left(\omega_{l}\right)\right)\geq2\pi\int_{0}^{1}f\left(u,\,\omega_{l}\right)du>0.$
For other values of $\left\{ l\right\} $ as $\omega_{l}\rightarrow0$,
the second term of \eqref{Eq. (E(I_T(wl))) Theorem B in Mikosh and Starica}
diverges to infinity. Thus, considering the behavior of  $\left\{ \mathbb{E}\left(I_{T}\left(\omega_{l}\right)\right)\right\} $
as $\omega_{l}\rightarrow0$, it generally takes  unbounded values
except for some $\omega_{l}$ for which $\mathbb{E}\left(I_{T}\left(\omega_{l}\right)\right)$
is bounded below by $2\pi\int_{0}^{1}f\left(u,\,\omega_{l}\right)du>0.$
 A SLS process with long memory has an unbounded local spectral density
$f\left(u,\,\omega\right)$ as $\omega\rightarrow0$ for some $u\in\left[0,\,1\right]$.
Since $f\left(\cdot,\,\cdot\right)$ cannot be negative, it follows
that $\int_{0}^{1}f\left(u,\,\omega\right)du$ is also unbounded as
$\omega\rightarrow0$. Theorem \ref{Theorem Periodogram Long Memory Effects}
suggests that nonstationarity consisting of time-varying first moment
results in a periodogram sharing features of a long memory series.

\section{\label{Supp Math App}Mathematical Appendix}

\subsection{Proofs of the Results in Section \ref{Section Low Freq Cont - Theory}
and \ref{Section Results on Low Frequency Contamination} }

\subsubsection{Proof of Theorem \ref{Theorem ACF Nonstat}}

Let $\overline{V}_{j}=\left(Tr_{j}\right)^{-1}\sum_{t=\left\lfloor T\lambda_{j-1}^{0}\right\rfloor +1}^{\left\lfloor T\lambda_{j}^{0}\right\rfloor }V_{t}$,
$\mu_{2,j}\left(u\right)=\mathbb{E}(V_{\left\lfloor Tu\right\rfloor })^{2}$
for $T_{j-1}^{0}\leq Tu\leq T_{j}^{0}$ and $\overline{\mu}_{2,j}=r_{j}^{-1}\int_{\lambda_{j-1}^{0}}^{\lambda_{j}^{0}}\mu_{2,j}\left(u\right)du$.
By Assumption \ref{Assumption Smothness of A (for HAC)}-\ref{Assumption A - Dependence}-(i),
the latter implying ergodicity, it follows for fixed $k\geq0$ that
 
\begin{align*}
\widehat{\Gamma}\left(k\right) & =\sum_{j=1}^{m_{0}+1}r_{j}\frac{1}{Tr_{j}}\sum_{t=\left\lfloor T\lambda_{j-1}^{0}\right\rfloor +1+k}^{\left\lfloor T\lambda_{j}^{0}\right\rfloor }V_{t}V_{t-k}-\left(\sum_{j=1}^{m_{0}+1}r_{j}\frac{1}{Tr_{j}}\sum_{t=\left\lfloor T\lambda_{j-1}^{0}\right\rfloor +1}^{\left\lfloor T\lambda_{j}^{0}\right\rfloor }V_{t}\right)^{2}\\
 & =\sum_{j=1}^{m_{0}+1}\int_{\lambda_{j-1}^{0}}^{\lambda_{j}^{0}}c\left(u,\,k\right)du+\sum_{j=1}^{m_{0}+1}r_{j}\frac{1}{Tr_{j}}\sum_{t=\left\lfloor T\lambda_{j-1}^{0}\right\rfloor +1+k}^{\left\lfloor T\lambda_{j}^{0}\right\rfloor }\mathbb{E}\left(V_{t}\right)\mathbb{E}\left(V_{t-k}\right)\\
 & \quad-\left(\sum_{j=1}^{m_{0}+1}r_{j}\frac{1}{Tr_{j}}\sum_{t=\left\lfloor T\lambda_{j-1}^{0}\right\rfloor +1}^{\left\lfloor T\lambda_{j}^{0}\right\rfloor }V_{t}\right)^{2}+O\left(T^{-1}\right)+o_{\mathrm{a.s.}}\left(1\right)\\
 & =\int_{0}^{1}c\left(u,\,k\right)du+\sum_{j=1}^{m_{0}+1}r_{j}\frac{1}{Tr_{j}}\sum_{t=\left\lfloor T\lambda_{j-1}^{0}\right\rfloor +1+k}^{\left\lfloor T\lambda_{j}^{0}\right\rfloor }\mathbb{E}\left(V_{t}\right)\mathbb{E}\left(V_{t-k}\right)\\
 & \quad-\left(\sum_{j=1}^{m_{0}+1}r_{j}\overline{V}_{j}\right)^{2}+O\left(T^{-1}\right)+o_{\mathrm{a.s.}}\left(1\right)\\
 & =\int_{0}^{1}c\left(u,\,k\right)du+\sum_{j=1}^{m_{0}+1}r_{j}\frac{1}{Tr_{j}}\sum_{t=\left\lfloor T\lambda_{j-1}^{0}\right\rfloor +1+k}^{\left\lfloor T\lambda_{j}^{0}\right\rfloor }\mu^{2}\left(t/T\right)-\left(\sum_{j=1}^{m_{0}+1}r_{j}\overline{V}_{j}\right)^{2}+O\left(T^{-1}\right)+o_{\mathrm{a.s.}}\left(1\right),
\end{align*}
where we have used $\mathbb{E}\left(V_{t-k}\right)-\mathbb{E}\left(V_{t}\right)=O\left(k/T\right)$
by local stationarity in the third equality. Note that by ergodicity
and an approximation to Riemann sums, we have 
\begin{align}
\sum_{j=1}^{m_{0}+1}r_{j}\overline{V}_{j}-\sum_{j=1}^{m_{0}+1}r_{j}\overline{\mu}_{j} & =\sum_{j=1}^{m_{0}+1}r_{j}\overline{V}_{j}-\sum_{j=1}^{m_{0}+1}r_{j}\mathbb{E}\left(\overline{V}_{j}\right)+\sum_{j=1}^{m_{0}+1}r_{j}\mathbb{E}\left(\overline{V}_{j}\right)-\sum_{j=1}^{m_{0}+1}r_{j}\overline{\mu}_{j}\nonumber \\
 & =o_{\mathrm{a.s.}}\left(1\right)+O\left(T^{-1}\right).\label{Eq. (1) for (5) Mikosh and Starica}
\end{align}
 Basic manipulations show that
\begin{align}
\sum_{j_{2}\neq j_{1}} & r_{j_{1}}r_{j_{2}}\left(\overline{\mu}_{j_{2}}-\overline{\mu}_{j_{1}}\right)^{2}\nonumber \\
 & =\sum_{j_{2}\neq j_{1}}r_{j_{1}}r_{j_{2}}\left(\overline{\mu}_{j_{2}}^{2}+\overline{\mu}_{j_{1}}^{2}-2\overline{\mu}_{j_{2}}\overline{\mu}_{j_{1}}\right)\nonumber \\
 & =\sum_{1\leq j_{2}\leq m_{0}+1}r_{j_{2}}\overline{\mu}_{j_{2}}^{2}\left(1-r_{j_{2}}\right)+\sum_{1\leq j_{1}\leq m_{0}+1}r_{j_{1}}\overline{\mu}_{j_{1}}^{2}\left(1-r_{j_{1}}\right)-2\sum_{j_{1}\neq j_{2}}r_{j_{1}}r_{j_{2}}\overline{\mu}_{j_{2}}\overline{\mu}_{j_{1}}\nonumber \\
 & =2\sum_{1\leq j\leq m_{0}+1}r_{j}\overline{\mu}_{j}^{2}-2\sum_{1\leq j\leq m_{0}+1}r_{j}^{2}\overline{\mu}_{j}^{2}-2\sum_{j_{1}\neq j_{2}}r_{j_{1}}r_{j_{2}}\overline{\mu}_{j_{2}}\overline{\mu}_{j_{1}}.\label{Eq. (2) open square for (5) in Mikosh and Starica}
\end{align}
 Note that 
\begin{align}
\left(Tr_{j}-k\right)\sum_{t=\left\lfloor T\lambda_{j-1}^{0}\right\rfloor +1+k}^{\left\lfloor T\lambda_{j}^{0}\right\rfloor }\mu^{2}\left(t/T\right)\geq\left(\sum_{t=\left\lfloor T\lambda_{j-1}^{0}\right\rfloor +1+k}^{\left\lfloor T\lambda_{j}^{0}\right\rfloor }\mu\left(t/T\right)\right)^{2} & .\label{Eq. CS Ineq Theorem 1}
\end{align}
 Thus, 
\begin{align}
\sum_{j=1}^{m_{0}+1}r_{j}\frac{1}{Tr_{j}}\sum_{t=\left\lfloor T\lambda_{j-1}^{0}\right\rfloor +1+k}^{\left\lfloor T\lambda_{j}^{0}\right\rfloor }\mu^{2}\left(t/T\right) & =\sum_{j=1}^{m_{0}+1}r_{j}\frac{1}{Tr_{j}\left(Tr_{j}-k\right)}\left(Tr_{j}-k\right)\sum_{t=\left\lfloor T\lambda_{j-1}^{0}\right\rfloor +1+k}^{\left\lfloor T\lambda_{j}^{0}\right\rfloor }\mu^{2}\left(t/T\right)\nonumber \\
 & \geq\sum_{j=1}^{m_{0}+1}r_{j}\frac{1}{Tr_{j}\left(Tr_{j}-k\right)}\left(\sum_{t=\left\lfloor T\lambda_{j-1}^{0}\right\rfloor +1+k}^{\left\lfloor T\lambda_{j}^{0}\right\rfloor }\mu\left(t/T\right)\right)^{2}\nonumber \\
 & =\sum_{1\leq j\leq m_{0}+1}r_{j}\overline{\mu}_{j}^{2}+o\left(1\right).\label{Eq. (CS) inequality for (5) in Mikosh and Starica}
\end{align}
Using \eqref{Eq. (1) for (5) Mikosh and Starica}-\eqref{Eq. (CS) inequality for (5) in Mikosh and Starica}
we have, 
\begin{align}
\widehat{\Gamma}\left(k\right) & =\int_{0}^{1}c\left(u,\,k\right)du+\sum_{j=1}^{m_{0}+1}r_{j}\frac{1}{Tr_{j}}\sum_{t=\left\lfloor T\lambda_{j-1}^{0}\right\rfloor +1+k}^{\left\lfloor T\lambda_{j}^{0}\right\rfloor }\mu^{2}\left(t/T\right)-\left(\sum_{j=1}^{m_{0}+1}r_{j}\overline{V}_{j}\right)^{2}+o_{\mathrm{a.s.}}\left(1\right)\nonumber \\
 & \geq\int_{0}^{1}c\left(u,\,k\right)du+\sum_{j=1}^{m_{0}+1}r_{j}\overline{\mu}_{2,j}-\left(\sum_{j=1}^{m_{0}+1}r_{j}\overline{V}_{j}\right)^{2}+O\left(T^{-1}\right)+o_{\mathrm{a.s.}}\left(1\right)\nonumber \\
 & =\int_{0}^{1}c\left(u,\,k\right)du+2^{-1}\sum_{j_{1}\neq j_{2}}r_{j_{1}}r_{j_{2}}\left(\overline{\mu}_{j_{2}}-\overline{\mu}_{j_{1}}\right)^{2}+O\left(T^{-1}\right)+o_{\mathrm{a.s.}}\left(1\right).\label{Eq. (5) Mikosh and Starica}
\end{align}
 The claim that $\widehat{\Gamma}\left(k\right)\geq d$ $\mathbb{P}$-a.s.
as $k\rightarrow\infty$ follows from Assumption \ref{Assumption A - Dependence}-(i)
since this implies that $c\left(u,\,k\right)\rightarrow0$ as $k\rightarrow\infty$
and from the fact that the second term on the right-hand side of \eqref{Eq. (5) Mikosh and Starica}
does not depend on $k.$ If in addition it holds that $\mu_{j}\left(t/T\right)=\mu_{j}$
for $j=1,\ldots,\,m_{0}+1$, then \eqref{Eq. CS Ineq Theorem 1} holds
with equality and the result follows as a special case of \eqref{Eq. (5) Mikosh and Starica}.
$\square$

\subsubsection{Proof of Theorem \ref{Theorem Periodogram Long Memory Effects}}
\begin{lem}
\label{Lemma Cross Products Periodogram}Assume that $\left\{ V_{t,T}\right\} $
satisfies Definition \ref{Definition Segmented-Locally-Stationary}.
Under Assumptions \ref{Assumption Smothness of A (for HAC)}-\ref{Assumption A - Dependence}
and \ref{Assumption Means for Periodogram}-(ii),
\begin{align*}
\sum_{j_{1}\neq j_{2}}\frac{1}{T}\sum_{t=\left\lfloor T\lambda_{j_{1}-1}^{0}\right\rfloor +1}^{\left\lfloor T\lambda_{j_{1}}^{0}\right\rfloor }\sum_{s=\left\lfloor T\lambda_{j_{2}-1}^{0}\right\rfloor +1}^{\left\lfloor T\lambda_{j_{2}}^{0}\right\rfloor }\mathbb{E}\left(\left(V_{t}-\mu\left(t/T\right)\right)\left(V_{s}-\mu\left(s/T\right)\right)\right)\exp\left(-i\omega_{l}\left(t-s\right)\right) & =o\left(1\right).
\end{align*}
\end{lem}
\noindent\textit{Proof. }Let $\overline{r}_{j_{1},j_{2}}=\max\left\{ r_{j_{1}},\,r_{j_{2}}\right\} $
and $\underline{r}_{j_{1},j_{2}}=\min\left\{ r_{j_{1}},\,r_{j_{2}}\right\} .$
We consider the case of adjacent regimes (i.e., $j_{2}=j_{1}+1$)
which also provides an upper bound for non-adjacent regimes due to
the short memory property. For any $k=s-t=1,\ldots,\,\bigl\lfloor T\underline{r}_{j_{1},j_{2}}\bigr\rfloor$
there are $k$ pairs in the above sum. The double sum above (over
$t$ and $s$) can be split into
\begin{align}
T^{-1} & \sum_{k=1}^{\left\lfloor CT^{\kappa}\right\rfloor }\left|\Gamma_{\left\{ 1:\left\lfloor CT^{\kappa}\right\rfloor \right\} }\left(\cdot,\,k\right)\right|+T^{-1}\sum_{k=\left\lfloor CT^{\kappa}\right\rfloor +1}^{\left\lfloor hT\right\rfloor }\left|\Gamma_{\left\{ \left\lfloor CT^{\kappa}\right\rfloor +1:\left\lfloor hT\right\rfloor \right\} }\left(\cdot,\,k\right)\right|\label{Eq. 4 sums for cross products}\\
 & \quad+T^{-1}\sum_{k=\left\lfloor hT\right\rfloor +1}^{\left\lfloor T\underline{r}_{j_{1},j_{2}}\right\rfloor -1}\left|\Gamma_{\left\{ \left\lfloor hT\right\rfloor +1:\left\lfloor T\underline{r}_{j_{1},j_{2}}\right\rfloor -1\right\} }\left(\cdot,\,k\right)\right|+T^{-1}\sum_{k=\left\lfloor T\underline{r}_{j_{1},j_{2}}\right\rfloor }^{\left\lfloor T\overline{r}_{j_{1},j_{2}}\right\rfloor }\left|\Gamma_{\left\{ \underline{r}_{j_{1},j_{2}}:\underline{r}_{j_{1},j_{2}}\right\} }\left(\cdot,\,k\right)\right|\nonumber 
\end{align}
where $C>0$, $0<h<1$ with $\left\lfloor hT\right\rfloor <\bigl\lfloor T\underline{r}_{j_{1},j_{2}}\bigr\rfloor-1$,
and $\Gamma_{S}\left(\cdot,\,k\right)$ is the sum of the autocovariances
at lag $k$ computed at the time points corresponding to $k\in S$.
Note that the term $\left|\exp\left(-i\omega_{l}\left(\pm k\right)\right)\right|$
can be bounded by some constant. The sums run over only $k>0$ because
by symmetry $\Gamma_{u}\left(k\right)=\Gamma_{u-k/T}\left(-k\right)$.
Consider the first sum in \eqref{Eq. 4 sums for cross products}.
This is of order $O\left(T^{-1}T^{2\kappa}\right)$ which goes to
zero given $\kappa<1/2.$ The second sum is also negligible using
the following arguments. By Assumption \ref{Assumption Means for Periodogram}-(ii),
$\left|\Gamma\left(u,\,k\right)\right|=C_{u,k}k^{-m}$ with $m>2$
and choosing $C$ large enough yields that the second sum of \eqref{Eq. 4 sums for cross products}
converges to zero. In the third sum, the number of summands grows
at rate $O\left(T\right)$ and for each lag $k$ there are $O\left(T\right)$
autocovariances. However, by Assumption \ref{Assumption Means for Periodogram}-(ii)
each autocovariance is $O\left(T^{-m}\right).$ Thus, the bound is
$O\left(T^{-1}T^{2-m}\right)$ which goes to zero as $T\rightarrow\infty$.
The difference between the arguments used for the third sum and fourth
sums is that now we do not have $O\left(T\right)$ autocovariances
for each lag $k$. Thus, the bound for the fourth sum cannot be greater
than the bound for the third sum. Thus, the fourth sum also converges
to zero. $\square$ 

\medskip{}

\noindent\textit{Proof of Theorem \ref{Theorem Periodogram Long Memory Effects}.
}We have,
\begin{align*}
I_{T}\left(\omega_{l}\right) & =\left|\frac{1}{\sqrt{T}}\sum_{j=1}^{m_{0}+1}\sum_{t=\left\lfloor T\lambda_{j-1}^{0}\right\rfloor +1}^{\left\lfloor T\lambda_{j}^{0}\right\rfloor }\exp\left(-i\omega_{l}t\right)V_{t}\right|^{2}\\
 & =\left|\frac{1}{\sqrt{T}}\sum_{j=1}^{m_{0}+1}\sum_{t=\left\lfloor T\lambda_{j-1}^{0}\right\rfloor +1}^{\left\lfloor T\lambda_{j}^{0}\right\rfloor }\left(X_{t}-\mu\left(t/T\right)\right)\exp\left(-i\omega_{l}t\right)+\frac{1}{\sqrt{T}}\sum_{j=1}^{m_{0}+1}\sum_{t=\left\lfloor T\lambda_{j-1}^{0}\right\rfloor +1}^{\left\lfloor T\lambda_{j}^{0}\right\rfloor }\mu\left(t/T\right)\exp\left(-i\omega_{l}t\right)\right|^{2}.
\end{align*}
 From Assumption \ref{Assumption Means for Periodogram}, 
\begin{align*}
\Biggl|\sum_{j=1}^{m_{0}+1} & \sum_{t=\left\lfloor T\lambda_{j-1}^{0}\right\rfloor +1}^{\left\lfloor T\lambda_{j}^{0}\right\rfloor }\mu\left(t/T\right)\exp\left(-i\omega_{l}t\right)\Biggr|^{2}\\
 & \geq\left|\sum_{j=1}^{m_{0}+1}B_{j}\sum_{t=\left\lfloor T\lambda_{j-1}^{0}\right\rfloor +1}^{\left\lfloor T\lambda_{j}^{0}\right\rfloor }\exp\left(-i\omega_{l}t\right)\right|^{2}\\
 & =\left|\sum_{j=1}^{m_{0}+1}B_{j}\exp\left(-i\omega_{l}\left(\left\lfloor T\lambda_{j-1}^{0}\right\rfloor +1\right)\right)\sum_{t=0}^{\left\lfloor T\lambda_{j}^{0}\right\rfloor -\left\lfloor T\lambda_{j-1}^{0}\right\rfloor -1}\exp\left(-i\omega_{l}t\right)\right|^{2}\\
 & =\left|\frac{\exp\left(-i\omega_{l}\right)}{1-\exp\left(-i\omega_{l}\right)}\sum_{j=1}^{m_{0}+1}B_{j}\exp\left(-i\omega_{l}\left(\left\lfloor T\lambda_{j-1}^{0}\right\rfloor \right)\right)\left(1-\exp\left(-i\omega_{l}\left(\left\lfloor T\lambda_{j}^{0}\right\rfloor -\left\lfloor T\lambda_{j-1}^{0}\right\rfloor \right)\right)\right)\right|^{2}\\
 & =\left|\frac{\exp\left(-i\omega_{l}\right)}{1-\exp\left(-i\omega_{l}\right)}\sum_{j=1}^{m_{0}+1}B_{j}\left(\exp\left(-i\omega_{l}\left(\left\lfloor T\lambda_{j-1}^{0}\right\rfloor \right)\right)-\exp\left(-i\omega_{l}\left\lfloor T\lambda_{j}^{0}\right\rfloor \right)\right)\right|^{2},
\end{align*}
using the formula for the first $n$-th terms of a geometric series
$\sum_{k=0}^{n-1}ar^{k}=a\sum_{k=0}^{n-1}r^{k}=a\left(1-r^{n}\right)/\left(1-r\right).$
Then, using summation by parts,
\begin{align*}
\frac{\exp\left(-i\omega_{j}\right)}{1-\exp\left(-i\omega_{j}\right)} & \sum_{j=1}^{m_{0}+1}B_{j}\left(\exp\left(-i\omega_{l}\left(\left\lfloor T\lambda_{j-1}^{0}\right\rfloor \right)\right)-\exp\left(-i\omega_{l}\left\lfloor T\lambda_{j}^{0}\right\rfloor \right)\right)\\
 & =\frac{\exp\left(-i\omega_{j}\right)}{1-\exp\left(-i\omega_{j}\right)}\left[B_{1}-B_{m_{0}+1}-\sum_{j=1}^{m_{0}}\left(B_{j}-B_{j+1}\right)\exp\left(-i\omega_{l}\left\lfloor T\lambda_{j}^{0}\right\rfloor \right)\right].
\end{align*}
 By Lemma \ref{Lemma Cross Products Periodogram}, it is sufficient
to consider the cross-products within each regime $j$, 
\begin{align*}
\mathbb{E}\left(I_{T}\left(\omega_{l}\right)\right) & \geq\sum_{j=1}^{m_{0}+1}r_{j}\frac{1}{Tr_{j}}\mathbb{E}\sum_{t=\left\lfloor T\lambda_{j-1}^{0}\right\rfloor +1}^{\left\lfloor T\lambda_{j}^{0}\right\rfloor }\sum_{s=\left\lfloor T\lambda_{j-1}^{0}\right\rfloor +1}^{\left\lfloor T\lambda_{j}^{0}\right\rfloor }\left(V_{t}-\mu\left(t/T\right)\right)\left(V_{s}-\mu\left(s/T\right)\right)\exp\left(-i\omega_{l}\left(t-s\right)\right)\\
 & \quad+\underset{j_{1}\neq j_{2}}{\sum\sum}\frac{1}{T}\mathbb{E}\sum_{t=\left\lfloor T\lambda_{j_{1}-1}^{0}\right\rfloor +1}^{\left\lfloor T\lambda_{j_{1}}^{0}\right\rfloor }\sum_{s=\left\lfloor T\lambda_{j_{2}-1}^{0}\right\rfloor +1}^{\left\lfloor T\lambda_{j_{2}}^{0}\right\rfloor }\left(V_{t}-\mu\left(t/T\right)\right)\left(V_{s}-\mu\left(s/T\right)\right)\exp\left(-i\omega_{l}\left(t-s\right)\right)\\
 & \quad+\left|\frac{1}{\sqrt{T}}\frac{\exp\left(-i\omega_{l}\right)}{1-\exp\left(-i\omega_{l}\right)}\sum_{j=1}^{m_{0}+1}B_{j}\left(\exp\left(-i\omega_{l}\left(\left\lfloor T\lambda_{j-1}^{0}\right\rfloor \right)\right)-\exp\left(-i\omega_{l}\left\lfloor T\lambda_{j}^{0}\right\rfloor \right)\right)\right|^{2}+o\left(1\right)\\
 & =\sum_{j=1}^{m_{0}+1}\left(\mathbb{E}\frac{1}{T}\sum_{t=\left\lfloor T\lambda_{j-1}^{0}\right\rfloor +1}^{\left\lfloor T\lambda_{j}^{0}\right\rfloor }\left(V_{t}-\mu\left(t/T\right)\right)^{2}+\frac{2}{Tr_{j}}\sum_{k=1}^{\left\lfloor Tr_{j}\right\rfloor -1}\sum_{t=\left\lfloor T\lambda_{j-1}^{0}\right\rfloor +k+1}^{\left\lfloor T\lambda_{j}^{0}\right\rfloor }\Gamma_{t/T}\left(k\right)\exp\left(-i\omega_{l}k\right)\right)\\
 & \quad+\left|\frac{1}{\sqrt{T}}\frac{\exp\left(-i\omega_{l}\right)}{1-\exp\left(-i\omega_{l}\right)}\sum_{j=1}^{m_{0}+1}B_{j}\left(\exp\left(-i\omega_{l}\left(\left\lfloor T\lambda_{j-1}^{0}\right\rfloor \right)\right)-\exp\left(-i\omega_{l}\left\lfloor T\lambda_{j}^{0}\right\rfloor \right)\right)\right|^{2}+o\left(1\right).
\end{align*}
Next, using the definition of $f\left(u,\,\omega_{l}\right),$ $e^{-2i\omega_{l}}=1$
by Euler's formula and letting $\omega_{l}\rightarrow0$ we have,
\begin{align}
\mathbb{E}\left(I_{T}\left(\omega_{l}\right)\right) & \geq\sum_{j=1}^{m_{0}+1}\left(\int_{\lambda_{j-1}^{0}}^{\lambda_{j}^{0}}c\left(u,\,0\right)du+2\sum_{k=1}^{\infty}\int_{\lambda_{j-1}^{0}}^{\lambda_{j}^{0}}c\left(u,\,k\right)\exp\left(-i\omega_{l}k\right)du\right)\nonumber \\
 & \quad+\frac{1}{T}\frac{1}{\left|1-\exp\left(-i\omega_{l}\right)\right|^{2}}\left|\left[B_{1}-B_{m_{0}+1}-\left(1+o\left(1\right)\right)\sum_{j=1}^{m_{0}}\left(B_{j}-B_{j+1}\right)\exp\left(-2\pi il\lambda_{j}^{0}\right)\right]\right|^{2}+o\left(1\right)\nonumber \\
 & =2\pi\sum_{j=1}^{m_{0}+1}\int_{\lambda_{j-1}^{0}}^{\lambda_{j}^{0}}f\left(u,\,\omega_{l}\right)du\nonumber \\
 & \quad+\frac{1}{T}\frac{1}{\left|1-\exp\left(-i\omega_{l}\right)\right|^{2}}\left|\left[B_{1}-B_{m_{0}+1}-\left(1+o\left(1\right)\right)\sum_{j=1}^{m_{0}}\left(B_{j}-B_{j+1}\right)\exp\left(-2\pi il\lambda_{j}^{0}\right)\right]\right|^{2}+o\left(1\right)\nonumber \\
 & =2\pi\int_{0}^{1}f\left(u,\,\omega_{l}\right)du+\frac{1}{T\omega_{l}^{2}}\left|\left[B_{1}-B_{m_{0}+1}-\sum_{j=1}^{m_{0}}\left(B_{j}-B_{j+1}\right)\exp\left(-2\pi il\lambda_{j}^{0}\right)\right]\right|^{2}+o\left(1\right).\label{Eq. (8) Mikosh and Starica}
\end{align}
 By Assumption \ref{Assumption Smothness of A (for HAC)}-(ii), the
first term of \eqref{Eq. (8) Mikosh and Starica} is bounded for all
frequencies $\omega_{j}$.  Since $B_{1},\ldots,\,B_{m_{0}+1}$
are fixed, if $T\omega_{l}^{2}\rightarrow0$ then the order of the
second term of \eqref{Eq. (8) Mikosh and Starica} is $O((T\omega_{l}^{2})^{-1})$.
Note that as $\omega_{l}\rightarrow0$ there are some values of $l$
for which the corresponding term involving $\left|\cdot\right|^{2}$
on the right-hand side of \eqref{Eq. (8) Mikosh and Starica} is equal
to zero {[}see the argument in \citeReferencesSupp{mikosh/starica:04}{]}.
In such a case, $\mathbb{E}\left(I_{T}\left(\omega_{l}\right)\right)\geq2\pi\int_{0}^{1}f\left(u,\,\omega_{l}\right)du>0.$
For the other values of $\left\{ l\right\} $ as $\omega_{l}\rightarrow0$,
the second term of \eqref{Eq. (8) Mikosh and Starica} diverges to
infinity. The outcome is that there are frequencies close to $\omega_{l}=0$
for which $\mathbb{E}\left(I_{T}\left(\omega_{l}\right)\right)\rightarrow\infty.$
$\square$

\subsubsection{Proof of Theorem \ref{Theorem Local ACF Nonstat}}

We consider the case $k\geq0$. The case $k<0$ follows similarly.
Consider any $u\in\left(0,\,1\right)$ such that $T_{j}^{0}\notin\mathbf{S}\left(u,\,k,\,n_{2,T}\right)$
for all $j=1,\ldots,\,m_{0}$.  Theorem S.B.3 in \citeReferencesSupp{casini_hac}
showed that 
\begin{align}
\mathbb{E}\left[\widehat{c}_{T}\left(u,\,k\right)\right] & =c\left(u_{0},\,k\right)+\frac{1}{2}\left(n_{2,T}/T\right)^{2}\left[\frac{\partial^{2}}{\partial^{2}u}c\left(u,\,k\right)\right]+o\left(\left(n_{2,T}/T\right)^{2}\right)+O\left(1/n_{2,T}\right).\label{eq. Bias Locally Stationary}
\end{align}
 Since $n_{2,T}\rightarrow\infty$ and $n_{2,T}/T\rightarrow0,$
$\mathbb{E}\left[\widehat{c}_{T}\left(u,\,k\right)\right]=c\left(u_{0},\,k\right)+o\left(1\right).$
 The same aforementioned theorem shows that $n_{2,T}\mathrm{Var}\left[\widehat{c}_{T}\left(u,\,k\right)\right]=O_{\mathbb{P}}\left(1\right)$.
This combined with \eqref{eq. Bias Locally Stationary} yields part
(i) of the theorem. 

Next, we consider case (ii-a) with $n_{j,L}\left(u,\,k,\,n_{2,T}\right)/n_{2,T}\rightarrow\gamma\in\left(0,\,1\right)$.
We have,
\begin{align}
\widehat{c}_{T}\left(u,\,k\right) & =n_{2,T}^{-1}\sum_{s=0}^{n_{2,T}}V_{\left\lfloor Tu\right\rfloor +k/2-n_{2,T}/2+s+1}V_{\left\lfloor Tu\right\rfloor +k/2-n_{2,T}/2+s+1-k}-\left(n_{2,T}^{-1}\sum_{s=0}^{n_{2,T}}V_{\left\lfloor Tu\right\rfloor -n_{2,T}/2+s+1}\right)^{2}\nonumber \\
 & =n_{2,T}^{-1}\sum_{s=0}^{T_{j}^{0}-\left(\left\lfloor Tu\right\rfloor +k/2-n_{2,T}/2+1\right)}V_{\left\lfloor Tu\right\rfloor +k/2-n_{2,T}/2+s+1}V_{\left\lfloor Tu\right\rfloor +k/2-n_{2,T}/2+s+1-k}\nonumber \\
 & \quad+n_{2,T}^{-1}\sum_{s=T_{j}^{0}-\left(\left\lfloor Tu\right\rfloor +k/2-n_{2,T}/2\right)}^{n_{2,T}}V_{\left\lfloor Tu\right\rfloor +k/2-n_{2,T}/2+s+1}V_{\left\lfloor Tu\right\rfloor +k/2-n_{2,T}/2+s+1-k}\nonumber \\
 & \quad-\biggl(n_{2,T}^{-1}\sum_{s=0}^{T_{j}^{0}-\left(\left\lfloor Tu\right\rfloor +k/2-n_{2,T}/2+1\right)}V_{\left\lfloor Tu\right\rfloor +k/2-n_{2,T}/2+s+1}\nonumber \\
 & \quad+n_{2,T}^{-1}\sum_{s=T_{j}^{0}-\left(\left\lfloor Tu\right\rfloor +k/2-n_{2,T}/2\right)}^{n_{2,T}}V_{\left\lfloor Tu\right\rfloor -n_{2,T}/2+s+1}\biggr)^{2}\nonumber \\
 & =n_{2,T}^{-1}\sum_{s=0}^{T_{j}^{0}-\left(\left\lfloor Tu\right\rfloor +k/2-n_{2,T}/2+1\right)}\biggl(V_{\left\lfloor Tu\right\rfloor +k/2-n_{2,T}/2+s+1}V_{\left\lfloor Tu\right\rfloor +k/2-n_{2,T}/2+s+1-k}\nonumber \\
 & \quad-\mathbb{E}\left(V_{\left\lfloor Tu\right\rfloor +k/2-n_{2,T}/2+s+1}\right)\mathbb{E}\left(V_{\left\lfloor Tu\right\rfloor +k/2-n_{2,T}/2+s+1-k}\right)\biggr)\nonumber \\
 & \quad+n_{2,T}^{-1}\sum_{s=T_{j}^{0}-\left(\left\lfloor Tu\right\rfloor +k/2-n_{2,T}/2\right)}^{n_{2,T}}\biggl(V_{\left\lfloor Tu\right\rfloor +k/2-n_{2,T}/2+s+1}V_{\left\lfloor Tu\right\rfloor +k/2-n_{2,T}/2+s+1-k}\nonumber \\
 & \quad-\mathbb{E}\left(V_{\left\lfloor Tu\right\rfloor +k/2-n_{2,T}/2+s+1}\right)\mathbb{E}\left(V_{\left\lfloor Tu\right\rfloor +k/2-n_{2,T}/2+s+1-k}\right)\biggr)\nonumber \\
 & \quad+n_{2,T}^{-1}\sum_{s=0}^{T_{j}^{0}-\left(\left\lfloor Tu\right\rfloor +k/2-n_{2,T}/2+1\right)}\mathbb{E}\left(V_{\left\lfloor Tu\right\rfloor +k/2-n_{2,T}/2+s+1}\right)\mathbb{E}\left(V_{\left\lfloor Tu\right\rfloor +k/2-n_{2,T}/2+s+1-k}\right)\nonumber \\
 & \quad+n_{2,T}^{-1}\sum_{s=T_{j}^{0}-\left(\left\lfloor Tu\right\rfloor +k/2-n_{2,T}/2\right)}^{n_{2,T}}\mathbb{E}\left(V_{\left\lfloor Tu\right\rfloor +k/2-n_{2,T}/2+s+1}\right)\mathbb{E}\left(V_{\left\lfloor Tu\right\rfloor +k/2-n_{2,T}/2+s+1-k}\right)\nonumber \\
 & \quad-\biggl(n_{2,T}^{-1}\sum_{s=0}^{T_{j}^{0}-\left(\left\lfloor Tu\right\rfloor +k/2-n_{2,T}/2+1\right)}V_{\left\lfloor Tu\right\rfloor -n_{2,T}/2+s+1}\label{Eq. E(chat) 1}\\
 & \quad+n_{2,T}^{-1}\sum_{s=T_{j}^{0}-\left(\left\lfloor Tu\right\rfloor +k/2-n_{2,T}/2\right)}^{n_{2,T}}V_{\left\lfloor Tu\right\rfloor -n_{2,T}/2+s+1}\biggr)^{2}+o_{\mathbb{P}}\left(1\right)\nonumber \\
 & \geq\gamma c\left(\lambda_{j}^{0},\,k\right)+\left(1-\gamma\right)c\left(u,\,k\right)+\gamma\mu_{j}\left(\lambda_{j}^{0}\right)^{2}+\left(1-\gamma\right)\mu_{j+1}\left(u\right)^{2}\nonumber \\
 & \quad-\left(\gamma\mu_{j}\left(\lambda_{j}^{0}\right)+\left(1-\gamma\right)\mu_{j+1}\left(u\right)\right)^{2}+o_{\mathbb{P}}\left(1\right)\nonumber \\
 & =\gamma c\left(\lambda_{j}^{0},\,k\right)+\left(1-\gamma\right)c\left(u,\,k\right)+\gamma\left(1-\gamma\right)\left(\mu_{j}\left(\lambda_{j}^{0}\right)-\mu_{j+1}\left(u\right)\right)^{2}+o_{\mathbb{P}}\left(1\right).\label{Eq. Gamma + d for Local Autocovariance}
\end{align}
Consider the case (ii-b) with $n_{j,L}\left(u,\,k,\,n_{2,T}\right)/n_{2,T}\rightarrow0$.
The other sub-case follows by symmetry. Eq. \eqref{Eq. E(chat) 1}
continues to hold. The first term, third term and the first summation
of the last term on the right-hand side of \eqref{Eq. E(chat) 1}
are negligible. Thus, using ergodicity, implied by Assumptions \ref{Assumption Smothness of A (for HAC)}-\ref{Assumption A - Dependence}-(i),
\begin{align*}
\widehat{c}_{T}\left(u,\,k\right) & =c\left(u,\,k\right)+n_{2,T}^{-1}\sum_{s=T_{j}^{0}-\left(\left\lfloor Tu\right\rfloor +k/2-n_{2,T}/2\right)}^{n_{2,T}}\mathbb{E}\left(V_{\left\lfloor Tu\right\rfloor +k/2-n_{2,T}/2+s+1}\right)\mathbb{E}\left(V_{\left\lfloor Tu\right\rfloor +k/2-n_{2,T}/2+s+1}\right)\\
 & \quad-\mu_{j+1}\left(u\right)^{2}+o_{\mathbb{P}}\left(1\right)\\
 & =c\left(u,\,k\right)+\mu_{j+1}\left(u\right)^{2}-\mu_{j+1}\left(u\right)^{2}+o_{\mathbb{P}}\left(1\right)=c\left(u,\,k\right)+o_{\mathbb{P}}\left(1\right),
\end{align*}
where we have used the smoothness of $\mathbb{E}(V_{t})$ implied
by local stationarity. The second claim of the lemma follows from
Assumption \ref{Assumption A - Dependence}-(i) since this implies
that $\sup_{u\in\left[0,\,1\right]}c\left(u,\,k\right)\rightarrow0$
as $k\rightarrow\infty$ and the fact that the third term on the right-hand
side of \eqref{Eq. Gamma + d for Local Autocovariance} does not depend
on $k.$ Thus, $\widehat{\Gamma}_{\mathrm{DK}}\left(k\right)\geq d_{T}^{*}+o_{\mathbb{P}}\left(1\right)$
where $d_{T}^{*}=\left(n_{2,T}/T\right)\gamma\left(1-\gamma\right)(\mu_{j}\left(\lambda_{j}^{0}\right)-\mu_{j+1}\left(u\right))^{2}>0$
and $d_{T}^{*}\rightarrow0$ since $n_{2,T}/T\rightarrow0$. The factor
$n_{2,T}/T$ in $d_{T}^{*}$ follows because the neighborhood $(\lambda_{j}^{0}-n_{2,T}/T,\,\lambda_{j}^{0}+n_{2,T}/T)$
includes $O(n_{2,T}/n_{T})$ blocks which are then averaged out. $\square$

\subsubsection{Proof of Theorem \ref{Theorem Local Periodogram Long Memory Effects}}

Consider first any $u\in\left(0,\,1\right)$ such that $T_{j}^{0}\notin\mathbf{S}\left(u,\,0,\,n_{T}\right)$
for all $j=1,\ldots,\,m_{0}$.  Theorem 3.3 in \citeReferencesSupp{casini:change-point-spectra}
shows that
\begin{align}
\mathbb{E}\left(I_{\mathrm{L},T}\left(u,\,\omega_{l}\right)\right) & =\left|\frac{1}{\sqrt{n_{T}}}\sum_{s=0}^{n_{T}-1}V_{\left\lfloor Tu\right\rfloor -n_{T}/2+s+1,T}\exp\left(-i\omega_{l}s\right)\right|^{2}\nonumber \\
 & =f\left(u,\,\omega_{l}\right)+\frac{1}{6}\left(\frac{n_{T}}{T}\right)^{2}\frac{\partial^{2}}{\partial u^{2}}f\left(u,\,\omega_{l}\right)+o\left(\left(\frac{n_{T}}{T}\right)^{2}\right)+O\left(\frac{\log\left(n_{T}\right)}{n_{T}}\right).\label{Eq. Expectation local periodogram}
\end{align}
 By Assumption \ref{Assumption Smothness of A (for HAC)} the absolute
value of the first term on the right-hand side is bounded for all
frequencies $\omega_{l}.$ By Assumption \ref{Assumption Means for Local Periodogram}-(iii)
$\left|\left(\partial^{2}/\partial u^{2}\right)f\left(u,\,\omega_{l}\right)\right|$
is bounded and, since $n_{T}/T\rightarrow0$, the second term converges
to zero. Similarly, the third and fourth terms are negligible. Thus,
$\mathbb{E}\left(I_{\mathrm{L},T}\left(u,\,\omega_{l}\right)\right)$
is bounded below by $f\left(u,\,\omega_{l}\right)>0$ as $\omega_{l}\rightarrow0$
which establishes part (i). Now we consider part (ii). We begin with
case (a). We only focus on the sub-case $n_{j,L}\left(u,\,0,\,n_{T}\right)/n_{T}$
$\rightarrow\gamma$ with $\gamma\in\left(0,\,1\right)$. We have
\begin{align}
I_{\mathrm{L,}T} & \left(\omega_{l}\right)=\nonumber \\
 & \Biggl|\frac{1}{\sqrt{n_{T}}}\left(\sum_{s=0}^{T_{j}^{0}-\left(\left\lfloor Tu\right\rfloor -n_{T}/2+1\right)}V_{\left\lfloor Tu\right\rfloor -n_{T}/2+s+1,T}\exp\left(-i\omega_{l}s\right)+\sum_{s=T_{j}^{0}-\left(\left\lfloor Tu\right\rfloor -n_{T}/2\right)}^{n_{T}-1}V_{\left\lfloor Tu\right\rfloor -n_{T}/2+s+1,T}\exp\left(-i\omega_{l}s\right)\Biggr|^{2}\right.\nonumber \\
 & =\frac{1}{n_{T}}\biggl|\sum_{s=0}^{T_{j}^{0}-\left(\left\lfloor Tu\right\rfloor -n_{T}/2+1\right)}\left(V_{\left\lfloor Tu\right\rfloor -n_{T}/2+s+1,T}-\mu\left(\left(\left\lfloor Tu\right\rfloor -n_{T}/2+s+1\right)/T\right)\right)\exp\left(-i\omega_{l}s\right)\nonumber \\
 & \quad+\sum_{s=T_{j}^{0}-\left(\left\lfloor Tu\right\rfloor -n_{T}/2\right)}^{n_{T}-1}\left(V_{\left\lfloor Tu\right\rfloor -n_{T}/2+s+1,T}-\mu\left(\left(\left\lfloor Tu\right\rfloor -n_{T}/2+s+1\right)/T\right)\right)\exp\left(-i\omega_{l}s\right)\nonumber \\
 & \quad+\sum_{s=0}^{n_{T}-1}\mu\left(\left(\left\lfloor Tu\right\rfloor -n_{T}/2+s+1\right)/T\right)\exp\left(-i\omega_{l}s\right)\biggr|^{2}.\label{Eq. Local periodogram decomposition}
\end{align}
Using Assumption \ref{Assumption Means for Local Periodogram}, we
have
\begin{align}
 & \left|\sum_{s=0}^{n_{T}-1}\mu\left(\left(\left\lfloor Tu\right\rfloor -n_{T}/2+s+1\right)/T\right)\exp\left(-i\omega_{l}s\right)\right|^{2}\geq\nonumber \\
 & \quad\left|B_{j}\sum_{s=0}^{T_{j}^{0}-\left(\left\lfloor Tu\right\rfloor -n_{T}/2+1\right)}\exp\left(-i\omega_{l}s\right)+B_{j+1}\sum_{s=T_{j}^{0}-\left(\left\lfloor Tu\right\rfloor -n_{T}/2\right)}^{n_{T}-1}\exp\left(-i\omega_{l}s\right)\right|^{2}.\label{Eq. inequality Bj > miut/T}
\end{align}
Note that 

\begin{align}
B_{j} & \sum_{s=0}^{T_{j}^{0}-\left(\left\lfloor Tu\right\rfloor -n_{T}/2+1\right)}\exp\left(-i\omega_{l}s\right)+B_{j+1}\sum_{s=T_{j}^{0}-\left(\left\lfloor Tu\right\rfloor -n_{T}/2\right)}^{n_{T}-1}\exp\left(-i\omega_{l}s\right)\nonumber \\
 & =B_{j}\sum_{s=0}^{T_{j}^{0}-\left(\left\lfloor Tu\right\rfloor -n_{T}/2+1\right)}\exp\left(-i\omega_{l}s\right)\label{Eq. (Bj+Bj+1)}\\
 & \quad+B_{j+1}\exp\left(-i\omega_{l}\left(T_{j}^{0}-\left(\left\lfloor Tu\right\rfloor -n_{T}/2\right)\right)\right)\sum_{s=0}^{n_{T}-1-\left(T_{j}^{0}-\left(\left\lfloor Tu\right\rfloor -n_{T}/2\right)\right)}\exp\left(-i\omega_{l}s\right).\nonumber 
\end{align}
Focusing on the second term on the right-hand side above,
\begin{align}
n_{T}^{-1} & \left|B_{j+1}\sum_{s=T_{j}^{0}-\left(\left\lfloor Tu\right\rfloor -n_{T}/2\right)}^{n_{T}-1}\exp\left(-i\omega_{l}s\right)\right|^{2}\nonumber \\
 & =n_{T}^{-1}\left|B_{j+1}\exp\left(-i\omega_{l}\left(T_{j}^{0}-\left(\left\lfloor Tu\right\rfloor -n_{T}/2\right)\right)\right)\sum_{s=0}^{n_{T}-1-\left(T_{j}^{0}-\left(\left\lfloor Tu\right\rfloor -n_{T}/2\right)\right)}\exp\left(-i\omega_{l}s\right)\right|^{2}\nonumber \\
 & =n_{T}^{-1}\left|B_{j+1}\exp\left(-i\omega_{l}\left(T_{j}^{0}-\left(\left\lfloor Tu\right\rfloor -n_{T}/2\right)\right)\right)\frac{1-\exp\left(-i\omega_{l}\left(n_{T}-\left(T_{j}^{0}-\left(\left\lfloor Tu\right\rfloor -n_{T}/2\right)\right)\right)\right)}{1-\exp\left(-i\omega_{l}\right)}\right|^{2}\nonumber \\
 & =n_{T}^{-1}\left|B_{j+1}\frac{\exp\left(-i\omega_{l}\left(T_{j}^{0}-\left(\left\lfloor Tu\right\rfloor -n_{T}/2\right)\right)\right)-\exp\left(-i\omega_{l}n_{T}\right)}{1-\exp\left(-i\omega_{l}\right)}\right|^{2}.\label{Eq. Right-hand side of Bj+1}
\end{align}
We show that the above equation diverges to infinity as $\omega_{l}\rightarrow0$
with $n_{T}\omega_{l}^{2}\rightarrow0$. If $n_{T}\omega_{l}\rightarrow a\in(0,\,\infty)$
then $\mathrm{Re}\left(\exp\left(-i\omega_{l}n_{T}\right)\right)\neq1$
and the order is determined by the denominator. As in the proof of
Theorem \ref{Theorem Periodogram Long Memory Effects}, $|1-\exp(-i\omega_{l})|^{2}=\omega_{l}^{2}.$
Since $n_{T}\omega_{l}^{2}\rightarrow0$, the right-hand side above
diverges. If $n_{T}\omega_{l}\rightarrow0,$ we apply L'H\^{o}pital's
rule to obtain 
\begin{align*}
n_{T}^{-1} & \left|B_{j+1}\frac{-i\left(T_{j}^{0}-\left(\left\lfloor Tu\right\rfloor -n_{T}/2\right)\right)+in_{T}}{i}\right|^{2}\\
 & =n_{T}^{-1}B_{j+1}^{2}\left(-\left(T_{j}^{0}-\left(\left\lfloor Tu\right\rfloor -n_{T}/2\right)\right)^{2}+n_{T}^{2}-\left(T_{j}^{0}-\left(\left\lfloor Tu\right\rfloor -n_{T}/2\right)\right)n_{T}\right)\\
 & =O\left(n_{T}^{2}/n_{T}\right)=O\left(n_{T}\right),
\end{align*}
which shows that the right-hand side of \eqref{Eq. Right-hand side of Bj+1}
diverges. A similar argument can be applied to the first term on the
right-hand side of \eqref{Eq. (Bj+Bj+1)} and to the product of the
latter term and the complex conjugate of the second term on the right-hand
side of \eqref{Eq. (Bj+Bj+1)}.

It remains to consider case (b) and the sub-case $n_{j,L}\left(u,\,0,\,n_{T}\right)/n_{T}\rightarrow0$.
The other sub-case follows by symmetry. We have \eqref{Eq. Local periodogram decomposition}
and \eqref{Eq. inequality Bj > miut/T}. Note that, 
\begin{align*}
\Biggl| & \frac{1}{\sqrt{n_{T}}}B_{j+1}\sum_{s=T_{j}^{0}-\left(\left\lfloor Tu\right\rfloor -n_{T}/2\right)}^{n_{T}-1}\exp\left(-i\omega_{l}s\right)\Biggr|^{2}\\
 & =\left|\frac{1}{\sqrt{n_{T}}}B_{j+1}\sum_{s=0}^{n_{T}-1}\exp\left(-i\omega_{l}s\right)-\frac{1}{\sqrt{n_{T}}}B_{j+1}\sum_{s=0}^{T_{j}^{0}-\left(\left\lfloor Tu\right\rfloor -n_{T}/2\right)-1}\exp\left(-i\omega_{l}s\right)\right|^{2}\\
 & =\left|-\frac{1}{\sqrt{n_{T}}}B_{j+1}\sum_{s=0}^{T_{j}^{0}-\left(\left\lfloor Tu\right\rfloor -n_{T}/2\right)-1}\exp\left(-i\omega_{l}s\right)\right|^{2}\rightarrow0.
\end{align*}
Thus, we have 
\begin{align*}
\mathbb{E}\left(I_{\mathrm{L}T}\left(\omega_{l}\right)\right) & =\frac{1}{n_{T}}\Biggl|\left(\sum_{s=0}^{T_{j}^{0}-\left(\left\lfloor Tu\right\rfloor -n_{T}/2+1\right)}\left(V_{\left\lfloor Tu\right\rfloor -n_{T}/2+s+1,T}-\mu\left(\left(\left\lfloor Tu\right\rfloor -n_{T}/2+s+1\right)/T\right)\right)\exp\left(-i\omega_{l}s\right)\right)\\
 & \quad+\sum_{s=T_{j}^{0}-\left(\left\lfloor Tu\right\rfloor -n_{T}/2\right)}^{n_{T}-1}\left(V_{\left\lfloor Tu\right\rfloor -n_{T}/2+s+1,T}-\mu\left(\left(\left\lfloor Tu\right\rfloor -n_{T}/2+s+1\right)/T\right)\right)\exp\left(-i\omega_{l}s\right)\Biggr|^{2}+o\left(1\right).
\end{align*}
 Note that the first sum above involves at most $C<\infty$ summands.
So the first term is negligible. The expectation of the product of
the first term and the conjugate of the second term is negligible
by using arguments similar to the proof in Lemma \ref{Lemma Cross Products Periodogram}
with $n_{T}$ in place of $T$. Thus, the limit of $\mathbb{E}\left(I_{T}\left(\omega_{l}\right)\right)$
is equal to the right-hand side of \eqref{Eq. Expectation local periodogram}
plus additional $o\left(1\right)$ terms. $\square$

\subsection{Proofs of the Results in Section \ref{Section Edgeworth-Expansions-for}}

We first introduce the multiple Fejér kernel as in \citeReferencesSupp{velasco/robinson:01},
\begin{align*}
\Psi_{T}^{\left(n\right)}\left(x_{1},\ldots,\,x_{n}\right) & =\frac{1}{\left(2\pi\right)^{n-1}T}\sum_{t_{1}\cdots t_{n}=1}^{T}\exp\left\{ i\sum_{j=1}^{n}t_{j}x_{j}\right\} ,
\end{align*}
with $x_{n}=-\sum_{j=1}^{n-1}x_{j}$.  \citeReferencesSupp{velasco/robinson:01}
discussed the following properties. $\Psi_{T}^{\left(n\right)}\left(x_{1},\ldots,\,x_{n}\right)$
is integrable in $\Pi^{n-1}$ and integrates to one for all $T$.
For $\delta>0$ and $T\geq1$, we have
\begin{align}
\int_{\mathbf{D}^{c}}\left|\Psi_{T}^{\left(n\right)}\left(x_{1},\ldots,\,x_{n}\right)\right|dx_{1}\ldots dx_{n-1} & =O\left(\frac{\log^{n-1}T}{T\sin\delta/2}\right),\label{Eq. (B.1) in VR}
\end{align}
 where $\mathbf{D}^{c}$ is the complement in $\Pi^{n-1}$ of the
set $\mathbf{D}=\{x\in\Pi^{n-1}:\,|x_{j}|\leq\delta,\,j=1,\ldots,\,n-1\}$.
For $j=1,\ldots,\,n-1$, 
\begin{align}
\int_{\Pi}\cdots\int_{\Pi}|x_{j}||\Psi_{T}^{\left(n\right)}\left(x_{1},\ldots,\,x_{n}\right)|dx_{1}\cdots dx_{n} & =O\left(T^{-1}\log^{n-1}T\right).\label{Eq. (B.2) in VR}
\end{align}
Recall that the Dirichlet kernel is defined as $D_{T}\left(x\right)=\sum_{t=1}^{T}\exp\left(itx\right)$.
It satisfies the following two relations, 
\begin{align}
\left|D_{T}\left(x\right)\right| & \leq\min\left\{ T,\,2\left|x\right|^{-1}\right\} ;\qquad\int_{\Pi}\left|D_{T}\left(x\right)\right|dx=O\left(\log T\right).\label{Eq. (B.4) in VR}
\end{align}
Eq. \eqref{Eq. (B.1) in VR}-\eqref{Eq. (B.2) in VR} follow from
\begin{align}
\left|\Psi_{T}^{\left(n\right)}\left(x_{1},\ldots,\,x_{n}\right)\right|\leq\frac{1}{\left(2\pi\right)^{n-1}T}\left|D_{T}\left(x_{1}\right)\right|\left|D_{T}\left(x_{2}\right)\right|\cdots\left|D_{T}\left(x_{n}\right)\right| & dx_{1}\cdots dx_{n}.\label{Eq. (B.3) VR}
\end{align}

\subsubsection{Preliminary Lemmas}
\begin{lem}
\label{Lemma Bhattacharya and Rao 1975}(Bhattacharya and Rao, 1975,
pp. 97-98, 113). Let $\mathbb{Q}_{1}$ and $\mathbb{Q}_{2}$ be probability
measures on $\mathbb{R}^{2}$ and $\mathscr{B}^{2}$ the class of
all Borel subsets of $\mathbb{R}^{2}$. Let $\phi$ be a positive
number. Then there exists a kernel probability measure $\mathbb{G}_{\phi}$
such that 
\begin{align*}
\sup_{\mathbf{B}\in\mathscr{B}^{2}} & \left|\mathbb{Q}_{1}\left(\mathbf{B}\right)-\mathbb{Q}_{2}\left(\mathbf{B}\right)\right|\leq\frac{2}{3}\left\Vert \left(\mathbb{Q}_{1}-\mathbb{Q}_{2}\right)\bullet\mathbb{G}_{\phi}\right\Vert +\frac{4}{3}\sup_{\mathbf{B}\in\mathscr{B}^{2}}\mathbb{Q}_{2}\left(\left(\partial\mathbf{B}\right)^{2\phi}\right),
\end{align*}
 where $\mathbb{G}_{\phi}$ satisfies 
\begin{align}
\mathbb{G}_{\phi}\left(\mathbf{B}\left(0,\,r\right)^{c}\right) & =O\left(\left(\frac{\phi}{r}\right)^{3}\right),\label{Eq. (B.19) in VR}
\end{align}
and its Fourier transform $\widehat{\mathbb{G}}_{\phi}$ satisfies
\begin{align}
\widehat{\mathbb{G}}_{\phi}\left(\mathbf{t}\right) & =0\qquad\mathrm{for}\quad\left\Vert \mathbf{t}\right\Vert \geq8\times2^{4/3}/\pi^{1/3}\phi.\label{Eq. (B.20) in VR}
\end{align}
 Here $\left(\partial\mathbf{B}\right)^{2\phi}$ is a neighborhood
of radius $2\phi$ of the boundary of $\mathbf{B}$, $\left\Vert \cdot\right\Vert $
is the variation norm, and $\bullet$ means convolution. 
\end{lem}
\begin{lem}
\label{Lemma: Proposition 1 in VR}Let Assumptions \ref{Assumption 1 in VR},
\ref{Assumption 3 VR}-\ref{Assumption 4 in VR} hold. For $s\geq2$
with $\epsilon_{T}\left(2s\right)\rightarrow0$, we have
\begin{align*}
\mathrm{Tr}\left(\left(\Sigma_{V}W_{b_{1}}\right)^{s}\right) & =T\left(2\pi\right)^{2s-1}\sum_{j=0}^{d_{f}}L_{j}\left(s\right)b_{1,T}^{1+j-s}+O\left(Tb_{1,T}^{1-s}\epsilon_{T}\left(2s\right)\right),
\end{align*}
where $\epsilon_{T}\left(2s\right)=(Tb_{1,T})^{-1}\log^{2s-1}T$,
$L_{j}\left(s\right)=\left(1/j\right)!\mu_{j}\left(K^{s}\right)\left(d^{j}/d\omega^{j}\right)\left(f\left(u,\,0\right)du\right)^{s}$
with $|L_{j}\left(s\right)|<\infty$ and  $L_{j}\left(s\right)$
differs from zero only for $j$ even $\left(j=0,\ldots,\,d_{f}\right)$. 
\end{lem}
\noindent\textit{Proof of Lemma \ref{Lemma: Proposition 1 in VR}.}
Let $r_{2s+1}=r_{1}$ and note that 
\begin{align}
\mathrm{Tr} & \left(\left(\Sigma_{V}W_{b_{1}}\right)^{s}\right)\nonumber \\
 & =\sum_{1\leq r_{1},\ldots,r_{2s}\leq T}\prod_{j=1}^{s}\mathbb{E}\left(V_{r_{2j-1}}V_{r_{2j}}\right)w\left(b_{1,T}\left(r_{2j}-r_{2j+1}\right)\right)\nonumber \\
 & =\sum_{1\leq r_{1},\ldots,r_{2s}\leq T}\prod_{j=1}^{s}\int_{\Pi}f\left(r_{2j-1}/T,\,\omega_{2j-1}\right)e^{i\left(r_{2j-1}-r_{2j}\right)\omega_{2j-1}}\int_{\Pi}\widetilde{K}_{b_{1}}\left(\omega_{2j}\right)e^{i\left(r_{2j}-r_{2j+1}\right)\omega_{2j}}d\omega\nonumber \\
 & =\sum_{k_{2},\,k_{4},\ldots,\,k_{2s}=-T+1}^{T-1}\sum_{r_{1}=|k_{2}|+1}^{T}\sum_{r_{3}=|k_{4}|+1}^{T}\cdots\sum_{r_{2s-1}=|k_{2s}|+1}^{T}\prod_{j=1}^{s}\int_{\Pi}f\left(r_{2j-1}/T,\,\omega_{2j-1}\right)e^{ik_{2j}\left(\omega_{2j-1}-\omega_{2j}\right)}\nonumber \\
 & \quad\times\int_{\Pi}\widetilde{K}_{b_{1}}\left(\omega_{2j}\right)e^{i\left(\left(-k_{2j}-k_{2j+2}\right)\omega_{2j}\right)}d\omega\nonumber \\
 & =\sum_{k_{2},\,k_{4},\ldots,\,k_{2s}=-T+1}^{T-1}\prod_{j=1}^{s}\left(T-|k_{2j}|\right)\int_{\Pi}\int_{0}^{1}f\left(u_{2j-1},\,\omega_{2j-1}\right)e^{ik_{2j}\left(\omega_{2j-1}-\omega_{2j}\right)}\nonumber \\
 & \quad\times\int_{\Pi}\widetilde{K}_{b_{1}}\left(\omega_{2j}\right)e^{i\left(\left(-k_{2j}-k_{2j+2}\right)\omega_{2j}\right)}dud\omega+O\left(T^{-1}\right)\nonumber \\
 & =\sum_{1\leq r_{1},\ldots,r_{2s}\leq T}\prod_{j=1}^{s}\left(T-|k_{2j}|\right)\int_{\Pi}\int_{0}^{1}f\left(u_{2j-1},\,\omega_{2j-1}\right)\int_{\Pi}\widetilde{K}_{b_{1}}\left(\omega_{2j}\right)\exp\left\{ i\sum_{j=1}^{2s}\omega_{j}\left(r_{j}-r_{j+1}\right)\right\} dud\omega+O\left(T^{-1}\right)\nonumber \\
 & =T\left(2\pi\right)^{2s-1}\int_{\Pi^{2s}}H_{b_{1}}\left(\omega,\,\mu\right)\widetilde{K}_{b_{1}}\left(\omega\right)\Psi_{T}^{\left(2s\right)}\left(\mu\right)d\omega d\mu+O\left(T^{-1}\right),\label{Eq. (B.5a)}
\end{align}
where $\Psi_{T}^{\left(2s\right)}\left(\mu\right)=\Psi_{T}^{\left(2s\right)}\left(\mu_{1},\ldots,\,\mu_{2s}\right),$
\begin{align*}
H_{b_{1}}\left(\omega,\,\mu\right) & =\int_{0}^{1}\cdots\int_{0}^{1}f\left(u_{1},\,\omega-\mu_{2}-\ldots-\mu_{2s}\right)\widetilde{K}_{b_{1}}\left(\omega-\mu_{3}-\ldots-\mu_{2s}\right)\\
 & \quad\times f\left(u_{3},\,\omega-\ldots-\mu_{2s}\right)\widetilde{K}_{b_{1}}\left(\omega-\mu_{4}-\ldots-\mu_{2s}\right)\ldots f\left(u_{2s-1},\,\omega-\mu_{2s}\right)du,
\end{align*}
 $d\mu=d\mu_{2},\ldots,\,d\mu_{2s},$ $d\omega=d\omega_{1},\ldots,\,\omega_{2s}$,
$du=du_{1},\,du_{3},\ldots,\,du_{2s-1}$, and we have made the change
in variables 
\begin{align*}
\begin{cases}
\mu_{1}=\omega_{1}-\omega_{2}\\
\mu_{2}=\omega_{2}-\omega_{1}\\
\cdots\\
\mu_{2s}=\omega_{2s}-\omega_{2s-1}
\end{cases} & \begin{cases}
\omega_{2s-1}=\omega-\mu_{2s}\\
\omega_{2s-2}=\omega-\mu_{2s}-\mu_{2s-1}\\
\cdots\\
\omega_{1}=\omega-\mu_{2s}-\ldots-\mu_{s}=\omega-\mu_{1}
\end{cases}
\end{align*}
with $\sum_{j=1}^{2s}\mu_{j}=0$, setting $\omega=\omega_{2s}$, and
expressing all the $\omega_{j}$ in terms of $\omega$ and $\mu_{j}$,
$j=2,\ldots,\,2s$. 

Let
\[
B=\left|\mathrm{Tr}\left(\left(\Sigma_{V}W_{b_{1}}\right)^{s}\right)-T\left(2\pi\right)^{2s-1}\int_{\Pi}\left(\int_{0}^{1}f\left(u,\,\omega\right)du\right)^{s}\widetilde{K}_{b_{1}}^{s-1}\left(\omega\right)d\omega\right|.
\]
Using \eqref{Eq. (B.5a)} we have
\begin{align}
B & \leq T\left(2\pi\right)^{2s-1}\int_{\Pi^{2s}}\left|H_{b_{1}}\left(\omega,\,\mu\right)-\left(\int_{0}^{1}f\left(u,\,\omega\right)du\right)^{s}\widetilde{K}_{b_{1}}^{s-1}\left(\omega\right)\right|\left|\widetilde{K}_{b_{1}}\left(\omega\right)\Psi_{T}^{\left(2s\right)}\left(\mu\right)\right|d\omega d\mu+O\left(T^{-1}\right).\label{Eq. (B.7) in VR}
\end{align}
 We split the integral in \eqref{Eq. (B.7) in VR} into two sets,
for small and for large $\mu_{j}$. Define the set $\mathbf{M}=\{\mu\in\Pi^{2s-1}:$
$\sup_{j}|\mu_{j}|\leq b_{1,T}/\left(2s\right)\}$. Since $K\left(\omega\right)$
takes small values for $\left|\omega\right|>\pi b_{1,T}$, for all
$u$ all functions $f\left(u,\,\omega\right)$ are boundedly differentiable
in $\omega$ in the set $\mathbf{M}$. We use the following inequality,
\begin{align}
\left|A_{1}\cdots A_{r}-B_{1}\cdots B_{r}\right| & \leq\sum_{q=0}^{r-1}\left|B_{1}\cdots B_{q}\right|\left|B_{q+1}-A_{q+1}\right|\left|A_{q+2}\cdots A_{r}\right|,\label{Eq. (B.8) in VR}
\end{align}
 and $\sup_{\omega}|\widetilde{K}_{b_{1}}\left(\omega\right)|=O(b_{1,T}^{-1})$
to bound the integral in \eqref{Eq. (B.7) in VR} over $\mathbf{M}$
by
\begin{align}
O & \left(Tb_{1,T}^{-s+1}\right)\sum_{q=0}^{s-1}\int_{\Pi}\int_{\mathbf{M}}\int_{0}^{1}\left|f\left(u_{2q+1},\,\omega-\mu_{2+2q}-\ldots-\mu_{2s}\right)-f\left(u_{2q+1},\,\omega\right)\right|\left|\widetilde{K}_{b_{1}}\left(\omega\right)\Psi_{T}^{\left(2s\right)}\left(\mu\right)\right|du_{2q+1}d\mu d\omega\label{Eq. (B.9) in VR}\\
+ & O\left(Tb_{1,T}^{-s+1}\right)\sum_{q=0}^{s-2}\int_{\Pi}\int_{\mathbf{M}}\left|\widetilde{K}_{b_{1}}\left(\omega-\mu_{3+2q}-\ldots-\mu_{2s}\right)-\widetilde{K}_{b_{1}}\left(\omega\right)\right|\left|\Psi_{T}^{\left(2s\right)}\left(\mu\right)\right|d\mu d\omega.\label{Eq. (B.10) in VR}
\end{align}
We apply the mean value theorem in \eqref{Eq. (B.9) in VR} to yield,
\begin{align*}
O\left(Tb_{1,T}^{1-s}\right) & \int_{\Pi}\left|\widetilde{K}_{b_{1}}\left(\omega\right)\right|d\omega\sum_{q=0}^{2s}\int_{\mathbf{M}}|\mu_{q}||\Psi_{T}^{\left(2s\right)}\left(\mu\right)|d\mu\\
 & \leq O\left(Tb_{1,T}^{1-s}\right)\int_{\Pi}\left|\widetilde{K}_{b_{1}}\left(\omega\right)\right|d\omega\sum_{q=0}^{2s}\int_{\Pi^{2s-1}}|\mu_{q}||\Psi_{T}^{\left(2s\right)}\left(\mu\right)|d\mu\\
 & =O\left(b_{1,T}^{1-s}\log^{2s-1}T\right),
\end{align*}
where the equality follows from \eqref{Eq. (B.2) in VR}. Using the
Lipschitz property of $K$ (cf. Assumption \ref{Assumption 4 in VR}),
the expression in \eqref{Eq. (B.10) in VR} is of order $O(b_{1,T}^{-s}\log^{2s-1}T)$. 

Let $\mathbf{M}^{c}$ denote the complement of $\mathbf{M}$ in $\Pi^{2s-1}.$
We now study the contribution to $B$ corresponding to the set $\mathbf{M}^{c}$.
This is bounded by
\begin{align}
T\left(2\pi\right)^{2s-1} & \int_{\Pi}\int_{\mathbf{M}^{c}}\left|H_{b_{1}}\left(\omega,\,\mu\right)\widetilde{K}_{b_{1}}\left(\omega\right)\right|\left|\Psi_{T}^{\left(2s\right)}\left(\mu\right)\right|d\omega d\mu\label{Eq. (B.11) in VR}\\
 & \quad+T\left(2\pi\right)^{2s-1}\int_{\Pi}\left|\left(\int_{0}^{1}f\left(u,\,\omega\right)du\right)^{s}\widetilde{K}_{b_{1}}^{s}\left(\omega\right)\right|d\omega\int_{\mathbf{M}^{c}}\left|\Psi_{T}^{\left(2s\right)}\left(\mu\right)\right|d\mu.\label{Eq. (B.12) in VR}
\end{align}
The expression in \eqref{Eq. (B.12) in VR} is $O(b_{1,T}^{-s}\log^{2s-1}T)$
using \eqref{Eq. (B.1) in VR} and 
\[
\int_{\Pi}\left|\left(\int_{0}^{1}f(u,\,\omega)du\right){}^{s}\widetilde{K}_{b_{1}}^{s}\left(\omega\right)\right|d\omega=O\left(b_{1,T}^{-s}\right).
\]
  Applying \eqref{Eq. (B.3) VR} the expression in \eqref{Eq. (B.11) in VR}
is bounded by 
\begin{align}
\int_{\mathbf{M}'}\prod_{j=1}^{s}\int_{0}^{1} & \left|f\left(u_{2j-1},\,\omega_{2j-1}\right)\widetilde{K}_{b_{1}}\left(\omega_{2j}\right)D_{T}\left(\omega_{2j}-\omega_{2j-1}\right)D_{T}\left(\omega_{2j+1}-\omega_{2j}\right)\right|du_{2j-1}d\omega_{2j}d\omega_{2j-1},\label{Eq. (B.13) in VR}
\end{align}
 where $\mathbf{M}'=\{\left|\omega_{2}-\omega_{1}\right|>\nu_{T}\}\cup\{\left|\omega_{3}-\omega_{2}\right|>\nu_{T}\}\cup\ldots\cup\{\left|\omega_{2s}-\omega_{2s-1}\right|>\nu_{T}\}$
with $\nu_{T}=b_{1,T}/\left(2s\right)$ and $2s+1$ is to be interpreted
as 1. Note that the integral in \eqref{Eq. (B.13) in VR} differs
from zero only if $\left|\omega_{2}\right|,\,\left|\omega_{4}\right|,\ldots,\,\left|\omega_{2s}\right|\leq b_{1,T}\pi$.
Without loss of generality, we consider only the case where just one
of the events in $\mathbf{M}'$ is satisfied, $|\omega_{2j}-\omega_{2j-1}|>\nu_{T}$,
say, the other cases can be handled similarly. 

From \eqref{Eq. (B.4) in VR} it follows that $|D_{T}(\omega_{2j}-\omega_{2j-1})|=O(b_{1,T}^{-1})$
since $|\omega_{2j}-\omega_{2j-1}|>\nu_{T}=b_{1,T}/\left(2s\right)$,
and $\int_{\Pi}|D_{T}(\omega_{2j}-\omega_{2j-1})\widetilde{K}_{b_{1}}(\omega_{2j})|d\omega_{2j}=O(b_{1,T}^{-1}\log T)$.
For $\epsilon>0,$ consider the following decomposition
\begin{align}
\int_{\Pi} & \int_{0}^{1}\left|f\left(u_{2j-1},\,\omega_{2j-1}\right)D_{T}\left(\omega_{2j-1}-\omega_{2j-2}\right)\right|du_{2j-1}d\omega_{2j-1}\label{Eq. (B.14) in VR}\\
 & =\int_{\left|\omega_{2j-1}\right|\leq\epsilon}\int_{0}^{1}\left|f\left(u_{2j-1},\,\omega_{2j-1}\right)D_{T}\left(\omega_{2j-1}-\omega_{2j-2}\right)\right|du_{2j-1}d\omega_{2j-1}\nonumber \\
 & \quad+\int_{\left|\omega_{2j-1}\right|>\epsilon}\int_{0}^{1}\left|f\left(u_{2j-1},\,\omega_{2j-1}\right)D_{T}\left(\omega_{2j-1}-\omega_{2j-2}\right)\right|du_{2j-1}d\omega_{2j-1}.\nonumber 
\end{align}
By Assumption \ref{Assumption 1 in VR} $f(u_{2j-1},\,\omega_{2j-1})$
is bounded if $|\omega_{2j-1}|\leq\epsilon$. Then, the integral
over $|\omega_{2j-1}|\leq\epsilon$ above is of order $O\left(\log T\right)$.
On the other hand, if $\left|\omega_{2j-1}\right|>\epsilon$ (and
recall that $|\omega_{2j-1}|\leq b_{1,T}\pi$), we yield as $T\rightarrow\infty$
$|\omega_{2j-1}-\omega_{2j-2}|>\epsilon/2$, say. Then, $|D_{T}(\omega_{2j-1}-\omega_{2j-2})|=O\left(1\right)$
by \eqref{Eq. (B.4) in VR}  and the second summand of \eqref{Eq. (B.14) in VR}
is finite in view of the integrability of $f\left(u,\,\omega\right)$
by Assumption \ref{Assumption: Assumption 2 in VR }. It follows that
\eqref{Eq. (B.14) in VR} is $O\left(\log T\right)$. There are other
$s-1$ integrals of this type that can be handled in the same way.
The remaining integral is of the form 
\begin{align*}
\int_{\Pi}\int_{\Pi}\int_{0}^{1}\left|\widetilde{K}_{b_{1}}\left(\omega_{2s}\right)f\left(u_{2s-1},\,\omega_{1}\right)D_{T}\left(\omega_{1}-\omega_{2s}\right)\right|du_{2s-1}d\omega_{1}d\omega_{2s} & =O\left(\log T\right),
\end{align*}
 where $\omega_{1}=\omega_{2s+1}$ and we have used the same argument
as in \eqref{Eq. (B.14) in VR} to show that the integral in $\omega_{1}$
is $O\left(\log T\right)$ for all $\omega_{2s}$ and that $\int_{\Pi}|\widetilde{K}_{b_{1}}(\omega_{2s})|d\omega_{2s}=O\left(1\right)$.
Thus, \eqref{Eq. (B.13) in VR} is $O(b_{1,T}^{-s}\log^{2s-1}T)$
and $B=O(b_{1,T}^{1-s}\log^{2s-1}T+b_{1,T}^{-s}\log^{2s-1}T+T^{-1})=O(Tb_{1,T}^{1-s}\epsilon_{T}\left(2s\right))$. 

Define $R_{b_{1}}\left(s\right)=\sum_{j=0}^{d_{f}}L_{j}\left(s\right)b_{1,T}^{1+j-s}$.
Using the Lipschitz property of $f^{\left(d_{f}\right)}\left(u,\,\omega\right)$
for all $u,$ 
\begin{align*}
\biggl|\int_{\Pi} & \widetilde{K}_{b_{1}}^{s}\left(\omega\right)\left(\int_{0}^{1}f\left(u,\,\omega\right)du\right)^{s}d\omega-R_{b_{1}}\left(s\right)\biggr|\\
 & \leq\int_{\Pi}\left|\widetilde{K}_{b_{1}}\left(\omega\right)\right|^{s-1}\left|\left(\int_{0}^{1}f\left(u,\,\omega\right)du\right)^{s}-\sum_{j=0}^{d_{f}}\frac{1}{j!}\left(\frac{d}{d\omega}\right)^{j}\left(\int_{0}^{1}f\left(u,\,0\right)du\right)^{s}\omega^{j}\right|\left|\widetilde{K}_{b_{1}}\left(\omega\right)\right|d\omega\\
 & =O\left(\sup_{\omega\in\Pi}\left|\widetilde{K}_{b_{1}}\left(\omega\right)\right|^{s-1}\left|\int_{\Pi}\left|\omega\right|^{d_{f}+\varrho}\right|\left|\widetilde{K}_{b_{1}}\left(\omega\right)\right|d\omega\right)=O\left(b_{1,T}^{d_{f}+\varrho-s+1}\right),
\end{align*}
 where we have used $\sup_{\omega\in\Pi}|\widetilde{K}_{b_{1}}\left(\omega\right)|=O(b_{1,T}^{-1})$.

Note that $L_{j}\left(s\right)$ differs from zero for $j$ even 
because $L_{j}\left(s\right)$ depends on $\mu_{j}(K^{s})$. $\square$ 
\begin{lem}
\label{Lemma: Proposition 2 in VR}Let Assumptions \ref{Assumption 1 in VR}
and \ref{Assumption 3 VR}-\ref{Assumption 4 in VR} hold. For $s\geq1$
with $\epsilon_{T}\left(2s+2\right)\rightarrow0$, we have 
\[
\mathbf{1}'\left(\Sigma_{V}W_{b_{1}}\right)^{s}\Sigma_{V}\mathbf{1}=T\left(2\pi\right)^{2s+1}\left(\int_{0}^{1}f\left(u,\,0\right)du\right)^{s+1}\left(\widetilde{K}_{b_{1}}\left(0\right)\right)^{s}+O\left(b_{1,T}^{-1-s}\log^{2s+1}T+T^{-1}\right).
\]
\end{lem}
\noindent\textit{Proof of Lemma \ref{Lemma: Proposition 2 in VR}.}
We first write $\mathbf{1}'(\Sigma_{V}W_{b_{1}})^{s}\Sigma_{V}\mathbf{1}$
using an argument similar to the one used to derive \eqref{Eq. (B.5a)},
the only difference being that we also have the summation over two
additional indexes. We write 
\begin{align}
\sum_{0\leq r_{1},\ldots,\,r_{2s+2}\leq T} & \mathbb{E}\left(V_{r_{2s+1}}V_{r_{2s+2}}\right)\Pi_{j=1}^{s}\left\{ \mathbb{E}\left(V_{r_{2j-1}}V_{r_{2j}}\right)w\left(b_{1,T}\left(r_{2j}-r_{2j+1}\right)\right)\right\} \nonumber \\
 & =\sum_{r}\int_{\Pi}f\left(r_{2s+1}/T,\,\omega_{2s+1}\right)e^{i\left(r_{2s+1}-r_{2s+2}\right)\omega_{2s+1}}\Pi_{j=1}^{s}\nonumber \\
 & \quad\times\left\{ f\left(r_{2j-1}/T,\,\omega_{2j-1}\right)e^{i\left(r_{2j-1}-r_{2j}\right)\omega_{2j-1}}\int_{\Pi}\widetilde{K}_{b_{1}}\left(\lambda_{2j}\right)e^{i\left(r_{2j}-r_{2j+1}\right)\lambda_{2j}}\right\} d\lambda d\omega\nonumber \\
 & =T\left(2\pi\right)^{2s+1}\int_{\Pi^{2s+1}}S_{b_{1}}\left(\mu\right)\Psi_{T}^{\left(2s+2\right)}\left(\mu\right)d\mu+O\left(T^{-1}\right),\label{Eq. (B.15) in VR}
\end{align}
using a change of variable, where $\Psi_{T}^{\left(2s+2\right)}\left(\mu\right)=\Psi_{T}^{\left(2s+2\right)}(\mu_{1},\ldots,\,\mu_{2s+1},\,-\sum_{j=1}^{2s+1}\mu_{j}),$
\begin{align*}
S_{b_{1}}\left(\mu\right) & =\int_{0}^{1}\cdots\int_{0}^{1}f\left(u_{1},\,\mu_{1}\right)\widetilde{K}_{b_{1}}\left(\mu_{1}+\mu_{2}\right)\ldots\widetilde{K}_{b_{1}}\left(\mu_{1}+\ldots+\mu_{2s}\right)f\left(u_{2s+1}\,,\mu_{1}+\ldots+\mu_{2s+1}\right)du,
\end{align*}
 and $d\mu=d\mu_{1}\ldots d\mu_{2s+1},\,du=du_{1}\ldots du_{2s+1}$
and $d\omega=d\omega_{1}\ldots d\omega_{2s+1}$. Proceeding as in
the proof of Lemma \ref{Lemma: Proposition 1 in VR}, we divide the
range of integration in \eqref{Eq. (B.15) in VR}, $\Pi^{2s+1}$,
into two sets, $\mathbf{M}$ and its complement $\mathbf{M}^{c}$,
where $\mathbf{M}=\{|\mu_{j}|\leq\pi b_{1,T}/\left(2s+2\right),\,j=1,\ldots,\,2s+1\}$.
We have 
\begin{align}
\biggl|\int_{\mathbf{M}} & S_{b_{1}}\left(\mu\right)\Psi_{T}^{\left(2s+2\right)}\left(\mu\right)d\mu-\int_{\mathbf{M}}\left(\int_{0}^{1}f\left(u,\,0\right)du\right)^{s+1}\widetilde{K}_{b_{1}}^{s}\left(0\right)\Psi_{T}^{\left(2s+2\right)}\left(\mu\right)d\mu\biggr|\nonumber \\
 & =O\left(b_{1,T}^{-s-1}\right)\int_{\Pi^{2s+1}}\sum_{j=2}^{2s}\left|\mu_{j}\right|\left|\Psi_{T}^{\left(2s+2\right)}\left(\mu\right)\right|d\mu\nonumber \\
 & =O\left(b_{1,T}^{-s-1}T^{-1}\log^{2s+1}T\right),\label{Eq. (B.16) in VR}
\end{align}
using \eqref{Eq. (B.2) in VR}, \eqref{Eq. (B.8) in VR}, Assumptions
\ref{Assumption 1 in VR} and \ref{Assumption 4 in VR}. On the other
hand, the contribution from $\mathbf{M}^{c}$ is less than or equal
to 
\begin{align}
\int_{\mathbf{M}^{c}}\left|S_{b_{1}}\left(\mu\right)\right|\left|\Psi_{T}^{\left(2s+2\right)}\left(\mu\right)\right|d\mu & +O\left(b_{1,T}^{-s-1}T^{-1}\log^{2s+1}T\right),\label{Eq. (B.17) in VR}
\end{align}
 where we have used \eqref{Eq. (B.1) in VR}. Using the same argument
used for \eqref{Eq. (B.13) in VR}, the integral in \eqref{Eq. (B.17) in VR}
is less than or equal to 
\begin{align}
\frac{1}{T\left(2\pi\right)^{2s+1}} & \int_{\mathbf{M}'}\prod_{j=1}^{s}\int_{0}^{1}\int_{0}^{1}\left[f\left(u_{2j-1},\,\omega_{2j-1}\right)\right.\widetilde{K}_{b_{1}}\left(\omega_{2j}\right)D_{T}\left(\omega_{2j}-\omega_{2j-1}\right)\label{Eq. (B.18) in VR}\\
 & \times D_{T}\left(\omega_{2j+1}-\omega_{2j}\right)\left.f\left(u_{2s+1},\,\omega_{2s+1}\right)D_{T}\left(\omega_{1}\right)D_{T}\left(-\omega_{2s-1}\right)\right]dud\omega,\nonumber 
\end{align}
where 
\[
\mathbf{M}'=\left\{ \left|\omega_{1}\right|>\pi b_{1,T}/\left(2s+2\right)\right\} \cup\left\{ \left|\omega_{2}-\omega_{1}\right|>\pi b_{1,T}/\left(2s+2\right)\right\} \cup\ldots\cup\left\{ \left|\omega_{2s-1}-\omega_{2s}\right|>\pi b_{1,T}/\left(2s+2\right)\right\} ,
\]
and \eqref{Eq. (B.18) in VR} is nonzero only if $\left|\omega_{2}\right|,\,\left|\omega_{4}\right|,\ldots,\,\left|\omega_{2s}\right|\leq\pi b_{1,T}$.

If $\left|\omega_{j+1}-\omega_{j}\right|>\pi b_{1,T}/\left(2s+2\right)$
for at least one index $j\in\left\{ 1,\ldots,\,2s\right\} $ we can
obtain a bound of order $(T^{-1}b_{1,T}^{-s-1}\log^{2s+1}T)$ for
\eqref{Eq. (B.18) in VR} as in Lemma \ref{Lemma: Proposition 1 in VR}.
The same bound is obtained for the case $\left|\omega_{1}\right|>\pi b_{1,T}/\left(2s+2\right)$
with a similar argument. Combining these results with \eqref{Eq. (B.15) in VR}-\eqref{Eq. (B.17) in VR}
concludes the proof. $\square$ 
\begin{lem}
\label{Lemma: Proposition 1 in VR DK-HAC}Let Assumptions \ref{Assumption 1 in VR},
\ref{Assumption 3 VR}-\ref{Assumption 4 in VR} and \ref{Assumption K2 and b2}-\ref{Assumption Lip of d2 f(u,w)}
hold. For $s\geq2$ with $\epsilon_{Tb_{2,T}}\left(2s\right)\rightarrow0$,
we have
\begin{align*}
\mathrm{Tr} & \left(\left(\Sigma_{\widetilde{V}}W_{b_{1}}\right)^{s}\right)=Tb_{2,T}\left(2\pi\right)^{2s-1}\left(\sum_{j=0}^{d_{f}}L_{j}\left(s\right)b_{1,T}^{1+j-s}+b_{2,T}^{2}\sum_{j=0}^{d_{f}}\left(\left(L_{2,j}\left(s\right)+L_{3,j}\left(s\right)\right)b_{1,T}^{1+j-s}\right)\right)\\
 & \quad+O\left(Tb_{2,T}b_{1,T}^{1-s}\epsilon_{Tb_{2,T}}\left(2s\right)+b_{1,T}^{-s}\frac{\log^{2s}\left(Tb_{2,T}\right)}{Tb_{2,T}}\right),
\end{align*}
where $\epsilon_{Tb_{2,T}}\left(2s\right)=(Tb_{2,T})^{-1}\log^{2s-1}\left(Tb_{2,T}\right)$,
 $L_{j}\left(s\right)=\left(1/j\right)!\mu_{j}(K^{s})\int_{0}^{1}K_{2}^{s}\left(x\right)dx\left(d^{j}/d\omega^{j}\right)(\int_{0}^{1}f\left(u,\,0\right)du)^{s}$
with $|L_{j}\left(s\right)|<\infty$, $L_{j}\left(s\right)$ differs
from zero only for $j$ even, $L_{2,j}\left(s\right)$ depends on
$\frac{\partial^{2}}{\partial u^{2}}\int_{\widetilde{\mathbf{C}}}f\left(u,\,\omega\right)du$,
$K_{2}$, $\widetilde{K}_{b_{1}}$ and $s$ with $|L_{2,j}\left(s\right)|<\infty$,
and $L_{3,j}\left(s\right)$ depends on $\Delta_{f}\left(\cdot\right)$,
$\widetilde{K}_{b_{1}}$ and $s$ with $|L_{3,j}\left(s\right)|<\infty$. 
\end{lem}
\noindent\textit{Proof of Lemma \ref{Lemma: Proposition 1 in VR DK-HAC}.}
Let $r_{2s+1}=r_{1}$ and note that 
\begin{align}
\mathrm{Tr}\left(\left(\Sigma_{\widetilde{V}}W_{b_{1}}\right)^{s}\right) & =\int_{0}^{1}\cdots\int_{0}^{1}\sum_{1\leq r_{1},\ldots,r_{2s}\leq T}\prod_{j=1}^{s}\mathbb{E}\left(\widetilde{V}_{r_{2j-1}}\left(u_{j}\right)\widetilde{V}_{r_{2j}}\left(u_{j}\right)\right)w\left(b_{1,T}\left(r_{2j}-r_{2j+1}\right)\right)du\nonumber \\
 & =\int_{0}^{1}\cdots\int_{0}^{1}\sum_{1\leq r_{1},\ldots,r_{2s}\leq T}\prod_{j=1}^{s}K_{2}\left(\frac{\left(Tu_{j}-\left(r_{2j-1}-\left(r_{2j}-r_{2j-1}\right)/2\right)\right)/T}{b_{2,T}}\right)\nonumber \\
 & \quad\times\int_{\Pi}f\left(r_{2j-1}/T,\,\omega\right)e^{i\left(r_{2j-1}-r_{2j}\right)\omega_{2j-1}}d\omega\int_{\Pi}\widetilde{K}_{b_{1}}\left(\omega_{2j}\right)e^{i\left(r_{2j}-r_{2j+1}\right)\omega_{2j}}d\omega du\nonumber \\
 & =\sum_{k_{2},\,k_{4},\ldots,\,k_{2s}=-\left\lfloor Tb_{2,T}\right\rfloor +1}^{\left\lfloor Tb_{2,T}\right\rfloor -1}\int_{0}^{1}\cdots\int_{0}^{1}\int_{\Pi^{2}}\prod_{j=1}^{s}\left(Tb_{2,T}-|k_{2j}|\right)f\left(u_{2j-1},\,\omega_{2j-1}\right)e^{i\left(\omega_{2j-1}-\omega_{2j}\right)k_{2j}}\nonumber \\
 & \quad\times\widetilde{K}_{b_{1}}\left(\omega_{2j}\right)e^{i\left(-k_{2j}-k_{2j+2}\right)\omega_{2j}}d\omega du+O\left(b_{2,T}^{2}\right)+O\left(\frac{\log\left(Tb_{2,T}\right)}{Tb_{2,T}}\right)\nonumber \\
 & =Tb_{2,T}\left(2\pi\right)^{2s-1}\int_{\Pi^{2s}}\left(H_{b_{1}}\left(\omega,\,\mu\right)\int_{0}^{1}K_{2}^{s}\left(x\right)dx+H_{2,b_{1}}\left(\omega,\,\mu\right)+H_{3,b_{1}}\left(\omega,\,\mu\right)\right)\label{Eq. (B.5a) DK-HAC}\\
 & \quad\times\widetilde{K}_{b_{1}}\left(\omega\right)\Psi_{Tb_{2,T}}^{\left(2s\right)}\left(\mu\right)d\omega d\mu+O\left(b_{2,T}^{2}b_{1,T}^{-s}\log^{2s-1}\left(Tb_{2,T}\right)\right)+O\left(b_{1,T}^{-s}\frac{\log^{2s}\left(Tb_{2,T}\right)}{Tb_{2,T}}\right),\nonumber 
\end{align}
where $H_{b_{1}}\left(\omega,\,\mu\right),$ $d\omega$ and $d\mu$
are defined as in \eqref{Eq. (B.5a)}, $\Psi_{Tb_{2,T}}^{\left(2s\right)}\left(\mu\right)=\Psi_{Tb_{2,T}}^{\left(2s\right)}\left(\mu_{1},\ldots,\,\mu_{2s}\right),$
\begin{align*}
H_{2,b_{1}}\left(\omega,\,\mu\right) & =b_{2,T}^{2}\left(\int_{0}^{1}x^{2}K_{2}\left(x\right)dx\right)\left(\int_{0}^{1}K_{2}^{s-1}\left(x\right)dx\right)\\
 & \quad\times\sum_{j\in\mathbf{J}}\frac{\partial^{2}}{\partial u_{j}^{2}}\int_{\widetilde{\mathbf{C}}}\cdots\int_{\widetilde{\mathbf{C}}}f\left(u_{1},\,\omega-\mu_{2}-\ldots-\mu_{2s}\right)\widetilde{K}_{b_{1}}\left(\omega-\mu_{3}-\ldots-\mu_{2s}\right)\\
 & \quad\times f\left(u_{3},\,\omega-\ldots-\mu_{2s}\right)\widetilde{K}_{b_{1}}\left(\omega-\mu_{4}-\ldots-\mu_{2s}\right)\ldots f\left(u_{2s-1},\,\omega-\mu_{2s}\right)du_{1}\cdots du_{2s-1},
\end{align*}
 with $\mathbf{J}=\left\{ 1,\,3,\ldots,\,2s-1\right\} $, and $H_{3,b_{1}}\left(\omega,\,\mu\right)$
depends on the discontinuity points, i.e.,
\begin{align*}
H_{3,b_{1}}\left(\omega,\,\mu\right) & =b_{2,T}^{2}\left(\int_{0}^{1}K_{2}^{s-1}\left(x\right)dx\right)\left(\mathbf{1}\left\{ u_{1}=\lambda_{j}^{0},\,j=1,\ldots,\,m_{0}\right\} \Delta_{f,j}\left(\omega-\mu_{2}-\ldots-\mu_{2s}\right)\right)\\
 & \quad\times\widetilde{K}_{b_{1}}\left(\omega-\mu_{3}-\ldots-\mu_{2s}\right)f\left(u_{3},\,\omega-\ldots-\mu_{2s}\right)\widetilde{K}_{b_{1}}\left(\omega-\mu_{4}-\ldots-\mu_{2s}\right)\ldots f\left(u_{2s-1},\,\omega-\mu_{2s}\right)\\
 & \quad\vdots\\
 & \quad+b_{2,T}^{2}\left(\int_{0}^{1}K_{2}^{s-1}\left(x\right)dx\right)f\left(u_{1},\,\omega-\mu_{2}-\ldots-\mu_{2s}\right)\widetilde{K}_{b_{1}}\left(\omega-\mu_{3}-\ldots-\mu_{2s}\right)\\
 & \quad\times f\left(u_{3},\,\omega-\ldots-\mu_{2s}\right)\widetilde{K}_{b_{1}}\left(\omega-\mu_{4}-\ldots-\mu_{2s}\right)\ldots\\
 & \quad\times\mathbf{1}\left\{ u_{2s-1}=\lambda_{j}^{0},\,j=1,\ldots,\,m_{0}\right\} \Delta_{f,j}\left(\omega-\mu_{2s}\right),
\end{align*}
 with 
\begin{align}
\Delta_{f,j}\left(\omega\right) & =\int_{0}^{1}\left(\frac{\partial}{\partial u_{-}}f\left(\lambda_{j}^{0},\,\omega\right)\int_{0}^{1-s}xK_{2}\left(x\right)dx+\frac{\partial}{\partial u_{+}}f\left(\lambda_{j}^{0},\,\omega\right)\int_{1-s}^{1}xK_{2}\left(x\right)dx\right)ds.\label{Eq. Delta,f,j}
\end{align}
Let
\[
B=\left|Tb_{2,T}\left(2\pi\right)^{2s-1}\int_{0}^{1}K_{2}^{s}\left(x\right)dx\int_{\Pi^{2s}}\left(H_{b_{1}}\left(\omega,\,\mu\right)\widetilde{K}_{b_{1}}\left(\omega\right)\Psi_{Tb_{2,T}}^{\left(2s\right)}\left(\mu\right)-\left(\int_{0}^{1}f\left(u,\,\omega\right)du\right)^{s}\widetilde{K}_{b_{1}}^{s}\left(\omega\right)\right)d\omega d\mu\right|.
\]
Using \eqref{Eq. (B.5a) DK-HAC} we have
\begin{align}
B & \leq Tb_{2,T}\left(2\pi\right)^{2s-1}\int_{0}^{1}K_{2}^{s}\left(x\right)dx\int_{\Pi^{2s}}\left|H_{b_{1}}\left(\omega,\,\mu\right)-\left(\int_{0}^{1}f\left(u,\,\omega\right)du\right)^{s}\widetilde{K}_{b_{1}}^{s-1}\left(\omega\right)\right|\left|\widetilde{K}_{b_{1}}\left(\omega\right)\Psi_{Tb_{2,T}}^{\left(2s\right)}\left(\mu\right)\right|d\omega d\mu.\label{Eq. (B.7) in VR DK-HAC}
\end{align}
We split the integral in \eqref{Eq. (B.7) in VR DK-HAC} into two
sets, for small and for large $\mu_{j}$. Define the set $\mathbf{M}=\{\mu\in\Pi^{2s-1}:$
$\thinspace\sup_{j}|\mu_{j}|\leq b_{1,T}/\left(2s\right)\}$. Proceeding
as in \eqref{Eq. (B.9) in VR}-\eqref{Eq. (B.10) in VR}, we have
\begin{align}
O\left(Tb_{2,T}b_{1,T}^{-s+1}\right) & \sum_{q=0}^{s-1}\int_{\Pi}\int_{\mathbf{M}}\int_{0}^{1}\left|f\left(u,\,\omega-\mu_{2+2q}-\ldots-\mu_{2s}\right)-f\left(u,\,\omega\right)\right|\left|\widetilde{K}_{b_{1}}\left(\omega\right)\Psi_{Tb_{2,T}}^{\left(2s\right)}\left(\mu\right)\right|dud\omega d\mu\label{Eq. (B.9) in VR DK-HAC}\\
+ & O\left(Tb_{2,T}b_{1,T}^{-s+1}\right)\sum_{q=0}^{s-2}\int_{\Pi}\int_{\mathbf{M}}\left|\widetilde{K}_{b_{1}}\left(\omega-\mu_{2+2q}-\ldots-\mu_{2s}\right)-\widetilde{K}_{b_{1}}\left(\omega\right)\right|\left|\Psi_{Tb_{2,T}}^{\left(2s\right)}\left(\mu\right)\right|d\omega d\mu.\label{Eq. (B.10) in VR DK-HAC}
\end{align}
We apply the mean value theorem in \eqref{Eq. (B.9) in VR DK-HAC}
and use \eqref{Eq. (B.2) in VR} to yield, 
\begin{align*}
O\left(Tb_{2,T}b_{1,T}^{-s+1}\right) & \int_{\Pi}\left|\widetilde{K}_{b_{1}}\left(\omega\right)\right|d\omega\sum_{q=0}^{2s}\int_{\mathbf{M}}\left|\mu_{q}\right|\left|\Psi_{Tb_{2,T}}^{\left(2s\right)}\left(\mu\right)\right|d\mu\\
 & \leq O\left(Tb_{2,T}b_{1,T}^{-s+1}\right)\int_{\Pi}\left|\widetilde{K}_{b_{1}}\left(\omega\right)\right|d\omega\sum_{q=0}^{2s}\int_{\Pi^{2s-1}}\left|\mu_{q}\right|\left|\Psi_{Tb_{2,T}}^{\left(2s\right)}\left(\mu\right)\right|d\mu\\
 & =O\left(b_{1,T}^{-s+1}\log^{2s-1}\left(Tb_{2,T}\right)\right).
\end{align*}
On the other hand, using the Lipschitz property of $K$ (cf. Assumption
\ref{Assumption 4 in VR}), the expression in \eqref{Eq. (B.10) in VR DK-HAC}
is of order $O(b_{1,T}^{-s}\log^{2s-1}(Tb_{2,T}))$. 

Let $\mathbf{M}^{c}$ denote the complement of $\mathbf{M}$ in $\Pi^{2s-1}.$
The contribution to $B$ corresponding to the set $\mathbf{M}^{c}$
is bounded by
\begin{align}
Tb_{2,T}\left(2\pi\right)^{2s-1} & \int_{\Pi}\int_{\mathbf{M}^{c}}\left|H_{b_{1}}\left(\omega,\,\mu\right)\widetilde{K}_{b_{1}}\left(\omega\right)\right|\left|\Psi_{Tb_{2,T}}^{\left(2s\right)}\left(\mu\right)\right|d\omega d\mu\label{Eq. (B.11) in VR DK-HAC}\\
 & \quad+Tb_{2,T}\left(2\pi\right)^{2s-1}\int_{\Pi}\left|\left(\int_{0}^{1}f\left(u,\,\omega\right)du\right)^{s}\widetilde{K}_{b_{1}}^{s}\left(\omega\right)\right|d\omega\int_{\mathbf{M}^{c}}\left|\Psi_{Tb_{2,T}}^{\left(2s\right)}\left(\mu\right)\right|d\mu.\label{Eq. (B.12) in VR DK-HAC}
\end{align}
The expression in \eqref{Eq. (B.12) in VR DK-HAC} is $O(b_{1,T}^{-s}\log^{2s-1}(Tb_{2,T}))$
using \eqref{Eq. (B.1) in VR} and
\[
\int_{\Pi}\left|\left(\int_{0}^{1}f\left(u,\,\omega\right)\right)^{s}\widetilde{K}_{b_{1}}^{s}\left(\omega\right)\right|d\omega=O\left(b_{1,T}^{-s}\right).
\]
  The expression in \eqref{Eq. (B.11) in VR DK-HAC} is bounded by
\begin{align}
\int_{\mathbf{M}'}\prod_{j=1}^{s}\int_{0}^{1} & \left|f\left(u_{2j-1},\,\omega_{2j-1}\right)\widetilde{K}_{b_{1}}\left(\omega_{2j}\right)D_{Tb_{2,T}}\left(\omega_{2j}-\omega_{2j-1}\right)D_{Tb_{2,T}}\left(\omega_{2j+1}-\omega_{2j}\right)\right|du_{2j-1}d\omega_{2j}d\omega_{2j-1},\label{Eq. (B.13) in VR DK-HAC}
\end{align}
 where $\mathbf{M}'$ is defined after \eqref{Eq. (B.13) in VR}.

From \eqref{Eq. (B.4) in VR} it follows that $|D_{Tb_{2,T}}\left(\omega_{2j}-\omega_{2j-1}\right)|=O(b_{1,T}^{-1})$
since $|\omega_{2j}-\omega_{2j-1}|>\nu_{T}=b_{1,T}/\left(2s\right)$,
and $\int_{\Pi}|D_{Tb_{2,T}}\left(\omega_{2j}-\omega_{2j+1}\right)\widetilde{K}_{b_{1}}\left(\omega_{2j}\right)|d\omega_{2j}=O(b_{1,T}^{-1}\log(Tb_{2,T}))$.
For $\epsilon>0,$ consider the following decomposition
\begin{align}
\int_{\Pi} & \int_{0}^{1}\left|f\left(u_{2j-1},\,\omega_{2j-1}\right)D_{Tb_{2,T}}\left(\omega_{2j-1}-\omega_{2j-2}\right)\right|du_{2j-1}d\omega_{2j-1}\label{Eq. (B.14) in VR DK-HAC}\\
 & =\int_{\left|\omega_{2j-1}\right|\leq\epsilon}\int_{0}^{1}\left|f\left(u_{2j-1},\,\omega_{2j-1}\right)D_{Tb_{2,T}}\left(\omega_{2j-1}-\omega_{2j-2}\right)\right|du_{2j-1}d\omega_{2j-1}\nonumber \\
 & \quad+\int_{\left|\omega_{2j-1}\right|>\epsilon}\int_{0}^{1}\left|f\left(u_{2j-1},\,\omega_{2j-1}\right)D_{Tb_{2,T}}\left(\omega_{2j-1}-\omega_{2j-2}\right)\right|du_{2j-1}d\omega_{2j-1}.\nonumber 
\end{align}
By Assumption \ref{Assumption 1 in VR} $f(u_{2j-1},\,\omega_{2j-1})$
is bounded if $\left|\omega_{2j-1}\right|\leq\epsilon$. Then the
integral over $|\omega_{2j-1}|\leq\epsilon$ above is of order $O(\log(Tb_{2,T}))$.
On the other hand, if $|\omega_{2j-1}|>\epsilon$   we have  $|D_{Tb_{2,T}}(\omega_{2j-1}-\omega_{2j-2})|=O\left(1\right)$
by \eqref{Eq. (B.4) in VR}  and the second summand of \eqref{Eq. (B.14) in VR DK-HAC}
is finite in view of the integrability of $f\left(u,\,\omega\right)$
by Assumption \ref{Assumption: Assumption 2 in VR }. It follows that
\eqref{Eq. (B.14) in VR DK-HAC} is $O(\log(Tb_{2,T}))$. There are
other $s-1$ integrals of this type that can be handled in the same
way. The remaining integral is of the form 
\begin{align*}
\int_{\Pi}\int_{\Pi}\int_{0}^{1}\left|\widetilde{K}_{b_{1}}\left(\omega_{2s}\right)f\left(u_{2s-1},\,\omega_{1}\right)D_{Tb_{2,T}}\left(\omega_{1}-\omega_{2s}\right)\right|du_{2s-1}d\omega_{1}d\omega_{2s} & =O\left(\log\left(Tb_{2,T}\right)\right),
\end{align*}
 where $\omega_{1}=\omega_{2s+1}$ and we have used the same argument
as in \eqref{Eq. (B.14) in VR DK-HAC} to show that the integral in
$\omega_{1}$ is $O(\log(Tb_{2,T}))$ for all $\omega_{2s}$ and that
$\int_{\Pi}|\widetilde{K}_{b_{1}}\left(\omega_{2s}\right)|d\omega_{2s}=O\left(1\right)$.
Thus, \eqref{Eq. (B.13) in VR DK-HAC} is $O(b_{1,T}^{-s}\log^{2s-1}Tb_{2,T})$
and $B=O(b_{1,T}^{1-s}\log^{2s-1}(Tb_{2,T})+b_{1,T}^{-s}\log^{2s-1}(Tb_{2,T}))=O(Tb_{2,T}b_{1,T}^{1-s}\epsilon_{Tb_{2,T}}\left(2s\right))$. 

Next, let
\begin{align*}
B_{2} & =Tb_{2,T}\left(2\pi\right)^{2s-1}\int_{\Pi^{2s}}\left|H_{2,b_{1}}\left(\omega,\,\mu\right)-b_{2,T}^{2}\Lambda_{2}\left(f'',\,\widetilde{\mathbf{C}},\,s\right)\widetilde{K}_{b_{1}}^{s-1}\left(\omega\right)\right|\left|\widetilde{K}_{b_{1}}\left(\omega\right)\Psi_{Tb_{2,T}}^{\left(2s\right)}\left(\mu\right)\right|d\omega d\mu,
\end{align*}
where $\Lambda_{2}(f'',\,\widetilde{\mathbf{C}},\,s)$ depends on
$f\left(u,\,\omega\right),$ the second partial derivative of $f\left(u,\,\omega\right)$
in $u$ at the continuity points in $\widetilde{\mathbf{C}}$ and
$s$. By Assumption \ref{Assumption Lip of d2 f(u,w)}, for $j\in\mathbf{J}$
and $u_{j}\in\widetilde{\mathbf{C}}$ $(\partial^{2}/\partial u_{j}^{2})f\left(u_{j},\,\omega_{j}\right)$
has similar smoothness properties in $\omega_{j}$ to those of $f\left(u_{j},\,\omega_{j}\right)$.
Thus, the proof used above to bound $B$ can be repeated which then
results in $B_{2}=O(Tb_{2,T}^{3}b_{1,T}^{1-s}\epsilon_{Tb_{2,T}}\left(2s\right))$. 

Let 
\begin{align*}
B_{3} & =Tb_{2,T}\left(2\pi\right)^{2s-1}\int_{\Pi^{2s}}\left|H_{3,b_{1}}\left(\omega,\,\mu\right)-b_{2,T}^{2}\Lambda_{3}\left(f',\,\left\{ \lambda_{j}^{0},\,j=1,\ldots,\,m_{0}\right\} ,\,s\right)\widetilde{K}_{b_{1}}^{s-1}\left(\omega\right)\right|\\
 & \quad\times\left|\widetilde{K}_{b_{1}}\left(\omega\right)\Psi_{Tb_{2,T}}^{\left(2s\right)}\left(\mu\right)\right|d\omega d\mu,
\end{align*}
 where $\Lambda_{3}(f',\,\{\lambda_{j}^{0},\,j=1,\ldots,\,m_{0}\},\,s)$
depends on $f\left(u,\,\omega\right),$ $\Delta_{f}\left(\cdot\right)$
and $s$. By Assumption \ref{Assumption Lip of d2 f(u,w)}, $\left(\partial/\partial u_{-}\right)f\left(u,\,\omega\right)$
and $\left(\partial/\partial u_{+}\right)f\left(u,\,\omega\right)$
for $u$ a discontinuity point have similar smoothness properties
in $\omega$ to those of $f\left(u,\,\omega\right)$. Thus, the proof
used above to bound $B$ can be repeated which then results in $B_{3}=O(Tb_{2,T}^{3}b_{1,T}^{1-s}$
$\epsilon_{Tb_{2,T}}\left(2s\right))$. 

The rest of the proof follows from the same arguments used in the
last part of the proof of Lemma \ref{Lemma: Proposition 1 in VR}.
$\square$ 
\begin{lem}
\label{Lemma: Proposition 2 in VR DK-HAC}Let Assumptions \ref{Assumption 1 in VR},
\ref{Assumption 3 VR}-\ref{Assumption 4 in VR} and \ref{Assumption K2 and b2}-\ref{Assumption Lip of d2 f(u,w)}
hold. For $s\geq1$ with $\epsilon_{T}\left(2s+2\right)\rightarrow0$,
we have 
\begin{align*}
\mathbf{1}'\left(\Sigma_{\widetilde{V}}W_{b_{1}}\right)^{s}\Sigma_{\widetilde{V}}\mathbf{1} & =Tb_{2,T}\left(2\pi\right)^{2s+1}\biggl(\left(\int_{0}^{1}f\left(u,\,0\right)du\right)^{s+1}\int_{0}^{1}K_{2}^{s+1}\left(x\right)dx\\
 & \quad+b_{2,T}^{2}\left(\widetilde{\Lambda}_{2}\left(f'',\,\widetilde{\mathbf{C}},\,s\right)+\widetilde{\Lambda}_{3}\left(f',\,\left\{ \lambda_{j}^{0},\,j=1,\ldots,\,m_{0}\right\} ,\,s\right)\right)\biggr)\left(\widetilde{K}_{b_{1}}\left(0\right)\right)^{s}\\
 & \quad+O\left(b_{1,T}^{1-s}\log^{2s+1}\left(Tb_{2,T}\right)+b_{1,T}^{-s}\frac{\log^{2s+1}\left(Tb_{2,T}\right)}{Tb_{2,T}}\right),
\end{align*}
 where $\Lambda_{2}(f'',\,\widetilde{\mathbf{C}},\,s)$ depends on
$f\left(u,\,\omega\right),$ the second partial derivative of $f\left(u,\,\omega\right)$
in $u$ at the continuity points in $\widetilde{\mathbf{C}}$ and
$s$, and $\widetilde{\Lambda}_{3}(f',\,\{\lambda_{j}^{0},\,j=1,\ldots,\,m_{0}\},\,s)$
depends on $f\left(u,\,\omega\right),$ $\Delta_{f}\left(\cdot\right)$
and $s$.
\end{lem}
\noindent\textit{Proof of Lemma \ref{Lemma: Proposition 2 in VR DK-HAC}.}
We first write $\mathbf{1}'(\Sigma_{\widetilde{V}}W_{b_{1}})^{s}\Sigma_{\widetilde{V}}\mathbf{1}$
using an argument similar to the one used to derive \eqref{Eq. (B.15) in VR},
  
\begin{align}
\int_{0}^{1} & \sum_{1\leq r_{1},\ldots,\,r_{2s+2}\leq T}\mathbb{E}\left(\widetilde{V}_{r_{2s+1}}\left(u_{s+1}\right)\widetilde{V}_{r_{2s+2}}\left(u_{s+1}\right)\right)\int_{0}^{1}\cdots\int_{0}^{1}\Pi_{j=1}^{s}\nonumber \\
 & \quad\times\left\{ \mathbb{E}\left(\widetilde{V}_{r_{2j-1}}\left(u_{j}\right)\widetilde{V}_{r_{2j}}\left(u_{j}\right)\right)w\left(b_{1,T}\left(r_{2j}-r_{2j+1}\right)\right)\right\} du\nonumber \\
 & =Tb_{2,T}\sum_{k_{2s+2}=-\left\lfloor Tb_{2,T}\right\rfloor +1}^{\left\lfloor Tb_{2,T}\right\rfloor -1}\int_{0}^{1}\int_{\Pi}f\left(u_{s+1}/T,\,\omega_{2s+1}\right)e^{-ik_{2s+2}\omega_{2s+1}}\Pi_{j=1}^{s}\int_{0}^{1}\cdots\int_{0}^{1}\nonumber \\
 & \quad\times\left\{ f\left(u_{2j-1}/T,\,\omega_{2j-1}\right)\sum_{k_{2},\,k_{4},\ldots,\,k_{2s}=-\left\lfloor Tb_{2,T}\right\rfloor +1}^{\left\lfloor Tb_{2,T}\right\rfloor -1}\frac{Tb_{2,T}-|k_{2j}|}{Tb_{2,T}}\int_{\Pi}\widetilde{K}_{b_{1}}\left(\omega_{2j}\right)e^{i\left(k_{2j}+k_{2j+1}\right)\omega_{2j}}\right\} d\omega du\nonumber \\
 & =Tb_{2,T}\left(2\pi\right)^{2s+1}\int_{\Pi^{2s+1}}\left(S_{b_{1}}\left(\mu\right)\int_{0}^{1}K_{2}^{s+1}\left(x\right)dx+S_{2,b_{1}}\left(\mu\right)+S_{3,b_{1}}\left(\mu\right)\right)\Psi_{Tb_{2,T}}^{\left(2s+2\right)}\left(\mu\right)d\mu\label{Eq. (B.15) in VR -DK-HAC}\\
 & \quad+O\left(b_{2,T}^{2}b_{1,T}^{-s}\log^{2s-1}\left(Tb_{2,T}\right)\right)+O\left(b_{1,T}^{-s}\frac{\log^{2s}\left(Tb_{2,T}\right)}{Tb_{2,T}}\right),\nonumber 
\end{align}
where  $\Psi_{Tb_{2,T}}^{\left(2s+2\right)}\left(\mu\right)$, 
$S_{b_{1}}\left(\mu\right)$ and $d\mu=d\mu_{1}\ldots d\mu_{2s+1}$
are defined as in \eqref{Eq. (B.15) in VR}, 
\begin{align*}
S_{2,b_{1}}\left(\mu\right) & =b_{2,T}^{2}\left(\int_{0}^{1}x^{2}K_{2}\left(x\right)dx\right)\int_{0}^{1}K_{2}^{s}\left(x\right)dx\sum_{j\in\mathbf{J}}\frac{\partial^{2}}{\partial u_{j}^{2}}\int_{\widetilde{\mathbf{C}}}\cdots\int_{\widetilde{\mathbf{C}}}f\left(u_{1},\,\mu_{1}\right)\widetilde{K}_{b_{1}}\left(\mu_{1}+\mu_{2}\right)\ldots\\
 & \quad\times\widetilde{K}_{b_{1}}\left(\mu_{1}+\ldots+\mu_{2s}\right)f\left(u_{2s+1}\,,\mu_{1}+\ldots+\mu_{2s+1}\right)du,
\end{align*}
with $\mathbf{J}=\left\{ 1,\,3,\ldots,\,2s+1\right\} $ and $S_{3,b_{1}}\left(\omega,\,\mu\right)$
depends on the discontinuity points, i.e.,
\begin{align*}
S_{3,b_{1}}\left(\mu\right) & =b_{2,T}^{2}\int_{0}^{1}K_{2}^{s}\left(x\right)dx\left(\mathbf{1}\left\{ u_{1}=\lambda_{j}^{0},\,j=1,\ldots,\,m_{0}\right\} \Delta_{f,j}\left(\mu_{1}\right)\right)\widetilde{K}_{b_{1}}\left(\mu_{1}+\mu_{2}\right)\\
 & \quad\ldots\widetilde{K}_{b_{1}}\left(\mu_{1}+\ldots+\mu_{2s}\right)f\left(u_{2s-1},\,\mu_{1}+\ldots+\mu_{2s+1}\right)\\
 & \quad\vdots\\
 & \quad+b_{2,T}^{2}\int_{0}^{1}K_{2}^{s}\left(x\right)dxf\left(u_{1},\,\omega-\mu_{2}-\ldots-\mu_{2s}\right)\widetilde{K}_{b_{1}}\left(\omega-\mu_{3}-\ldots-\mu_{2s}\right)\\
 & \quad\times\widetilde{K}_{b_{1}}\left(\mu_{1}+\ldots+\mu_{2s}\right)\mathbf{1}\left\{ u_{2s-1}=\lambda_{j}^{0},\,j=1,\ldots,\,m_{0}\right\} \Delta_{f,j}\left(\mu_{1}+\ldots+\mu_{2s+1}\right),
\end{align*}
 with $\Delta_{f,j}\left(\omega\right)$ defined in \eqref{Eq. Delta,f,j}.
Proceeding as in the proof of Lemma \ref{Lemma: Proposition 2 in VR},
we divide the range of integration of the integral involving $S_{b_{1}}\left(\mu\right)$
in \eqref{Eq. (B.15) in VR -DK-HAC}, $\Pi^{2s+1}$, into two sets,
$\mathbf{M}$ and its complement $\mathbf{M}^{c}$, where $\mathbf{M}=\{|\mu_{j}|\leq\pi b_{1,T}/\left(2s+2\right),\,j=1,\ldots,\,2s+1\}$.
We have 
\begin{align}
\biggl|\int_{\mathbf{M}} & S_{b_{1}}\left(\mu\right)\Psi_{Tb_{2,T}}^{\left(2s+2\right)}\left(\mu\right)d\mu-\int_{\mathbf{M}}\left(\int_{0}^{1}f\left(u,\,0\right)du\right)^{s+1}\widetilde{K}_{b_{1}}^{s}\left(0\right)\Psi_{Tb_{2,T}}^{\left(2s+2\right)}\left(\mu\right)d\mu\biggr|\nonumber \\
 & =O\left(b_{1,T}^{-s-1}\right)\int_{\Pi^{2s+1}}\sum_{j=2}^{2s}\left|\mu_{j}\right|\left|\Psi_{Tb_{2,T}}^{\left(2s+2\right)}\left(\mu\right)\right|d\mu\nonumber \\
 & =O\left(b_{1,T}^{-s-1}\left(Tb_{2,T}\right)^{-1}\log^{2s+1}\left(Tb_{2,T}\right)\right),\label{Eq. (B.16) in VR DK-HAC}
\end{align}
using \eqref{Eq. (B.2) in VR}, \eqref{Eq. (B.8) in VR}, Assumptions
\ref{Assumption 1 in VR} and \ref{Assumption 4 in VR}. On the other
hand, the contribution from $\mathbf{M}^{c}$ is less than or equal
to 
\begin{align}
Tb_{2,T}\left(2\pi\right)^{2s+1}\int_{\mathbf{M}^{c}}\left|S_{b_{1}}\left(\mu\right)\right|\left|\Psi_{Tb_{2,T}}^{\left(2s+2\right)}\left(\mu\right)\right|d\mu & +O\left(b_{1,T}^{-s}\log^{2s+1}\left(Tb_{2,T}\right)\right),\label{Eq. (B.17) in VR DK-HAC}
\end{align}
where we have used \eqref{Eq. (B.1) in VR}. Using the same argument
used for \eqref{Eq. (B.13) in VR DK-HAC}, the expression in \eqref{Eq. (B.17) in VR DK-HAC}
is less than or equal to 
\begin{align}
\int_{\mathbf{M}'} & \prod_{j=1}^{s}\int_{0}^{1}\int_{0}^{1}\left|f\left(u_{2j-1},\,\lambda_{2j-1}\right)\widetilde{K}_{b_{1}}\left(\lambda_{2j}\right)D_{Tb_{2,T}}\left(\lambda_{2j}-\lambda_{2j-1}\right)\right.\label{Eq. (B.18) in VR-1}\\
 & \quad\times D_{Tb_{2,T}}\left(\lambda_{2j+1}-\lambda_{2j}\right)f\left(u_{2s+1},\,\lambda_{2s+1}\right)\left.D_{Tb_{2,T}}\left(\lambda_{1}\right)D_{Tb_{2,T}}\left(-\lambda_{2s-1}\right)\right|du_{2s+1}du_{2j-1}d\lambda,\nonumber 
\end{align}
where $\mathbf{M}'=\left\{ \left|\lambda_{1}\right|>\pi b_{1,T}/\left(2s+2\right)\right\} \cup\left\{ \left|\lambda_{2}-\lambda_{1}\right|>\pi b_{1,T}/\left(2s+2\right)\right\} \cup\ldots\cup\left\{ \left|\lambda_{2s-1}-\lambda_{2s}\right|>\pi b_{1,T}/\left(2s+2\right)\right\} $
and \eqref{Eq. (B.18) in VR-1} is nonzero only if $\left|\lambda_{2}\right|,\,\left|\lambda_{4}\right|,\ldots,\,\left|\lambda_{2s}\right|\leq\pi b_{1,T}$.

If $\left|\lambda_{j+1}-\lambda_{j}\right|>\pi b_{1,T}/\left(2s+2\right)$
for at least one index $j\in\left\{ 1,\ldots,\,2s\right\} $ we can
obtain a bound of order $((Tb_{2,T})^{-1}b_{1,T}^{-s-1}\log^{2s+1}(Tb_{2,T}))$
for \eqref{Eq. (B.18) in VR-1} as in Lemma \ref{Lemma: Proposition 1 in VR DK-HAC}.

Next, we have 
\begin{align}
Tb_{2,T} & \left(2\pi\right)^{2s+1}\biggl|\int_{\Pi^{2s}}\left(S_{b_{2}}\left(\mu\right)+S_{b_{3}}\left(\mu\right)\right)\Psi_{Tb_{2,T}}^{\left(2s+2\right)}\left(\mu\right)d\mu\label{Eq. Diff Sb2 Sb3}\\
 & -b_{2,T}^{2}\int_{\Pi^{2s}}\left(\widetilde{\Lambda}_{2}\left(f'',\,\widetilde{\mathbf{C}},\,s\right)+\widetilde{\Lambda}_{3}\left(f',\,\left\{ \lambda_{j}^{0},\,j=1,\ldots,\,m\right\} ,\,s\right)\right)\widetilde{K}_{b_{1}}^{s}\left(0\right)\Psi_{Tb_{2,T}}^{\left(2s+2\right)}\left(\mu\right)d\mu\biggr|.\nonumber 
\end{align}
 By Assumption \ref{Assumption Lip of d2 f(u,w)}, $(\partial^{2}/\partial u^{2})f\left(u,\,\omega\right)$
for $u\in\widetilde{\mathbf{C}}$, $\left(\partial/\partial u_{-}\right)f\left(u,\,\omega\right)$
and $\left(\partial/\partial u_{+}\right)f\left(u,\,\omega\right)$
for $u$ a discontinuity point have similar smoothness properties
in $\omega$ to those of $f\left(u,\,\omega\right)$. Thus, the proof
used above to bound \eqref{Eq. (B.16) in VR DK-HAC} can be repeated
which then results in \eqref{Eq. Diff Sb2 Sb3} being $O(b_{2,T}^{2}b_{1,T}^{-s-1}\log^{2s+1}(Tb_{2,T})$.
$\square$ 
\begin{lem}
\label{Lemma: Lemma 16 in VR}Let Assumptions \ref{Assumption 1 in VR},
\ref{Assumption: Assumption 2 in VR } $\left(p>1\right)$, \ref{Assumption 3 VR}-\ref{Assumption 4 in VR}
and \ref{Assumption 7 VR} $\left(0<q<1\right)$ hold. Then, $||\Sigma_{V}W_{b_{1}}||\leq C_{1}\nu_{2,T}$
where $C_{1}$ depends on $f\left(\cdot,\,\cdot\right)$ and $K$,
$0<C_{1}<\infty$ and $\nu_{2,T}=\max\{b_{1,T}^{-1}\log^{2}T,\,T{}^{\left(2-p\right)/2p}b_{1,T}^{-1/2}\log^{2}T)\}$
$\rightarrow\infty.$ 
\end{lem}
\noindent\textit{Proof of Lemma \ref{Lemma: Lemma 16 in VR}. }We
have  
\begin{align}
\left\Vert \Sigma_{V}W_{b_{1}}\right\Vert  & =\sup_{\left\Vert x\right\Vert =1}\left|\sum_{j,h=1}^{T}x_{j}x_{h}\sum_{t=1}^{T}\sum_{s=1}^{T}\int_{\Pi^{2}}f\left(t/T,\,\lambda\right)\widetilde{K}_{b_{1}}\left(\omega\right)e^{it\lambda}e^{-is\omega}e^{i\left(h\omega-j\lambda\right)}d\lambda d\omega\right|+O\left(T^{-1}\right)\nonumber \\
 & =\sup_{\left\Vert x\right\Vert =1}\left|\sum_{t=1}^{T}f\left(t/T,\,\lambda\right)e^{it\lambda}\sum_{j,h}x_{j}x_{h}\int_{\Pi^{2}}\widetilde{K}_{b_{1}}\left(\omega\right)D_{T}\left(-\omega\right)e^{i\left(h\omega-j\lambda\right)}d\lambda d\omega\right|+O\left(T^{-1}\right)\nonumber \\
 & \leq\sup_{\left\Vert x\right\Vert =1}\left|\int_{\omega\leq\epsilon}\int_{\lambda}\sum_{t=1}^{T}f\left(t/T,\,\lambda\right)e^{it\lambda}D_{T}\left(-\omega\right)\sum_{j,h}x_{j}x_{h}\widetilde{K}_{b_{1}}\left(\omega\right)e^{i\left(h\omega-j\lambda\right)}d\lambda d\omega\right|\nonumber \\
 & \quad+\sup_{\left\Vert x\right\Vert =1}\left|\int_{\omega>\epsilon}\int_{\lambda}\sum_{t=1}^{T}f\left(t/T,\,\lambda\right)e^{it\lambda}D_{T}\left(-\omega\right)\sum_{j,h}x_{j}x_{h}\widetilde{K}_{b_{1}}\left(\omega\right)e^{i\left(h\omega-j\lambda\right)}d\lambda d\omega\right|+O\left(T^{-1}\right)\nonumber \\
 & \triangleq A_{1}+o\left(1\right)+O\left(T^{-1}\right).\label{Eq. (B.28) in VR}
\end{align}
 Let $L_{2,T}:\,\mathbb{R}\rightarrow\mathbb{R}$ be the periodic
extension with period $2\pi$ of 
\begin{align*}
L_{2,T}\left(\omega\right) & =\begin{cases}
T, & \left|\omega\right|\leq1/T,\\
1/|\omega|, & 1/T\leq\left|\omega\right|\leq|\pi|.
\end{cases}
\end{align*}
Lemma S.A.1-2 in \citeReferencesSupp{casini:change-point-spectra}
showed that
\begin{align}
\left|\sum_{t=1}^{T}f\left(t/T,\,\lambda\right)e^{-it\lambda}\right| & \leq L_{2,T}\left(\lambda\right),\label{Eq. Lemma A.S.2. CP}
\end{align}
and $\int_{\Pi}L_{2,T}\left(\lambda\right)d\lambda\leq C_{L}\log T$
for $T>1$ and $C_{L}>0$ being a constant independent of $T$. Let
$X_{T}\left(\omega\right)=\sum_{j=1}^{T}x_{j}e^{ij\omega}$. Then,
the contribution to $A_{1}$ from $\left|\lambda\right|\leq\epsilon$
is bounded by 

\begin{align}
\sup_{\left\Vert x\right\Vert =1} & \int_{\omega\leq\epsilon}\int_{\lambda}\left|\sum_{t=1}^{T}f\left(t/T,\,\lambda\right)e^{it\lambda}\right|\left|D_{T}\left(-\omega\right)\right|\left|X_{T}\left(\omega\right)\right|\left|X_{T}\left(\lambda\right)\right|\left|\widetilde{K}_{b_{1}}\left(\omega\right)\right|d\lambda d\omega\nonumber \\
 & \leq\sup_{\left\Vert x\right\Vert =1}b_{1,T}^{-1}\sup_{\omega\in\Pi}|K\left(\omega\right)|\int_{\Pi}L_{2,T}\left(\lambda\right)\left(\int_{\Pi}\left|D_{T}\left(-\omega\right)\right|\left|X_{T}\left(\omega\right)\right|\left|X_{T}\left(\lambda\right)\right|\right)d\lambda d\omega\nonumber \\
 & \leq\sup_{\left\Vert x\right\Vert =1}b_{1,T}^{-1}\sup_{\omega\in\Pi}|K\left(\omega\right)|\left(\int_{\Pi}L_{2,T}\left(\lambda\right)^{2}d\lambda\right)^{1/2}\left(\int_{\Pi}\left|X_{T}\left(\lambda\right)\right|^{2}d\lambda\right)^{1/2}\nonumber \\
 & \quad\times\left(\int_{\Pi}\left|D_{T}\left(-\omega\right)\right|^{2}d\omega\right)^{1/2}\left(\int_{\Pi}\left|X_{T}\left(\omega\right)\right|^{2}d\omega\right)^{1/2}\nonumber \\
 & \leq2\pi C_{2}b_{1,T}^{-1}\sup_{\omega\in\Pi}|K\left(\omega\right)|\log^{2}T,\label{Eq. (B.29) in VR}
\end{align}
where $0<C_{2}<\infty$ and we have used $\sup_{\omega\in\Pi}|K\left(\omega\right)|=O(b_{1,T}^{-1})$,
$(\int_{\omega}|X_{T}\left(\omega\right)|^{2}d\omega)=2\pi$ and \eqref{Eq. Lemma A.S.2. CP}.
For $\left|\lambda\right|>\epsilon$ the contribution to $A_{1}$
is bounded by 

\begin{align}
\sup_{\left\Vert x\right\Vert =1} & \int_{\omega\leq\epsilon}\sum_{t=1}^{T}\left(\int_{\Pi}\left(f\left(t/T,\,\lambda\right)\right)^{p}d\lambda\right)^{1/p}\left(\int_{\Pi}|e^{it\lambda}X_{T}\left(\lambda\right)|^{\frac{p}{p-1}}d\lambda\right)^{\left(p-1\right)/p}\left|D_{T}\left(-\omega\right)X_{T}\left(\omega\right)\widetilde{K}_{b_{1}}\left(\omega\right)\right|d\omega d\omega\nonumber \\
 & \leq C_{2}\sup_{\left\Vert x\right\Vert =1}\sum_{t=1}^{T}\left(\int_{\Pi}|e^{it\lambda}X_{T}\left(\lambda\right)|^{\frac{p}{p-1}}d\lambda\right)^{\left(p-1\right)/p}\int_{\omega\leq\epsilon}\left|D_{T}\left(-\omega\right)X_{T}\left(\omega\right)\widetilde{K}_{b_{1}}\left(\omega\right)\right|d\omega\nonumber \\
 & \leq C_{2}\sup_{\left\Vert x\right\Vert =1}\sum_{t=1}^{T}\left(\int_{\Pi}|e^{it\lambda}|^{\frac{p}{p-1}}d\lambda\right)^{\left(p-1\right)/p}\int_{\omega\leq\epsilon}\left(\int_{\Pi}|X_{T}\left(\lambda\right)|^{\frac{p}{p-1}}d\lambda\right)^{\left(p-1\right)/p}\nonumber \\
 & \quad\times\left(\int_{\Pi}\left|D_{T}\left(-\omega\right)\right|d\omega\right)\left(\int_{\Pi}\left|X_{T}\left(\omega\right)\right|^{2}d\omega\right)^{1/2}\left(\int_{\Pi}\left|\widetilde{K}_{b_{1}}\left(\omega\right)\right|^{2}d\omega\right)^{1/2}\nonumber \\
 & \leq\sqrt{2\pi}C_{2}\left(\sup_{\omega}|K\left(\omega\right)|\right)^{1/2}\left\Vert K\right\Vert _{1}\left(2\pi\right)^{\left(p-1\right)/p}T^{\frac{2-p}{2p}}b_{1,T}^{-1}\log^{2}T,\label{Eq. (B.30) in VR}
\end{align}
 where $0<C_{2}<\infty$ and we have used $\sup_{x,\lambda}\left|X_{T}\left(\lambda\right)\right|\leq\sqrt{T}$
and 
\begin{align*}
\left(\int_{\Pi}|X_{T}\left(\lambda\right)|^{\frac{p}{p-1}}d\lambda\right)^{\left(p-1\right)/p} & =\left(\int_{\Pi}|X_{T}\left(\lambda\right)|^{2+\frac{2-p}{p-1}}d\lambda\right)^{\left(p-1\right)/p}\\
 & =\left(\int_{\Pi}|X_{T}\left(\lambda\right)|^{2}|X_{T}\left(\lambda\right)|^{\frac{2-p}{p-1}}d\lambda\right)^{\left(p-1\right)/p}\\
 & \leq\left(\int_{\Pi}|X_{T}\left(\lambda\right)|^{2}T^{\frac{1}{2}\left(\frac{2-p}{p-1}\right)}d\lambda\right)^{\left(p-1\right)/p}\\
 & \leq\left(2\pi\right)^{\left(p-1\right)/p}T^{\frac{2-p}{2p}}.
\end{align*}
 From \eqref{Eq. (B.29) in VR}-\eqref{Eq. (B.30) in VR} we have
$A_{1}\leq C_{1}\nu_{2,T}$ for some $C_{1}$ such that $0<C_{1}<\infty$.
$\square$
\begin{lem}
\label{Lemma: Lemma 15 in VR}Let Assumptions \ref{Assumption 1 in VR},
\ref{Assumption: Assumption 2 in VR } (for some $p>1)$, \ref{Assumption 3 VR},
\ref{Assumption 4 in VR} and $b_{1,T}+T^{-1}b_{1,T}^{-1}\log^{3}T\rightarrow0$
hold. Then, there exists $c_{2}>0$ such for $\left\Vert \mathbf{t}\right\Vert >c_{1}m_{T}$
with $c_{1}>0$ we have $\left|\psi\left(\mathbf{t}\right)\right|\leq\exp\left\{ -c_{2}m_{T}^{2}\right\} ,$
where $m_{T}=\min\{(Tb_{1,T})^{-1/2}\log T,\,T{}^{\left(p-1\right)/p}\}\rightarrow\infty$. 
\end{lem}
\noindent\textit{Proof of Lemma \ref{Lemma: Lemma 15 in VR}. }The
proof is similar to the proof of Lemma 15 in \citeReferencesSupp{velasco/robinson:01}
with the difference that reference to Lemma 16 there is changed to
reference to Lemma \ref{Lemma: Lemma 16 in VR}.\textit{ $\square$ }
\begin{lem}
\label{Lemma: Lemma 16 DK-HAC} Let Assumptions \ref{Assumption 1 in VR},
\ref{Assumption: Assumption 2 in VR } $\left(p>1\right)$, \ref{Assumption 3 VR}-\ref{Assumption 4 in VR},
\ref{Assumption 7 VR} $\left(0<q<1\right)$ and \ref{Assumption K2 and b2}-\ref{Assumption Lip of d2 f(u,w)}
hold. Then, $||\Sigma_{\widetilde{V}}W_{b_{1}}||\leq C_{1}\nu_{2,T}$
where $C_{1}$ depends on $f\left(u,\,\omega\right)$ and $K$, $0<C_{1}<\infty$
and $\nu_{2,T}=\max\{b_{1,T}^{-1}\log\left(Tb_{2T}\right),\,(Tb_{2,T}){}^{\left(2-p\right)/2p}$
$b_{1,T}^{-1/2})\}\rightarrow\infty$. 
\end{lem}
\noindent\textit{Proof of Lemma \ref{Lemma: Lemma 16 DK-HAC}. }The
proof is similar to the proof of Lemma \ref{Lemma: Lemma 16 in VR}.
$\square$
\begin{lem}
\label{Lemma: Lemma 15 in VR DK-HAC}Let Assumptions \ref{Assumption 1 in VR},
\ref{Assumption: Assumption 2 in VR } $(p>1)$, \ref{Assumption 3 VR}-\ref{Assumption 4 in VR},
\ref{Assumption K2 and b2}-\ref{Assumption Lip of d2 f(u,w)} and
$b_{1,T}+(Tb_{1,T}b_{2,T})^{-1}\log^{3}T\rightarrow0$ hold. Then,
there exists a $c_{4}>0$ such for $\left\Vert \mathbf{t}\right\Vert >c_{3}m_{2,T}$
with $c_{3}>0$ we have $\left|\psi\left(t_{1},\,t_{2}\right)\right|\leq\exp(-c_{4}m_{2,T}^{2}),$
where $m_{2,T}=\min\{(Tb_{2,T}b_{1,T})^{1/2}/\log(Tb_{2,T}),\,(Tb_{2,T}){}^{\left(p-1\right)/p}\}\rightarrow\infty$.
\end{lem}
\noindent\textit{Proof of Lemma \ref{Lemma: Lemma 15 in VR DK-HAC}.
}Following \citeReferencesSupp{bentkus/rudzkis:1982} and \citeReferencesSupp{velasco/robinson:01}
we first study the characteristic function of $\widehat{J}_{\mathrm{DK},T}$.
Define $\tau\left(t_{2}\right)=\mathbb{E}(\exp(it_{2}v_{2}))=\tau'\left(t_{2}\right)\exp(-it_{2}\Upsilon_{2,T})$,
where 
\begin{align*}
\tau'\left(t_{2}\right) & =\left|I-\frac{2it_{2}}{\sqrt{Tb_{2,T}/b_{1,T}}\mathsf{V}_{2,T}J_{T}}\Sigma_{\widetilde{V}}W_{b_{1}}\right|^{-1/2}=\prod_{j=1}^{T}\left(1-2it_{2}\frac{\widetilde{\lambda}_{j}}{\sqrt{Tb_{2,T}/b_{1,T}}\mathsf{V}_{2,T}J_{T}}\right)^{-1/2},
\end{align*}
and $\widetilde{\lambda}_{j}$ are the eigenvalues of $\Sigma_{\widetilde{V}}W_{b_{1}}$.
Note that 
\begin{align*}
1= & \mathrm{Var}\left(v_{2}\right)=\frac{b_{1,T}}{Tb_{2,T}}\frac{1}{\mathsf{V}_{2,T}^{2}J_{T}^{2}}2\mathrm{Tr}\left[(\Sigma_{\widetilde{V}}W_{b_{1}})^{2}\right]=\frac{b_{1,T}}{Tb_{2,T}}\frac{2}{\mathsf{V}_{2,T}^{2}J_{T}^{2}}\sum_{j=1}^{T}\widetilde{\lambda}_{j}^{2},
\end{align*}
where we have used the normality of $\left\{ V_{t}\right\} $ and
the relationship between the trace and the eigenvalues. Rearranging
yields $\sum_{j=1}^{T}\widetilde{\lambda}_{j}^{2}=2^{-1}b_{1,T}^{-1}Tb_{2,T}\mathsf{V}_{2,T}^{2}J_{T}^{2}=O(b_{1,T}^{-1}Tb_{2,T}).$
Further, we have $\max_{j}|\widetilde{\lambda}_{j}|=\sup_{\left\Vert x\right\Vert =1}|\Sigma_{\widetilde{V}}W_{b_{1}}x,\,x|=||\Sigma_{\widetilde{V}}W_{b_{1}}||.$
We can apply Lemma \ref{Lemma: Lemma 16 DK-HAC} to yield
\begin{align*}
\max_{j}\left|\widetilde{\lambda}_{j}\right| & \leq C_{1}\nu_{2,T},\qquad\nu_{2,T}=\max\left\{ b_{1,T}^{-1}\log\left(Tb_{2T}\right),\,(Tb_{2,T}){}^{\left(2-p\right)/2p}b_{1,T}^{-1/2}\right\} \rightarrow\infty,
\end{align*}
where $C_{1}>0$ is such that $C_{1}<\infty$. Let $g_{j}=\widetilde{\lambda}_{j}(C_{1}\nu_{2,T})^{-1}$
and note that for $T$ large enough we have $\left|g_{j}\right|\leq1.$
Using $\sum_{j=1}^{T}g_{j}^{2}=(2C_{1}^{2}\nu_{2,T}^{2})\mathsf{V}_{2,T}^{2}J_{T}^{2}b_{1,T}^{-1}Tb_{2,T}$
we yield  
\begin{align*}
\left|\tau\left(t_{2}\right)\right| & \leq\prod_{j=1}^{T}\left(1+4t^{2}\frac{C_{1}^{2}\nu_{2,T}^{2}}{b_{1,T}^{-1}Tb_{2,T}\mathsf{V}_{2,T}^{2}J_{T}^{2}}\right)^{-(1/4)g_{j}^{2}}\\
 & =\left(1+t_{2}^{2}\frac{\nu_{2,T}^{2}}{b_{1,T}^{-1}Tb_{2,T}}\frac{4C_{1}^{2}}{\mathsf{V}_{2,T}^{2}J_{T}^{2}}\right)^{-(1/8)C_{1}^{-2}\mathsf{V}_{2,T}^{2}J_{T}^{2}b_{1,T}^{-1}Tb_{2,T}\nu_{2,T}^{-2}}\\
 & =\left(1+t_{2}^{2}\frac{\nu_{2,T}^{2}}{b_{1,T}^{-1}Tb_{2,T}}\left[C_{2}+O\left(b_{1,T}^{2}+\epsilon_{Tb_{2,T}}\left(2\right)\right)\right]\right)^{-(1/2)\left(C_{2}^{-1}+O\left(b_{1,T}^{2}+\epsilon_{Tb_{2,T}}\left(2\right)\right)\right)Tb_{2,T}b_{1,T}^{-1}\nu_{2,T}^{-2}},
\end{align*}
 where $C_{2}=C_{1}^{2}/(\pi^{3}4(\int_{0}^{1}f\left(u,\,0\right)du)^{2}\left\Vert K\right\Vert _{2}^{2}\left\Vert K_{2}\right\Vert _{2}^{2})$
 and we have applied $\left(1+at\right)\geq\left(1+t\right)^{a}$
which is valid for $t\geq0$ and $0\leq a\leq1$.  Thus, for all
$\eta>0$, we have
\begin{align}
\left|\tau\left(t_{2}\right)\right| & \leq\left(1+\eta_{1}^{2}\right)^{-\eta_{2}\left(Tb_{2,T}b_{1,T}^{-1}\nu_{2,T}^{-2}\right)},\label{Eq. (B.25) in VR DK-HAC}
\end{align}
 for $\left|t_{2}\right|>\eta\sqrt{Tb_{2,T}b_{1,T}^{-1}}\nu_{2,T}^{-1}$
and for $\eta_{1}>0$ and $\eta_{2}>0$ depending on $\eta$. 

Next, we consider the joint characteristic function $\psi_{T}\left(t_{1},\,t_{2}\right)$.
Its modulus is equal to 
\begin{align}
\left|\psi_{T}\left(t_{1},\,t_{2}\right)\right| & =\left|\tau\left(t_{2}\right)\right|\exp\left(-\frac{1}{2}t_{1}^{2}\xi'_{2,T}\mathscr{R}\left(I-2it_{2}\Sigma_{\widetilde{V}}Q_{2,T}\right)^{-1}\Sigma_{\widetilde{V}}\xi_{2,T}\right),\label{Eq. (B.26) in VR DK-HAC}
\end{align}
where $\mathscr{R}\left(A\right)$ stands for the real part of $A$.
From \nociteReferencesSupp{anderson:1958}\textcolor{MyBlue}{Anderson (1958, p. 161)}
$\mathscr{R}(\Sigma_{\widetilde{V}}^{-1}-2it_{2}Q_{2,T})^{-1}=\mathscr{R}(I-2it_{2}Q_{2,T})^{-1}\Sigma_{\widetilde{V}}$
is positive definite since $t_{2}Q_{2,T}$ is real. Then $\xi'_{2,T}\mathscr{R}(I-2it_{2}\Sigma_{\widetilde{V}}Q_{2,T})^{-1}\Sigma_{\widetilde{V}}\xi_{2,T}>0$
for all $t_{2}\in\mathbb{R}$. Thus, $|t_{2}|\leq d\sqrt{Tb_{2,T}b_{1,T}^{-1}}/\nu_{2,T}$
for all $d>0$ and $\xi'_{2,T}\mathscr{R}(I-2it_{2}\Sigma_{\widetilde{V}}Q_{2,T})^{-1}\Sigma_{\widetilde{V}}\xi_{2,T}>\epsilon$
for some $\epsilon>0$ depending on $d$ because $||\Sigma_{\widetilde{V}}Q_{2,T}||=O(Tb_{2,T}b_{1,T}^{-1})^{-1/2}||\Sigma_{\widetilde{V}}W_{b_{1}}||=(O(Tb_{2,T}b_{1,T}^{-1})^{-1/2}\nu_{2,T})$,
and $||\xi_{2,T}||=(\sqrt{Tb_{2,T}J_{T}})^{-1}\sqrt{1^{2}+1^{2}+\ldots+1^{2}}=1/\sqrt{b_{2,T}J_{T}}$,
with $J_{T}\rightarrow2\pi\int_{0}^{1}f\left(u,\,0\right)du,$ $0<f\left(u,\,0\right)<\infty$
for all $u$ by Assumption \ref{Assumption 1 in VR}. Then, for $\left|t_{1}\right|\sqrt{2}>d_{1}\sqrt{Tb_{2,T}b_{1,T}^{-1}}/\nu_{2,T}$
and $\left|t_{2}\right|\sqrt{2}\leq d_{1}\sqrt{Tb_{2,T}b_{1,T}^{-1}}/\nu_{2,T}$
and some $\epsilon_{1}>0$ depending on $d_{1}$, 
\begin{align}
\exp\left(-\frac{1}{2}t_{1}^{2}\xi'_{2,T}\mathscr{R}\left(I-2it_{2}\Sigma_{\widetilde{V}}Q_{2,T}\right)^{-1}\Sigma_{\widetilde{V}}\xi_{2,T}\right) & \leq\exp\left(-\frac{1}{2}t_{1}^{2}\epsilon_{1}\right)\leq\exp\left(-\frac{1}{4}d_{1}^{2}\epsilon_{1}\frac{Tb_{2,T}b_{1,T}^{-1}}{\nu_{2,T}^{2}}\right).\label{Eq. (B.27) in VR DK-HAC}
\end{align}
From \eqref{Eq. (B.25) in VR DK-HAC}-\eqref{Eq. (B.27) in VR DK-HAC},
there exists a $d_{2}>0$ such that $|\psi_{T}\left(\mathbf{t}\right)|\leq\exp(-d_{2}(Tb_{2,T}b_{1,T}^{-1}/\nu_{2,T}^{2}))$
for $\{\mathbf{t}:\,||\mathbf{t}||>d_{1}\sqrt{Tb_{2,T}b_{1,T}^{-1}}/\nu_{2,T}\}\subset\mathbf{B}_{1}\cup\mathbf{B}_{2}$
where $\mathbf{B}_{1}=\{\mathbf{t}\in\mathbb{R}^{2}:\,\left|t_{2}\right|>(d_{1}/\sqrt{2})\sqrt{Tb_{2,T}b_{1,T}^{-1}}/\nu_{2,T}\}$
and $\mathbf{B}_{2}=\{\mathbf{t}\in\mathbb{R}^{2}:\,\left|t_{2}\right|\leq(d_{1}/\sqrt{2})\sqrt{Tb_{2,T}b_{1,T}^{-1}}/\nu_{2,T}$
and $\left|t_{1}\right|>(d_{1}/\sqrt{2})\sqrt{Tb_{2,T}b_{1,T}^{-1}}/\nu_{2,T}\}$,
and the lemma follows because    $Tb_{2,T}b_{1,T}^{-1}/\nu_{2,T}^{2}=m_{2,T}^{2}\rightarrow\infty$.\textit{
$\square$ }

\subsubsection{Additional Lemmas Used for the Proofs of Theorem \ref{Theorem: Theorem 1 in VR}-\ref{Theorem: Theorem 2 in VR}}

We first present a result about the limit of $J_{T}$ and a result
about the bias of $\widehat{J}_{\mathrm{HAC,}T}$. 
\begin{lem}
\label{Lemma 4.1 in VR}Let Assumption \ref{Assumption 1 in VR} with
$d_{f}=1$ and $\varrho=0$ hold. Then, $J_{T}-2\pi\int_{0}^{1}f\left(u,\,0\right)du=O\left(T^{-1}\log T\right)$.
If in addition Assumption \ref{Assumption A - Dependence}-(i) holds,
then the order is $O(T^{-1})$.
\end{lem}
\begin{lem}
\label{Lemma 2 in VR }Let Assumptions \ref{Assumption 1 in VR},
\ref{Assumption 3 VR}, \ref{Assumption 5 VR}, and \ref{Assumption 6 VR}
hold. Then, 
\begin{align*}
\mathbb{E}\left(\widehat{J}_{\mathrm{HAC,}T}\right)-2\pi\int_{0}^{1}f\left(u,\,0\right)du-2\pi\frac{\int_{0}^{1}f^{\left(d_{f}\right)}\left(u,\,0\right)du}{d_{f}!}\mu_{d_{f}}\left(K\right)b_{1,T}^{d_{f}} & =O\left(T^{-1}\log T+b_{1,T}^{d_{f}+\varrho}\right).
\end{align*}
\end{lem}
We now study the cumulants of the normalized spectral estimate $h_{2}$. 
\begin{lem}
\label{Lemma 3 in VR}Let Assumptions \ref{Assumption 1 in VR}, \ref{Assumption 3 VR}-\ref{Assumption 4 in VR}
hold. For $s>2$ with $\epsilon_{T}\left(s\right)=b_{1,T}^{d_{f}+\varrho}+T^{-1}b_{1,T}\log^{2s-1}T\rightarrow0$,
we have
\begin{align*}
\overline{\kappa}_{T}\left(0,\,s\right) & \triangleq\kappa_{T}\left(0,\,s\right)\left(\frac{T}{b_{1,T}}\right)^{\left(s-2\right)/2}=\sum_{j=0}^{d_{f}}\Xi_{j}\left(0,\,s\right)b_{1,T}^{j}+O\left(\epsilon_{T}\left(s\right)\right),
\end{align*}
where $\Xi_{j}\left(0,\,s\right)$ is bounded and depends on  $K$
and $f^{\left(j\right)}\left(u,\,0\right)$ $(j=0,\ldots,\,d_{f})$.
\end{lem}
A few examples of $\Xi_{j}\left(0,\,s\right)$ are $\Xi_{0}\left(0,\,s\right)=\left(4\pi\right)^{(s-2)/2}\left(s-1\right)!\int_{\Pi}K^{s}\left(\omega\right)d\omega\left\Vert K\right\Vert _{2}^{-s}$
and $\Xi_{1}(2,\,s)=0.$ If $(\partial/\partial\omega)(\int_{0}^{1}f\left(u,\,\omega\right)du)|_{\omega=0}=0$
then $\Xi_{j}(0,\,s)=0$ for $j\geq1$.  In order to develop an Edgeworth
expansion to approximate the distribution of $\mathbf{h}$, we need
to study the cross-cumulants of $\mathbf{h}.$ 
\begin{lem}
\label{Lemma: Lemma 4 in VR}Let Assumptions \ref{Assumption 1 in VR}
and \ref{Assumption 3 VR}-\ref{Assumption 4 in VR} hold. For $s>0$
with $\epsilon_{T}\left(s+2\right)\rightarrow0$, we have
\begin{align*}
\overline{\kappa}_{T}(2,\,s) & \triangleq\kappa_{T}(2,\,s)\left(Tb_{1,T}\right)^{s/2}=\sum_{j=0}^{d_{f}}\Xi_{j}(2,\,s)b_{1,T}^{j}+O\left(\epsilon_{T}\left(s+2\right)\right),
\end{align*}
 where $\Xi_{j}(2,\,s)$ is bounded and depends on $K$ and $f^{\left(j\right)}\left(u,\,0\right)$
$(j=0,\ldots,\,d_{f})$.
\end{lem}
For example, we have $\Xi_{0}(2,\,s)=\left(4\pi\right)^{s/2}s!K^{s}\left(0\right)\left\Vert K\right\Vert _{2}^{-s}$
and $\Xi_{1}(2,\,s)=0$. Using Lemmas \ref{Lemma 3 in VR}-\ref{Lemma: Lemma 4 in VR}
we can substitute out $\mathsf{B}_{T}$ and $\mathsf{V}_{T}$ in 
$Z_{T}$ and, by only focusing on the leading terms, we define the
following linear stochastic approximation,
\begin{align*}
\widetilde{Z}_{T} & \triangleq h_{1}\left(1-2^{-1}\overline{c}_{1}b_{1,T}^{d_{f}}-2^{-1}\sqrt{4\pi}\left\Vert K_{2}\right\Vert h_{2}\left(Tb_{1,T}\right)^{-1/2}\right).
\end{align*}

\begin{lem}
\label{Lemma: Lemma 5 in VR}Let Assumptions \ref{Assumption 1 in VR},
\ref{Assumption: Assumption 2 in VR } $\left(p>1\right)$, \ref{Assumption 3 VR}-\ref{Assumption 5 VR}
and \ref{Assumption 7 VR} $(q=1/(1+2d_{f}))$ hold. Then, $Z_{T}$
has the same Edgeworth expansion as $\widetilde{Z}_{T}$ uniformly
for convex Borel sets up to order $O((Tb_{1,T})^{-1/2})$. 
\end{lem}
 Note that the condition $q=1/(1+2d_{f})$ is sufficient for the
consistency of $\widehat{J}_{\mathrm{HAC},T}$. Indeed, for $d_{f}=2$
it implies that $b_{1,T}=T^{-1/5}$ which coincides with the MSE-optimal
bandwidth choice for the quadratic spectral kernel {[}cf. \citet{andrews:91}{]}.\footnote{Note that the MSE bounds under nonstationarity in Section 8 in \citet{andrews:91},
which are used to determine the optimal bandwidth, are not correctly
stated {[}cf. \citet{casini_comment_andrews91}{]}.} 

\subsubsection{Proof of Lemma \ref{Lemma 4.1 in VR}}

Note that $J_{T}=\sum_{k=-T+1}^{T-1}\Gamma_{T}\left(k\right)$ where
$\Gamma_{T}\left(k\right)=T^{-1}\sum_{t=|k|+1}^{T}\mathbb{E}(V_{t}V_{t-|k|})$.
We have
\begin{align*}
J_{T} & =\sum_{k=-T+1}^{T-1}\frac{1}{T}\sum_{t=|k|+1}^{T}\int_{\Pi}f\left(t/T,\,\omega\right)e^{ik\omega}d\omega\\
 & =\sum_{k=-T+1}^{T-1}\frac{T-|k|}{T}\int_{|k|/T}^{1}\int_{\Pi}f\left(u,\,\omega\right)e^{ik\omega}d\omega du+O\left(T^{-1}\right)\\
 & =2\pi\int_{0}^{1}\int_{\Pi}f\left(u,\,\omega\right)\Psi_{T}^{\left(2\right)}\left(\omega\right)d\omega du+O\left(T^{-1}\right).
\end{align*}
Since $\int_{\Pi}\Psi_{T}^{\left(2\right)}\left(\omega\right)d\omega=1$,
we can apply the mean value theorem for $f\left(u,\,\omega\right)$
in a small interval $\left[-\epsilon,\,\epsilon\right],\,\epsilon>0$,
for some $|\eta|\leq1$ depending on $\omega$, 
\begin{align*}
\left|J_{T}-2\pi\int_{0}^{1}f\left(u,\,0\right)du\right| & \leq2\pi\left(\int_{\left|\omega\right|\leq\epsilon}+\int_{\left|\omega\right|>\epsilon}\right)\int_{0}^{1}\int_{\Pi}\left|f\left(u,\,\omega\right)-f\left(u,\,0\right)\right|\left|\Psi_{T}^{\left(2\right)}\left(\omega\right)\right|d\omega du+O\left(T^{-1}\right)\\
 & =O\left(\int_{\left|\omega\right|\leq\epsilon}\int_{0}^{1}\left|\omega\right||f^{\left(1\right)}\left(u,\,\omega\eta\right)|\left|\Psi_{T}^{\left(2\right)}\left(\omega\right)\right|dud\omega\right.\\
 & \quad\left.+\left(\int_{0}^{1}\left(||f\left(u,\,\omega\right)||_{1}+f\left(u,\,0\right)\right)du\right)T^{-1}\right)+O\left(T^{-1}\right)\\
 & =O\left(T^{-1}\log T\right)+O\left(T^{-1}\right),
\end{align*}
where we have used Assumption \ref{Assumption 1 in VR}, 
\begin{align*}
\left|\Psi_{T}^{\left(2\right)}\left(\omega\right)\right| & \leq\frac{1}{2\pi T}\left|D_{T}\left(\omega\right)\right|\left|D_{T}\left(-\omega\right)\right|\leq\frac{1}{\pi T}\left|\omega^{-2}\right|,
\end{align*}
 from \eqref{Eq. (B.4) in VR}-\eqref{Eq. (B.3) VR} and $|\Psi_{T}^{\left(2\right)}\left(\omega\right)|\leq O\left((T)^{-1}\right)$
if $\left|\omega\right|>\epsilon$. 

For the second result in the lemma, note that 
\begin{align*}
J_{T}=\sum_{k=-T+1}^{T-1}T^{-1}\sum_{t=|k|+1}^{T}\mathbb{E}\left(V_{t}V_{t-|k|}\right)= & -\sum_{k=-T+1}^{T-1}T^{-1}\sum_{t=1}^{|k|}\mathbb{E}\left(V_{t}V_{t-|k|}\right)+\sum_{k=-T+1}^{T-1}T^{-1}\sum_{t=1}^{T}\mathbb{E}\left(V_{t}V_{t-|k|}\right).
\end{align*}
 Then, 
\begin{align*}
\left|J_{T}-2\pi\int_{0}^{1}f\left(u,\,0\right)du\right| & \leq\left|\sum_{k=-T+1}^{T-1}T^{-1}\sum_{t=1}^{T}\mathbb{E}\left(V_{t}V_{t-|k|}\right)-2\pi\int_{0}^{1}f\left(u,\,0\right)du\right|+\left|\sum_{k=-T+1}^{T-1}T^{-1}\sum_{t=1}^{k}\mathbb{E}\left(V_{t}V_{t-|k|}\right)\right|,\\
 & =O\left(T^{-1}\right),
\end{align*}
using Assumption \ref{Assumption A - Dependence}-(i). $\square$

\subsubsection{Proof of Lemma \ref{Lemma 2 in VR }}

We can write $\widehat{J}_{\mathrm{HAC},T}=2\pi\int_{\Pi}\widetilde{K}_{b_{1}}\left(\omega\right)I_{T}\left(\omega\right)d\omega.$
Note that 
\begin{align*}
\mathbb{E}\left(I_{T}\left(\omega\right)\right) & =\int_{0}^{1}\int_{\Pi}f\left(u,\,\lambda\right)\Psi_{T}^{\left(2\right)}\left(\omega-\lambda\right)d\lambda du+O\left(T^{-1}\right).
\end{align*}
Thus, we obtain
\begin{align*}
\mathbb{E}\left(\widehat{J}_{\mathrm{HAC},T}\right) & =2\pi\int_{\Pi}\widetilde{K}_{b_{1}}\left(\omega\right)\int_{0}^{1}\int_{\Pi}f\left(u,\,\alpha+\omega\right)\Psi_{T}^{\left(2\right)}\left(\alpha\right)d\alpha dud\omega+O\left(T^{-1}\right).
\end{align*}
Then, using $\int_{\Pi}\Psi_{T}^{\left(2\right)}\left(\omega\right)d\omega=1$
and $\int_{\Pi}\widetilde{K}_{b_{1}}\left(\omega\right)d\omega=1$
we have 
\begin{align*}
\mathbb{E}\left(\widehat{J}_{\mathrm{HAC},T}\right) & -2\pi\int_{0}^{1}f\left(u,\,0\right)du-2\pi b_{1,T}^{d_{f}}\mu_{d_{f}}\left(K\right)\int_{0}^{1}\frac{f^{\left(d_{f}\right)}\left(u,\,0\right)}{d_{f}!}du\\
 & =2\pi\int_{\Pi}\widetilde{K}_{b_{1}}\left(\omega\right)\int_{0}^{1}\int_{\Pi}\Psi_{T}^{\left(2\right)}\left(\alpha\right)\left(f\left(u,\,\omega+\alpha\right)-f\left(u,\,\omega\right)\right)d\alpha dud\omega\\
 & \quad+\int_{\Pi}\widetilde{K}_{b_{1}}\left(\omega\right)\int_{0}^{1}\left[f\left(u,\,\omega\right)-f\left(u,\,0\right)-b_{1,T}^{d_{f}}\mu_{d_{f}}\left(K\right)\frac{f^{\left(d_{f}\right)}\left(u,\,0\right)}{d_{f}!}\right]dud\omega+O\left(T^{-1}\right)\\
 & \triangleq A_{1}+A_{2}+O\left(T^{-1}\right).
\end{align*}
For $\epsilon>0,$ we introduce the sets $\mathbf{A}=\left\{ \left|\alpha\right|,\,\left|\omega\right|\leq\epsilon/2\right\} $
and its complement $\mathbf{A}^{c}$, both defined in $\Pi^{2}$.
Let $A_{11}$ and $A_{12}$ be the contributions to $A_{1}$ corresponding
to $\mathbf{A}$ and $\mathbf{A}^{c}$, respectively. Then, applying
the mean value theorem we have  
\begin{align*}
|A_{11}| & =2\pi\int_{\left|\omega\right|\leq\epsilon/2}\left|\widetilde{K}_{b_{1}}\left(\omega\right)d\omega\right|d\omega\int_{\left|\alpha\right|\leq\epsilon/2}\left|\Psi_{T}^{\left(2\right)}\left(\alpha\right)\right|\left|\alpha\right|d\alpha\int_{0}^{1}\sup_{\left|\omega\right|\leq\epsilon}\left|f^{\left(1\right)}\left(u,\,\omega\right)\right|du\\
 & =O\left(T^{-1}\log T\right),
\end{align*}
 where we have used \eqref{Eq. (B.4) in VR}-\eqref{Eq. (B.3) VR}
and Assumption \ref{Assumption 1 in VR}.  Let $\mathbf{B}_{1}=\left\{ \left|\alpha\right|>\epsilon/2\right\} $
and $\mathbf{B}_{2}=\left\{ \left|\omega\right|>\epsilon/2,\,\left|\alpha\right|\leq\epsilon/2\right\} $
and note that $\mathbf{A}^{c}\subset\{\mathbf{B}_{1}\cup\mathbf{B}_{2}\}$.
The contribution to $A_{12}$ from $\mathbf{B}_{1}$ is  
\begin{align}
\left|\int_{\left|\alpha\right|>\epsilon/2}\right. & \left.\Psi_{T}^{\left(2\right)}\left(\alpha\right)\int_{\Pi}\widetilde{K}_{b_{1}}\left(\omega\right)\int_{0}^{1}\left(f\left(u,\,\omega+\alpha\right)-f\left(u,\,\omega\right)\right)dud\omega d\alpha\right|\nonumber \\
 & =O\left(T^{-1}\int_{\Pi^{2}}\int_{0}^{1}\left|\widetilde{K}_{b_{1}}\left(\omega\right)\left(f\left(u,\,\omega+\alpha\right)-f\left(u,\,\omega\right)\right)\right|dud\omega d\alpha\right)\nonumber \\
 & =O\left(T^{-1}\left(1+\int_{\left|\omega\right|\leq\epsilon}\int_{0}^{1}\left|\widetilde{K}_{b_{1}}\left(\omega\right)f\left(u,\,\omega\right)\right|dud\omega\right)\right)\nonumber \\
 & =O\left(T^{-1}\int_{\Pi}\left|\widetilde{K}_{b_{1}}\left(\omega\right)\right|d\omega\right),\label{Eq. (B.5) in VR}
\end{align}
using \eqref{Eq. (B.4) in VR}-\eqref{Eq. (B.3) VR} and Assumption
\ref{Assumption 1 in VR}. Since $\widetilde{K}_{b_{1}}\left(\omega\right)$
is of reduced magnitude for $\omega>\epsilon/2$, the contribution
to $A_{12}$ from $\mathbf{B}_{2}$ is, for large $T$,
\begin{align}
\left|\int_{\left|\omega\right|>\epsilon/2}\int_{\left|\alpha\right|\leq\epsilon/2}\widetilde{K}_{b_{1}}\left(\omega\right)\Psi_{T}^{\left(2\right)}\left(\alpha\right)\int_{0}^{1}\left(f\left(u,\,\omega+\alpha\right)-f\left(u,\,\omega\right)\right)dud\alpha d\omega\right| & =0,\label{Eq. (B.6) VR}
\end{align}
This implies that $A_{12}=O\left(T^{-1}\right).$ 

As for $A_{2}$ we apply a Taylor's expansion of $f\left(u,\,\omega\right)$
around $\omega=0$ and we split the integral into two parts for $\left|\omega\right|\leq\epsilon$
and $\left|\omega\right|>\epsilon$, denoted as $A_{21}$ and $A_{22}$,
respectively. We have for $\left|\eta\right|\leq1$ depending on $\omega$,
\begin{align*}
A_{21} & =\int_{\left|\omega\right|\leq\epsilon}\widetilde{K}_{b_{1}}\left(\omega\right)\int_{0}^{1}\left(\sum_{j=1}^{d_{f}-1}f^{\left(j\right)}\left(u,\,0\right)\frac{\omega^{j}}{j!}+f^{\left(d_{f}\right)}\left(u,\,\eta\omega\right)\frac{\omega^{d_{f}}}{d_{f}!}-\frac{f^{\left(d_{f}\right)}\left(u,\,0\right)}{d_{f}!}\mu_{d_{f}}\left(K\right)b_{1,T}^{d_{f}}\right)dud\omega\\
 & =\sum_{j=1}^{d_{f}-1}\int_{\Pi}\omega^{j}\widetilde{K}_{b_{1}}\left(\omega\right)d\omega\int_{0}^{1}f^{\left(j\right)}\left(u,\,0\right)\frac{1}{j!}du\\
 & \quad+d_{f}^{-1}\int_{\left|\omega\right|\leq b_{1,T}\pi}\omega^{d_{f}}\widetilde{K}_{b_{1}}\left(\omega\right)\int_{0}^{1}\left(f^{\left(d_{f}\right)}\left(u,\,\eta\omega\right)-f^{\left(d_{f}\right)}\left(u,\,0\right)\right)dud\omega\\
 & =O\left(\int_{\left|\omega\right|\leq b_{1,T}\pi}\left|\widetilde{K}_{b_{1}}\left(\omega\right)\right|\left|\omega\right|^{d_{f}+\varrho}d\omega\right)=O\left(b_{1,T}^{d_{f}+\varrho}\right),
\end{align*}
where we have used Assumption \ref{Assumption 5 VR} and the fact
that as $b_{1,T}\rightarrow0$ the integration is within $\left[-\epsilon,\,\epsilon\right]$
and that by Assumption \ref{Assumption 1 in VR} $f^{\left(d_{f}\right)}\left(u,\,0\right)$
is Lipschitz continuous of order $\varrho$ for all $u\in\left[0,\,1\right]$.
 We can use the same argument used for $A_{12}$ to show that $A_{22}=0$.
$\square$ 

\subsubsection{Proof of Lemma \ref{Lemma 3 in VR}}

From the definition of $Q_{T}$, we have 
\[
\kappa_{T}(0,\,s)=2^{s-1}\left(s-1\right)!\left(\mathsf{V}_{T}J_{T}\right)^{-s}\left(T/b_{1,T}\right)^{-s/2}\mathrm{Tr}((\Sigma_{V}W_{b_{1}})^{s}),
\]
 for $s>1$. By Lemma \ref{Lemma: Proposition 1 in VR}, 
\begin{align}
\overline{\kappa}_{T}(0,\,s) & =\kappa_{T}(0,\,s)\left(b_{1,T}T\right)^{\left(s-2\right)/2}=\frac{2^{s-1}\left(s-1\right)!\left(2\pi\right)^{2s-1}}{\left(\mathsf{V}_{T}J_{T}\right)^{s}}\left(\sum_{j=0}^{d_{f}}L_{j}\left(s\right)b_{1,T}^{j}+O\left(\epsilon_{T}\left(2s\right)\right)\right).\label{Eq. (A.1) in VR}
\end{align}
 Using again Lemma \ref{Lemma: Proposition 1 in VR} with $s=2$ to
evaluate $\mathsf{V}_{T}^{2}$ yields 
\begin{align*}
\mathsf{V}_{T}^{2}\frac{J_{T}^{2}}{4\pi^{2}} & =\frac{1}{4\pi^{2}}Tb_{1,T}\mathrm{Var}\left(\widehat{J}_{\mathrm{HAC},T}\right)=\frac{1}{4\pi^{2}}Tb_{1,T}\mathrm{Var}\left(\mathbf{V}'\frac{W_{b_{1}}}{T}\mathbf{V}\right)\\
 & =\frac{2b_{1,T}}{4\pi^{2}T}\mathrm{Tr}\left(W_{b_{1}}^{2}\Sigma_{V}^{2}\right)=\frac{2b_{1,T}}{4\pi^{2}T}\left(T\left(2\pi\right)^{3}\sum_{j=0}^{d_{f}}L_{j}\left(2\right)b_{1,T}^{j-1}+Tb_{1,T}^{-1}\epsilon_{T}\left(2\right)\right)\\
 & =4\pi\sum_{j=0}^{d_{f}}L_{j}\left(2\right)b_{1,T}^{j}+\epsilon_{T}\left(2\right),
\end{align*}
 where we have use the normality of $V_{t}$.  Lemma \ref{Lemma: Proposition 1 in VR}
implies that $0<L_{0}\left(2\right)<\infty$ and $L_{j}\left(2\right)$
are fixed constants independent of $T$. Then
\begin{align}
\left(\mathsf{V}_{T}\frac{J_{T}}{2\pi}\right)^{-s} & =(4\pi)^{-s/2}\sum_{j=0}^{d_{f}}H_{j}\left(s\right)b_{1,T}^{j}+O\left(\epsilon_{T}\left(s\right)\right),\label{Eq. (A.2) in VR}
\end{align}
 where  $H_{0}\left(s\right)=L_{0}\left(2\right)^{-s/2}$ and so
on. Denoting $c\left(0,\,s\right)=\left(4\pi\right)^{\left(s-2\right)/2}(s-1)!$
and using \eqref{Eq. (A.1) in VR}-\eqref{Eq. (A.2) in VR} we yield
the following expression for the cumulants,  $\overline{\kappa}_{T}(0,\,s)=c\left(0,\,s\right)\sum_{j=0}^{d_{f}}P_{j}\left(s\right)b_{1,T}^{j}+O\left(\epsilon_{T}\left(s\right)\right)$,
where $P_{j}\left(s\right)=\sum_{t=0}^{j}H_{t}\left(s\right)L_{j-t}\left(s\right)$
are constants not depending on $T$ with $P_{1}\left(s\right)=0$,
$P_{2}\left(s\right)=H_{0}\left(s\right)L_{2}\left(s\right)+J_{2}\left(s\right)L_{0}\left(s\right)$,
and so on. Setting $\Xi_{j}\left(0,\,s\right)=c\left(0,\,s\right)P_{j}\left(s\right)$
the lemma follows. $\square$ 

\subsubsection{Proof of Lemma \ref{Lemma: Lemma 4 in VR}}

Note that for $s>0$ we have 
\begin{align*}
\kappa_{T}(2,\,s)=2^{s}s!\xi'_{T}\left(\Sigma_{V}Q_{T}\right)^{s}\Sigma_{V}\xi_{T} & =2^{s}s!\frac{1}{TJ_{T}}\frac{b_{1,T}^{s/2}}{T^{s/2}\mathsf{V}_{T}^{s}J_{T}^{s}}\mathbf{1}'\left(W_{b_{1}}\Sigma_{V}\right)^{s}\Sigma_{V}\mathbf{1}.
\end{align*}
From Lemma \ref{Lemma: Proposition 2 in VR}, 
\begin{align*}
\overline{\kappa}_{T}(2,\,s) & =\left(Tb_{1,T}\right)^{s/2}2^{s}s!\frac{1}{TJ_{T}}\frac{b_{1,T}^{s/2}}{T^{s/2}\mathsf{V}_{T}^{s}J_{T}^{s}}\mathbf{1}'\left(W_{b_{1}}\Sigma_{V}\right)^{s}\Sigma_{V}\mathbf{1}\\
 & =\left(Tb_{1,T}\right)^{s/2}2^{s}s!\frac{1}{TJ_{T}}\frac{b_{1,T}^{s/2}}{T^{s/2}\mathsf{V}_{T}^{s}J_{T}^{s}}\left(T\left(2\pi\right)^{2s+1}\right.\left(\int_{0}^{1}f\left(u,\,0\right)du\right)^{s+1}\left(\widetilde{K}_{b_{1}}\left(0\right)\right)^{s}\\
 & \quad+\left.O\left(b_{1,T}^{-1-s}\log^{2s+1}T\right)\right)\\
 & =\left(\frac{2\pi}{J_{T}\mathsf{V}_{T}}\right)^{s}\frac{2\pi\int_{0}^{1}f\left(u,\,0\right)du}{J_{T}}\left(4\pi\right)^{s}s!\left(\int_{0}^{1}f\left(u,\,0\right)du\right)^{s}K\left(0\right)^{s}+O\left(\epsilon_{T}\left(s+2\right)\right),
\end{align*}
where we have used the fact that $\widetilde{K}_{b_{1}}\left(0\right)=b_{1,T}^{-1}K\left(0\right)$.
Using Lemma \ref{Lemma 4.1 in VR} and eq. \eqref{Eq. (A.2) in VR},
we yield
\begin{align*}
\overline{\kappa}_{T}(2,\,s) & =\left(\frac{2\pi}{J_{T}\mathsf{V}_{T}}\right)^{s}\left(1+O\left(T^{-1}\log T\right)\right)\left(4\pi\right)^{s}s!\left(\int_{0}^{1}f\left(u,\,0\right)du\right)^{s}K\left(0\right)^{s}+O\left(\epsilon_{T}\left(s+2\right)\right)\\
 & =\left(4\pi\right)^{-s/2}\left(4\pi\right)^{s}s!\left(\int_{0}^{1}f\left(u,\,0\right)du\right)^{s}K\left(0\right)^{s}\sum_{j=0}^{d_{f}}H_{j}\left(s\right)b_{1,T}^{j}+O\left(\epsilon_{T}\left(s+2\right)\right),
\end{align*}
where the $H_{s}\left(j\right)$ are as in the proof of Lemma \ref{Lemma 3 in VR}.
The lemma follows by setting $\Xi_{j}(2,\,s)=\left(4\pi\right)^{-s/2}\left(4\pi\right)^{s}s!$
$(\int_{0}^{1}f\left(u,\,0\right)du)^{s}K\left(0\right)^{s}H_{j}\left(s\right)$.
$\square$ 

\subsubsection{Proof of Theorem \ref{Theorem: Theorem 1 in VR}}

We first construct the approximation for $\psi_{T}\left(\mathbf{t}\right)$.
It follows from \citeReferencesSupp{velasco/robinson:01} and \citeReferencesSupp{taniguchi/puri:1996}
that only the cumulants $\kappa_{T}(0,\,s)$ and $\kappa_{T}(2,\,s)$
are nonzero, and that the cumulant generating function is given by
\begin{align}
\log\psi_{T}\left(\mathbf{t}\right) & =\frac{1}{2}\left\Vert i\mathbf{t}\right\Vert ^{2}+\sum_{s=3}^{\tau+1}\frac{\left(Tb_{1,T}\right)^{\left(2-s\right)/2}}{s!}\sum_{|\mathbf{r}|=s}\frac{s!}{r_{1}!r_{2}!}\overline{\kappa}_{T}(r_{1},\,r_{2})\left(it_{1}\right)^{r_{1}}\left(it_{2}\right)^{r_{2}}+R_{T}\left(\tau\right),\label{Eq. (A.3) in VR}
\end{align}
where $\mathbf{r}=\left(r_{1},\,r_{2}\right)'$ with $r_{1}\in\left\{ 0,\,2\right\} $
and $|\mathbf{r}|=r_{1}+r_{2}$, and 

\begin{align*}
R_{T}\left(\tau\right) & =\left(Tb_{1,T}\right)^{-\tau/2}\left(R_{0,\tau+2}\left(it_{2}\right)^{\tau+2}+R_{2,\tau}\left(it_{1}\right)^{2}\left(it_{2}\right)^{\tau}\right),\qquad\qquad\qquad\tau\,\mathrm{even},\\
R_{T}\left(\tau\right) & =\left(Tb_{1,T}\right)^{-\tau/2}\frac{1}{\left(\tau+2\right)!}\left(\overline{\kappa}_{T}(0,\,\tau+2)\left(it_{2}\right)^{\tau+2}+\frac{\left(\tau+2\right)\left(\tau+1\right)}{2}\overline{\kappa}_{T}(2,\,\tau)\left(it_{1}\right)^{2}\left(it_{2}\right)^{\tau}\right)\\
 & \qquad+\left(Tb_{1,T}\right)^{-\tau/2}\left(R_{0,\tau+3}\left(it_{2}\right)^{\tau+3}+R_{2,\tau+1}\left(it_{1}\right)^{2}\left(it_{2}\right)^{\tau+1}\right),\qquad\tau\,\mathrm{odd},
\end{align*}
where the $R_{0,j}$ and $R_{2,j}$ are bounded. Using Lemmas \ref{Lemma 3 in VR}-\ref{Lemma: Lemma 4 in VR},
we have
\begin{align*}
\log\psi_{T}\left(\mathbf{t}\right) & =\frac{1}{2}\left\Vert i\mathbf{t}\right\Vert ^{2}+\sum_{s=3}^{\tau+1}\frac{\left(Tb_{1,T}\right)^{\left(2-s\right)/2}}{s!}\left(\overline{\kappa}_{T}(0,\,s)\left(it_{2}\right)^{s}+\frac{s\left(s-1\right)}{2}\overline{\kappa}_{T}(2,\,s-2)\left(it_{1}\right)^{2}\left(it_{2}\right)^{s-2}\right)+R_{T}\left(\tau\right)\\
 & =\frac{1}{2}\left\Vert i\mathbf{t}\right\Vert ^{2}+\sum_{s=3}^{\tau+1}\left(Tb_{1,T}\right)^{\left(2-s\right)/2}\left(B_{T}\left(s,\,\mathbf{t}\right)+\left\{ \left(it_{2}\right)^{s}+\left(it_{1}\right)^{2}\left(it_{2}\right)^{s-2}\right\} O\left(\epsilon_{T}\left(s\right)\right)\right)+R_{T}\left(\tau\right),
\end{align*}
where
\begin{align*}
B_{T}\left(s,\,\mathbf{t}\right) & =\frac{1}{s!}\sum_{j=0}^{d_{f}}b_{1,T}^{j}\left\{ \Xi_{j}(0,\,s)\left(it_{2}\right)^{s}+\frac{s\left(s-1\right)}{2}\Xi_{j}(2,\,s-2)\left(it_{1}\right)^{2}\left(it_{2}\right)^{s-2}\right\} .
\end{align*}
The approximation of the characteristic function of $\mathbf{u}$
using its cumulant generating function is
\begin{align*}
\mathcal{A}_{T}\left(\mathbf{t},\,\tau\right) & =\exp\left\{ \frac{1}{2}\left\Vert i\mathbf{t}\right\Vert ^{2}\right\} \left[1+\sum_{j=3}^{\tau+1}\left(Tb_{1,T}\right)^{\left(2-j\right)/2}\sum_{\mathbf{r}}\prod_{n=3}^{\tau+1}\left[B_{T}\left(n,\,\mathbf{t}\right)\right]^{r_{n}}\frac{1}{r_{3}!\cdots r_{\tau+1}!}\right],
\end{align*}
where $\mathbf{r}=\left(r_{3},\ldots,\,r_{\tau+1}\right)'$, $r_{n}\in\left\{ 0,\,1,\ldots\right\} $,
and the summation is over all $\mathbf{r}$ satisfying $\sum_{n=3}^{\tau+1}\left(n-2\right)r_{n}=j-2$.
To obtain a second-order Edgeworth expansion we set $\tau=2$ and
we include in $\mathcal{A}_{T}\left(\mathbf{t},\,2\right)$ terms
up to order $(Tb_{1,T})^{-1/2}$, 
\begin{align}
\mathcal{A}_{T}\left(\mathbf{t},\,2\right) & =\exp\left\{ \frac{1}{2}\left\Vert i\mathbf{t}\right\Vert ^{2}\right\} \left(1+\overline{B}_{T}\left(3,\,\mathbf{t}\right)\left(Tb_{1,T}\right)^{-1/2}\right),\label{Eq. (A.4) in VR}
\end{align}
 where in $\overline{B}_{T}\left(3,\,\mathbf{t}\right)$ includes
only the leading term in $b_{1,T}^{j}$ $\left(j=0\right)$ in the
expansion for the cumulant of order three. Note that the characteristic
function of $\mathbb{Q}_{T}^{\left(2\right)}(\cdot)$ is $\mathcal{A}_{T}\left(\mathbf{t},\,2\right)$.

The rest of the proof consists of studying the distance between the
true distribution and its Edgeworth approximation. Lemma \ref{Lemma: Lemma 14 in VR}
studies the Edgeworth approximation for the characteristic function
for $\left\Vert \mathbf{t}\right\Vert \leq c_{1}\sqrt{Tb_{1,T}}$,
whereas Lemma \ref{Lemma: Lemma 15 in VR} analyzes its tail behavior.
The desired result follows from the same steps as in Theorem 1 of
\citeReferencesSupp{velasco/robinson:01} which relies on Lemma \ref{Lemma Bhattacharya and Rao 1975}.
$\square$ 
\begin{lem}
\label{Lemma: Lemma 14 in VR}Let Assumptions \ref{Assumption 1 in VR},
\ref{Assumption 3 VR}, \ref{Assumption 4 in VR} and $b_{1,T}+\left(Tb_{1,T}\right)^{-1}\log^{5}T\rightarrow0$
hold. There exists $\delta_{1}>0$ such that, for $\left\Vert \mathbf{t}\right\Vert \leq\delta_{1}\sqrt{Tb_{1,T}}$
and a number $d_{1}>0$, 
\begin{align*}
\left|\psi_{T}\left(\mathbf{t}\right)-A_{T}\left(\mathbf{t},\,2\right)\right| & \leq\exp\left\{ -d_{1}\left\Vert \mathbf{t}\right\Vert ^{2}\right\} \widetilde{F}\left(\left\Vert \mathbf{t}\right\Vert \right)O\left(\left(Tb_{1,T}\right)^{-1/2}\left(b_{1,T}^{2}+\epsilon_{T}\left(3\right)\right)+\frac{1}{Tb_{1,T}}\right),
\end{align*}
where $\widetilde{F}\left(\left\Vert \mathbf{t}\right\Vert \right)$
is a polynomial in $\mathbf{t}$ with bounded coefficients and $\mathcal{A}_{T}\left(\mathbf{t},\,2\right)$
is defined as in \eqref{Eq. (A.4) in VR}. 
\end{lem}
\noindent\textit{Proof of Lemma \ref{Lemma: Lemma 14 in VR}. }It
is similar to the proof of Lemma 14 in \citeReferencesSupp{velasco/robinson:01}.
\textit{$\square$ }

\subsubsection{Proof of Lemma \ref{Lemma: Lemma 5 in VR}}

It is similar to the proof of Lemma 5 in \citeReferencesSupp{velasco/robinson:01}.
$\square$

\subsubsection{Proof of Theorem \ref{Theorem: Theorem 2 in VR}}

Consider the transformation $\mathbf{s}=\left(s_{1},\,s_{2}\right)'=(\widetilde{Z}_{T}(h_{1},\,h_{2}),\,h_{2})'=\Delta_{T}\left(\mathbf{h}\right)$
say, and its inverse $\mathbf{h}=\Delta_{T}^{-1}\left(\mathbf{s}\right)=(h_{1}^{\dagger}(s_{1},\,s_{2}),\,s_{2})'$.
Let $\mathbf{L}_{T}=\{\mathbf{h}:\,\left|h_{i}\right|<l_{1}T^{\gamma},\,0<\gamma<d_{f}/(3(1+2d_{f})),\,i=1,\,2\}$,
where $l_{i}$ are some fixed constants. Using $(1+x)^{-1}=1-x+x^{2}-x^{3}+\ldots$
for $\left|x\right|<1$, we have uniformly in the set $\mathbf{L}_{T}$,
\begin{align*}
h_{1}^{\dagger}\left(\mathbf{s}\right) & =s_{1}\left[1+\frac{1}{2}\overline{c}_{1}b_{1,T}^{d_{f}}+\frac{1}{2}\sqrt{4\pi}\left\Vert K_{2}\right\Vert s_{2}\left(Tb_{1,T}\right)^{-1/2}\right]+o\left(\left(Tb_{1,T}\right)^{-1/2}\right).
\end{align*}
 We have $\mathbb{P}(Z_{T}\in\mathbf{C})=\mathbb{P}(\mathbf{h}\in\Delta_{T}^{-1}\left(\mathbf{C}\times\mathbb{R}\right))$
and from Theorem \ref{Theorem: Theorem 1 in VR},  
\[
\sup_{\mathbf{C}}\left|\mathrm{\mathbb{P}\left(\mathbf{h}\in\Delta_{\mathit{T}}^{-1}\left(\mathbf{C}\times\mathbb{R}\right)\right)-}\mathbb{Q}_{T}^{\left(2\right)}\left(Z_{T}^{-1}\left(\mathbf{C}\times\mathbb{R}\right)\right)\right|=o\left(\left(Tb_{1,T}\right)^{-1/2}\right)+\mathrm{cost}\,\sup_{\mathbf{C}}\mathbb{Q}_{T}^{\left(2\right)}\left(\left(\partial\Delta_{T}^{-1}\left(\mathbf{C}\times\mathbb{R}\right)\right)^{2\phi_{T}}\right),
\]
 where $\phi_{T}=(Tb_{1,T})^{-\varpi}$ with $1/2<\varpi<1$. The
rest of the proof is similar to the proof of Theorem 2 in \citeReferencesSupp{velasco/robinson:01}.
$\square$ 

\subsection{Additional Lemmas Used for the Proofs of Theorem \ref{Theorem: Theorem 1 in VR DK-HAC}-\ref{Theorem: Theorem 2 in VR DK-HAC}}
\begin{lem}
\label{Lemma 2 in VR DK-HAC}Let Assumptions \ref{Assumption 1 in VR},
\ref{Assumption 3 VR}, \ref{Assumption 5 VR}-\ref{Assumption 6 VR}
and \ref{Assumption K2 and b2}-\ref{Assumption Lip of d2 f(u,w)}
hold. Then, 
\begin{align*}
\mathbb{E}\left(\widehat{J}_{\mathrm{DK},T}^{*}\right) & -2\pi\int_{0}^{1}f\left(u,\,0\right)du-2\pi\frac{\int_{0}^{1}f^{\left(d_{f}\right)}\left(u,\,0\right)du}{d_{f}!}\mu_{d_{f}}\left(K\right)b_{1,T}^{d_{f}}\\
 & \quad-\pi b_{2,T}^{2}\int_{0}^{1}x^{2}K_{2}\left(x\right)dx\int_{\widetilde{\mathbf{C}}}\frac{\partial^{2}}{\partial u^{2}}f\left(u,\,0\right)du-2\pi b_{2,T}^{2}\Delta_{f}\left(0\right)\\
 & =O\left(b_{1,T}^{d_{f}+\varrho}+\left(Tb_{2,T}\right)^{-1}\log\left(Tb_{2,T}\right)\right)+o\left(b_{2,T}^{2}\right).
\end{align*}
\end{lem}
The term $2\pi b_{2,T}^{2}\Delta_{f}\left(0\right)$ in Lemma \ref{Lemma 2 in VR DK-HAC}
is the contribution to the bias due to the local time-smoothing in
the neighborhoods involving a discontinuity point.

We now consider the cumulants of the normalized spectral estimate
$v_{2}$. 
\begin{lem}
\label{Lemma 3 in VR DK-HAC}Let Assumptions \ref{Assumption 1 in VR},
\ref{Assumption 3 VR}-\ref{Assumption 4 in VR} and \ref{Assumption K2 and b2}-\ref{Assumption Lip of d2 f(u,w)}
hold. For $s>2$ with $\epsilon_{Tb_{2,T}}\left(s\right)=b_{1,T}^{d_{f}+\varrho}+(Tb_{2,T}b_{1,T})^{-1}$
$\log^{2s-1}(Tb_{2,T})\rightarrow0$, we have
\begin{align*}
\overline{\kappa}_{2,T}\left(0,\,s\right) & \triangleq\kappa_{2,T}\left(0,\,s\right)\left(Tb_{1,T}b_{2,T}\right)^{\left(s-2\right)/2}\\
 & =\sum_{j=0}^{d_{f}}\Xi_{2,j}\left(0,\,s\right)b_{1,T}^{j}+b_{2,T}^{2}\sum_{j=0}^{d_{f}}\left(\widetilde{\Xi}_{2,j}\left(0,\,s\right)+\widetilde{\Xi}_{3,j}\left(0,\,s\right)\right)b_{1,T}^{j}+O\left(\epsilon_{Tb_{2,T}}\left(s\right)\right),
\end{align*}
where $\Xi_{2,j}\left(0,\,s\right)$ is bounded and depends on  $K,\,K_{2}$
and on $f^{\left(j\right)}\left(u,\,0\right)$ $(j=0,\ldots,\,d_{f})$,
$\widetilde{\Xi}_{2,j}\left(0,\,s\right)$ is bounded and depends
on $K,\,K_{2}$, $f^{\left(j\right)}\left(u,\,0\right)$ and $\left(\partial^{2}/\partial u^{2}\right)f\left(u,\,\omega\right)$
and $\widetilde{\Xi}_{3,j}\left(0,\,s\right)$ is bounded and depends
on $K,\,K_{2}$, $f^{\left(j\right)}\left(u,\,0\right)$ and $\Delta_{f}\left(\omega\right)$.
\end{lem}
We now consider the cross-cumulants of $\mathbf{v}.$ 
\begin{lem}
\label{Lemma: Lemma 4 in VR DK-HAC}Let Assumptions \ref{Assumption 1 in VR},
\ref{Assumption 3 VR}-\ref{Assumption 4 in VR} and \ref{Assumption K2 and b2}-\ref{Assumption Lip of d2 f(u,w)}
hold. For $s>0$ with $\epsilon_{Tb_{2,T}}\left(s+2\right)\rightarrow0$,
\begin{align*}
\overline{\kappa}_{2,T}(2,\,s) & \triangleq\kappa_{2,T}(2,\,s)\left(Tb_{2,T}b_{1,T}\right)^{s/2}=\sum_{j=0}^{d_{f}}\left(\Xi_{2,j}(2,\,s)+b_{2,T}^{2}\left(\widetilde{\Xi}_{2,j}(2,\,s)+\widetilde{\Xi}_{3,j}(2,\,s)\right)\right)b_{1,T}^{j}\\
 & \quad+O\left(\epsilon_{Tb_{2,T}}\left(s+2\right)\right),
\end{align*}
 where $\Xi_{2,j}(2,\,s)$ is bounded and depends on $K,\,K_{2}$
and $f^{\left(j\right)}\left(u,\,0\right)$ $(j=0,\ldots,\,d_{f})$,
$\widetilde{\Xi}_{2,j}\left(2,\,s\right)$ is bounded and depends
on $K,\,K_{2},$ $f^{\left(j\right)}\left(u,\,0\right)$ and $\left(\partial^{2}/\partial u^{2}\right)f\left(u,\,\omega\right)$,
and $\widetilde{\Xi}_{3,j}\left(2,\,s\right)$ is bounded and depends
on $K,\,K_{2}$, $f^{\left(j\right)}\left(u,\,0\right)$ and $\Delta_{f}\left(\omega\right)$.
\end{lem}

\subsubsection{Proof of Lemma \ref{Lemma 2 in VR DK-HAC}}

For $r\in\widetilde{\mathbf{C}},$ using a second-order Taylor's expansion
as in the proof of Theorem 7.3 in \citeReferencesSupp{casini:change-point-spectra},
we yield 
\begin{align*}
\mathbb{E}\left(\widetilde{I}_{T}\left(r,\,\omega\right)\right) & =\mathbb{E}\left(\frac{1}{2\pi Tb_{2,T}}\left|\sum_{t=1}^{T}\exp\left(-i\omega t\right)\widetilde{V}_{t}\left(r\right)\right|^{2}\right)\\
 & =\frac{1}{2\pi}\frac{1}{Tb_{2,T}}\sum_{k=-\left\lfloor Tb_{2,T}\right\rfloor +1}^{\left\lfloor Tb_{2,T}\right\rfloor -1}\sum_{t=|k|+1}^{T}\int_{\Pi}K_{2}\left(\frac{\left(Tr-\left(t-k/2\right)\right)/T}{b_{2,T}}\right)f\left((t+k/2)/T,\,\lambda\right)e^{ik\left(\omega-\lambda\right)}d\lambda\\
 & \quad+O\left(\left(Tb_{2,T}\right)^{-1}\log\left(Tb_{2,T}\right)\right)\\
 & =\int_{\Pi}f\left(r,\,\lambda\right)\Psi_{Tb_{2,T}}^{\left(2\right)}\left(\omega-\lambda\right)d\lambda\\
 & \quad+\frac{b_{2,T}^{2}}{2}\int_{0}^{1}x^{2}K_{2}\left(x\right)dx\frac{\partial^{2}}{\partial u^{2}}f\left(u,\,\omega\right)|_{u=r}+o\left(b_{2,T}^{2}\right)+O\left(\left(Tb_{2,T}\right)^{-1}\log\left(Tb_{2,T}\right)\right).
\end{align*}
In a neighborhood of a break point $\lambda_{j}^{0}$, let  $r=\lambda_{j}^{0}+sb_{2,T}$
for some $s\in(0,\,1)$. Then, 
\begin{align*}
\mathbb{E}\left(\widetilde{I}_{T}\left(r,\,\omega\right)\right) & =\int_{\Pi}f\left(r,\,\lambda\right)\Psi_{Tb_{2,T}}^{\left(2\right)}\left(\omega-\lambda\right)d\lambda\\
 & \quad+b_{2,T}\left(\int_{0}^{1-s}xK_{2}\left(x\right)dx\frac{\partial}{\partial u_{-}}f\left(\lambda_{j}^{0},\,\omega\right)+\int_{1-s}^{1}xK_{2}\left(x\right)dx\frac{\partial}{\partial u_{+}}f\left(\lambda_{j}^{0},\,\omega\right)\right).
\end{align*}
When integrating the last term above over $r$ we have 
\begin{align*}
b_{2,T}^{2} & \sum_{j=1}^{m_{0}}\int_{0}^{1}\left(\frac{\partial}{\partial u_{-}}f\left(\lambda_{j}^{0},\,\omega\right)\int_{0}^{1-s}xK_{2}\left(x\right)dx+\frac{\partial}{\partial u_{+}}f\left(\lambda_{j}^{0},\,\omega\right)\int_{1-s}^{1}xK_{2}\left(x\right)dx\right)ds.
\end{align*}
 Thus, we obtain
\begin{align*}
\mathbb{E}\left(\widehat{J}_{\mathrm{DK},T}^{*}\right) & =2\pi\int_{\Pi}\widetilde{K}_{b_{1}}\left(\omega\right)\int_{0}^{1}\int_{\Pi}f\left(u,\,\alpha+\omega\right)\Psi_{T}^{\left(2\right)}\left(\alpha\right)d\lambda dud\omega\\
 & \quad+\pi b_{2,T}^{2}\int_{0}^{1}x^{2}K_{2}\left(x\right)dx\int_{\Pi}\widetilde{K}_{b_{1}}\left(\omega\right)\int_{\widetilde{\mathbf{C}}}\frac{\partial^{2}}{\partial u^{2}}f\left(u,\,\omega\right)dud\omega\\
 & \quad+2\pi b_{2,T}^{2}\int_{\Pi}\widetilde{K}_{b_{1}}\left(\omega\right)\Delta_{f}\left(\omega\right)d\omega+o\left(b_{2,T}^{2}\right)+O\left(\left(Tb_{2,T}\right)^{-1}\log\left(Tb_{2,T}\right)\right).
\end{align*}
Then, using $\int_{\Pi}\Psi_{T}^{\left(2\right)}\left(\omega\right)d\omega=1$,
$\int_{\Pi}\widetilde{K}_{b_{1}}\left(\omega\right)d\omega=1$, Assumption
\ref{Assumption Lip of d2 f(u,w)} and similar arguments as in the
proof of Lemma \ref{Lemma 2 in VR } applied to the terms involving
$\frac{\partial^{2}}{\partial u^{2}}f\left(u,\,\omega\right)$ and
$\Delta_{f}\left(\omega\right)$, we have 
\begin{align*}
\mathbb{E}\left(\widehat{J}_{\mathrm{DK},T}^{*}\right) & -2\pi\int_{0}^{1}f\left(u,\,0\right)du-2\pi b_{1,T}^{d_{f}}\mu_{d_{f}}\left(K\right)\int_{0}^{1}\frac{f^{\left(d_{f}\right)}\left(u,\,0\right)}{d_{f}!}du\\
 & \quad-\pi b_{2,T}^{2}\int_{0}^{1}x^{2}K_{2}\left(x\right)dx\int_{\widetilde{\mathbf{C}}}\frac{\partial^{2}}{\partial u^{2}}f\left(u,\,0\right)du-2\pi b_{2,T}^{2}\Delta_{f}\left(0\right)\\
 & =2\pi\int_{\Pi}\widetilde{K}_{b_{1}}\left(\omega\right)\int_{0}^{1}\int_{\Pi}\Psi_{T}^{\left(2\right)}\left(\alpha\right)\left(f\left(u,\,\omega+\alpha\right)-f\left(u,\,\omega\right)\right)d\alpha dud\omega\\
 & \quad+2\pi\int_{\Pi}\widetilde{K}_{b_{1}}\left(\omega\right)\int_{0}^{1}\left[f\left(u,\,\omega\right)-f\left(u,\,0\right)-b_{1,T}^{d_{f}}\mu_{d}\left(K\right)\frac{f^{\left(d_{f}\right)}\left(u,\,0\right)}{d_{f}!}\right]dud\omega\\
 & \quad+o\left(b_{2,T}^{2}\right)+O\left(\left(Tb_{2,T}\right)^{-1}\log\left(Tb_{2,T}\right)\right)+o\left(b_{2,T}^{2}b_{1,T}^{q_{2}}\right)\\
 & \triangleq A_{1}+A_{2}+o\left(b_{2,T}^{2}\right)+O\left(\left(Tb_{2,T}\right)^{-1}\log\left(Tb_{2,T}\right)\right).
\end{align*}
To conclude the proof, note that by Lemma \ref{Lemma 2 in VR } we
have $|A_{1}|+|A_{2}|=O\left(T^{-1}\log T\right)+O(b_{1,T}^{d_{f}+\varrho})$.
$\square$ 

\subsubsection{Proof of Lemma \ref{Lemma 3 in VR DK-HAC}}

 We have 
\[
\kappa_{2,T}(0,\,s)=2^{s-1}\left(s-1\right)!\left(\mathsf{V}_{2,T}J_{T}\right)^{-s}\left(Tb_{2,T}/b_{1,T}\right)^{-s/2}\mathrm{Tr}((\Sigma_{\widetilde{V}}W_{b_{1}})^{s}),
\]
 for $s>1$. By Lemma \ref{Lemma: Proposition 1 in VR DK-HAC}, 
\begin{align}
\overline{\kappa}_{2,T}(0,\,s) & =\kappa_{2,T}(0,\,s)\left(Tb_{1,T}b_{2,T}\right)^{\left(s-2\right)/2}\label{Eq. (A.1) in VR-1}\\
 & =\frac{2^{s-1}\left(s-1\right)!\left(2\pi\right)^{2s-1}}{\left(\mathsf{V}_{2,T}J_{T}\right)^{s}}\left(\sum_{j=0}^{d_{f}}L_{j}\left(s\right)b_{1,T}^{j}+b_{2,T}^{2}\sum_{j=0}^{d_{f}}\left(\left(L_{2,j}\left(s\right)+L_{3,j}\left(s\right)\right)b_{1,T}^{j}\right)+O\left(\epsilon_{Tb_{2,T}}\left(s\right)\right)\right).\nonumber 
\end{align}
 Using Lemma \ref{Lemma: Proposition 1 in VR DK-HAC}  to evaluate
$\mathsf{V}_{2,T}^{2}$ yields 
\begin{align*}
\mathsf{V}_{2,T}^{2}\frac{J_{T}^{2}}{4\pi^{2}} & =\frac{1}{4\pi^{2}}Tb_{1,T}b_{2,T}\mathrm{Var}\left(\widehat{J}_{\mathrm{DK},T}^{*}\right)=Tb_{1,T}b_{2,T}\mathrm{Var}\left(\int_{0}^{1}\mathbf{\widetilde{V}}\left(r\right)'\frac{W_{b_{1}}}{Tb_{2,T}}\mathbf{\widetilde{V}}\left(r\right)dr\right)\\
 & =\frac{2b_{1,T}}{4\pi^{2}Tb_{2,T}}\mathrm{Tr}\left(W_{b_{1}}^{2}\Sigma_{\widetilde{V}}^{2}\right)\\
 & =\frac{2b_{1,T}}{4\pi^{2}}\left(2\pi\right)^{3}\left(\sum_{j=0}^{d_{f}}L_{j}\left(2\right)b_{1,T}^{j-1}+b_{2,T}^{2}\sum_{j=0}^{d_{f}}\left(\left(L_{2,j}\left(s\right)+L_{3,j}\left(s\right)\right)b_{1,T}^{j-1}\right)\right)+Tb_{2,T}b_{1,T}^{-1}O\left(\epsilon_{Tb_{2,T}}\left(2\right)\right)\\
 & =4\pi\left(\sum_{j=0}^{d_{f}}L_{j}\left(2\right)b_{1,T}^{j}+b_{2,T}^{2}\sum_{j=0}^{d_{f}}\left(\left(L_{2,j}\left(s\right)+L_{3,j}\left(s\right)\right)b_{1,T}^{j}\right)\right)+O\left(\epsilon_{Tb_{2,T}}\left(2\right)\right),
\end{align*}
 where we have use the normality of $\{V_{t}\}$.  Since Lemma \ref{Lemma: Proposition 1 in VR DK-HAC}
implies that $0<L_{0}\left(2\right)<\infty$ and $L_{j}\left(2\right)$
are fixed constants independent of $T$, we then have
\begin{align}
\left(\mathsf{V}_{2,T}\frac{J_{T}}{2\pi}\right)^{-s} & =(4\pi)^{-s/2}\sum_{j=0}^{d_{f}}H_{j}\left(2\right)b_{1,T}^{j}+O\left(\epsilon_{Tb_{2,T}}\left(2\right)\right),\label{Eq. (A.2) in VR-1}
\end{align}
 where  $H_{0}\left(s\right)=L_{0}\left(2\right)^{-s/2}$ and so
on. Using \eqref{Eq. (A.1) in VR-1}-\eqref{Eq. (A.2) in VR-1} we
yield  
\[
\overline{\kappa}_{2,T}(0,\,s)=c\left(0,\,s\right)\left(\sum_{j=0}^{d_{f}}P_{2,j}\left(s\right)b_{1,T}^{j}+b_{2,T}^{2}\sum_{j=0}^{d_{f}}\left(\left(\widetilde{P}_{2,j}\left(s\right)+\widetilde{P}_{3,j}\left(s\right)\right)b_{1,T}^{j}\right)\right)+O\left(\epsilon_{Tb_{2,T}}\left(2\right)\right),
\]
where $c\left(0,\,s\right)=\left(4\pi\right)^{\left(s-2\right)/2}(s-1)!$,
$P_{2,j}\left(s\right)=\sum_{t=0}^{j}H_{t}\left(s\right)L_{j-t}\left(s\right)$
are constants not depending on $T$ with $P_{2,1}\left(s\right)=0$,
$P_{2,2}\left(s\right)=H_{0}\left(s\right)L_{2}\left(s\right)+H_{2}\left(s\right)L_{0}\left(s\right)$
and so on, and $\widetilde{P}_{2,j}\left(s\right)=\sum_{t=0}^{j}H_{t}\left(s\right)L_{2,j-t}\left(s\right)$
and $\widetilde{P}_{3,j}\left(s\right)=\sum_{t=0}^{j}H_{t}\left(s\right)L_{3,j-t}\left(s\right)$.
The lemma follows from setting $\Xi_{2,j}(0,\,s)=c\left(0,\,s\right)P_{2,j}\left(s\right)$,
$\widetilde{\Xi}_{2,j}(0,\,s)=c\left(0,\,s\right)\widetilde{P}_{2,j}\left(s\right)$
and $\widetilde{\Xi}_{2,j}(0,\,s)=c\left(0,\,s\right)\widetilde{P}_{3,j}\left(s\right)$.
$\square$ 

\subsubsection{Proof of Lemma \ref{Lemma: Lemma 4 in VR DK-HAC}}

For $s>0$ we have  
\begin{align*}
\kappa_{2,T}(2,\,s)=2^{s}s!\xi'_{T}\left(\Sigma_{\widetilde{V}}Q_{2,T}\right)^{s}\Sigma_{\widetilde{V}}\xi_{T} & =2^{s}s!\frac{1}{Tb_{2,T}J_{T}}\frac{b_{1,T}^{s/2}}{\left(Tb_{2,T}\right){}^{s/2}\mathsf{V}_{2,T}^{s}J_{T}^{s}}\mathbf{1}'\left(W_{b_{1}}\Sigma_{\widetilde{V}}\right)^{s}\Sigma_{\widetilde{V}}\mathbf{1}.
\end{align*}
From Lemma \ref{Lemma: Proposition 2 in VR DK-HAC}, we have 
\begin{align*}
\overline{\kappa}_{2,T}(2,\,s) & =\left(Tb_{1,T}b_{2,T}\right)^{s/2}2^{s}s!\frac{1}{Tb_{2,T}J_{T}}\frac{b_{1,T}^{s/2}}{\left(Tb_{2,T}\right){}^{s/2}\mathsf{V}_{2,T}^{s}J_{T}^{s}}\mathbf{1}'\left(W_{b_{1}}\Sigma_{\widetilde{V}}\right)^{s}\Sigma_{\widetilde{V}}\mathbf{1}\\
 & =\left(Tb_{1,T}b_{2,T}\right)^{s/2}2^{s}s!\frac{1}{Tb_{2,T}J_{T}}\frac{b_{1,T}^{s/2}}{\left(Tb_{2,T}\right){}^{s/2}\mathsf{V}_{2,T}^{s}J_{T}^{s}}\\
 & \quad\times\Biggl(Tb_{2,T}\left(2\pi\right)^{2s+1}\biggl(\left(\int_{0}^{1}f\left(u,\,0\right)du\right)^{s+1}\int_{0}^{1}K_{2}^{s+1}\left(x\right)dx+b_{2,T}^{2}\widetilde{\Lambda}_{2}\left(f'',\,\widetilde{\mathbf{C}},\,s\right)\\
 & \quad+b_{2,T}^{2}\widetilde{\Lambda}_{3}\left(f',\,\left\{ \lambda_{j}^{0},\,j=1,\ldots,\,m_{0}\right\} ,\,s\right)\biggr)\left(\widetilde{K}_{b_{1}}\left(0\right)\right)^{s}\\
 & \quad+O\left(b_{1,T}^{1-s}\log^{2s+1}\left(Tb_{2,T}\right)+b_{1,T}^{-s}\frac{\log^{2s+1}\left(Tb_{2,T}\right)}{Tb_{2,T}}\right)\Biggr)\\
 & =\left(\frac{2\pi}{J_{T}\mathsf{V}_{2,T}}\right)^{s}\frac{2\pi\int_{0}^{1}f\left(u,\,0\right)du}{J_{T}}\left(4\pi\right)^{s}s!\left(\left(\int_{0}^{1}f\left(u,\,0\right)du\right)^{s}\int_{0}^{1}K_{2}^{s+1}\left(x\right)dx+b_{2,T}^{2}\left(\widetilde{\Lambda}_{2}^{*}+\widetilde{\Lambda}_{3}^{*}\right)\right)K\left(0\right)^{s}\\
 & \quad+O\left(\epsilon_{Tb_{2,T}}\left(s+2\right)\right),
\end{align*}
where $\widetilde{\Lambda}_{2}^{*}$ and $\widetilde{\Lambda}_{3}^{*}$
are equal to $\widetilde{\Lambda}_{2}$ and $\widetilde{\Lambda}_{3}$,
respectively, without the factor $\int_{0}^{1}f\left(u,\,0\right)du$,
and we have used $\widetilde{K}_{b_{1}}\left(0\right)=b_{1,T}^{-1}K\left(0\right)$.
Using Lemma \ref{Lemma 4.1 in VR} and \eqref{Eq. (A.2) in VR-1},
we yield
\begin{align*}
\overline{\kappa}_{2,T}(2,\,s) & =\left(\frac{J_{T}\mathsf{V}_{2,T}}{2\pi}\right)^{-s}\left(1+O\left(\left(Tb_{2,T}\right)^{-1}\log(Tb_{2,T})\right)\right)\\
 & \quad\times\left(4\pi\right)^{s}s!\left(\left(\int_{0}^{1}f\left(u,\,0\right)du\right)^{s}\int_{0}^{1}K_{2}^{s+1}\left(x\right)dx+b_{2,T}^{2}\left(\widetilde{\Lambda}_{2}^{*}+\widetilde{\Lambda}_{3}^{*}\right)\right)K\left(0\right)^{s}+O\left(\epsilon_{Tb_{2,T}}\left(s+2\right)\right)\\
 & =\left(4\pi\right)^{-s/2}\left(4\pi\right)^{s}s!\left(\left(\int_{0}^{1}f\left(u,\,0\right)du\right)^{s}\int_{0}^{1}K_{2}^{s+1}\left(x\right)dx+b_{2,T}^{2}\left(\widetilde{\Lambda}_{2}^{*}+\widetilde{\Lambda}_{3}^{*}\right)\right)K\left(0\right)^{s}\sum_{j=0}^{d_{f}}H_{j}\left(s\right)b_{1,T}^{j}\\
 & \quad+O\left(\epsilon_{Tb_{2,T}}\left(s+2\right)\right),
\end{align*}
where the $H_{j}\left(s\right)$ are as in \eqref{Eq. (A.2) in VR-1}.
Letting
\begin{align*}
\Xi_{2,j}(2,\,s) & =\left(4\pi\right)^{-s/2}\left(4\pi\right)^{s}s!\left(\int_{0}^{1}f\left(u,\,0\right)du\right){}^{s}K\left(0\right)^{s}\int_{0}^{1}K_{2}^{s+1}\left(x\right)dxH_{j}\left(s\right)\\
\widetilde{\Xi}_{2,j}(2,\,s) & =\left(4\pi\right)^{-s/2}\left(4\pi\right)^{s}s!\widetilde{\Lambda}_{2}^{*}K\left(0\right)^{s}\int_{0}^{1}K_{2}^{s}\left(x\right)dxH_{j}\left(s\right)\\
\widetilde{\Xi}_{3,j}(2,\,s) & =\left(4\pi\right)^{-s/2}\left(4\pi\right)^{s}s!\widetilde{\Lambda}_{3}^{*}K\left(0\right)^{s}\int_{0}^{1}K_{2}^{s}\left(x\right)dxH_{j}\left(s\right),
\end{align*}
the lemma follows. $\square$ 

\subsubsection{Proof of Theorem \ref{Theorem: Theorem 1 in VR DK-HAC}}

It follows from \citeReferencesSupp{velasco/robinson:01} and \citeReferencesSupp{taniguchi:1987}
that only the cumulants $\kappa_{2,T}(0,\,s)$ and $\kappa_{2,T}(2,\,s)$
are nonzero, and that the cumulant generating function is given by
\begin{align}
\log\psi_{T}\left(\mathbf{t}\right) & =\frac{1}{2}\left\Vert i\mathbf{t}\right\Vert ^{2}+\sum_{s=3}^{\tau+1}\frac{\left(Tb_{1,T}b_{2,T}\right)^{\left(2-s\right)/2}}{s!}\sum_{|\mathbf{r}|=s}\frac{s!}{r_{1}!r_{2}!}\overline{\kappa}_{2,T}(r_{1},\,r_{2})\left(it_{1}\right)^{r_{1}}\left(it_{2}\right)^{r_{2}}+R_{T}^{*}\left(\tau\right),\label{Eq. (A.3) in VR DK-HAC}
\end{align}
where $\mathbf{r}=\left(r_{1},\,r_{2}\right)'$, with $r_{1}\in\left\{ 0,\,2\right\} $
and $|\mathbf{r}|=r_{1}+r_{2}$, and 

\begin{align*}
R_{T}^{*}\left(\tau\right) & =\left(Tb_{1,T}b_{2,T}\right)^{-\tau/2}\left[R'_{0,\tau+2}\left(it_{2}\right)^{\tau+2}+R'_{2,\tau}\left(it_{1}\right)^{2}\left(it_{2}\right)^{\tau}\right],\qquad\qquad\qquad\tau\,\,\mathrm{even},\\
R_{T}^{*}\left(\tau\right) & =\left(Tb_{1,T}b_{2,T}\right)^{-\tau/2}\frac{1}{\left(\tau+2\right)!}\left[\overline{\kappa}_{2,T}(0,\,\tau+2)\left(it_{2}\right)^{\tau+2}+\frac{\left(\tau+2\right)\left(\tau+1\right)}{2}\overline{\kappa}_{2,T}(2,\,\tau)\left(it_{1}\right)^{2}\left(it_{2}\right)^{\tau}\right]\\
 & \qquad+\left(Tb_{1,T}b_{2,T}\right)^{-\tau/2}\left[R'_{0,\tau+3}\left(it_{2}\right)^{\tau+3}+R'_{2,\tau+1}\left(it_{1}\right)^{2}\left(it_{2}\right)^{\tau+1}\right],\qquad\quad\tau\,\mathrm{odd},
\end{align*}
where the $R'_{0,j}$ and $R_{2,j}$ are bounded. Using Lemmas \ref{Lemma 3 in VR DK-HAC}-\ref{Lemma: Lemma 4 in VR DK-HAC},
we have
\begin{align*}
\log\psi_{T}\left(\mathbf{t}\right) & =\frac{1}{2}\left\Vert i\mathbf{t}\right\Vert ^{2}+\sum_{s=3}^{\tau+1}\frac{\left(Tb_{1,T}b_{2,T}\right)^{\left(2-s\right)/2}}{s!}\left(\overline{\kappa}_{2,T}(0,\,s)\left(it_{2}\right)^{s}+\frac{s\left(s-1\right)}{2}\overline{\kappa}_{2,T}(2,\,s-2)\left(it_{1}\right)^{2}\left(it_{2}\right)^{s-2}\right)\\
 & \quad+R_{T}^{*}\left(\tau\right)\\
 & =\frac{1}{2}\left\Vert i\mathbf{t}\right\Vert ^{2}+\sum_{s=3}^{\tau+1}\left(Tb_{1,T}b_{2,T}\right)^{\left(2-s\right)/2}\left[B_{2,T}\left(s,\,\mathbf{t}\right)+\left\{ \left(it_{2}\right)^{s}+\left(it_{1}\right)^{2}\left(it_{2}\right)^{s-2}\right\} O\left(\epsilon_{T}\left(s\right)\right)\right]+R_{T}^{*}\left(\tau\right),
\end{align*}
where
\begin{align*}
B_{2,T}\left(s,\,\mathbf{t}\right) & =\frac{1}{s!}\sum_{j=0}^{d_{f}}b_{1,T}^{j}\biggl\{\left(\Xi_{2,j}(0,\,s)+b_{2,T}^{2}\left(\widetilde{\Xi}_{2,j}(0,\,s)+\widetilde{\Xi}_{3,j}(0,\,s)\right)\right)\left(it_{2}\right)^{s}\\
 & \quad+\frac{s\left(s-1\right)}{2}\left(\Xi_{2,j}(2,\,s-2)+b_{2,T}^{2}\left(\widetilde{\Xi}_{2,j}(2,\,s-2)+\widetilde{\Xi}_{3,j}(2,\,s-2)\right)\right)\left(it_{1}\right)^{2}\left(it_{2}\right)^{s-2}\biggr\}.
\end{align*}
The approximation of the characteristic function of $\mathbf{v}$
using its cumulant generating function is
\begin{align*}
\mathcal{A}_{2,T}\left(\mathbf{t},\,\tau\right) & =\exp\left(\frac{1}{2}\left\Vert i\mathbf{t}\right\Vert ^{2}\right)\left[1+\sum_{j=3}^{\tau+1}\left(Tb_{1,T}b_{2,T}\right)^{\left(2-j\right)/2}\sum_{\mathbf{r}}\prod_{n=3}^{\tau+1}\left(B_{2,T}\left(n,\,\mathbf{t}\right)\right)^{r_{n}}\frac{1}{r_{3}!\ldots r_{\tau+1}!}\right],
\end{align*}
where $\mathbf{r}=\left(r_{3},\ldots,\,r_{\tau+1}\right)'$, $r_{n}\in\left\{ 0,\,1,\ldots\right\} $,
and the summation is over all $\mathbf{r}$ satisfying $\sum_{n=3}^{\tau+1}\left(n-2\right)r_{n}=j-2$.
To obtain a second-order Edgeworth expansion we set $\tau=2$ and
we include in $\mathcal{A}_{2,T}\left(\mathbf{t},\,2\right)$ the
terms up to order $(Tb_{1,T}b_{2,T})^{-1/2}$, 
\begin{align}
\mathcal{A}_{2,T}\left(\mathbf{t},\,2\right) & =\exp\left(\frac{1}{2}\left\Vert i\mathbf{t}\right\Vert ^{2}\right)\left[1+\overline{B}_{2,T}\left(3,\,\mathbf{t}\right)\left(Tb_{1,T}b_{2,T}\right)^{-1/2}\right],\label{Eq. (A.4) in VR DK-HAC}
\end{align}
 where $\overline{B}_{2,T}\left(3,\,\mathbf{t}\right)$ includes only
the leading term  in $b_{1,T}^{j}$ $\left(j=0\right)$ in the expansion
for the cumulant of order three. Note that the characteristic function
of $\mathbb{Q}_{2,T}^{\left(2\right)}(\cdot)$ is $\mathcal{A}_{2,T}\left(\mathbf{t},\,2\right)$.
We use Lemma \ref{Lemma Bhattacharya and Rao 1975} with kernel $\mathbb{G}$
to bound the distance between $\mathbb{P}_{T}$ and  $\mathbb{Q}_{2,T}^{\left(2\right)}$.
First, 
\begin{align*}
\left\Vert \left(\mathbb{P}_{T}-\mathbb{Q}_{2,T}^{\left(2\right)}\right)\bullet\mathbb{G}_{\phi_{T}}\right\Vert _{\mathrm{TV}} & \leq2\sup_{\mathbf{B}\subset\mathbf{B}\left(0,\,r_{T}\right)}\left|\left(\mathbb{P}_{T}-\mathbb{Q}_{2,T}^{\left(2\right)}\right)\bullet\mathbb{G}_{\phi_{T}}\right|+2\sup_{\mathbf{B}\subset\mathbf{B}\left(0,\,r_{T}\right)^{c}}\left|\left(\mathbb{P}_{T}-\mathbb{Q}_{2,T}^{\left(2\right)}\right)\bullet\mathbb{G}_{\phi_{T}}\right|,
\end{align*}
 where $\mathbf{B}\left(0,\,r_{T}\right)$ is a neighborhood around
0 with radius $r_{T}$, $r_{T}=\left(Tb_{1,T}b_{2,T}\right)^{a}$
with $a>0$, and $\left\Vert \cdot\right\Vert _{\mathrm{TV}}$ denotes
the total variation norm. For $\mathbf{B}\subset\mathbf{B}\left(0,\,r_{T}\right)^{c}$
we have uniformly
\begin{align*}
\left|\left(\mathbb{P}_{T}-\mathbb{Q}_{2,T}^{\left(2\right)}\right)\bullet\mathbb{G}_{\phi_{T}}\right| & \leq\left|\mathbb{P}_{T}\bullet\mathbb{G}_{\phi_{T}}\right|+\left|\mathbb{Q}_{2,T}^{\left(2\right)}\bullet\mathbb{G}_{\phi_{T}}\right|\\
 & \leq\mathbb{P}\left(\left\Vert \mathbf{v}\right\Vert \geq r_{T}/2\right)+2\mathbb{G}_{\phi_{T}}\left(\mathbf{B}\left(0,\,r_{T}/2\right)^{c}\right)+2\mathbb{Q}_{2,T}^{\left(2\right)}\left(\mathbf{B}\left(0,\,r_{T}/2\right)^{c}\right).
\end{align*}
By definition of $q_{2,T}^{\left(2\right)}\left(\mathbf{v}\right)$
it follows that $\mathbb{Q}_{2,T}^{\left(2\right)}(\mathbf{B}\left(0,\,r_{T}/2\right)^{c})=o((Tb_{1,T}b_{2,T})^{-1/2})$.
 In view of the definition of $v_{2}$, we have $\mathbb{P}\{\left\Vert \mathbf{v}\right\Vert \geq r_{T}/2\}=o((Tb_{1,T}b_{2,T})^{-1/2})$.
By Lemma \ref{Lemma Bhattacharya and Rao 1975},
\begin{align*}
\mathbb{G}_{\phi_{T}}\left(\mathbf{B}\left(0,\,r_{T}/2\right)^{c}\right) & =O\left(\left(\phi_{T}/r_{T}\right)^{3}\right)=O\left(\left(Tb_{1,T}b_{2,T}\right)^{-3\left(\varpi+a\right)}\right)=o\left(\left(Tb_{1,T}b_{2,T}\right)^{-1/2}\right).
\end{align*}
For $\mathbf{B}\subset\mathbf{B}\left(0,\,r_{T}\right)$ we have
by Fourier inversion 
\begin{align}
\left|\left(\mathbb{P}_{T}-\mathbb{Q}_{2,T}^{\left(2\right)}\right)\bullet\mathbb{G}_{\phi_{T}}\right| & \leq\left(2\pi\right)^{-1}\pi r_{T}^{2}\int\left|\left(\widehat{\mathbb{P}}_{T}-\mathbb{\widehat{Q}}_{2,T}^{\left(2\right)}\right)\left(\mathbf{t}\right)\widehat{\mathbb{G}}_{\phi_{T}}\left(\mathbf{t}\right)\right|d\mathbf{t},\label{Eq. (A.5) in VR DK-HAC}
\end{align}
 where $\widehat{\mathbb{P}}_{T}$ denotes the characteristic function
of $\mathbb{P}_{T}$ (i.e., $\widehat{\mathbb{P}}_{T}=\psi_{T}\left(\mathbf{t}\right)$)
and $\mathbb{\widehat{Q}}_{2,T}^{\left(2\right)}=\mathcal{A}_{2,T}\left(\mathbf{t},\,2\right)$.
 Let $a'=8\times2^{4/3}\pi^{-1/3}$. Using Lemma \ref{Lemma: Lemma 14 in VR DK-HAC},
a bound for \eqref{Eq. (A.5) in VR DK-HAC} is given by 
\begin{align}
O\left(\left(Tb_{1,T}b_{2,T}\right)^{2a-1/2}\right) & \left[b_{1,T}^{2}+\epsilon_{Tb_{2,T}}\left(3\right)\right]\int_{\left\Vert \mathbf{t}\right\Vert \leq c_{2}\sqrt{Tb_{1,T}b_{2,T}}}\left|e^{-d_{2}\left\Vert \mathbf{t}\right\Vert ^{2}}F\left(\left\Vert \mathbf{t}\right\Vert \right)\right|\left|\widehat{\mathbb{G}}_{\phi_{T}}\left(\left\Vert \mathbf{t}\right\Vert \right)\right|d\mathbf{t}\label{Eq. (A.6) in VR DK-HAC}\\
 & +O\left(Tb_{1,T}b_{2,T}\right)^{2a}\int_{c_{2}\sqrt{Tb_{1,T}b_{2,T}}<\left\Vert \mathbf{t}\right\Vert \leq a'\left(Tb_{1,T}b_{2,T}\right)^{\varpi}}\int\left|\left(\widehat{\mathbb{P}}_{T}-\mathbb{\widehat{Q}}_{2,T}^{\left(2\right)}\right)\left(\mathbf{t}\right)\widehat{\mathbb{G}}_{\phi_{T}}\left(\mathbf{t}\right)\right|d\mathbf{t}.\label{Eq. (A.7) in VR DK-HAC}
\end{align}
The integral over $\left\Vert \mathbf{t}\right\Vert >a'\left(Tb_{1,T}b_{2,T}\right)^{\varpi}$
is equal to zero from \eqref{Eq. (B.20) in VR}. Choosing $a\leq1/4$
\eqref{Eq. (A.6) in VR DK-HAC} is $o(\left((Tb_{1,T}b_{2,T})\right)^{-1/2})$. 

By Lemma \ref{Lemma: Lemma 15 in VR DK-HAC}, for $c_{2}m_{2,T}<\left\Vert \mathbf{t}\right\Vert $
the expression in \eqref{Eq. (A.7) in VR DK-HAC} is bounded by 
\begin{align*}
O\left(\left(Tb_{1,T}b_{2,T}\right)^{2a}\right) & \int_{c_{2}\sqrt{Tb_{1,T}b_{2,T}}<\left\Vert \mathbf{t}\right\Vert \leq a'\left(Tb_{1,T}b_{2,T}\right)^{\varpi}}e^{-d_{3}m_{2,T}^{2}}d\mathbf{t}+o\left(\left(Tb_{1,T}b_{2,T}\right)^{-1/2}\right),
\end{align*}
for some $d_{3}>0$. This implies that \eqref{Eq. (A.7) in VR DK-HAC}
is bounded by $O(((Tb_{1,T}b_{2,T})^{2\left(\varpi+a\right)})e^{-d_{3}m_{2,T}^{2}})+o((Tb_{1,T}b_{2,T})^{-1/2})$
 since by Assumptions \ref{Assumption 7 VR}-\ref{Assumption K2 and b2}
it holds $m_{2,T}\geq\epsilon(Tb_{2,T})^{\epsilon}$ for some $\epsilon>0$
depending on $q$ and $p$. $\square$ 
\begin{lem}
\label{Lemma: Lemma 14 in VR DK-HAC}Let Assumptions \ref{Assumption 1 in VR},
\ref{Assumption 3 VR}-\ref{Assumption 4 in VR}, \ref{Assumption K2 and b2}-\ref{Assumption Lip of d2 f(u,w)}
and $b_{1,T}+\left(Tb_{1,T}b_{2,T}\right)^{-1}\log^{5}(Tb_{2,T})\rightarrow0$
hold. Then there exists a $c_{2}>0$ such that, for $\left\Vert \mathbf{t}\right\Vert \leq c_{2}\sqrt{Tb_{1,T}b_{2,T}}$
and a $d_{2}>0$, 
\begin{align*}
\left|\psi_{T}\left(\mathbf{t}\right)-\mathcal{A}_{2,T}\left(\mathbf{t},\,2\right)\right| & \leq\exp\left(-d_{2}\left\Vert \mathbf{t}\right\Vert ^{2}\right)\widetilde{F}\left(\left\Vert \mathbf{t}\right\Vert \right)O\left(\left(Tb_{1,T}b_{2,T}\right)^{-1/2}\left(b_{1,T}^{2}+\epsilon_{Tb_{2,T}}\left(3\right)\right)+\frac{1}{Tb_{1,T}b_{2,T}}\right),
\end{align*}
where $\widetilde{F}\left(\left\Vert \mathbf{t}\right\Vert \right)$
is a polynomial in $\mathbf{t}$ with bounded coefficients and $\mathcal{A}_{2,T}\left(\mathbf{t},\,2\right)$
is defined in \eqref{Eq. (A.4) in VR DK-HAC}. 
\end{lem}
\noindent\textit{Proof of Lemma \ref{Lemma: Lemma 14 in VR DK-HAC}.
}From \nociteReferencesSupp{feller:1971}Feller (1971, p. 535) for
complex $\alpha$ and $\beta$ it holds that $|e^{a}-1-b|\leq e^{\gamma}(\left|a-b\right|+\left|b\right|^{2}/2)$,
where $\gamma=\max\{\left|a\right|,\,\left|b\right|\}.$ We set
\begin{align*}
a & =\log\psi\left(\mathbf{t}\right)-\frac{1}{2}\left\Vert i\mathbf{t}\right\Vert ^{2}=\left(Tb_{1,T}b_{1,T}\right)^{-1/2}\sum_{|\mathbf{r}|=3}\frac{s!}{r_{1}!r_{2}!}\overline{\kappa}_{2,T}(r_{1},\,r_{2})\left(it_{1}\right)^{r_{1}}\left(it_{2}\right)^{r_{2}}+R_{T}^{*}\left(2\right),
\end{align*}
where the right-hand side follows from \eqref{Eq. (A.3) in VR DK-HAC}.
Let $b=(Tb_{1,T}b_{1,T})^{-1/2}\overline{B}_{2,T}\left(3,\,\mathbf{t}\right)$
where $\overline{B}_{2,T}\left(3,\,\mathbf{t}\right)$ is defined
after \eqref{Eq. (A.4) in VR DK-HAC}. Using Lemmas \ref{Lemma 3 in VR DK-HAC}-\ref{Lemma: Lemma 4 in VR DK-HAC}
for $s=3$ we have
\begin{align}
\left|a-b\right| & \leq\Biggl|\left(Tb_{1,T}b_{1,T}\right)^{-1/2}O\left(b_{1,T}^{2}+\epsilon_{Tb_{2,T}}\left(3\right)\right)\left(\left(it_{2}\right)^{3}+\left(it_{1}\right)^{2}\left(it_{2}\right)\right)\label{Eq. (B.21) VR DK-HAC}\\
 & \quad+\frac{1}{Tb_{1,T}b_{2,T}}\left(R'_{0,4}\left(it_{2}\right)^{4}+R'_{2,2}\left(it_{1}\right)^{2}\left(it_{1}\right)^{2}\right)\Biggr|\nonumber \\
 & \leq P_{1}\left(\left\Vert \mathbf{t}\right\Vert \right)O\left(\left(Tb_{1,T}b_{1,T}\right)^{-1/2}\left(b_{1,T}^{2}+\epsilon_{Tb_{2,T}}\left(3\right)\right)+\frac{1}{Tb_{1,T}b_{2,T}}\right),\nonumber 
\end{align}
 where $P_{1}$ is a polynomial of degree of $4$. Note that $\left|b\right|^{2}/2\leq P_{2}\left(\left\Vert \mathbf{t}\right\Vert \right)O(Tb_{1,T}b_{1,T})^{-1})$
where $P_{2}$ is a polynomial of degree 6. Then, for some polynomial
$P$
\begin{align*}
\left|a-b\right|+\frac{\left|b\right|^{2}}{2} & \leq P\left(\left\Vert \mathbf{t}\right\Vert \right)O\left(\left(Tb_{1,T}b_{1,T}\right)^{-1/2}\left(b_{1,T}^{2}+\epsilon_{Tb_{2,T}}\left(3\right)\right)+\frac{1}{Tb_{1,T}b_{2,T}}\right).
\end{align*}
 Next, we need to find a bound for $\gamma=\max\left\{ \left|a\right|,\,\left|b\right|\right\} $.
For  $\left\Vert \mathbf{t}\right\Vert \leq c_{b}\sqrt{Tb_{1,T}b_{2,T}}$
with $c_{b}>0$ we have
\begin{align}
\left|b\right| & =\left|\left(Tb_{1,T}b_{1,T}\right)^{-1/2}\overline{B}_{2,T}\left(3,\,\mathbf{t}\right)\right|\leq\left\Vert \mathbf{t}\right\Vert ^{2}\left\{ \frac{1}{3!}\left(Tb_{1,T}b_{1,T}\right)^{-1/2}\left[\left|\Xi_{2,0}(0,\,3)\right|+3\left|\Xi_{2,0}(2,\,1)\right|\left\Vert \mathbf{t}\right\Vert \right]\right\} \label{Eq. (B.22) VR DK-HAC}\\
 & \leq\left\Vert \mathbf{t}\right\Vert ^{2}\left\{ \frac{c_{b}}{3!}\left(\left|\Xi_{2,0}(0,\,3)\right|+3\left|\Xi_{2,0}(2,\,1)\right|\right)\right\} \leq\left\Vert \mathbf{t}\right\Vert ^{2}T_{b},\nonumber 
\end{align}
 where $0<T_{b}<1/4$ by choosing $c_{b}$ sufficiently small. For
a given $a$ we can choose a $c_{a}>0$ sufficiently small such that,
for $\left\Vert \mathbf{t}\right\Vert \leq c_{a}\sqrt{Tb_{1,T}b_{1,T}}$,
\begin{align}
\left|a\right| & \leq\left\Vert \mathbf{t}\right\Vert ^{2}\Biggl\{\frac{1}{3!}\left(Tb_{1,T}b_{1,T}\right)^{-1/2}\left[\left|\Xi_{2,0}(0,\,3)\right|+3\left|\Xi_{2,1}(2,\,1)\right|+O\left(b_{1,T}^{2}+\epsilon_{Tb_{2,T}}\left(3\right)\right)\right]\label{Eq. (B.23) in DK-HAC}\\
 & \quad\times\left\Vert \mathbf{t}\right\Vert +\left(Tb_{1,T}b_{1,T}\right)^{-1}\left[\left|R'_{0,4}\right|+\left|R'_{2,2}\right|\right]\left\Vert \mathbf{t}\right\Vert ^{2}\Biggr\}\nonumber \\
 & \leq\left\Vert \mathbf{t}\right\Vert ^{2}\left\{ \frac{c_{a}}{3!}\left[\left|\Xi_{2,0}(0,\,3)\right|+3\left|\Xi_{2,0}(2,\,1)\right|+O\left(b_{1,T}^{2}+\epsilon_{Tb_{2,T}}\left(3\right)\right)\right]+c_{a}^{2}\left[\left|R'_{0,4}\right|+\left|R'_{2,2}\right|\right]\right\} \nonumber \\
 & \leq\left\Vert \mathbf{t}\right\Vert ^{2}\left\{ \frac{1}{4}+O\left(b_{1,T}^{2}+\epsilon_{Tb_{2,T}}\left(3\right)\right)\right\} .\nonumber 
\end{align}
 From \eqref{Eq. (B.22) VR DK-HAC}-\eqref{Eq. (B.23) in DK-HAC}
we have for $\left\Vert \mathbf{t}\right\Vert \leq c_{2}\sqrt{Tb_{1,T}b_{1,T}}$
with $c_{2}=\min\left\{ c_{a},\,c_{b}\right\} $, 
\[
\exp\left(\gamma\right)\leq\exp\left\{ \left\Vert \mathbf{t}\right\Vert ^{2}\left[\frac{1}{4}+O\left(b_{1,T}^{2}+\epsilon_{Tb_{2,T}}\left(3\right)\right)\right]\right\} ,
\]
 or 
\begin{align}
\exp\left\{ -\frac{1}{2}\mathbf{t}^{2}+\gamma\right\}  & \leq\exp\left\{ \left\Vert \mathbf{t}\right\Vert ^{2}\left[-\frac{1}{4}+O\left(b_{1,T}^{2}+\epsilon_{Tb_{2,T}}\left(3\right)\right)\right]\right\} \leq\exp\left\{ -d_{2}\left\Vert \mathbf{t}\right\Vert ^{2}\right\} ,\label{Eq. (B.24) DK-HAC}
\end{align}
for some $d_{2}>0$. Note that $\psi\left(\mathbf{t}\right)=\exp\{\frac{1}{2}\left\Vert i\mathbf{t}\right\Vert ^{2}+a\}$
and $\mathcal{A}_{2,T}\left(\mathbf{t},\,2\right)=\exp\{\frac{1}{2}\left\Vert i\mathbf{t}\right\Vert ^{2}\}(1+b)$.
Using \eqref{Eq. (B.21) VR DK-HAC}-\eqref{Eq. (B.24) DK-HAC} the
result of the lemma follows. $\square$ 

\subsubsection{Proof of Theorem \ref{Theorem: Theorem 2 in VR DK-HAC}}

Consider the following linear stochastic approximation to $U_{T}$,
 
\begin{align}
\widetilde{U}_{T} & \triangleq v_{1}\left(1-\frac{1}{2}\overline{c}_{1}b_{1,T}^{d_{f}}-\frac{1}{2}\sqrt{4\pi}\left\Vert K\right\Vert _{2}\left\Vert K_{2}\right\Vert _{2}v_{2}\left(Tb_{1,T}b_{2,T}\right)^{-1/2}-\frac{1}{2}\overline{c}_{2}b_{2,T}^{2}\right).\label{Eq. U tilde T}
\end{align}
 Consider the transformation $\mathbf{s}=\left(s_{1},\,s_{2}\right)'=(\widetilde{U}_{T}\left(h_{1},\,v_{2}\right),\,v_{2})'=\Delta_{T}\left(\mathbf{v}\right)$
say, and its inverse $\mathbf{v}=\Delta_{T}^{-1}\left(\mathbf{s}\right)=(h_{1}^{\dagger}\left(s_{1},\,s_{2}\right),\,s_{2})'$.
Let $\gamma>0$ be such that 
\begin{align*}
\frac{T^{3\gamma}}{\left(Tb_{1,T}b_{2,T}\right)^{3/2}} & \rightarrow0,
\end{align*}
and define $\mathbf{L}_{T}=\{\mathbf{v}:\,\left|v_{i}\right|<l_{i}T^{\gamma},\,i=1,\,2\}$,
where $l_{i}$ are some fixed constants. Using $\left(1+x\right)^{-1}=1-x+x^{2}-x^{3}+\ldots$
for $\left|x\right|<1$, we have uniformly in the set $\mathbf{L}_{T}$,
\begin{align*}
h_{1}^{\dagger}\left(\mathbf{s}\right) & =s_{1}\left[1+\frac{1}{2}\overline{c}_{1}b_{1,T}^{d_{f}}+\frac{1}{2}\sqrt{4\pi}\left\Vert K_{2}\right\Vert \left\Vert K_{2}\right\Vert _{2}s_{2}\left(Tb_{1,T}b_{2,T}\right)^{-1/2}+\frac{1}{2}\overline{c}_{2}b_{2,T}^{2}\right]+o\left(\left(Tb_{1,T}b_{2,T}\right)^{-1/2}\right).
\end{align*}
 We have $\mathbb{P}(U_{T}\in\mathbf{C})=\mathbb{P}(\mathbf{v}\in\Delta_{T}^{-1}\left(\mathbf{C}\times\mathbb{R}\right))$
and from Theorem \ref{Theorem: Theorem 1 in VR},  
\begin{align}
\sup_{\mathbf{C}} & \left|\mathrm{\mathbb{P}\left(\mathbf{v}\in\Delta_{\mathit{T}}^{-1}\left(\mathbf{C}\times\mathbb{R}\right)\right)-}\mathbb{Q}_{2,T}^{\left(2\right)}\left(\Delta_{T}^{-1}\left(\mathbf{C}\times\mathbb{R}\right)\right)\right|\nonumber \\
 & \quad=o\left(\left(Tb_{1,T}b_{2,T}\right)^{-1/2}\right)+\mathrm{cost}\,\sup_{\mathbf{C}}\mathbb{Q}_{2,T}^{\left(2\right)}\left(\left(\partial\Delta_{T}^{-1}\left(\mathbf{C}\times\mathbb{R}\right)\right)^{2\phi_{T}}\right),\label{Eq. (A.12) VR}
\end{align}
 where $\phi_{T}=(Tb_{1,T}b_{2,T})^{-\rho},$ $1/2<\rho<1$. From
the continuity of $\Delta_{T}$, we can obtain, for some $c>0$, 
\begin{align}
Q_{2,T}^{\left(2\right)}\left(\left(\partial\Delta_{T}^{-1}\left(\mathbf{C}\times\mathbb{R}\right)\right)^{2\phi_{T}}\right) & \leq Q_{2,T}^{\left(2\right)}\left(\Delta_{T}^{-1}\left(\partial\mathbf{C}\right)^{c\phi_{T}}\times\mathbb{R}\right),\label{Eq. (A.13) in VR DK-HAC}
\end{align}
 and 
\begin{align*}
Q_{2,T}^{\left(2\right)}\left(\Delta_{T}^{-1}\left(\mathbf{C}\times\mathbb{R}\right)\right) & =\int_{\mathbf{L}_{T}\cap\Delta_{T}^{-1}\left(\mathbf{C}\times\mathbb{R}\right)}\varphi_{2}\left(\mathbf{x}\right)q_{2,T}^{\left(2\right)}\left(\mathbf{x}\right)d\mathbf{x}+o\left(\left(Tb_{1,T}b_{2,T}\right)^{-1/2}\right)\\
 & =\int_{\mathbf{L}_{T}^{*}\cap\{\mathbf{C}\times\mathbb{R}\}}\varphi_{2}\left(\Delta_{T}^{-1}\left(\mathbf{s}\right)\right)q_{2,T}^{\left(2\right)}\left(\Delta_{T}^{-1}\left(\mathbf{s}\right)\right)\left|\mathcal{J}\right|d\mathbf{s}+o\left(\left(Tb_{1,T}b_{2,T}\right)^{-1/2}\right),
\end{align*}
 where $\varphi_{2}\left(\cdot\right)$ is the bivariate standard
normal density, $\mathbf{L}_{T}^{*}=\Delta_{T}\left(\mathbf{L}_{T}\right)$,
and $\left|\mathcal{J}\right|$ is the Jacobian of the transformation.
Neglecting the terms that contribute $o((Tb_{1,T}b_{2,T})^{-1/2})$
to the integrals, we yield 
\begin{align}
\varphi_{2}\left(\Delta_{T}^{-1}\left(\mathbf{s}\right)\right) & =\varphi\left(s_{1}\right)\varphi\left(s_{2}\right)\left(1-\frac{1}{2}s_{1}^{2}\left[\overline{c}_{1}b_{1,T}^{d_{f}}+\frac{1}{2}\sqrt{4\pi}\left\Vert K\right\Vert _{2}\left\Vert K_{2}\right\Vert _{2}s_{2}\left(Tb_{1,T}b_{2,T}\right)^{-1/2}+\frac{1}{2}\overline{c}_{2}b_{2,T}^{2}\right]\right),\label{Eq.A.13a VR}
\end{align}
 and 
\begin{align}
q_{2,T}^{\left(2\right)}\left(\mathbf{v}\right) & =1+\frac{1}{3!}\left(Tb_{1,T}b_{2,T}\right)^{-1/2}\left(\Xi_{2,0}\left(0,\,3\right)\mathcal{H}_{3}\left(v_{2}\right)+\Xi_{2,0}\left(2,\,1\right)\mathcal{H}_{2}\left(h_{1}\right)\mathcal{H}_{1}\left(v_{2}\right)\right),\label{Eq. (A.13b) VR}
\end{align}
where 
\[
\left|\mathcal{J}\right|=1+\frac{1}{2}\overline{c}_{1}b_{1,T}^{d_{f}}+\frac{1}{2}\sqrt{4\pi}\left\Vert K_{2}\right\Vert \left\Vert K_{2}\right\Vert _{2}s_{2}\left(Tb_{1,T}b_{2,T}\right)^{-1/2}+\frac{1}{2}\overline{c}_{2}b_{2,T}^{2}.
\]
For $j=1,\,2,\,3$ let $p_{j}\left(\mathbf{s}\right)$ denote polynomials
not depending on $T$. We have
\begin{align}
Q_{2,T}^{\left(2\right)}\left(\Delta_{T}^{-1}\left(\mathbf{C}\times\mathbb{R}\right)\right) & =\int_{\mathbf{C}}\varphi\left(s_{1}\right)\left\{ \int_{\mathbb{R}}\left[1+p_{1}\left(\mathbf{s}\right)\left(Tb_{1,T}b_{2,T}\right)^{-1/2}+p_{2}\left(\mathbf{s}\right)b_{1,T}^{d_{f}}+p_{3}\left(\mathbf{s}\right)b_{2,T}^{2}\right]\varphi\left(s_{2}\right)ds_{2}\right\} ds_{1}\label{Eq. (A.13c) VR}\\
 & \quad+o\left(\left(Tb_{1,T}b_{2,T}\right)^{-1/2}\right)\nonumber \\
 & =\int_{\mathbf{C}}\varphi\left(s_{1}\right)\left[1+r_{1}\left(s_{1}\right)\left(Tb_{1,T}b_{2,T}\right)^{-1/2}+r_{2}\left(s_{1}\right)b_{1,T}^{d_{f}}+r_{3}\left(s_{1}\right)b_{2,T}^{2}\right]ds_{1}\nonumber \\
 & \quad+o\left(\left(Tb_{1,T}b_{2,T}\right)^{-1/2}\right),\nonumber 
\end{align}
 where $r_{j}\left(s_{1}\right)$ are polynomials in $s_{1}$ for
$j=1,\,2,\,3$ with bounded coefficients. Integration with respect
to $s_{2}$ in $\mathbb{R}$ yields $r_{1}\left(x\right)=0$, $r_{2}\left(x\right)=-2^{-1}\overline{c}_{1}\left(x^{2}-1\right)$
and $r_{3}\left(x\right)=-2^{-1}\overline{c}_{2}\left(x^{2}-1\right)$.
Using \eqref{Eq. (A.12) VR}-\eqref{Eq. (A.13c) VR} provides the
second-order Edgeworth expansion for the linear stochastic approximation
$\widetilde{U}_{T}$. Since Lemma \ref{Lemma: Lemma 5 in VR DK-HAC}
below shows that $\widetilde{U}_{T}$ and $U_{T}$ have the same Edgeworth
expansion, the proof is concluded. $\square$
\begin{lem}
\label{Lemma: Lemma 5 in VR DK-HAC}Let Assumptions \ref{Assumption 1 in VR},
\ref{Assumption: Assumption 2 in VR } $\left(p>1\right)$ and \ref{Assumption 3 VR}-\ref{Assumption 5 VR},
\ref{Assumption K2 and b2}-\ref{Assumption condition Repalce Assumption 7}
hold. Then, $U_{T}$ has the same Edgeworth expansion as $\widetilde{U}_{T}$
uniformly for convex Borel sets up to the order $O((Tb_{1,T}b_{2,T})^{-1/2})$. 
\end{lem}
\noindent\textit{Proof of Lemma \ref{Lemma: Lemma 5 in VR DK-HAC}.
}We first expand $U_{T}\left(\mathbf{v}\right)$ around $\mathbf{0}$
in $\mathbf{L}_{T}$ with $|\eta_{2}|\leq1$,  
\begin{align}
U_{T} & =d_{T}h_{1}-\frac{1}{2}d_{T}^{3}\mathsf{V}_{2,T}h_{1}v_{2}\left(Tb_{1,T}b_{2,T}\right)^{-1/2}+U_{1,T}^{*}\left(Tb_{1,T}b_{2,T}\right)^{-1},\label{Eq. (A.8) in VR DK-HAC}
\end{align}
 where $d_{T}=(1+\mathrm{\mathsf{B}}_{2,T})^{-1/2}$ and 
\begin{align*}
U_{1,T}^{*} & =\frac{3}{8}\left(1+\mathrm{\mathsf{B}}_{2,T}+\eta_{2}\mathsf{V}_{2,T}v_{2}\left(Tb_{1,T}b_{2,T}\right)^{-1/2}\right)^{-5/2}\mathsf{V}_{2,T}^{2}h_{1}v_{2}^{2}.
\end{align*}
 We now express $U_{T}$ in terms of $\widetilde{U}_{T}$ where the
latter is defined in \eqref{Eq. U tilde T}. Substituting for $\mathrm{\mathsf{B}}_{2,T}$
and $\mathsf{V}_{2,T}$ in \eqref{Eq. (A.8) in VR DK-HAC}, we yield
$U_{T}=\widetilde{U}_{T}+U_{T}^{*}\left(Tb_{1,T}b_{2,T}\right)^{-1}$
where $U_{T}^{*}=\sum_{i=1}^{3}U_{i,T}^{*}$, 
\[
U_{2,T}^{*}=h_{1}\left(O\left(\left(b_{1,T}b_{2,T}\right){}^{-1}\log T+Tb_{2,T}b_{1,T}^{1+d_{f}+\varrho}\right)+o\left(Tb_{2,T}^{3}b_{1,T}\right)\right)
\]
and 
\[
U_{3,T}^{*}=h_{1}v_{2}O\left(\left(Tb_{1,T}b_{2,T}\right)^{1/2}\left(b_{1,T}^{2}+\epsilon_{T}\left(2\right)\right)\right).
\]
We now show that $U_{T}^{*}(Tb_{1,T}b_{2,T})^{-1}$ can be neglected
with error $o((Tb_{1,T}b_{2,T})^{1/2})$. This follows from Theorem
2 in \citeReferencesSupp{chibisov:1972} provided that the following
condition holds, 
\begin{align}
\mathbb{P}\left(|U_{T}^{*}|>\gamma_{T}\sqrt{Tb_{1,T}b_{2,T}}\right) & \leq\sum_{i=1}^{3}\mathbb{P}\left(\left|U_{i,T}^{*}\right|>\frac{1}{3}\gamma_{T}\sqrt{Tb_{1,T}b_{2,T}}\right)=o\left(\left(Tb_{1,T}b_{2,T}\right)^{-1/2}\right),\label{Eq. (A.9) VR DK-HAC}
\end{align}
for some positive sequence $\left\{ \gamma_{T}\right\} $ such that
$\gamma_{T}\rightarrow0$ and $\gamma_{T}\sqrt{Tb_{1,T}b_{2,T}}\rightarrow\infty$.
Note that 
\begin{align*}
\left(Tb_{1,T}b_{2,T}\right)^{-1/2}U_{2,T}^{*} & =h_{1}O\left(\left(Tb_{2,T}\right)^{1/2}b_{1,T}^{-3/2}\left(Tb_{2,T}\right)^{-1}\log T+\left(Tb_{2,T}b_{1,T}\right)^{1/2}b_{1,T}^{d_{f}+\varrho}\right).
\end{align*}
 By Assumption \ref{Assumption condition Repalce Assumption 7} the
right-hand side above is $O((Tb_{2,T}b_{1,T})^{-\upsilon})$ for some
$\upsilon>0$. Further, 
\begin{align*}
\left(Tb_{1,T}b_{2,T}\right)^{-1/2}U_{3,T}^{*} & =h_{1}v_{2}O\left(b_{1,T}^{2}+\epsilon_{T}\left(2\right)\right)=O((Tb_{2,T}b_{1,T})^{-\upsilon}),
\end{align*}
for some $\upsilon>0.$ Since $h_{1}$ and $v_{2}$ have finite moments
of all orders, we can take $\gamma_{T}=1/\log T$ and apply Chebyshev's
inequality to establish $\mathbb{P}(|U_{i,T}^{*}|>3^{-1}\gamma_{T}\sqrt{Tb_{1,T}b_{2,T}})=o((Tb_{1,T}b_{2,T})^{-1/2})$
for $i=2,\,3$. 

It remains to show $\mathbb{P}(|U_{1,T}^{*}|>3^{-1}\gamma_{T}\sqrt{Tb_{1,T}b_{2,T}})=o((Tb_{1,T}b_{2,T})^{-1/2})$.
We have 
\begin{align*}
\mathbb{P} & \left(\left|U_{1,T}^{*}\right|>\frac{1}{3}\gamma_{T}\sqrt{Tb_{1,T}b_{2,T}}\right)\\
 & <\mathbb{P}\left(\left|\frac{3}{8}\mathsf{V}_{2,T}^{2}h_{1}v_{2}^{2}\right|\left(Tb_{1,T}b_{2,T}\right)^{-1/4}>\gamma_{T}^{1/2}\right)\\
 & \quad+\mathbb{P}\left(\left|1+\mathsf{B}_{2,T}+\eta_{2}\mathsf{V}_{2,T}v_{2}\left(Tb_{1,T}b_{2,T}\right)^{-1/2}\right|\left(Tb_{1,T}b_{2,T}\right)^{-1/4}>\gamma_{T}^{1/2}\right).\\
 & \triangleq A_{1}+A_{2}.
\end{align*}
Using Chebyshev\textquoteright s inequality $A_{1}=o((Tb_{1,T}b_{2,T})^{-1/2})$.
Using $\left(Tb_{1,T}b_{2,T}\right)^{-1/10}\gamma_{T}^{-1/5}\rightarrow0$
we yield 
\begin{align*}
A_{2} & <C_{2}\mathbb{P}\left(\left|v_{2}\left(Tb_{1,T}b_{2,T}\right)^{-1/2}\right|>c_{2}\right)=o\left(\left(Tb_{1,T}b_{2,T}\right)^{-1/2}\right),
\end{align*}
 where $C_{2}$ and $c_{2}$ are some positive constants and we have
used Chebyshev\textquoteright s inequality. $\square$ 

\subsection{Proof of the Results of Section \ref{Section Consequences for HAR}}

\subsubsection{Proof of Theorem \ref{Theorem Power DM HAR Tests}}

Consider first the numerator of $t_{\mathrm{DM},i}$.  We have 
\begin{align*}
T_{n}^{1/2}\overline{d}_{L} & =\delta^{2}O_{\mathbb{P}}\left(T_{n}^{1/2}T_{n}^{-1}n_{\delta}\right)+O_{\mathbb{P}}\left(T_{n}^{1/2}T_{n}^{-1}\left(T_{n}-n_{\delta}\right)^{1/2}\right)\mathscr{N}\left(0,\,J_{\mathrm{DM}}\right)\\
 & =\delta^{2}O_{\mathbb{P}}\left(T_{n}^{-1/2}n_{\delta}\right)+O_{\mathbb{P}}\left(1\right),
\end{align*}
 for some $J_{\mathrm{DM}}\in\left(0,\,\infty\right)$ where $n_{\delta}$
depends on the length of the segment where the mean of $x_{t}^{(2)}$
shifts by $\delta$. The factor $\delta^{2}$ follows from the quadratic
loss.

Next, we focus on the expansion of the denominator of $t_{\mathrm{DM},i}$
which hinges on which LRV estimator is used. We begin with part (i).
Under Assumption \ref{Assumption 6 VR} $b_{1,T}\rightarrow0$ as
$T\rightarrow\infty$. Using Theorem \ref{Theorem ACF Nonstat}, 
\begin{align*}
\widehat{J}_{d_{L},\mathrm{NW87},T} & =\sum_{k=-\left\lfloor b_{T}^{-1}\right\rfloor }^{\left\lfloor b_{T}^{-1}\right\rfloor }\left(1-\left|b_{1,T}k\right|\right)\widehat{\Gamma}\left(k\right)\\
 & =\sum_{k=-\left\lfloor b_{1,T}^{-1}\right\rfloor }^{\left\lfloor b_{1,T}^{-1}\right\rfloor }\left(1-\left|b_{1,T}k\right|\right)\int_{0}^{1}c\left(u,\,k\right)du\\
 & \quad+\sum_{k=-\left\lfloor b_{1,T}^{-1}\right\rfloor }^{\left\lfloor b_{1,T}^{-1}\right\rfloor }\left(1-\left|b_{1,T}k\right|\right)\left(2^{-1}\left(\frac{T_{b}-T_{m}-1}{T_{n}}\right)\left(\frac{T_{n}-T_{b}-2}{T_{n}}\right)\delta^{4}+o_{\mathbb{P}}\left(1\right)\right)\\
 & =CJ_{\mathrm{DM}}+\sum_{k=-\left\lfloor b_{1,T}^{-1}\right\rfloor }^{\left\lfloor b_{1,T}^{-1}\right\rfloor }\left(1-\left|b_{1,T}k\right|\right)\left(2^{-1}\left(\frac{T_{b}-T_{m}-1}{T_{n}}\right)\left(\frac{T_{n}-T_{b}-2}{T_{n}}\right)\delta^{4}+o_{\mathbb{P}}\left(1\right)\right),
\end{align*}
for some $C>0$ such that $C<\infty$. By Exercise 1.7.12 in \citet{brillinger:75},
\begin{align*}
\sum_{k=-\left\lfloor b_{1,T}^{-1}\right\rfloor }^{\left\lfloor b_{1,T}^{-1}\right\rfloor }\left(1-\left|b_{1,T}k\right|\right)\exp\left(-i\omega k\right) & =b_{1,T}\left(\frac{\sin\frac{\left\lfloor b_{1,T}^{-1}\right\rfloor \omega}{2}}{\sin\frac{\omega}{2}}\right)^{2}.
\end{align*}
 Evaluating the expression above at $\omega=0$ and applying L'Hôpital's
rule we yield,
\begin{align*}
\sum_{k=-\left\lfloor b_{1,T}^{-1}\right\rfloor }^{\left\lfloor b_{1,T}^{-1}\right\rfloor }\left(1-\left|b_{1,T}k\right|\right) & =b_{1,T}\left(\frac{\frac{\left\lfloor b_{1,T}^{-1}\right\rfloor }{2}}{\frac{1}{2}}\right)^{2}=\left\lfloor b_{1,T}^{-1}\right\rfloor .
\end{align*}
 Therefore, $\widehat{J}_{d_{L},\mathrm{NW87},T}=CJ_{\mathrm{DM}}+\delta^{4}O_{\mathbb{P}}\left(b_{1,T}^{-1}\right)$
and 

\begin{align}
\left|t_{\mathrm{DM},\mathrm{NW87}}\right| & \leq\frac{\delta^{2}O_{\mathbb{P}}\left(T_{n}^{-1/2}n_{\delta}\right)+O_{\mathbb{P}}\left(1\right)}{\left(\delta^{4}O\left(b_{1,T}^{-1}\right)\right)^{1/2}}\label{Eq. t-test NW87}\\
 & =\frac{\delta^{2}O\left(T_{n}^{\zeta}\right)}{\delta^{2}O\left(b_{1,T}^{-1/2}\right)}=O\left(T_{n}^{\zeta}b_{1,T}^{1/2}\right),\nonumber 
\end{align}
 which implies $\mathbb{P}_{\delta}(|t_{\mathrm{DM},\mathrm{NW87}}|>z_{\alpha})\rightarrow0$. 

Under Assumption \ref{Assumption 7 VR} with $q=1/3$, similar derivations
yield $|t_{\mathrm{DM},\mathrm{NW87}}|=O(T_{n}^{\zeta-1/6})$ and
$\mathbb{P}_{\delta}(|t_{\mathrm{DM},\mathrm{NW87}}|>z_{\alpha})\rightarrow0$. 

In part (ii), $b_{1,T}=T^{-1}$. Proceeding as in \eqref{Eq. t-test NW87}
we have $|t_{\mathrm{DM},\mathrm{KVB}}|=O(T_{n}^{\zeta-1})$ and
$\mathbb{P}_{\delta}(|t_{\mathrm{DM},\mathrm{KVB}}|>z_{\alpha})\rightarrow0$
since $T_{n}^{\zeta-1}\rightarrow0.$

Finally, we consider part (iii). Using Theorem \ref{Theorem Local ACF Nonstat},
we have 
\begin{align*}
\widehat{J}_{d_{L},\mathrm{DK},T} & =\sum_{k=-T_{n}+1}^{T_{n}-1}K_{1}\left(\widehat{b}_{1,T}k\right)\frac{n_{T}}{T_{n}}\sum_{r=1}^{\left\lfloor T_{n}/n_{T}\right\rfloor }\widehat{c}_{\mathrm{DK},T}\left(rn_{T}/T,\,k\right)\\
 & =\sum_{k=-T_{n}+1}^{T_{n}-1}K_{1}\left(\widehat{b}_{1,T}k\right)\frac{n_{T}}{T_{n}}\sum_{r=1}^{\left\lfloor T_{n}/n_{T}\right\rfloor }\biggl(c\left(rn_{T}/T,\,k\right)\\
 & \quad+\delta^{2}\mathbf{1}\left\{ \left(|rn_{T}+k/2+n_{2,T}/2+1)-T_{j}^{0}|/n_{2,T}\right)\in\left(0,\,1\right)\right\} \biggr)+o_{\mathbb{P}}\left(1\right)\\
 & =J_{\mathrm{DM}}+\delta^{2}O_{\mathbb{P}}\left(\widehat{b}_{1,T}^{-1}\frac{T\widehat{\overline{b}}_{2,T}}{n_{T}}\frac{n_{T}}{T_{n}}\right)+o_{\mathbb{P}}\left(1\right).
\end{align*}
 It follows that 
\begin{align*}
\left|t_{\mathrm{DM},\mathrm{DK}}\right| & =\frac{\delta^{2}O_{\mathbb{P}}\left(T_{n}^{-1/2}n_{\delta}\right)+O_{\mathbb{P}}\left(1\right)}{\left(J_{\mathrm{DM}}+\delta^{2}O_{\mathbb{P}}\left(b_{1,T}^{-1}\widehat{\overline{b}}_{2,T}\right)\right)^{1/2}}\\
 & =\delta^{2}O\left(T_{n}^{\zeta}\right),
\end{align*}
and so $\mathbb{P}_{\delta}(|t_{\mathrm{DM},\mathrm{DK}}|>z_{\alpha})\rightarrow1$
since $T_{n}^{\zeta}\rightarrow\infty$. $\square$ 

\pagebreak{}

\section{Figures\label{Section Figures Supp}}

\setcounter{figure}{0}
\renewcommand{\thefigure}{S.\arabic{figure}}

\begin{center}
\begin{figure}[H]
\includegraphics[width=18cm,height=17cm]{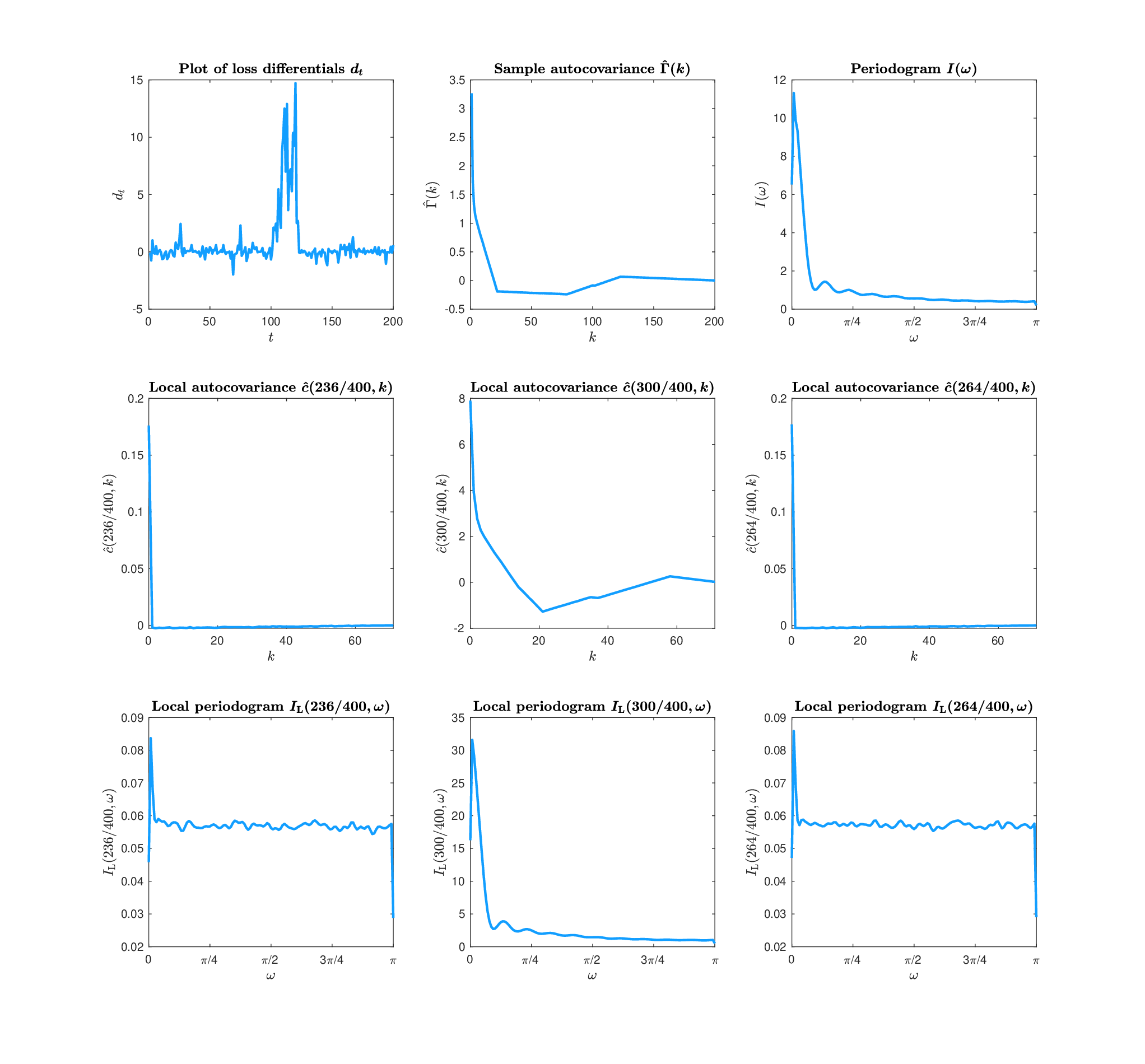}

{\footnotesize{}\caption{{\scriptsize{}\label{Fig_2_ET}Plots of loss differentials $d_{t}$,
sample autocovariance $\widehat{\Gamma}\left(k\right)$, periodogram
$I\left(\omega\right)$, sample local autocovariance $\widehat{c}(u,\,k)$
and local periodogram $I_{\mathrm{L}}(u,\,\omega)$. In all panels
$\delta=2.$}}
}{\footnotesize\par}
\end{figure}
\end{center}

\begin{center}
\begin{figure}[H]
\includegraphics[width=18cm,height=17cm]{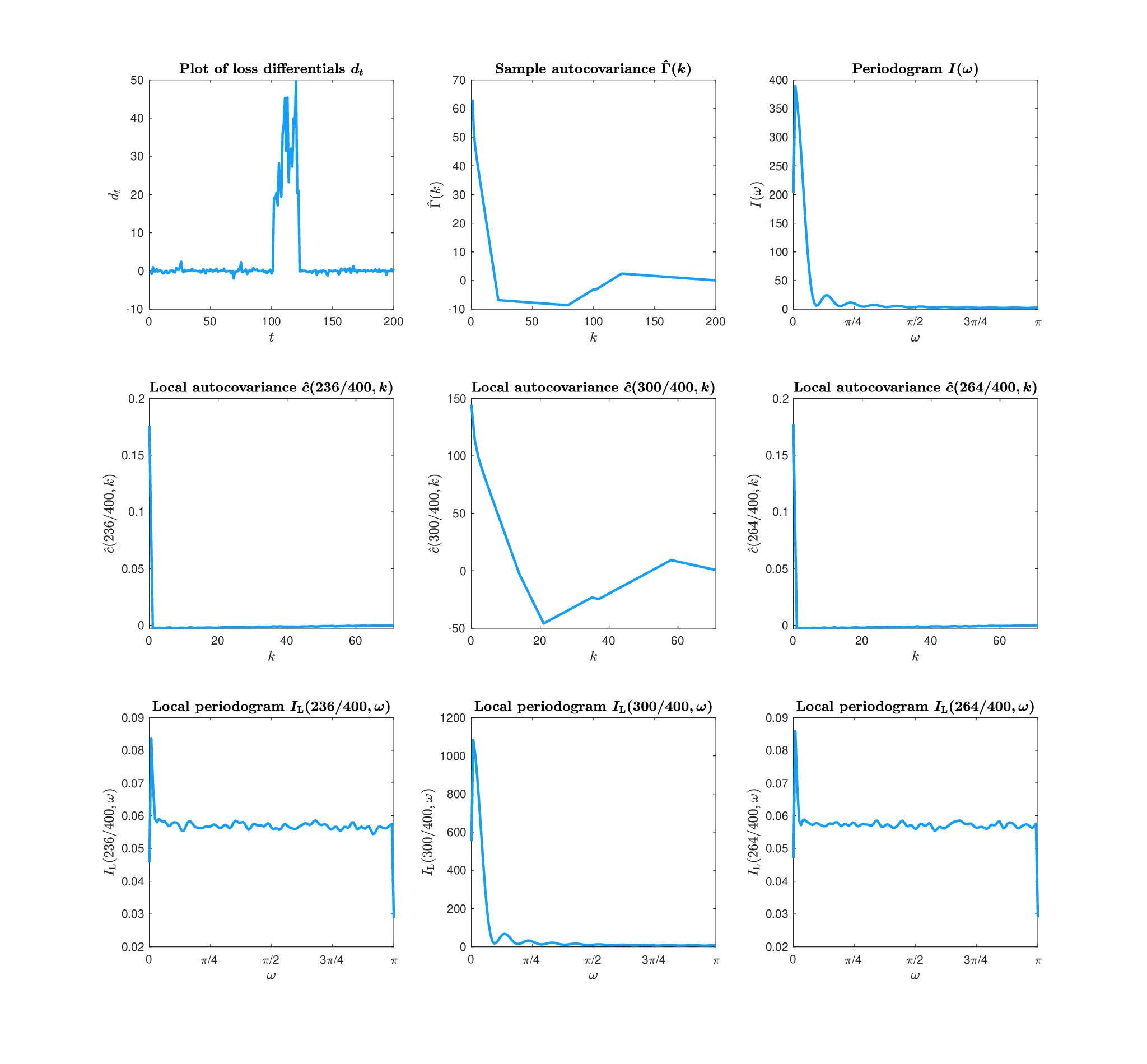}

{\footnotesize{}\caption{{\scriptsize{}\label{Fig_3_ET}Plots of loss differentials $d_{t}$,
sample autocovariance $\widehat{\Gamma}\left(k\right)$, periodogram
$I\left(\omega\right)$, sample local autocovariance $\widehat{c}(u,\,k)$
and local periodogram $I_{\mathrm{L}}(u,\,\omega)$. In all panels
$\delta=5.$}}
}{\footnotesize\par}
\end{figure}
\end{center}

\bibliographystyleReferencesSupp{elsarticle-harv}  
\bibliographyReferencesSupp{References_Supp}

\clearpage{}

\end{singlespace} 
\end{document}